\documentclass[preprint]{aastex}

\newcommand{\kms}{km s$^{-1}$}

\newcommand{\water}{H$_{2}$O}
\newcommand{\methanol}{CH$_{3}$OH}
\newcommand{\ammonia}{NH$_{3}$}

\usepackage{xcolor}
\newif\ifmod
\modtrue
\newif\ifdel
\delfalse

\shorttitle{Water and Methanol Maser Survey}
\shortauthors{Kim et al.}

\begin{document}

\title{Simultaneous Survey of Water and Class~I Methanol Masers toward 
Red MSX Sources}

\author{Chang-Hee Kim\altaffilmark{1,2}, Kee-Tae Kim\altaffilmark{1}, and Yong-Sun Park\altaffilmark{2}}

\altaffiltext{1}{Korea Astronomy and Space Science Institute, Yuseong-gu, Daejeon 34055, Korea: ktkim@kasi.re.kr}
\altaffiltext{2}{Department of Physics and Astronomy, Seoul National University, Gwanak-gu, Seoul 08826, Korea: chkim@astro.snu.ac.kr}

\begin{abstract}

 We report simultaneous single-dish surveys of 22 GHz \water ~and 44 and 95 GHz class I \methanol ~masers toward 299 Red MSX Sources in the protostellar stage.
 The detection rates are 45\% at 22~GHz, 28\% at 44~GHz, and 23\% at 95~GHz. There are 15, 53, and 51 new discoveries at 22, 44, and 95~GHz, respectively. 
We detect high-velocity ($>$30~\kms) features in 27 \water\ maser sources.
The 95~GHz maser emission is detected only in 44~GHz maser sources. The two transitions show strong correlations in the peak velocity, peak flux density, and isotropic maser luminosity, 
indicating that they are likely generated in the same sites by the same mechanisms. 
The 44~GHz masers have much narrower distributions than 22~GHz masers in the relative peak velocity and velocity range,
while 6.7~GHz class~II \methanol\ masers have distributions intermediate between the two.
The maser luminosity significantly correlates with the parental clump mass,
while it correlates well with the bolometric luminosity of the central protostar
only when data of the low-mass regime from the literature are added.
Comparison with the results of previous maser surveys toward massive star-forming regions suggests that the detection rates of 22 and 44~GHz masers tend to increase as the central objects evolve. This is  contrary to the trends found in low- and intermediate-mass star-forming regions. Thus
the occurrence of both masers might depend on the surrounding environments as well as on the evolution of the central object.

\end{abstract}


\section{Introduction}

Interstellar masers of \water, \methanol, and OH are considered excellent probes of high-mass star formation regions (SFRs).
Because at radio frequencies, these masers are quite common, intense, 
and rarely affected by dust extinction in massive SFRs, they allow us to investigate such regions.
The different maser species favor different physical conditions for their production and survival, which has led to the proposition of tracing different evolutionary phases of massive star formation \citep[e.g.,][]{ell07,bre10a,fon10}.
Since many sources of massive SFRs show emission from multiple maser species or transitions, there must be significant overlap for the evolutionary stage that is traced by the most common types of masers \citep{bre10a}.
\water\ masers are known to occur in both low- and high-mass SFRs \citep[e.g.,][]{fur03,szy05} and are thus an important signpost of ongoing star formation.
The evolutionary stage associated with 22~GHz \water\ maser emission has been investigated through extensive single-dish and interferometric observations \citep[e.g.,][]{tor97,urq11}, which have indicated that this type of maser is closely associated with high-mass young stellar objects (YSOs) in various evolutionary stages.
Although it has been claimed in a few cases that \water\ masers could trace circumstellar disks around (proto)stellar objects \citep[e.g.,][]{tor97,seth02}, connected-array and VLBI observations toward high-mass YSOs have shown that \water\ masers are predominantly associated with jets and outflows \citep[e.g.,][]{tor01,cod04}.

Methanol masers are commonly observed in massive SFRs. They have been empirically classified into two categories: class I and class II. The initial classification was based on the sources toward which the different transitions were detected \citep{bat88,men91}.
Class I \methanol\ masers (e.g., the 7$_{0}$--6$_{1}$ A$^{+}$ and 8$_{0}$--7$_{1}$ A$^{+}$ at 44 and 95 GHz, respectively) are often observed significantly offset from high-mass YSOs \citep[e.g.,][]{kur04,vor06,cyg09}.
In contrast, class II \methanol\ masers (e.g., the 5$_{1}$--6$_{0}$ A$^+$ and 2$_0$--3$_1$ E at 6.7 and 12.2 GHz, respectively) are usually very close to high-mass YSOs  \citep[e.g.,][]{cyg09,fuj14}.
These observational findings were supported by early theoretical models of \methanol\ masers, which suggest that the class~I masers are pumped by collisions with molecular hydrogen, while the class II masers are pumped by external far infrared radiation \citep{crag92}.
Compared to class II masers, class I \methanol\ masers are relatively poorly studied and understood. There have only been several large surveys of class I masers (mainly at 44 and 95 GHz), primarily undertaken with single-dish telescopes \citep[e.g.,][]{has90,bac90,sly94,valt00,ell05,chen11,chen12,chen13,gan13,kang15,kang16}  along with interferometric searches \citep[e.g,][]{kur04,cyg09,vor14,mat14,jor15,gom16,rod17}. The catalog from these surveys currently comprises of about 500 class I \methanol\ maser sources \citep[e.g,][]{bay12}. 
  
  \citet{ell07} suggested that the common maser species (\water, class I and II \methanol, and OH masers) may help identify the evolutionary phase of the central objects, and proposed a possible evolutionary sequence of these maser species. This proposed sequence has been improved upon by \citet{bre10b} (in their Figure 6). However, there remains significant uncertainty about where in SFRs the different maser species arise and which evolutionary phase they are associated with. 
  
  In this study, we performed simultaneous 22 GHz \water\ and 44 and 95~GHz class~I \methanol\ maser surveys toward high-mass 
protostellar objects (HMPOs) using single-dish telescopes.
  Our aim is to find new maser sources and to investigate the relationship between the different maser species and between the maser activity and the properties of the central objects and the natal clumps.  
 In Section 2 we describe the source selection and the observations. In Section 3 we present the results of the survey. A discussion of the analyses is given in Section 4, followed by a summary in Section 5.

\section{Source Selection and Observations}

\subsection{Source Selection}

\citet{lum02} identified
about 2000 candidates for young massive stars with 21~$\mu$m flux densities of $>$2.5~Jy and infrared colors consistent with known massive YSOs,
using the color-color plots of the Midcourse Space Experiment (MSX) and Two Micron All Sky Survey (2MASS) data
\citep[see also][]{lum13}.
They carried out follow-up observations and incorporated the complementary data from other surveys to distinguish between
HMPOs and ultracompact HII regions (UCHIIs) 
\citep[e.g.,][]{urq14a}.
From this so-called Red MSX Source (RMS) catalog\footnote{http://rms.leeds.ac.uk/cgi-bin/public/RMS\_DATABASE.cgi.}, 
we selected 299 HMPO candidates at the beginning of this study in 2011 with two criteria: bolometric luminosities of $> 10^3 L_{\odot}$ and declinations of $>-30^{\circ}$. These sources are very likely under active accretion and so have not yet developed detectable UCHIIs.
As of 2018 February, however, 
the online catalog contains 353 HMPOs satisfying the criteria.
Fifty eight sources have been added because they were re-classified as HMPOs with new data,
while four sources were re-classified as UCHII regions and so now are excluded.
Table~\ref{tbl_source} lists their information as extracted from the RMS catalog, including their RMS ID, MSX name, equatorial coordinates, source type, distance, bolometric luminosity, and systemic velocity. The 4 UCHIIs are marked by daggers  in the table.

\subsection{Observations}

We surveyed the 299 RMS sources in three maser transitions: H$_2$O $6_{16} - 5_{23}$ \citep[22.235080 GHz,][]{lov04}, CH$_3$OH $7_{0} - 6_{1}$ A$^{+}$ \citep[44.069430 GHz,][]{pick98}, and $8_{0} - 7_{1}$ A$^{+}$ \citep[95.169463 GHz,][]{mul04}.
The observations were undertaken using the Korean VLBI Network (KVN) 
21~m telescopes from 2011 April to 2014 May. 
The telescopes were equipped with multi-frequency receiving systems
which made it possible to observe in the 22, 43, 86, and 129~GHz bands
simultaneously \citep{han08}.
The first-epoch observations were made only at 22 and 44 GHz in 2011
because the 86 and 129~GHz receivers were not then installed, 
while the second-epoch observations were conducted at 22, 44, and 95~GHz in 2012.
Supplementary observations were performed toward 15 sources in 2013 April and May and 2015 May to confirm their detections.
We observed dual polarization
both at 22 and 44~GHz in the first epoch, and single polarization at 22 and 44~GHz and dual polarization at 95~GHz in the second epoch. The backend was a digital spectrometer
that provided 4096 channels and 64~MHz bandwidth for each stream. The velocity coverage and the spectral resolution for each transition are summarized in Table~\ref{OBS_summary}. The central velocity of each source was taken from the RMS catalog (see also Table~\ref{tbl_source})

The full widths at half maximum (FWHMs)
of the telescopes were about 130\arcsec\ at 22 GHz, 65\arcsec\ at 44 GHz, and 32\arcsec\ at 95 GHz.
The telescope pointing and focus were checked every $\sim$2~hr
by observing strong SiO maser sources at 43 or 86 GHz. The pointing accuracy
was better than 5$\arcsec$.
Each spectrum was obtained in position-switching mode usually with an offset of +2~minutes in the right ascension. The total (ON+OFF) integration time was 30 minutes for each source, which yielded typical 1 $\sigma$ noise levels of about
0.6 Jy, 0.7 Jy, and 0.8 Jy at about 0.2~\kms\ resolution after smoothing for 22, 44, and 95~GHz, respectively. 
The data were calibrated with the standard chopper wheel method to provide 
the line intensity on the $T_{\rm A}^*$ scale, which were converted to flux density using the conversion factors in Table~\ref{OBS_summary}. 
The KVN telescopes are of shaped Cassegraign type and hence have quite high first-sidelobe levels of $\sim$14~dB (4\%) at  $\sim$ 1.5 times FWHM away from the pointing center \citep{kim11,lee11}. We thus mapped each detected source at the same frequency to investigate contamination from nearby bright maser sources. The mapping area was 1.5 FWHM $\times$ 1.5 FWHM around a source with half-beam spacing at each frequency. The typical rms noise level was 5~Jy at 0.2~\kms. Seven and four sources turned out to be detected by the first sidelobe at 22 and 44~GHz, respectively (see Table~\ref{tbl_source}).
They were not counted as detected sources except RMS 3308 and 3555 for which maser emission was detected not only by the sidelobe but also by the main beam at different velocities. 

We also surveyed 10 sources in our sample in the $^{13}$CO~J=1--0 and HCO$^+$~J=1--0 lines in 2013 June. The $^{13}$CO line observations were conducted with the Taeduk Radio Astronomy Observatory (TRAO) 14~m telescope, while the HCO$^+$ line observations were made with the KVN telescope at the Tamna station.
The observed sources were \water\ maser sources with maser features largely offset from the systemic velocities (see Section~3.3.1). These observations aimed to search for dense molecular cores associated with those features. 
The TRAO telescope was equipped with the 15-beam receiver, 
QUARRY (QUabbin ARRaY) \citep{eri92}.
The backends were autocorrelators, each of which had 427 channels and a bandwidth of 100~MHz. We utilized the position switching mode. 
Table~\ref{OBS_summary} summarizes the observational details. All the spectral data were 
reduced and analyzed with the GILDAS/CLASS package.

\section{Result}

\subsection{Detection Statistics}

\begin{figure}
\epsscale{0.8}
\plotone{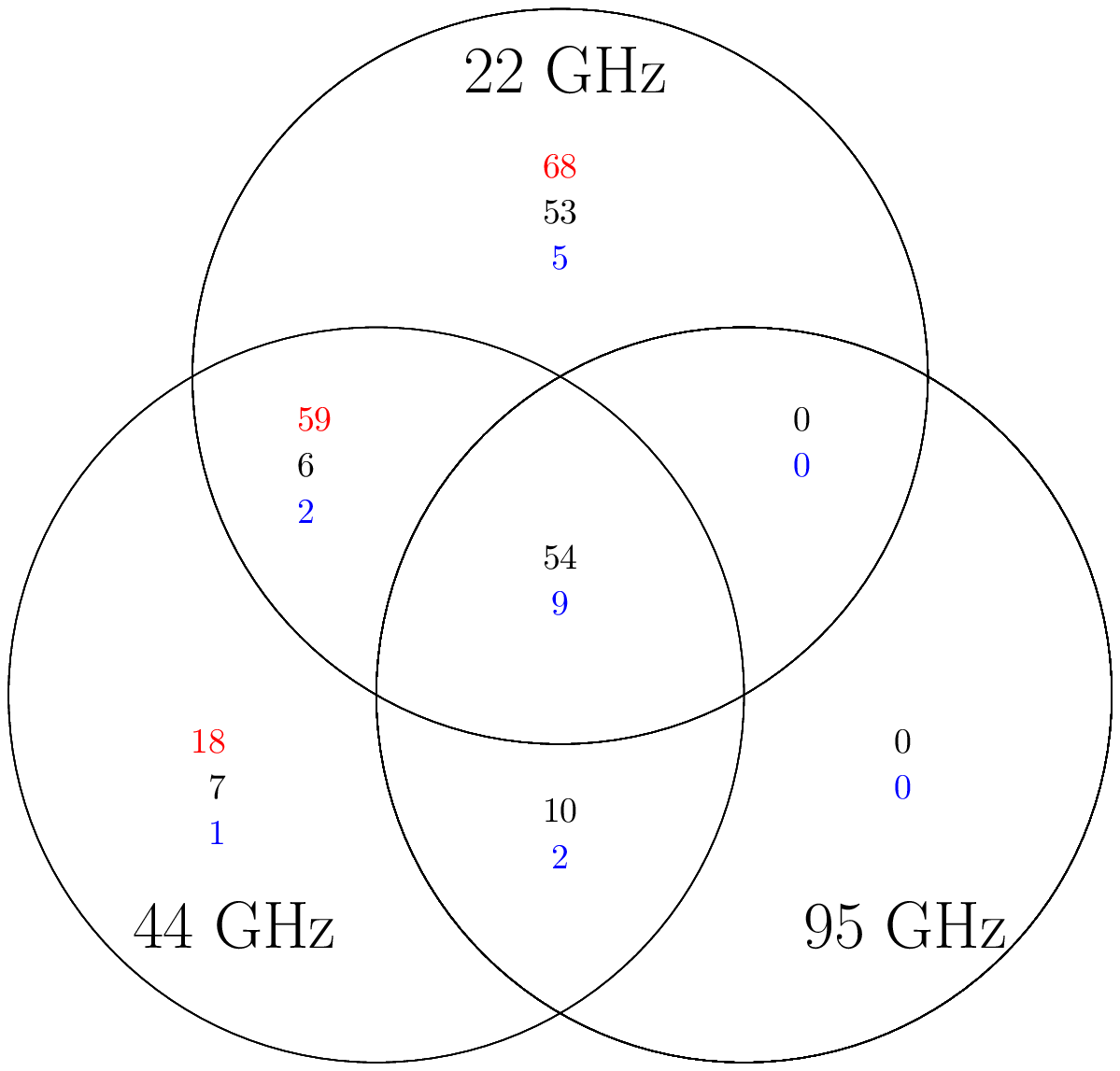}
\caption{
Venn diagram showing the numbers of detected 22~GHz \water\ and 44 and 95 GHz class~I \methanol\ maser sources in each epoch. The numbers in red and black represent the first and second epochs, respectively. 
Note that 95 GHz \methanol\ maser was not observed in the first epoch.
The numbers in blue are for a subsample of 38 sources with 6.7~GHz class~II \methanol\ masers (see Section~3.1 for details).
\label{fig_venn}}
\end{figure}

Our search toward the 299 RMS sources resulted in the detection of 151 (51\%) sources at least in one of the three maser transitions.
Table~\ref{tbl_source} presents the detection summary: detection (y), nondetection (n), and new detection (Y).
All the detected maser lines were limited to signals stronger than the 3$\sigma$ rms noise levels, which were typically 1.8, 2.1, and 2.4 Jy for 22, 44, and 95 GHz, respectively.
Table~\ref{tbl_DR} lists the numbers of observed and detected sources and the corresponding detection rates for each transition and epoch (see also Figure~\ref{fig_venn}).
We detected 22, 44, and 95~GHz maser emission in 135 (45\%), 83 (28\%), and 68 (23\%) sources, respectively.
The detection rates (42\% and 37\%) of \water\ maser emission are a little different in the first and second epochs,
although the difference is not statistically significant, while those (25\% and 27\%) of 44~GHz \methanol\ maser emission are practically the same. 
All detected maser spectra are shown in Figures~\ref{fig_224495}$-$\ref{fig_44}: sources detected in all three masers in Figure~\ref{fig_224495}; sources detected in 22 and 44 GHz masers in Figure~\ref{fig_2244}; sources detected in 44 and 95 GHz masers in Figure~\ref{fig_4495}; sources detected only in 22 GHz maser in Figure~\ref{fig_22}; sources detected only in 44 GHz maser in Figure~\ref{fig_44}.

There are several maser surveys of HMPO candidates with various sensitivities. 
 \citet{srid02} detected 22~GHz \water\ maser emission toward 42 \%\ of 69 HMPOs at a rms noise level of $\sim$0.4~Jy.
 \citet{urq11} searched for \water\ maser emission toward 597 RMS sources with a detection rate of $\sim$52\% for 275 HMPOs.
 The survey had a mean rms noise level of $\sim$0.12 Jy at a velocity resolution of $\sim$0.33 \kms.
If it had had sensitivity similar to ours, $\sim$0.6 Jy, neglecting the difference in the velocity resolution, the detection rate would be 42\%.
 \citet{fon10} surveyed 88 HMPOs in the 44 GHz \methanol~maser with a twice better sensitivity to ours, and obtained a detection rate of 31 \%.
 \citet{gan13} detected 95 GHz maser emission in 22\% (62)  of 288 outflow sources. 
\citet{valt07} also found that 24 \% of the outflow sources are associated with class~I \methanol~maser sources, including 36, 44, and 95~GHz masers, within 2\arcmin.
Thus the detection rates of the three masers in this study seem to be comparable to those of the previous surveys.
On the other hand, \citet{chen11} and \citet{chen13} detected 95~GHz maser emission toward 55\% of 192 and 71\% of 52 extended green objects (EGOs), respectively, at comparable detection limit (1.6~Jy) with ours.
Because EGOs are known to trace shocked gas in high-mass protostellar outflows, which are closely associated with class I \methanol~masers, the high detection rates may be due to selection effects.

One hundred seventy sources in our sample are distributed in the Methanol Multibeam (MMB) survey area of 6.7~GHz
class~II \methanol\ maser with the Parkes 64~m telescope \citep{gre10,gre12,bre15}.
Among them, 54 (32\%), 51 (30\%), and 45 (26\%) sources are associated with 22, 44, and 95~GHz masers, respectively.
We found that 38 of the 170 corresponds to 6.7~GHz maser sources.
Here we used a matching radius of 2\arcsec, because
the 6.7 GHz \methanol~maser spots are spread typically within 2\arcsec \citep{cas09} and the astrometric accuracy of RMS sources is better than about 2\arcsec \citep{lum13}. 
Of the 38 sources, 16 (42\%), 14 (37\%), and 11 (29\%) are related to 22, 44, and 95~GHz masers, respectively. 
For comparison, \citet{kang15} searched 22, 44, and 95 GHz masers toward seventy seven 6.7 GHz maser sources at similar detection limits to this survey using the same telescopes, and achieved detection rates of 51\%, 32\%, and 25\% for 22, 44, and 95 GHz masers, respectively.
\citet{szy05} surveyed only \water\ maser emission toward seventy nine 6.7 GHz maser sources at a comparable detection limit of $\sim$1.5~Jy with the Effelsberg 100~m (FWHM$\simeq$40$''$). The detection rate was 52\%.
 \citet{tit14,tit16} conducted interferometric observations (the synthesized beam size $\simeq$10$''$) of \water\ maser emission toward 323 6.7~GHz maser sources found in the MMB survey with detection limits of $\sim$0.2~Jy at 0.5~\kms\ resolution, and obtained a detection of 48\%.
Despite much higher sensitivity, this detection rate similar to those of the aforementioned single-dish surveys may be in part due to their much smaller matching radius of 3\arcsec.

\subsection{New Detections}

We compared our results with the previous surveys of \water\ and class I \methanol\ masers to identify newly detected sources. In this comparison we used a search radius of a HWHM at each frequency.
For 22~GHz \water\ masers, we examined the catalogs of \citet{han95}, \citet{han98}, \citet{vald01}, \citet{sun07}, \citet{urq09}, \citet{bre11}, \citet{urq11}, and \citet{tit14}. 
In particular, \citet{urq11} observed $\sim$600 RMS sources using the Green Bank Telescope (GBT) in 2009 and 2010 and detected 308 sources ($\sim$50\%). Their sample contains 257 sources in common with our sample, 145 (56\%) of which were detected. 
There are 110 common detections with this survey.
Fifteen  and thirty five sources were detected only in this survey and \citet{urq11}, respectively (see Table~\ref{tbl_source}).
We identified 15, 53, and 51 new maser sources at 22, 44, and 95 GHz, respectively. 
All of the new 22~GHz maser sources have been observed but not detected by \citet{urq09} (2) and \citet{urq11} (13) with better sensitivities of 0.1 and 0.25 Jy (1~$\sigma$), respectively. Thus these non-detections were very likely to be due to significant variability in flux density.
Two of the new 95 GHz maser sources have previously been searched for emission in this transition. 
RMS 2996 was observed by \citet{ell05} at a detection limit of 4.2 Jy (3~$\sigma$) using the Mopra~22~m. Considering the measured peak flux density of 7.9 Jy, their non-detection could be caused by flux variability. RMS 3314 was observed by \citet{chen11} at a detection limit of 1.7 Jy (3~$\sigma$) using the same telescope, which has a FWHM of 36\arcsec. The observed position was about 40\arcsec\ away from RMS 3314. Since the measured flux density is 3.4 Jy, their non-detection might be due to the large positional offset.

\subsection{Maser Properties}

We determined the line parameters of the observed maser spectra, including the peak velocity, the peak flux density, the integrated flux density, and the minimum and maximum values of the velocity range over which the maser lines are distributed.
Table~\ref{tbl_linepara} summarizes the measurements.
Table~\ref{tbl_line_summary} presents the mean, median, minimum, and maximum values of the peak flux density ($S_{\rm p}$), the relative peak velocity with respect to the systemic velocity ($V_{\rm rel}$), and the velocity range ($V_{\rm range}$) of 22, 44, and 95~GHz masers for the detected sources. The mean and median values of 44 and 95~GHz \methanol ~masers are very similar for all the three parameters, while they are much lower than those of the 22~GHz \water\ masers (see also Figure~\ref{fig_Vrel_Vrange_hist}).

The isotropic luminosity of maser emission can be estimated from the integrated flux density using the following equations:

\begin{equation}
L_{22} =
 2.30\times10^{-8}~{L_\odot}
 \biggl( \frac{\int S_{22}dv}{\rm Jy~km~s^{-1}} \biggr)
 \biggl( {\frac{D}{\rm kpc} \biggr)^2},
\end{equation}

\begin{equation}
L_{44} =
 4.60\times10^{-8}~{L_\odot}
 \biggl( \frac{\int S_{44}dv}{\rm Jy~km~s^{-1}} \biggr)
 \biggl( {\frac{D}{\rm kpc} \biggr)^2},
\end{equation}

\begin{equation}
L_{95} =
 9.92\times10^{-8}~{L_\odot}
 \biggl( \frac{\int S_{95}dv}{\rm Jy~km~s^{-1}} \biggr)
 \biggl( {\frac{D}{\rm kpc} \biggr)^2}.
\end{equation}

\begin{equation}
L_{6.7} =
 6.95\times10^{-9}~{L_\odot}
 \biggl( \frac{\int S_{6.7}dv}{\rm Jy~km~s^{-1}} \biggr)
 \biggl( {\frac{D}{\rm kpc} \biggr)^2},
\end{equation}

\noindent
where $D$ is the distance to the source. We adopted the distances from \citet{urq14a}, who gathered well-determined distances from the literature or derived kinematic distances using available molecular and HI line data for the individual sources. The integrated flux densities given by \citet{bre15} were used for 6.7 GHz \methanol\ masers.
The derived isotropic maser luminosities are listed in the 11th column of Table~\ref{tbl_linepara}.

\subsubsection{\water~maser}

 We detected 22~GHz \water\ maser emission toward 135 sources in our sample: 126 in the first epoch and 112 in the second epoch. The median of the peak flux densities is 15.5~Jy with the minimum and maximum values of 0.8 and 4552~Jy, respectively.
 The vast majority (78\%) of the detected sources  have velocity ranges smaller than 20~\kms. RMS~3555 exhibits the largest velocity range, 230~\kms. 
  Twenty seven sources (19\%) have one or more high-velocity features, which are defined to be offset from the systemic velocity by more than 30~\kms\ as in \citet{bre10a}.
These high-velocity sources fall into three categories: eighteen with only blueshifted features, three with both blue- and redshifted ones, and six with only redshifted ones. 
 For comparison, \citet{bre10a} detected high-velocity features toward 33\% (77) of 223 detected \water\ maser sources in their sample. This higher detection rate might be due to their sensitivity, which is nearly six times better than ours. 
The median value of $V_{\rm rel}$ is slightly blueshifted, --0.9 \kms, for all the 135 sources. 
  
\citet{cas08} reported four so-called dominant blueshifted \water\ maser sources, which show blueshifted high-velocity maser features with no or very weak features around the systemic velocities, and suggested that the features can be generated by pole-on jets. We found 9 blue and 2 red dominant \water\ maser source candidates. To examine whether the dominant blueshifted maser feature candidates are associated with other YSOs in the same lines of sight, we observed 10 of them in the $^{13}$CO~J=1--0 line 
and found that 3 of them have weak $^{13}$CO lines around the candidate features. For the remaining one (RMS~3936), no $^{13}$CO~J=2--1 line emission near the candidate feature was confirmed using the JCMT archive. We observed
the 3 candidates in the HCO$^{\rm +}$~J=1--0 line, which is known to be a good tracer of massive star-forming cores \citep{pur06}, and detected the line emission toward 2 of them (RMS 2716, 3360). 
Thus these two appear to originate from YSOs other than the target source which happen to lie along a nearby line of sight,
although we cannot exclude the possibility that they are related to high-velocity outflows from targets. 
The remaining 9 are very likely to be dominant shifted \water\ maser sources.
Table~\ref{tab_HV} summarizes the survey results and Figure~\ref{fig_HV} shows the detected molecular spectra together with the 22 and 44~GHz maser spectra.
\citet{urq11} also observed 8 of the 9 candidates and detected \water\ maser emission in 7 sources. However, only two (RMS~2547, 2584) of them showed dominant blueshifted features with similar velocity offsets. The others (RMS~3158, 3587, 3766, 3911, 3936) showed maser emission only around the systemic velocities except RMS~3911, which showed high-velocity features, as well. Most of these sources are discussed in more detail in Section~3.4.

\subsubsection{Class I \methanol~maser}
 
 We detected 44 and 95~GHz class~I \methanol\ maser emission toward 83 and 68 sources, respectively, in the two epochs.
It should be noted that 95~GHz maser emission was detected only in 44~GHz maser sources (Figure.~\ref{fig_venn}).
The peak flux densities range from 1.1 to 253~Jy at 44~GHz and between 1.1 and 147~Jy at 95~GHz. 
The peak velocities are always very close to the systemic velocities within $\pm$5~\kms\ (Figure~\ref{fig_Compare_4495}). The medians of the relative peak velocities are between $-$0.1 and 0.1~\kms\ for both masers (Table~\ref{tbl_line_summary}).
The measured velocity ranges limited by 3 $\sigma$ noise levels are 0.2--15.5~\kms\ at 44~GHz and 0.2--13.4 \kms\ at 95~GHz.
The median values are 2.3 and 2.4~\kms\ for 44 and 95~GHz masers, respectively. 
With the same criterion (FWHM = 1~\kms) as in \citet{ell05}, we found by Gaussian fitting to the spectrum that  the broad emission features appear in 48\% and 56\% of detected 44 and 95~GHz maser sources.
Such features have been reported by previous surveys of class~I \methanol\ masers \citep[e.g.,][]{ell05,chen11}.
They might include quasi-thermal emission, which means maser emission that is blended with the thermal component.

Figure~\ref{fig_Compare_4495} compares 44~GHz transition with 95~GHz transition in the relative peak velocity, the peak flux density, and the isotropic maser luminosity. There appear to be strong correlations between the two transitions. The least-squares fitting results are as follows:
  \begin{eqnarray}
    V_{\rm rel,95}        = (0.98 \pm 0.05)~(V_{\rm rel,44})     - (0.01 \pm 0.10)        \ (\rho = 0.73) , \\
   log (S_{\rm p,95})  = (0.98 \pm 0.06)~log (S_{\rm p,44})- (0.12 \pm 0.07)        \ (\rho = 0.91) , \\
   log (L_{95})    = (0.96 \pm 0.05)~log (L_{44})  - (0.09 \pm 0.26)        \ (\rho = 0.90) .
 \end{eqnarray}

\noindent
Four sources (RMS 2445, 3963, 3982, and 3998) show differences of $>$1~\kms\ between $V_{\rm rel,44}$ and $V_{\rm rel,95}$. They all have two or more velocity components and/or broad emission features. Thus the peak velocity measurements could be affected by the combination of the positional offsets from the pointing center and the different beam sizes at 44 and 95~GHz. 
If the four are excluded, the correlation between $V_{\rm rel,44}$ and $V_{\rm rel,95}$
becomes stronger with a correlation coefficient of 0.96.
\citet{valt00} and \citet{jor15} estimated the peak flux density ratio of 44 and 95~GHz transitions ($S_{\rm p,95}$/$ S_{\rm p,44}$) to be 0.31$-$0.32, 
while \citet{kang15} recently derived the ratio to be about 0.71$\pm$0.08.
Our result of best fit in linear scale plot is $S_{\rm p,95}=(0.56 \pm 0.08) \times S_{\rm p,44}+(2.16 \pm 0.91)$ ($\rho=0.89$) 
(Figure~\ref{fig_Sp_4495_lin}), and this is consistent with the estimate of \citet{kang15}.
It is worth noting that the former two studies established datasets using different telescopes in epochs separated by several years
while this study and  \citet{kang15} utilized the datasets simultaneously taken from the same telescopes at the same velocity resolution.
We also found a strong correlation between $L_{44}$ and $L_{95}$.
However, there is no correlation in the isotropic maser luminosity between 44~GHz masers and 22~GHz \water\ or 6.7 GHz class II \methanol\ masers.

We compared the first-epoch spectrum with the second-epoch one for each of seventy four 44~GHz \methanol\ maser sources that were detected in both epochs. 
Using a simple equation of $S_{\rm p,2nd} $/$S_{\rm p,1st}$, variability in the peak flux density was quantitatively examined for each source. The values range between 0.35 and 2.1 with a mean of 0.92$\pm$0.28. We also compared the peak velocities of the first- and second-epoch spectra and found the differences range from $-$0.85 to 0.64 \kms\ with a mean of 0.03$\pm$0.18~\kms.   
There is a trend that weaker sources are more variable in the peak flux density. 
If we focus on bright sources, i.e., 30 sources with signal-to-noise ratios $>$10, the variations significantly decrease both in the peak flux and velocity: 0.53 to 1.24 with a mean of 0.91$\pm$0.17 and $-$0.02 to 0.42~\kms\ with 
a mean of 0.03$\pm$0.09.
The first- and second-epoch spectra show very similar line profiles for almost all sources.

\subsubsection{Class II \methanol~maser}

 As mentioned in \S~3.1, 38 of the 170 sources in the MMB survey area are associated with 6.7~GHz class~II \methanol\ maser emission. 
\citet{gre10,gre12} and \citet{bre15} measured the line parameters of the 6.7~GHz maser spectra. The peak flux densities range from 0.9 to 517~Jy.
Table~\ref{tbl_line_summary} presents the mean, median, minimum, and maximum values of $S_{\rm p}$, $V_{\rm rel}$, and $V_{\rm range}$ of four maser species for the maser-detected sources in the MMB survey area.
The median value of $S_{\rm p}$ of 6.7~GHz masers is comparable to that of 22~GHz masers,
while the two are (2$-$3) times larger than those of 44 and 95~GHz masers. The median values of $V_{\rm rel}$ and $V_{\rm range}$ for 6.7~GHz masers are in the middle of the values of 22 and 44 (or 95)~GHz masers. These trends are well displayed in Figure~\ref{fig_Vrel_Vrange_hist}. We could \textit{not} find any correlation in the isotropic maser luminosity between 6.7~GHz masers and the other maser species.

\subsection{Comments on Individual Sources of Interest}

\textit{RMS 121.}
Strong \water\ maser emissions were detected in the first and second epochs with peak flux densities of 53 and 25 Jy, respectively. During the observing period of 2 years, the peak velocity changed from redshifted (7.2~\kms) to blueshifted ($-$3.1~\kms) relative to the systemic velocity (3.1 \kms). While there is a considerable variation in the \water\ maser line, the 44~GHz \methanol\ maser lines are invariable in peak flux density and peak velocity.

\textit{RMS 145.}
As noted in \S~3.3.1, this new \water\ maser source can be classified as a dominant redshifted source (Table~6; Figure~\ref{fig_HV}). A high-velocity feature appears at 125~\kms\ without any feature near the systemic velocity, 46.6 \kms. Both $^{13}$CO (1--0) and HCO$^+$ (1--0) lines were detected only around the systemic velocity. The peak flux density practically remained constant, $\sim$8~Jy, between the two epochs. Neither 44~GHz nor 95~GHz \methanol\ maser emission was detected toward this object.

\textit{RMS 2490.}
This is located in the W33A massive star-forming region.  \water\ maser lines were detected near the systemic velocity of 35.8 \kms\ in both epochs. The peak flux density of them increased from 11.4 Jy in the first epoch to 15.6~Jy in the second epoch. A high-velocity maser feature was also detected at $-$7.9~\kms\ in the second epoch. The peak flux was 15.8~Jy.  
\citet{urq11} detected only maser features around the systemic velocity.

\textit{RMS 2547.} 
This is a dominant blueshifted \water\ maser source (Table~6; Figure~\ref{fig_HV}). High-velocity features were detected at $-$70.9 and $-$65.5~\kms\ in the first and second epochs, respectively, with no emission around the systemic velocity of 20.9~\kms. 
\citet{urq11} also detected this source in 2009--2010. 
The peak flux density rapidly increased from 13.5~Jy in 2009--2010 to 128~Jy in 2011 April and 945~Jy in 2012 June,
while the velocity range remained very similar over the three epochs. 
Weak 44~GHz \methanol\ maser emission was detected near the systemic velocity in both epochs.

\textit{RMS 2584.}
This source showed strong dominant blue-shifted \water\ maser features in both epochs (Table~6; Figure~\ref{fig_HV}). \citet{urq11} also detected them. The peak flux density rapidly faded out from 165~Jy in 2009--2010 to 75~Jy in 2011 March and 5~Jy in 2012 June, while the velocity range appeared similar in three epochs.

\textit{RMS 3236.}
Two weak 44 GHz \methanol\ maser features were detected toward this source.
One is at the systemic velocity, while the other is offset by $-$29 \kms.
From the Galactic Ring Survey (GRS) $^{13}$CO J=1$-$0 line data archive \citep{jack06}, we examined whether there are $^{13}$CO lines associated with both maser features or not. As a result, we found that the latter feature in the offset is related to another YSO in the same line of sight.

\textit{RMS 3546.}
This source was observed in both epochs with the same OFF position, i.e., +1$^{\rm m}$ offset in R.A., (19:24:26.61, 14:40:16.9). An absorption feature was detected at $\sim$57~\kms\ only in the second epoch (Figure~5). It can be explained by the appearance of \water\ maser source, which did not appear in the first epoch, at the OFF position in the second epoch. This source is located in the active star-forming complex W51. \cite{kang09} identified 737 YSO candidates from the Spitzer data, and several of them lie within 60$''$ of the OFF position, including No. 608 that is 15$''$ away. One of them might emit \water\ maser emission in the second epoch, although there is no previous report of the maser detection toward the position.

\textit{RMS 3555.}
This source is located within G49.5--0.4 in the W51 A complex.
\water\ maser emission toward this is the broadest (188 and  230 \kms) in velocity range at each epoch. The peak flux density was about 3600~Jy at 57.0 \kms\ in the second epoch. The emission of \methanol\ maser was detected near the systemic velocity (60.1~\kms) at 44 and 95 GHz. Interestingly, weak 44 GHz maser emission was also detected at 50 \kms\ far from the systemic velocity in both epochs. The grid mapping result suggests that this emission  may result from W51 IRS 1 in a sidelobe.

\textit{RMS 3587.}
In the first epoch, \water ~maser emission showed dominant blueshifted features (Table~6; Figure~\ref{fig_HV}). High-velocity features were detected around --44~\kms\ in both epochs, although the peak flux density decreased from 4.2~Jy in 2011 March to 2.2~Jy in 2012 May by a factor of 2. 
The \water ~maser emission near the systemic velocity (16.1~\kms) emerged in the second epoch with a peak flux of 6.7~Jy.
\citet{urq11} detected multiple features only near the systemic velocity.

\textit{RMS 3735.}
The emission of \methanol\ maser was marginally detected only at 44 GHz in the second epoch, with a peak flux density of 1.4~Jy near the systemic velocity. As for the \water\ maser,  a single velocity component was detected in both epochs, and the peak flux density increased by a factor of 6 in a period of two years : from 5 Jy in 2011 April to 30 Jy in 2013 April. There is no significant change in velocity position of peak flux density.

\textit{RMS 3749.}
The emission of \water\ maser was detected in both epochs. The peak flux density varied from 73 Jy in 2011 April to 266 Jy in 2012 October. Also, the velocity of peak flux density moved from blueshifted (1.7~\kms) to redshifted (15.6~\kms) relative to the systemic velocity (7.9 \kms). \citet{urq11} detected a peak flux density of 175 Jy at $-$2.4 \kms.

\textit{RMS 3766.}
A dominant blueshifted \water\ maser feature with a peak flux of 5.0~Jy was detected at --25.7 \kms\ in the first epoch (Table~6; Figure~\ref{fig_HV}), while a 1.2~Jy feature solely appeared around the systemic velocity of 5.6 \kms\ in the second epoch. 
In both epochs 44 GHz \methanol\ maser emission ($\sim$2~Jy) was detected near the systemic velocity.

\textit{RMS 3846.}
In both epochs, 44 GHz \methanol\ maser spectra consisted mainly of two strong features of about 25 Jy at --5.2 \kms\ and --3.1 \kms. These lines appeared to have a symmetrical profile with respect to the systemic velocity ($-$4.4 \kms). Contrasting with the 44 GHz \methanol\ maser, 95 GHz maser spectrum showed a red-skewed profile in a similar velocity range of 44 GHz maser. 

\textit{RMS 3911.}
A dominant high-velocity \water\ maser feature was detected at $-$75.0~\kms\ and $-$65.8~\kms\ in the first and second epochs, respectively (Table~6; Figure~\ref{fig_HV}).
The peak flux density decreased by a factor of 2 from 6.4 Jy to 3.2 Jy.
In comparison, \citet{urq11} detected multiple features with a peak flux density of 5.8 Jy around the systemic velocity, $-$46.6~\kms, as well as a high-velocity feature around $-$75.0~\kms. 

\textit{RMS 3936}
Weak \water\ maser features  were detected between $-$75 and $-$55~\kms\ in the first epoch, while a dominant redshifted feature was detected at $-$10.1~\kms\ with a peak flux of 3.2~Jy in the second epoch (Table~6; Figure~\ref{fig_HV}).
\citet{urq11} detected the peak flux density of $\sim$19 Jy at the systemic velocity ($-$87.5~\kms) and multiple high-velocity features from $-$88.6 to $-$27.6~\kms. No class I \methanol ~maser emission was detected in our survey.

\textit{RMS 3958.}
Dominant blueshifted \water ~maser features were detected around $-$96.4~\kms\ without any feature near the systemic velocity ($-$51.2~\kms) in the first epoch, while no maser emission was detected in the second epoch (Table~6; Figure~\ref{fig_HV}).
 The peak flux density was 11.7~Jy.
\citet{urq11} detected \water ~maser features only near the systemic velocity. The 44 GHz and 95 GHz \methanol ~maser emission were detected at the systemic velocity.

\section{Analysis and Discussion}

\subsection{Masers and the Central Objests}
  In Figure~\ref{fig_lbol_lmaser}, the isotropic luminosities of four maser species are plotted against the bolometric luminosity ($L_{\rm bol}$) of the central protostar. 
 In our sample, as marked in Table~\ref{tbl_source}, 62 sources have one or two additional RMS sources within 20$''$ of each of them.
By considering the FWHMs of the KVN telescopes, we integrated the bolometric luminosities of the individual members and use the total luminosity for this comparison.
 The total bolometric luminosity is less than 5 times larger than the luminosity of a target source for 90\% of them.
 Thus this integration may not significantly affect the derived relationship here.
The maser luminosity tends to increase with the bolometric luminosity for all maser species, although the correlations seem to be weak. Linear least-square fittings result in the following relations with correlation coefficients of 0.27--0.50:
 \begin{eqnarray}
  log(L_{\rm 22})  = (1.40 \pm 0.11)~log(L_{\rm bol}) -(11.00 \pm 0.46)  \ (\rho = 0.50) , \\
  log(L_{\rm 44})  = (1.25 \pm 0.10)~log(L_{\rm bol}) -(10.31 \pm 0.41)  \ (\rho = 0.40) , \\
  log(L_{\rm 95})  = (1.17 \pm 0.10)~log(L_{\rm bol}) -~(9.70 \pm 0.39)  \ (\rho = 0.34) , \\
  log(L_{\rm 6.7}) = (1.27 \pm 0.18)~log(L_{\rm bol}) -(11.06 \pm 0.84)  \ (\rho = 0.27) .
 \end{eqnarray}	  

The relationship between the isotropic maser luminosity and the bolometric luminosity has been investigated in several previous studies of \water\ and \methanol\ masers in SFRs.
  For \water\ masers, the correlation between $L_{\rm H_2O}$ and $L_{\rm bol}$ has been proposed by \citet{fur03} for low-mass YSOs ($L_{\rm bol} \sim 0.1-10^2 L_{\odot}$), \citet{bae11} for intermediate-mass YSOs ($L_{\rm bol} \sim 10^2-10^4 L_{\odot}$), and \citet{fel92} for low- to high-mass YSOs ($L_{\rm bol} \sim 10-10^6 L_{\odot}$).
  The slopes of their fit lines are close to 1 (0.81 $-$ 1.02). For comparison, we plot our data together with \water\ maser data of low- and intermediate-mass YSOs from \citet{fur03} and \citet{bae11}.
  We obtain a relation of  $\rm log(L_{22}) = (1.07 \pm 0.05) \times log(L_{\rm bol})-(9.21 \pm 0.20)$ with a much higher correlation coefficient ($\rho = 0.76$), which is consistent with previous studies (Figure~\ref{fig_lbol_lmaser_wide}$a$).

  In the case of 44 \methanol\ masers, the relation between $L_{\rm CH_3OH}$ and $L_{\rm bol}$ has been investigated by \citet{kal13} for low-mass YSOs ($L_{\rm FIR} \sim 0.1-10^2 L_{\odot}$) and \citet{bae11} for intermediate-mass YSOs ($L_{\rm bol} \sim 10^2-10^3 L_{\odot}$).  We combine our data with the data of \citet{kal13} and \citet{bae11} to widen the range of $L_{\rm bol}$.
  The best-fit line of 44 GHz data is $\rm log(L_{44}) = (1.12 \pm 0.07)\times log(L_{\rm bol})-(9.77 \pm 0.27)$ ($\rho = 0.71$) (Figure~\ref{fig_lbol_lmaser_wide}$b$).
  For the 95 GHz \methanol\ masers, \citet{gan13} presented a relation of $\rm log(L_{95})  =
   (0.51 \pm 0.08)\times log(L_{\rm bol})-(7.77 \pm 0.33)$ ($\rho = 0.66$) from low to high-mass YSOs ($L_{\rm bol} \sim 10^1-10^6 L_{\odot}$).

\subsection{Masers and the Parental Clumps}

  We search for dust clumps associated with our target sources using the 870~$\mu$m continuum data of the APEX Telescope Large Area Survey of the Galaxy
(ATLASGAL) \citep{sch09,con13}. 
We find that 135 sources are located in the survey area and that 116 (86\%) of them are associated with the ATLASGAL clumps within a search radius of 30$\arcsec$, in which $\sim$90\% of ATLASGAL$-$RMS matches are distributed \citep{urq14b}.
The maser detection rates of the 116 sources are 35\%, 36\%, and 31\% at 22, 44, and 95~GHz, respectively, in the second epoch. These values are higher than the maser detection rates of the remaining 19 sources: 26\%, 26\%, and 10\%.
Assuming that dust emission is optically thin at 870 $\mu$m,
we derive the clump mass in a similar way to \citet{urq14b} using the following equation

  \begin{equation}
 M_{\rm clump} =
 \frac{S_{\rm int}D^2}{k_\nu B_{\nu}(T_{\rm d}) R_{\rm d}}.
  \end{equation}

\noindent
Here $S_{\rm int}$ is the integrated flux density, $D$ is the distance to the  source, $k_\nu$ is the mass absorption coefficient per unit mass of dust, $B_\nu(T)$ is the Planck function, $T_{\rm d}$ is the dust temperature, and $R_{\rm d}$ is the dust$-$to$-$gas mass ratio. We take $S_{\rm int}$ from the ATLASGAL catalog and adopt $k_\nu$=1.85 cm$^2$ g$^{-1}$, $T_{\rm d}$=20~K, and $R_{\rm d}$=0.01 as in \citet{urq14b}. The effective radii of the associated clumps range from 7\arcsec\ to 154\arcsec\ with a median of 31\arcsec.

  Figure~\ref{fig_M_Lmaser} shows the maser luminosity versus the estimated clump mass for the detected maser sources.
The maser luminosity appears to correlate with the clump mass for all four maser species, especially 44 and 95~GHz \methanol\ masers. Linear least-squares fittings result in the following relations with correlation coefficients of 0.57 to 0.77.
 \begin{eqnarray}
 log(L_{\rm 22}) = (1.66 \pm 0.25)~log(M_{\rm clump}) -(10.11 \pm 0.81)   \ (\rho = 0.57) ,\\
 log(L_{\rm 44}) = (1.30 \pm 0.10)~log(M_{\rm clump}) -~(9.16 \pm 0.36)   \ (\rho = 0.69) ,\\
 log(L_{\rm 95}) = (1.38 \pm 0.13)~log(M_{\rm clump}) -~(9.10 \pm 0.44)   \ (\rho = 0.77) ,\\
 log(L_{\rm 6.7})= (1.63 \pm 0.27)~log(M_{\rm clump}) -(10.67 \pm 0.92)   \ (\rho = 0.58) .
 \end{eqnarray}

  \noindent
 \citet{chen12} also found a strong ($\rho$=0.84) correlation between $L_{95}$ and the clump mass for the Bolocam Galactic Plane Survey (BGPS) sources. However, the slope was much lower, 0.81. \citet{chen11} and \citet{gan13} obtained even lower (0.5$-$0.6) slopes of the best linear fits between $L_{95}$ and $M_{\rm clump}$ for the BGPS clumps associated with EGOs and molecular outflows, respectively.
 In our sample, twenty 95~GHz maser sources are associated both with the ATLASGAL clumps and BGPS clumps. After deriving the mass of the BGPS clumps as in \citet{chen12}, we examine the $M_{\rm clump}$$-$$L_{95}$ relations for the two groups and find no significant difference. The slopes of the best linear fits are 1.3 and 1.2 for the ATLASGAL and BGPS clumps with correlation coefficients of 0.89 and 0.74, respectively.

	The peak H$_2$ column density can be calculated from the peak flux density of each clump using the equation
	
  \begin{equation}
 N_{\rm H_2} =
 \frac{S_{\rm p}}{\Omega_{\rm b} \mu~ m_{\rm H} k_\nu B_\nu(T_{\rm d}) R_{\rm d} } ,
  \end{equation}

\noindent
  where $S_{\rm p}$ is the peak flux density, $\Omega_{\rm b}$ is the beam solid angle, $\mu$ is the mean molecular weight, and $m_{\rm H}$ is the mass of hydrogen atom. We adopt $\Omega_{\rm b}$=9.8 $\times~10^{-9}$~Sr \citep{con13} and $\mu$=2.37. Figure~\ref{fig_NH2_Lmaser} plots the maser luminosity against $N_{\rm H_2}$ for four maser species. No correlation appears for 22 and 6.7~GHz masers while very weak correlations exist with $\rho \simeq$0.3 for 44 and 95~GHz masers. 
Figure~\ref{fig_NH2_hist} shows the histograms of $N_{\rm H_2}$ for detected and undetected sources in each maser transition.  The median values of  $N_{\rm H_2}$'s are 11.1, 10.7, 11.7, and 6.7 in units of $10^{22}$~cm$^{-2}$ for 22, 44, 95, and 6.7 GHz maser sources, respectively. 
The two distributions are significantly different for 22, 44, and 95 GHz masers by Kolmogorov-Smirnov(K-S) test ($p<10^{-7}$), while they are statistically similar for 6.7~GHz maser ($p =0.67$).
This difference can be caused by different pumping mechanisms of the two groups. The occurrence of 6.7~GHz class~II masers, which are radiatively pumped, appears to be less dependent on the ambient physical conditions than \water\ and class~I \methanol\ masers that are collisionally pumped \citep[see also][]{bre14}.

Figure~\ref{fig_NH2_Smint} plots the integrated flux density of each maser ($\int S_{\nu} dv$) versus $N_{\rm H_2}$. The relationship might be meaningful for the future maser survey because the two parameters are independent of the distance and other intrinsic physical parameters \citep{chen12}.
  We perform least-squares fit and obtain the following relations:
 \begin{eqnarray}
 log(\int S_{22}~dv)  = (2.00 \pm 0.42)~log(N_{\rm H_2})-(44.20 \pm 9.65)  \ (\rho = 0.42) , \\
 log(\int S_{44}~dv)  = (1.69 \pm 0.16)~log(N_{\rm H_2})-(37.72 \pm 3.56)  \ (\rho = 0.71) , \\
 log(\int S_{95}~dv)  = (1.80 \pm 0.20)~log(N_{\rm H_2})-(40.35 \pm 4.53)  \ (\rho = 0.71) , \\
 log(\int S_{6.7}~dv) = (1.81 \pm 0.39)~log(N_{\rm H_2})-(40.10 \pm 8.93)  \ (\rho = 0.35) .
 \end{eqnarray}

 The slopes of linear fits are similar (1.7$-$2.0) for all four maser species. However, there is a significant difference in the correlation coefficient in that 44 and 95 GHz masers show quite high correlation coefficients of 0.71, while 22 and 6.7 GHz masers have considerably lower coefficients of $\sim$0.4. The derived correlation coefficient of 95~GHz masers is consistent with the value (0.69) of \citet{chen12}, although they used 40\arcsec ~beam-averaged column density of the BGPS data.

\subsection{NH$_3$ Line width and Kinetic Temperature}

 \citet{urq11} surveyed about 600 RMS sources in the NH$_3$ as well as 22 GHz \water~maser lines, and detected the \ammonia~line emission toward 479 ($\sim$80\%) sources. They derived some line and physical parameters using the NH$_3$ data, including the line width (FWHM) of each transition and the kinetic temperature of gas.
As mentioned in Section~3.2, their sample includes 257 sources in our sample. NH$_3$ line emission was detected in 218 (85\%) of them.   For these NH$_3$-detected sources,
the line widths range from 0.49 to 8.23~\kms\ with mean and median values of 1.85 and 1.73 \kms, respectively. The mean and median are similar to those (1.7 and 1.6~\kms) of infrared dark clouds (IRDCs) measured by 
\citet{chi13} from the NH$_3$ line observations, while the mean is significantly smaller than that (2.1~\kms) of UCHIIs measured by \citet{urq11}. 
We compare the line widths of four subsamples: 99 non-maser, 44 only 22~GHz, 21 only 44~GHz, and 54 both (22 and 44~GHz) maser-detected sources.
The median values are 1.5, 1.8, 1.7, and 2.2~\kms~for non-maser, only 22~GHz, only 44~GHz, and both maser-detected sources, respectively. The line width tends to increase from non-maser to only 22 or 44~GHz to both maser-detected subsamples. This tendency is more clearly displayed in Figure~\ref{fig_NH3_cumul}, which presents cumulative probabilities of the line width for the four subsamples. The difference between non-maser and both maser-detected subsamples seems to be distinct. We perform a K-S test for combinations of the four subsamples in order to examine the statistical difference in their distributions. The K-S test shows that both maser-detected subsample has a significantly different distribution from the other three subsamples, namely, the p-values being 1.2$\times$10$^{-10}$, 0.001, and 0.02 for non-maser, only 22 GHz, and only 44 GHz maser-detected subsamples, respectively.

  The estimated kinetic temperatures are in the range of 12$-$45 K with mean and median values of 21.6 and 21.1~K for the 218 NH$_3$-detected sources. 
 The mean and median are significantly higher than those (18 and 16~K) of IRDCs \citep{chi13}, while the mean is significantly lower than that (24.6~K) of UCHIIs \citep{urq11}. The median values are 19.2, 23.0, 21.3, and 23.1~K for no masers, only 22~GHz, only 44~GHz, and both maser-detected sources, respectively. The kinetic temperature tends to increase from no masers to only 44~GHz to only 22~GHz and both maser-detected subsamples. Figure~\ref{fig_NH3_cumul} shows this tendency. The K-S test also reveals that both maser-detected sources have a different distribution from no masers (p=1.3$\times$10$^{-7}$) and only 44~GHz (p=0.005) maser-detected sources. 
  Therefore, non-maser and both maser-detected sources have similar NH$_3$ line widths and kinetic temperatures to IRDCs and UCHIIs, resepctively. This may suggest that both maser-detected sources are more evolved than non-maser sources although they are all classified as HMPOs.

\subsection{The Virial Parameters}

  We estimate the virial masses of the molecular clumps associated with our RMS sources using the \ammonia\ line widths. Following \citet{ful92} (see also Urquhart et al. 2015), we first derive the average line width of the total column of gas from the observed \ammonia\ line width using the equation below
   \begin{equation}
   \Delta v_{\rm avg}^2 =
      \Delta v_{\rm T}^2 + \Delta v_{\rm NT}^2 =
   \Delta v_{\rm corr}^2 + 8~ln2 \ \frac{k_{\rm b} T_{\rm kin}}{m_{\rm H}}
   \left(\frac{1}{\mu_{\rm p}}-\frac{1}{\mu_{\rm NH_3}}\right).
   \end{equation}
 Here  $\Delta v_{\rm corr}$ is the observed line width corrected for the resolution of the spectrometer,
$k_b$ is the Boltzmann constant, $T_{\rm kin}$ is the kinetic temperature of gas, taken from ammonia analysis in Section 4.3, $\mu_{\rm p}$ and $\mu_{\rm NH_3}$ are the mean molecular weights of hydrogen and ammonia molecules taken as 2.37 and 17, respectively.

Assuming the associated clumps are self-gravitating, we calculate the virial mass from
  \begin{eqnarray}
    M_{\rm vir} = \frac{5}{\rm 8ln2 G} \frac{R_{eff} \Delta v_{\rm avg}^2}{a_1 a_2}
    = 210M_{\odot}\frac{(R_{eff}/\rm pc)(\Delta v_{\rm avg}/\rm km s^{-1})^2}{a_1 a_2}, \\
    a_1 = \frac{1-p/3}{1-2p/5} \rm  ~for ~p < 2.5,~~\ a_2 = y \frac{\rm arcsinh (y^2-1)^{\frac{1}{2}}}{(y^2-1)^{\frac{1}{2}}},
  \end{eqnarray}
  where $R_{eff}$ is the effective radius of the clump, $G$ is the gravitational constant, $a_1$ is the correction for the power-law density distribution $\rho(r) \sim r^{-p}$, $a_2$ accounts for the effect of the clump ellipticity, and $y$ is the aspect ratio of the clump \citep{ber92}. We adopt the mean power value determined by \citet{mul02}, $<$p$>$=1.8 (i.e. $a_1$=1.43), from modeling the 350~$\mu$m continuum emission maps of 31 massive star-forming clumps, and the mean aspect ratio of the 116 associated ATLASGAL clumps, $y$=1.55 (i.e., $a_2 \sim$1.3).

  Figure~\ref{fig_virial} shows the virial parameter $\alpha$ ($\equiv$ $M_{\rm vir}$/$M_{\rm clump}$) versus $M_{\rm clump}$ for the clumps with and without maser emission. The virial parameter tends to decrease with increasing clump mass. This implies that more massive clumps are more gravitationally unstable.
  Linear regression fittings to the clumps with and without maser emission give slopes of about --0.54$\pm$0.08 and --0.61$\pm$0.07 respectively.
  These values are similar to the slopes from $-$0.37 to $-$0.79 reported by \citet{kauf13} for 260 other massive star-forming clumps.
In addition, the maser-detected clumps appear to have higher $\alpha$'s than the non-detected clumps with similar masses, which suggests that the former are more stable than the latter. This may be because of more active feedback in the clumps with maser emission.

\subsection{Maser Occurrence and Evolutionary Stage} 

  IRDCs are generally believed to be the best candidates of birthplaces for high-mass stars and clusters \citep[e.g.,][]{rath06}. UCHIIs are produced by young massive stars that have reached the main sequence stage. Thus IRDCs, HMPOs, and UCHIIs represent the evolutionary sequence of massive star formation. We investigate how the occurrence rates of \water\ and class~I \methanol\ masers vary with the evolution of the central objects. \citet{wang06} surveyed \water\ maser emission toward 140 IRDC cores with the Very Large Array (VLA)
 at a typical rms noise level of $\sim$0.1~Jy, and detected the emission in 12\% of them. \citet{cham09} also observed the 140 and 50 more IRDC cores using the GBT
 with a twice lower noise level ($\sim$0.05~Jy) and obtained a detection rate of 35\% . 
As mentioned earlier, the previous surveys of \water\ maser emission toward HMPOs showed a detection rate of 42\% at a similar sensitivity to ours \citep{srid02}, and a slightly higher rate of $\sim$ 52\% with a better sensitivity ($\sim$0.12~Jy) \citep{urq11}. Although \citet{urq11} obtained the same detection rate of \water\ maser emission toward UCHIIs as that for HMPOs, other surveys showed significantly higher detection rates for UCHIIs. \citet{chur90} surveyed 84 UCHIIs and detected \water\ maser emission in 67\% of them, and \cite{kim17} also observed 103 UCHIIs and obtained a similar detection rate with the KVN telescopes at a similar detection limit to this study. Therefore, the detection rate of \water\ maser emission tends to increase as the central objects evolve.

As for 44~GHz \methanol\ masers, as mentioned in \S~3.1,
\citet{fon10} detected the emission in 31\% of 88 HMPO candidates at a rms noise level of about 0.3~Jy.  \cite{kim17} observed 103 UCHIIs in the 44~GHz maser tranisition at a similar sensitivity to this study, and obtained a significantly higher detection rate of 48\%. In actual, \citet{fon10} also acquired the same value for the $high$ group of their sample, which has similar $IRAS$ colors to UCHIIs, but a considerably lower detection rate (17\%) for the $low$ group, which may be in an earlier evolutionary phase than the $high$ group. Thus the detection rates of 44~GHz \methanol\ maser emission also appear to increase as the central objects evolve. Moreover, \citet{fon10} found a similar evolutionary trend for 95~GHz masers of which the detection rate increase from the $low$ group (6\%) to the $high$ group (20\%), although the number of the detected sources is just 11.

These trends are in contrast to the findings in low- and intermediate-mass star-forming regions. \citet{fur03} found in low-mass YSOs that the detection rate of \water\ masers dramatically decreases from Class 0 to Class I to Class II objects. \citet{bae11} also found similar trends in the detection rates of both \water\ and 44~GHz \methanol\ masers toward intermediate-mass YSOs. \citet{bae11} suggested that this difference can be caused by the different environments
of low- and high-mass star-forming regions. UCHIIs are still deeply embedded in the natal molecular clouds although the central stars have already reached the main sequence stage. On the contrary, low- and intermediate-mass stars in the pre-main sequence phase are usually visible.  Therefore, the occurrence of 22~GHz \water\ and 44~GHz class~I \methanol\ masers is closely related to the surrounding environments as well as the evolutionary stage of the central objects.
\citet{bre14} also suggested  from their investigation of \water\ masers associated with 12.2~GHz  class~II \methanol\ masers that the occurrence of 22 GHz \water\ masers depends more on the surrounding environments than class~II \methanol\ and OH masers, which are produced by radiative pumping.

\section{Summary}

We have simultaneously surveyed 22 GHz \water\ and 44 and 95 GHz class I \methanol\ masers toward 299 HMPOs in the RMS catalog. The main results are summarized as follows.

1. The overall detection rates are 45\%, 28\%, and 23\% for 22, 44, and 95 GHz masers, respectively.
We found 123 new maser sources: 15 at 22 GHz, 56 at 44 GHz, and 51 at 95 GHz. In our sample, 170 sources are distributed in the MMB survey area  and 38 of them are associated with 6.7~GHz class~II \methanol\ maser emission. We detected high-velocity ($>$30~\kms) features in 27 \water\ maser sources. Nine of them are very likely to be dominant shifted \water\ maser outflow sources, 7 blueshifted and 2 redshifted ones.

2. The 44 and 95~GHz class~I \methanol\ masers have very similar properties. The 95~GHz maser emission was detected only in 44~GHz maser sources. The two transitions have strong correlations with each other in the peak velocity, the peak flux density, and the isotropic luminosity.  The peak flux density ratio ($S_{\rm p, 95}$/$S_{\rm p, 44}$) was estimated to be 0.56. Both of these masers always have the peak velocities within 5~\kms\ from the systemic velocities.  Therefore, 44 and 95~GHz masers  might be produced by the same mechanisms in the same sites. This is consistent with the prediction of some modeling that these two transitions can be mased in similar physical conditions \citep[e.g.,][]{mce14}.
On the other hand, they show no significant correlation with 22~GHz \water\ or 6.7~GHz class~II \methanol\ masers in the isotropic luminosity. 

3.  The 44~GHz class~I \methanol\ masers have much narrower distributions than 22~GHz \water\ masers in the peak velocity relative to the systemic velocity and the velocity range. The 6.7~GHz class~II \methanol\ masers have intermediate distributions between the two maser species. As for the peak flux density, 44~GHz masers have a significantly smaller median value than those of 6.7 and 22~GHz masers, which are comparable. 
 
4. We investigated for 22, 44, 95, and 6.7~GHz masers whether the isotropic luminosity of each maser species correlates with the physical properties of the central objects and the parental clumps. 
The maser luminosity shows significant correlations with the clump mass  for all the four maser species, but does \textit{not} correlate with the peak H$_2$ column density. We found weak correlation between the maser luminosity and the bolometric luminosity only for our sample. However, quite strong correlations appeared between the two parameters for 22 and 44~GHz masers in the case where the data points of low- and intermediate-mass YSOs from the previous studies were added.

5. The line width and kinetic temperature of \ammonia\ line emission tend to increase from non-maser to only 22~GHz or 44~GHz to both maser-detected sources. 
This may suggest that both maser-detected sources are more evolved than non-maser sources although they are all classified as HMPOs.
The investigation of the virial mass suggests that the ATLASGAL clumps with any maser emission could be more gravitationally stable than those associated with no maser emission.

6. The detection rates of 22 GHz \water\ and 44 GHz \methanol\ maser emission appear to increase as the central objects evolve in massive star-forming regions.
This is contrary to the results of low- and intermediate-mass cases. Therefore, the occurrence of both masers might depend on the encompassing environments as well as on the evolutionary stage of the central objects.

\acknowledgments
We thank the anonymous referee for many constructive comments  and  suggestions.
We are grateful to all staff members in KVN who helped to operate the array and to correlate the data. The KVN is a facility operated by KASI (Korea Astronomy and Space Science Institute). The KVN operations are supported by KREONET (Korea Research Environment Open NETwork) which is managed and operated by KISTI (Korea Institute of Science and Technology Information).
This paper made use of information from the Red MSX Source survey database at $http://rms.leeds.ac.uk/cgi-bin/public/RMS\_DATABASE.cgi$ which was constructed with support from the Science and Technology Facilities Council of the UK.

\clearpage

\begin{figure}
\epsscale{1}
\begin{minipage}{0.1cm}
\rotatebox{90}{\textcolor{black}{Flux density (Jy)}}
\end{minipage}
\begin{tabular}{cccccc}
&
\includegraphics[width=50mm]{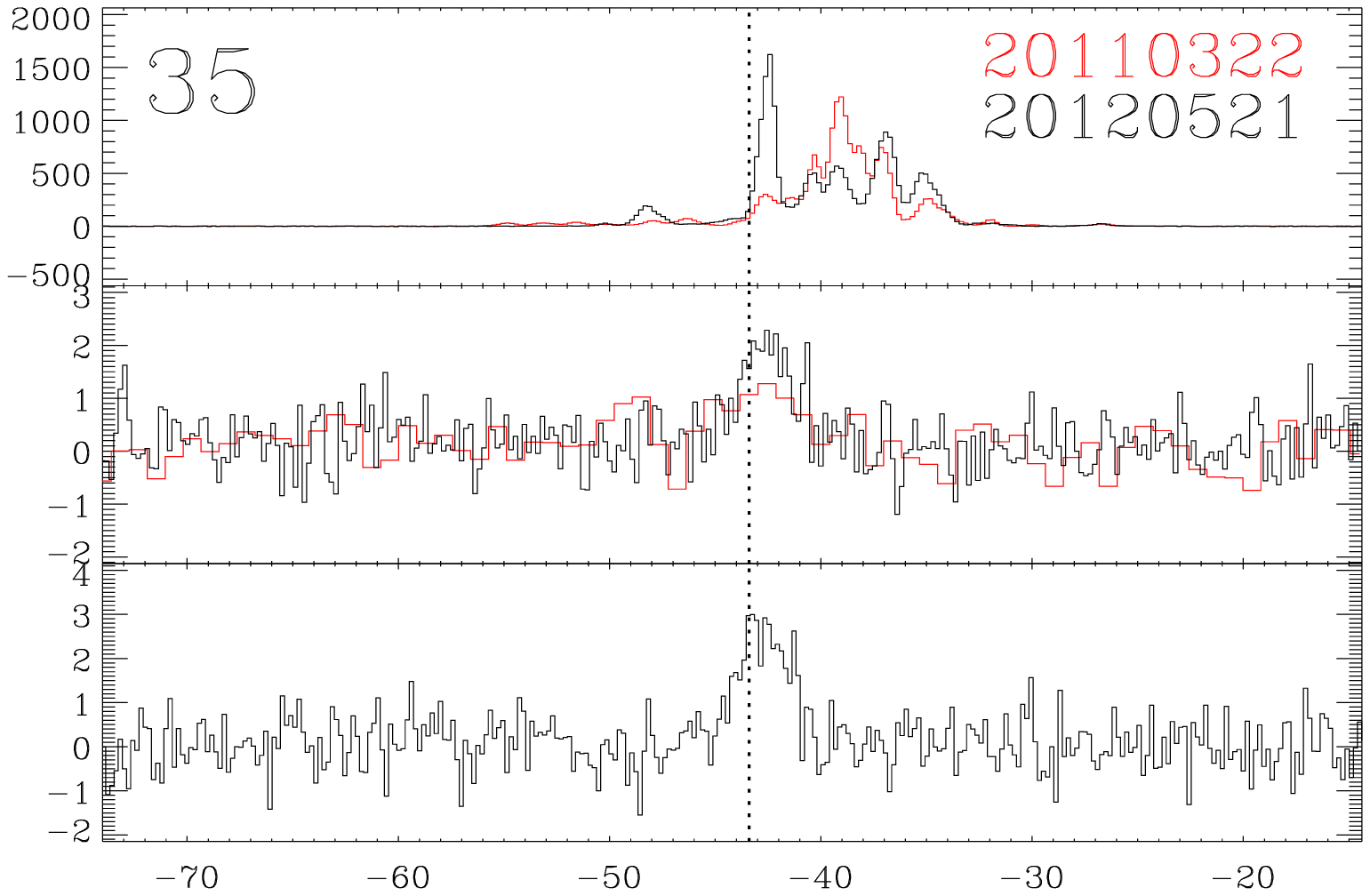} 
&
&
\includegraphics[width=50mm]{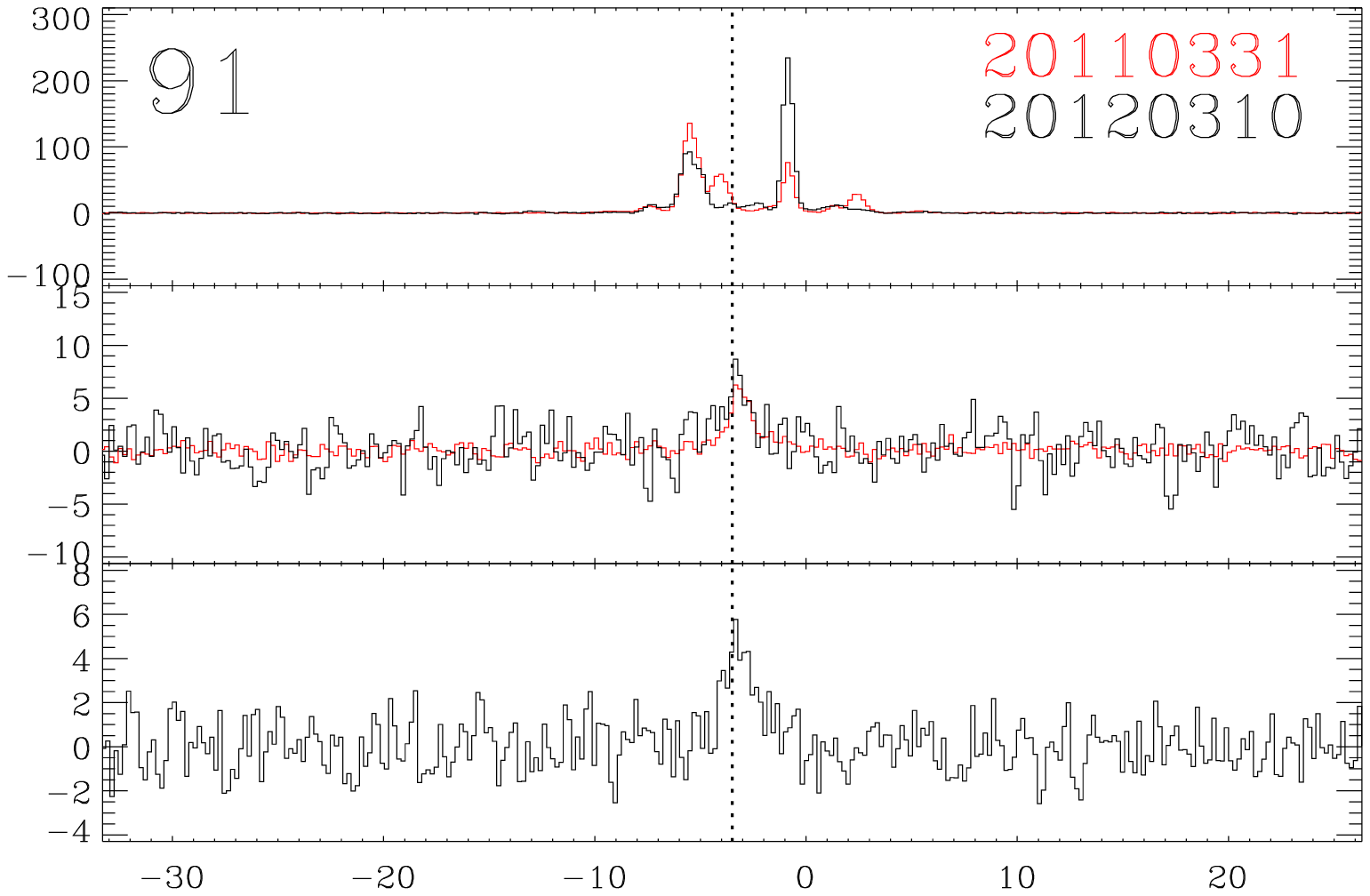} 
&
& 
\includegraphics[width=50mm]{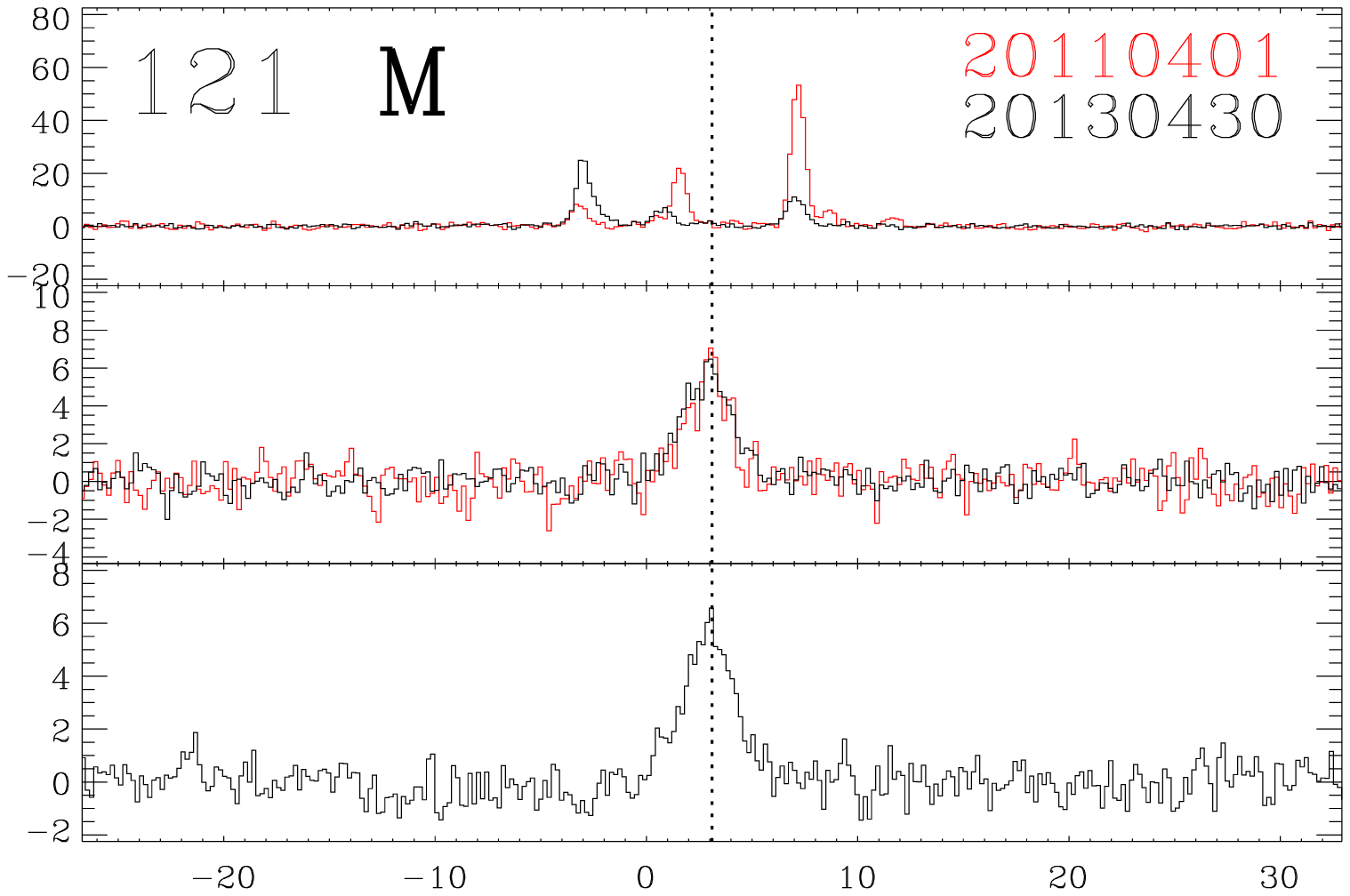} 
\\
&
\includegraphics[width=50mm]{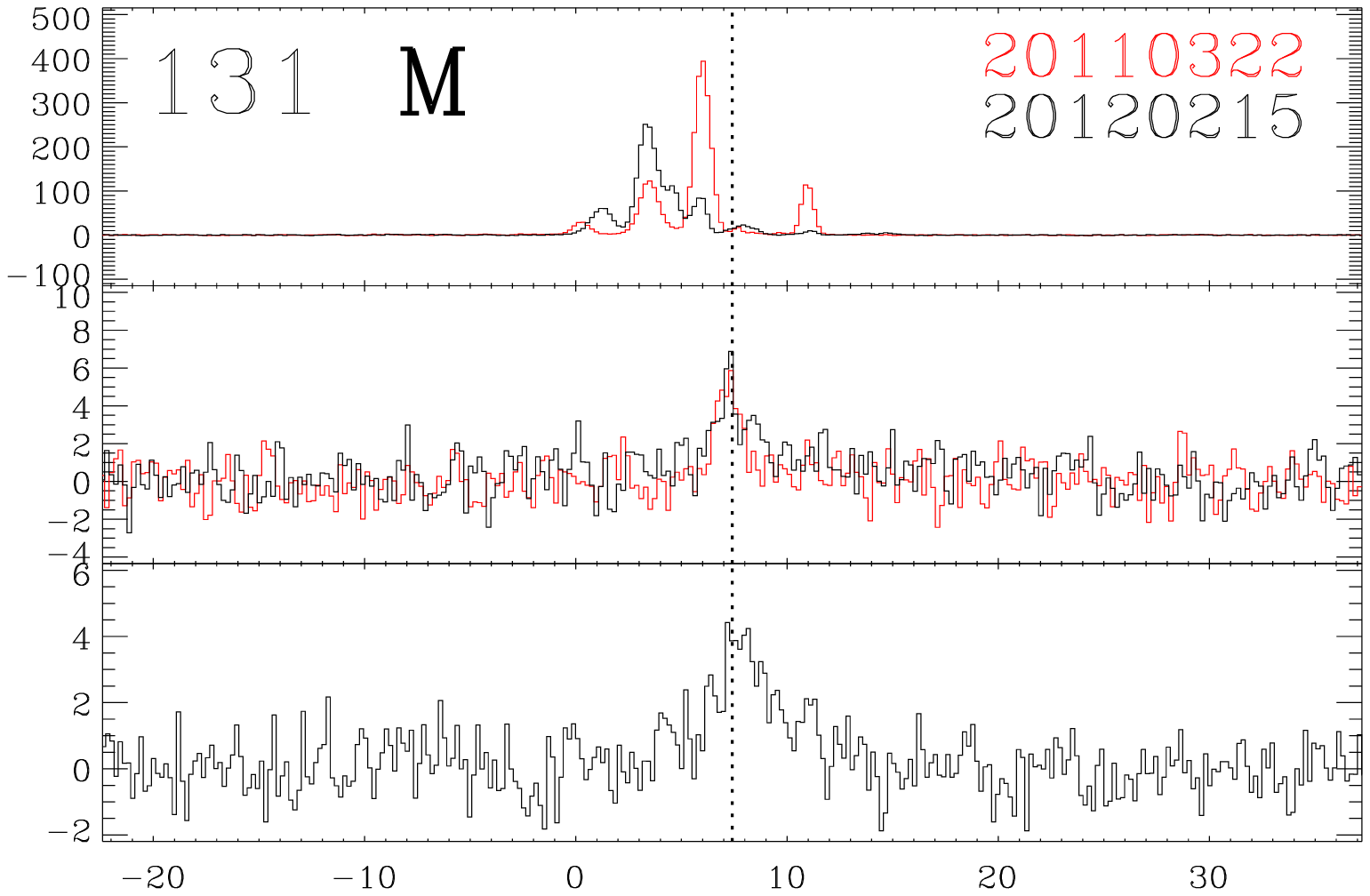} 
&
& 
\includegraphics[width=50mm]{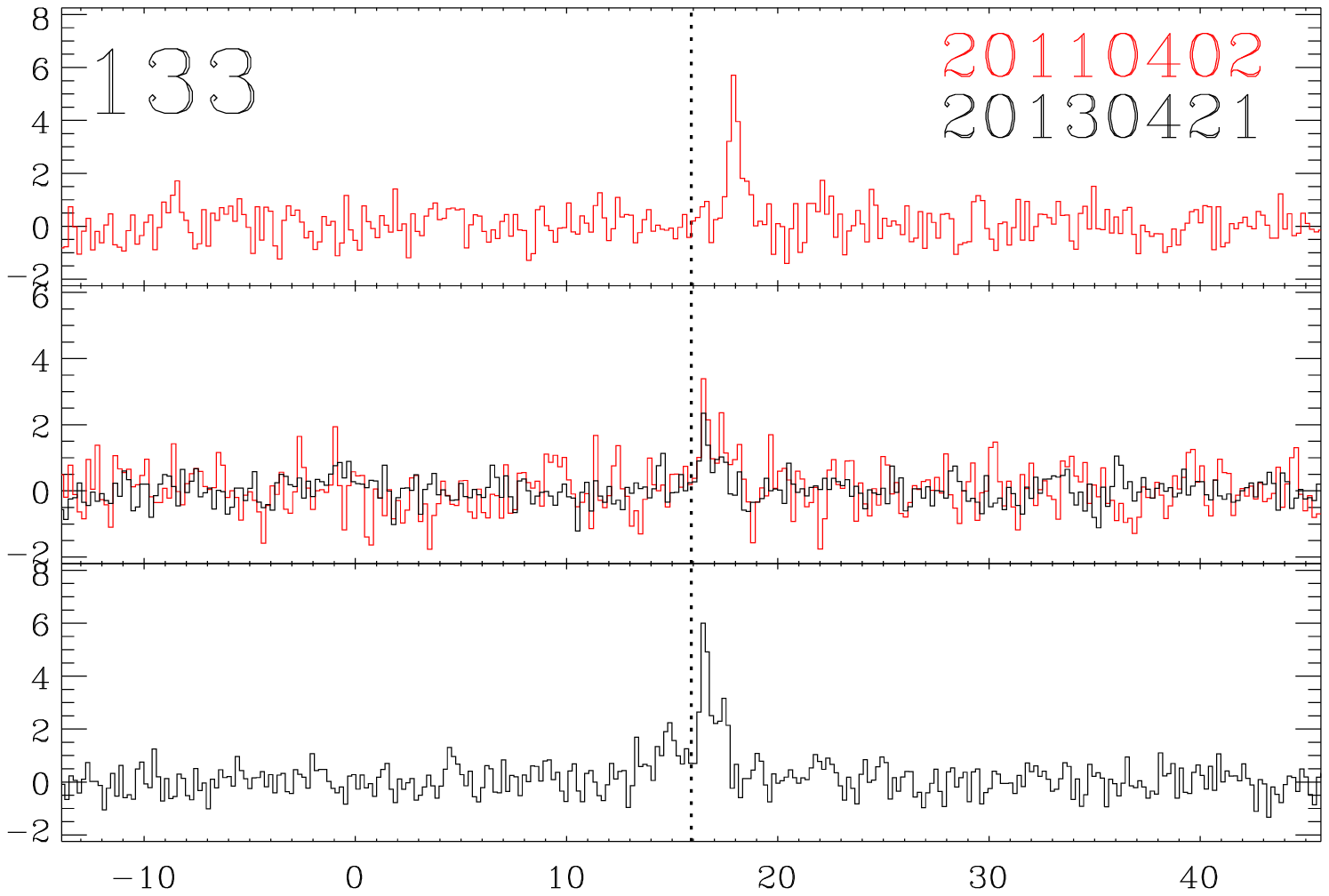} 
&
&
\includegraphics[width=50mm]{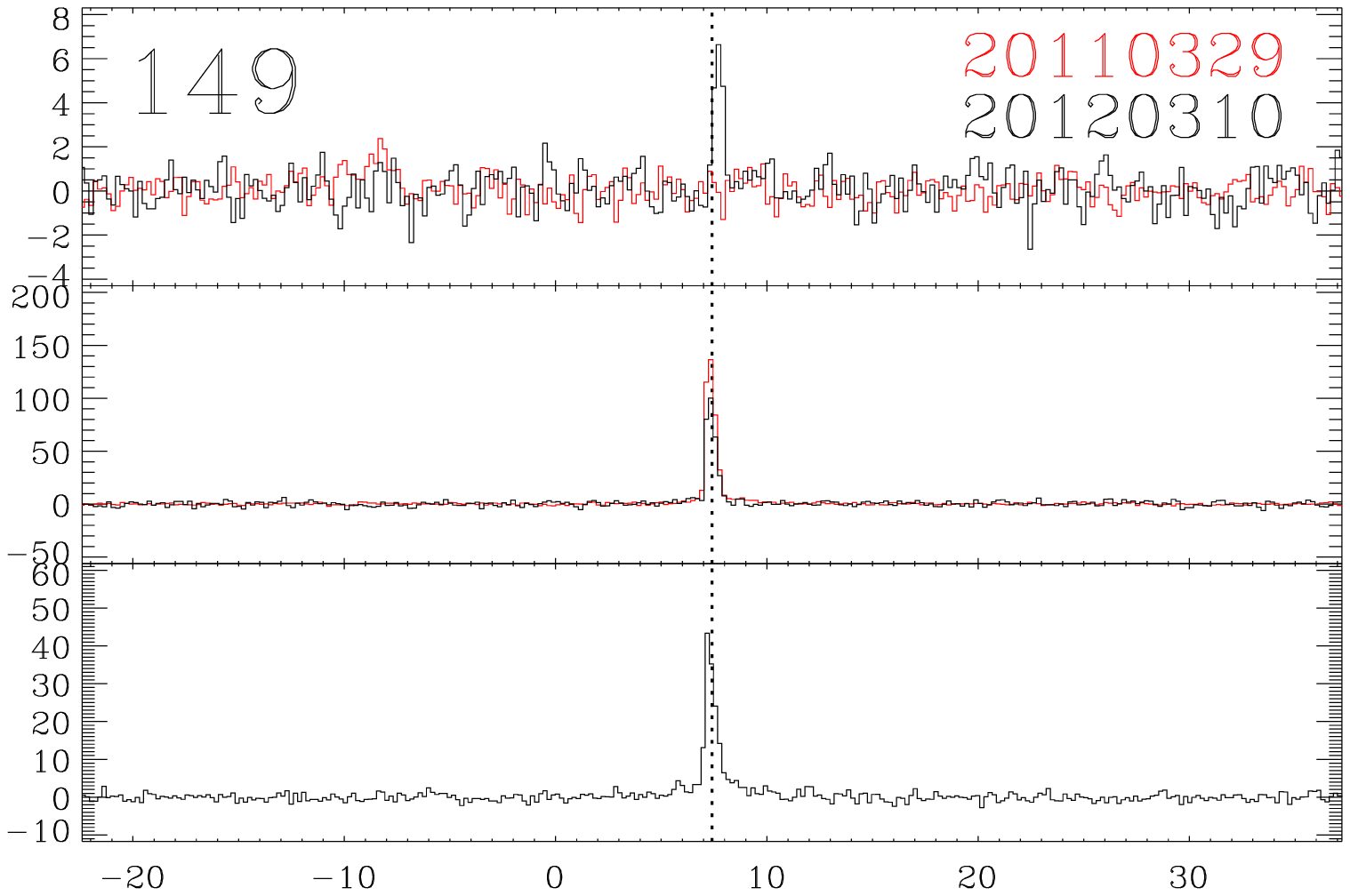} 
\\
&
\includegraphics[width=50mm]{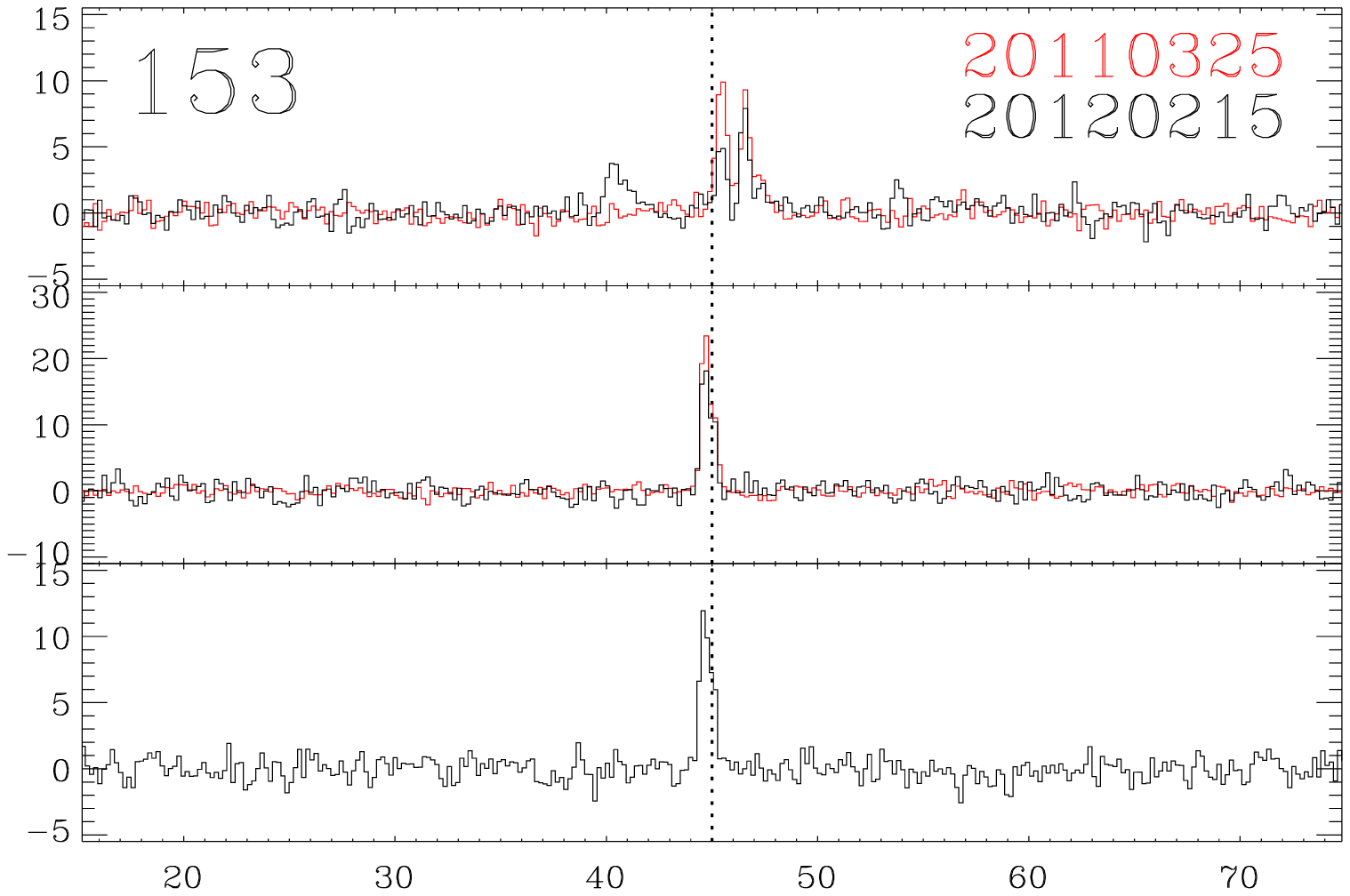} 
&
&  
\includegraphics[width=50mm]{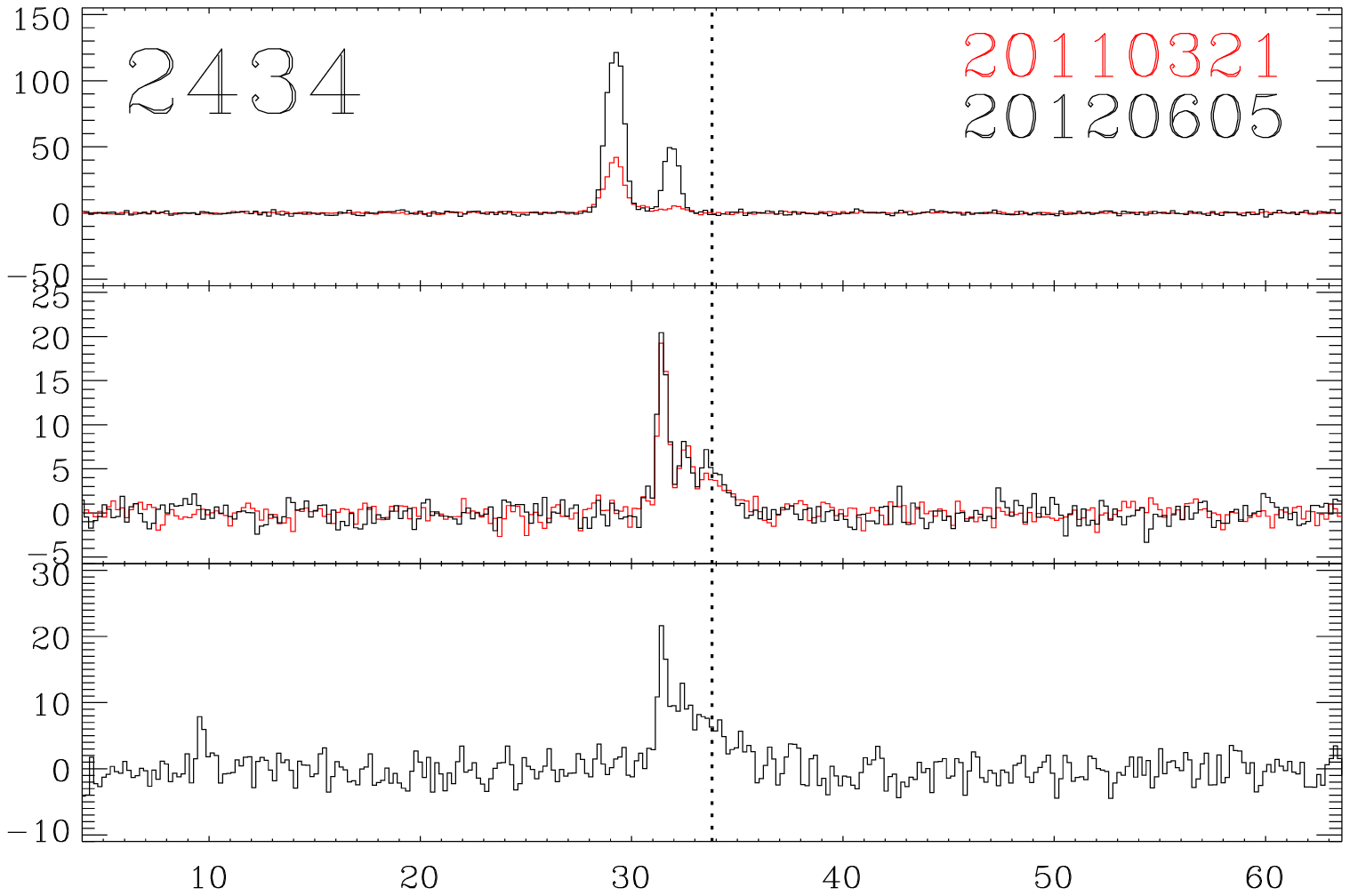} 
&
&
\includegraphics[width=50mm]{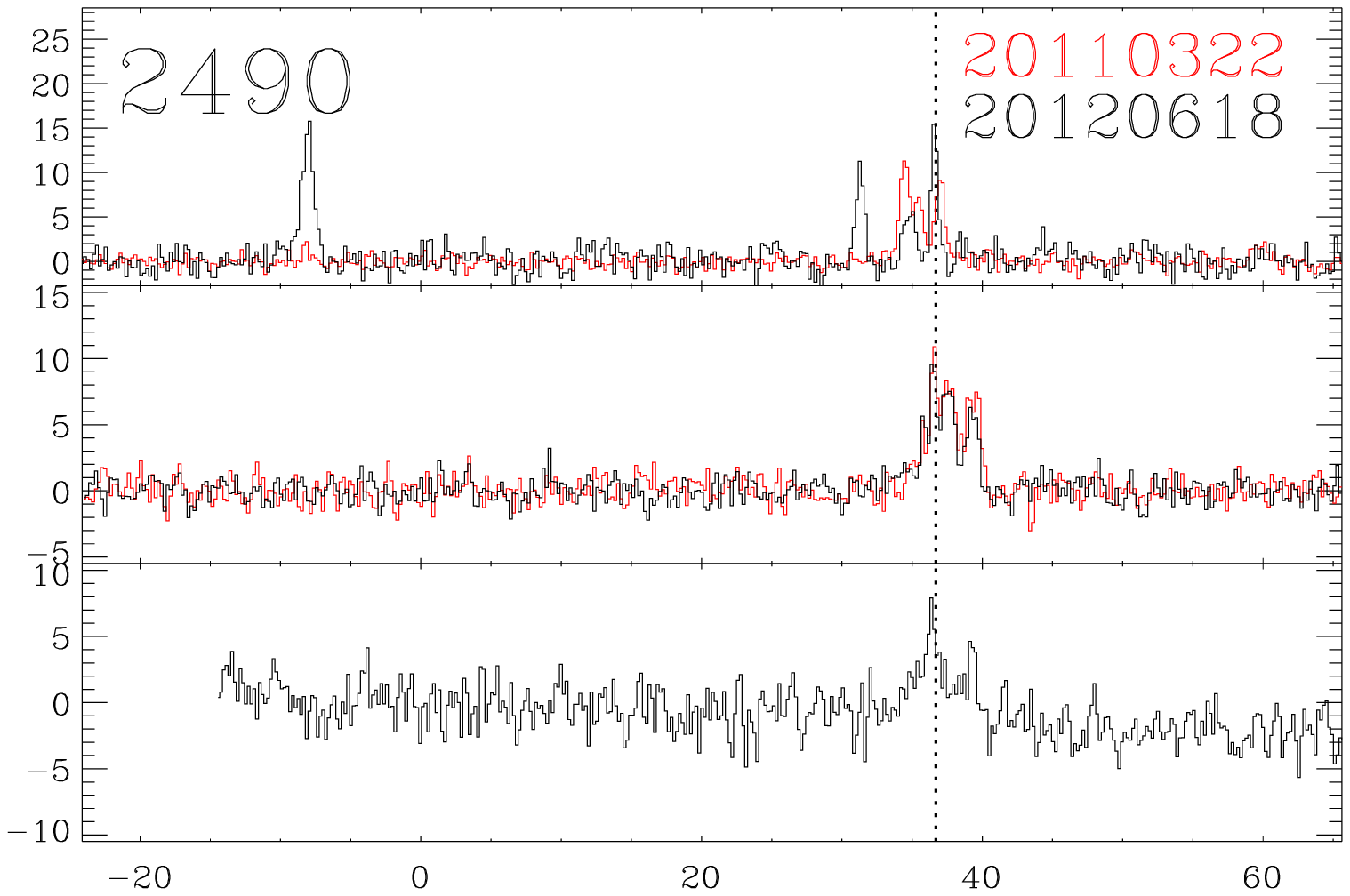} 
\\
&
\includegraphics[width=50mm]{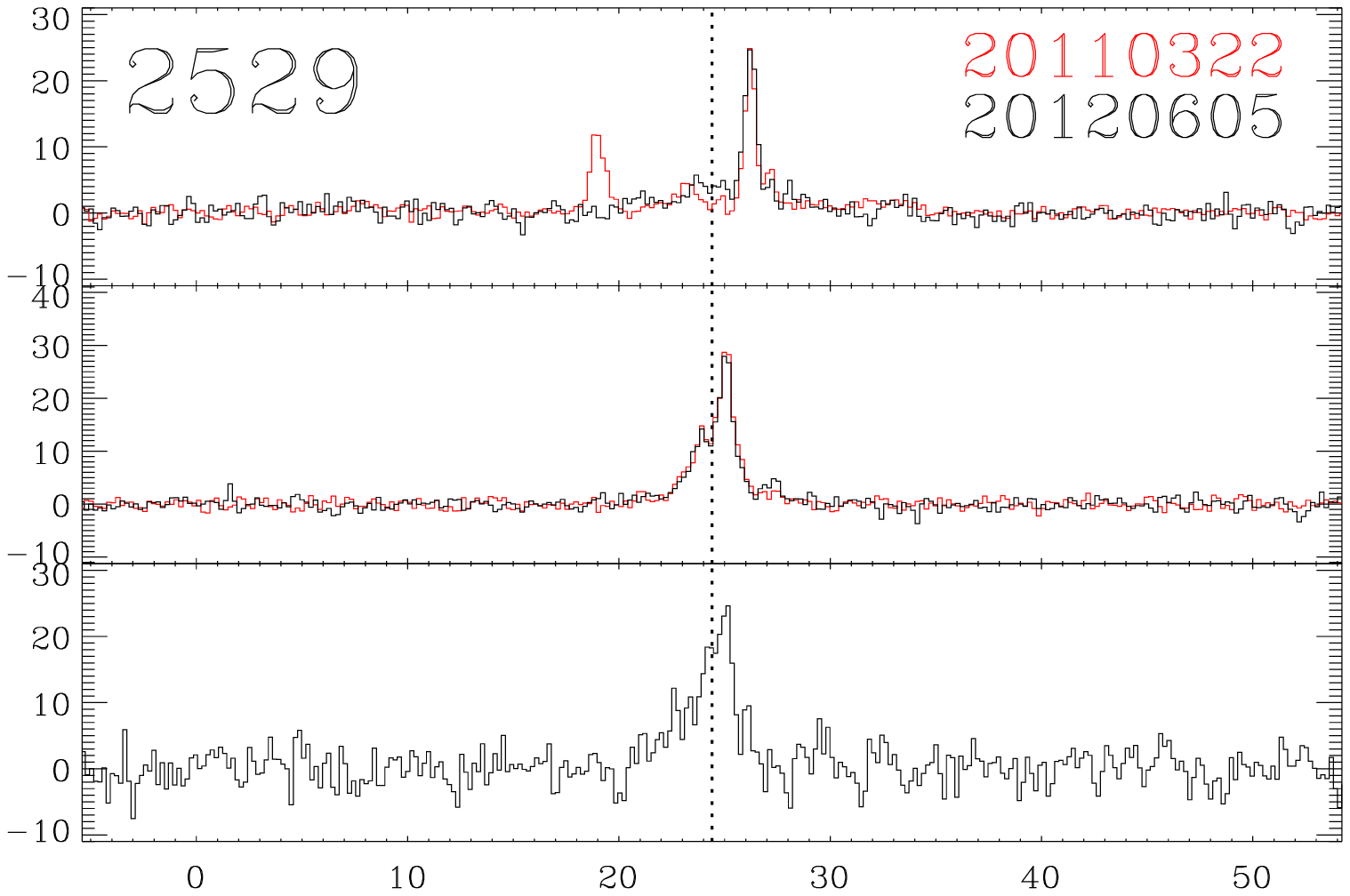} 
&
&  
\includegraphics[width=50mm]{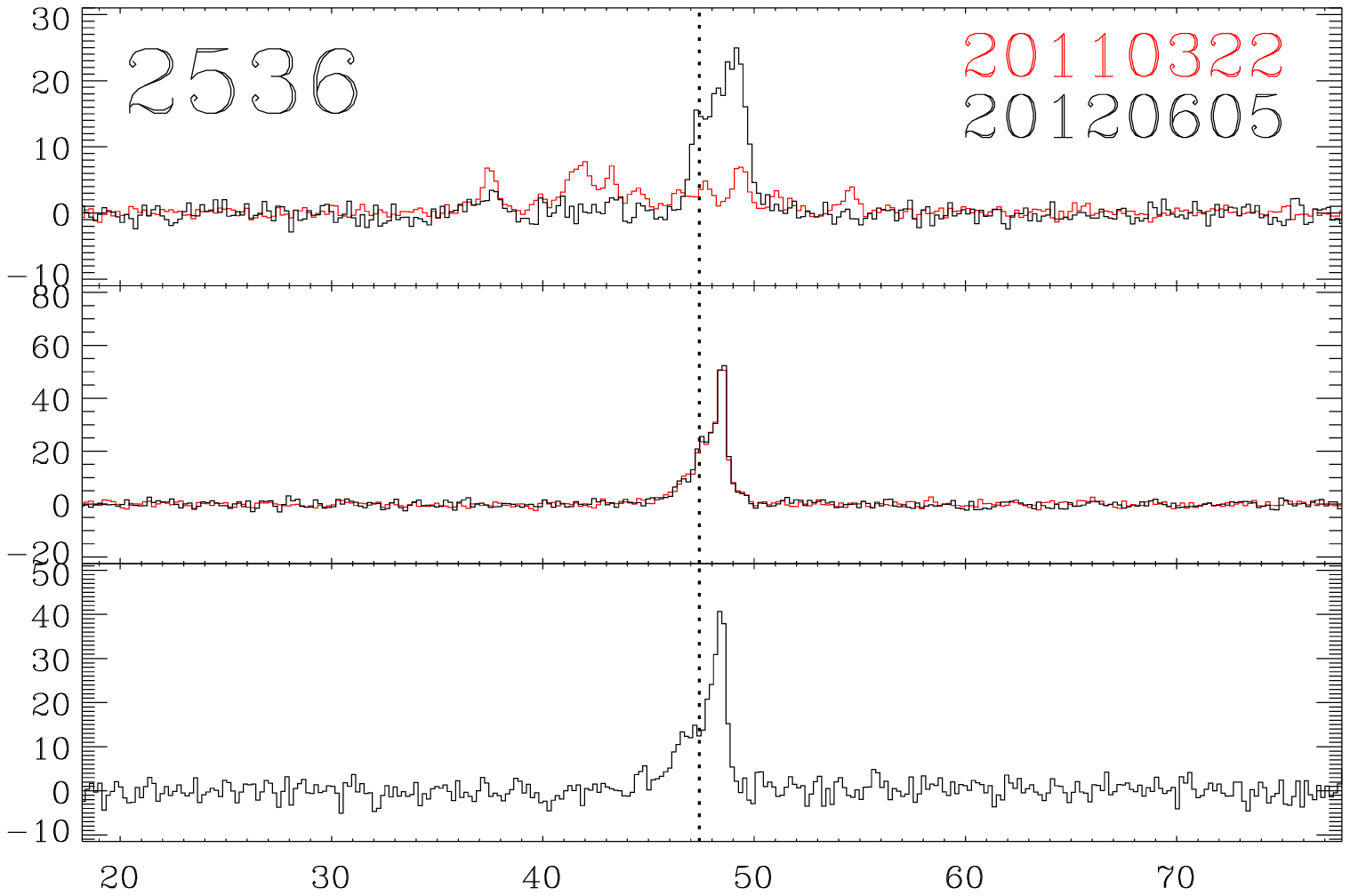} 
&
&
\includegraphics[width=50mm]{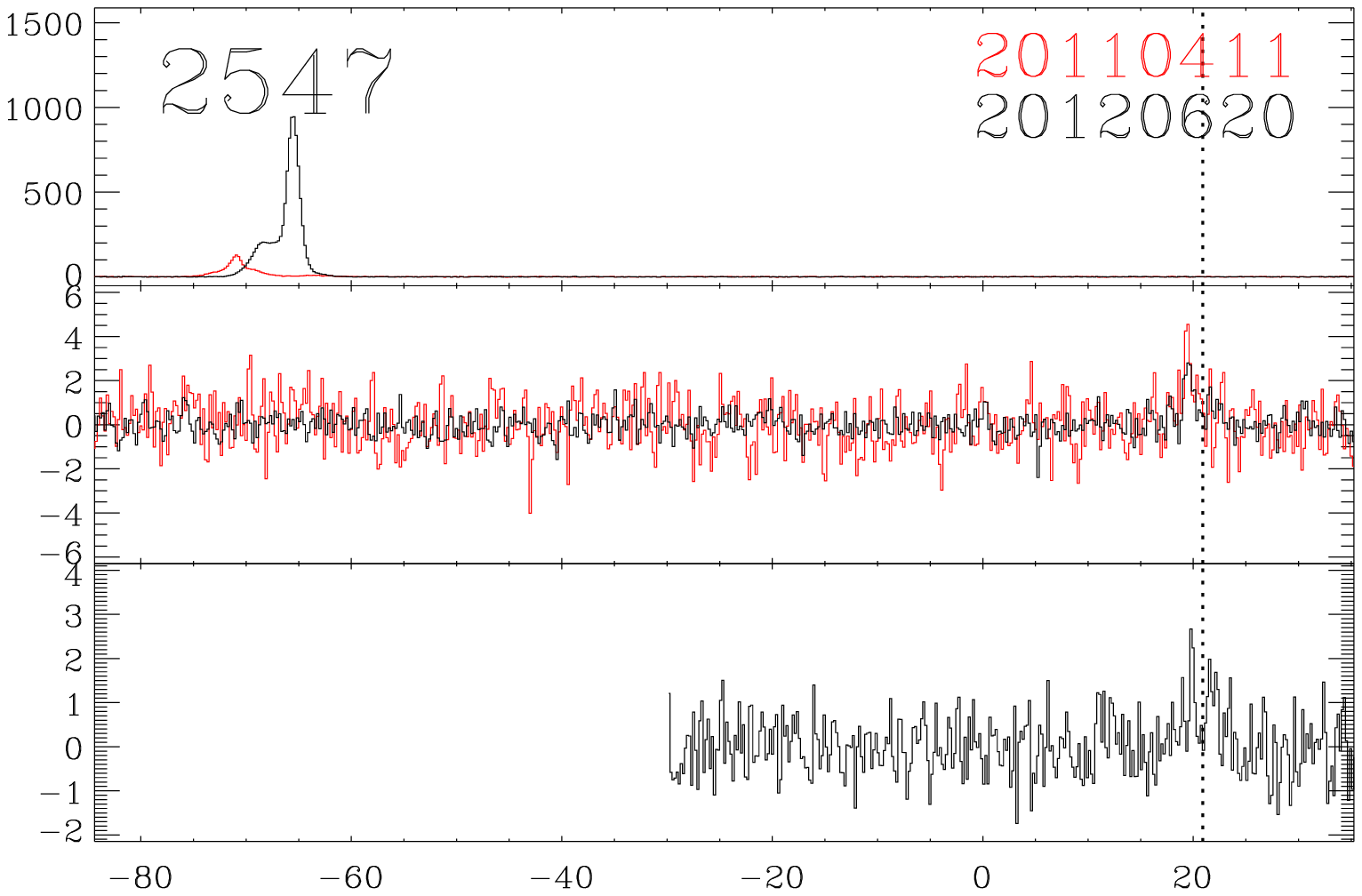}
\\
&
\includegraphics[width=50mm]{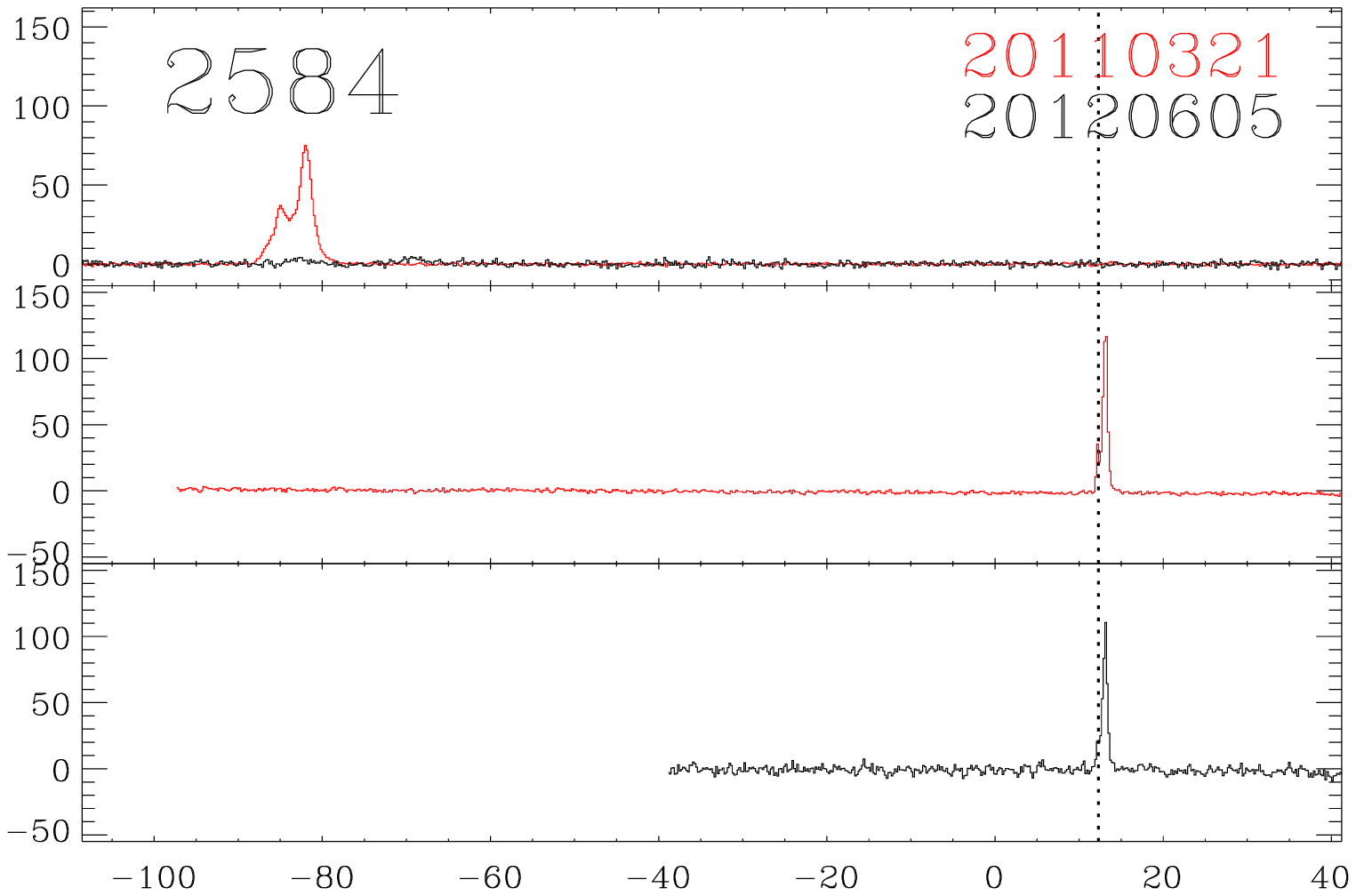} 
&
&  
\includegraphics[width=50mm]{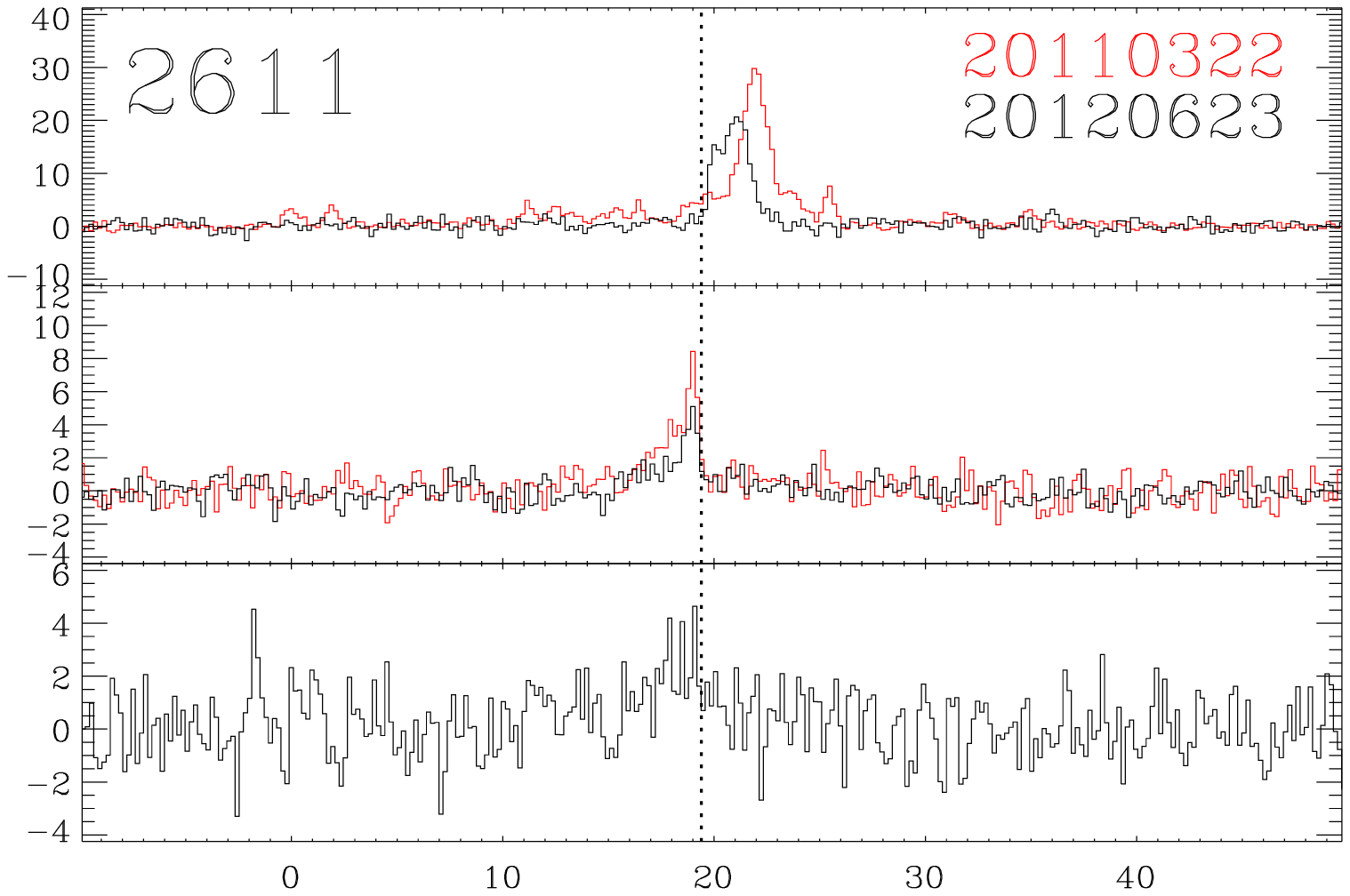} 
&
&
\includegraphics[width=50mm]{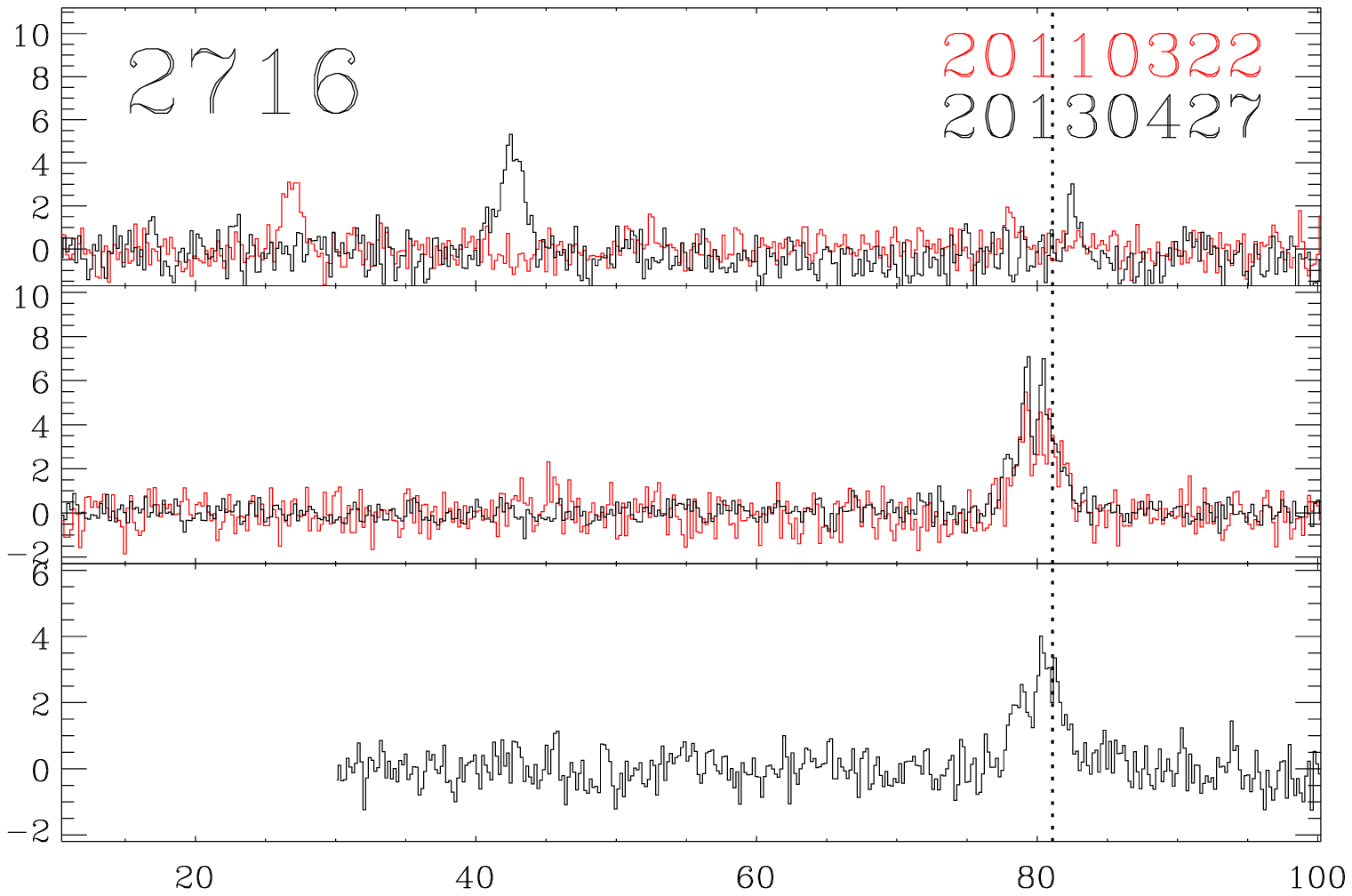}  
\end{tabular}
\begin{minipage}{\linewidth}
\center{LSR velocity ( \kms\ )}
\end{minipage}
\caption{(Upper panel) The detected 22 GHz \water, (Middle panel) 44 GHz \methanol, and (Lower panel) 95 GHz \methanol\ maser spectra of the sources detected in all of 3 transitions.  Red and Black colors represent the first and second epochs, respectively. 
The source name is given in the top-left corner of the upper panel and `M' is attached for the source associated with 6.7 GHz \methanol\ maser emission (see Section~3.1). The observing dates are shown in the top-right corner. The vertical dotted line indicates the systemic velocity, which was mostly determined by the NH$_3$ line observations (Table~\ref{tbl_source}).
\label{fig_224495}}
\end{figure}
\clearpage

\begin{figure}
\figurenum{2}
\epsscale{1}
\begin{minipage}{0.1cm}
\rotatebox{90}{\textcolor{black}{Flux density (Jy)}}
\end{minipage}
\begin{tabular}{cccccc}
&
\includegraphics[width=50mm]{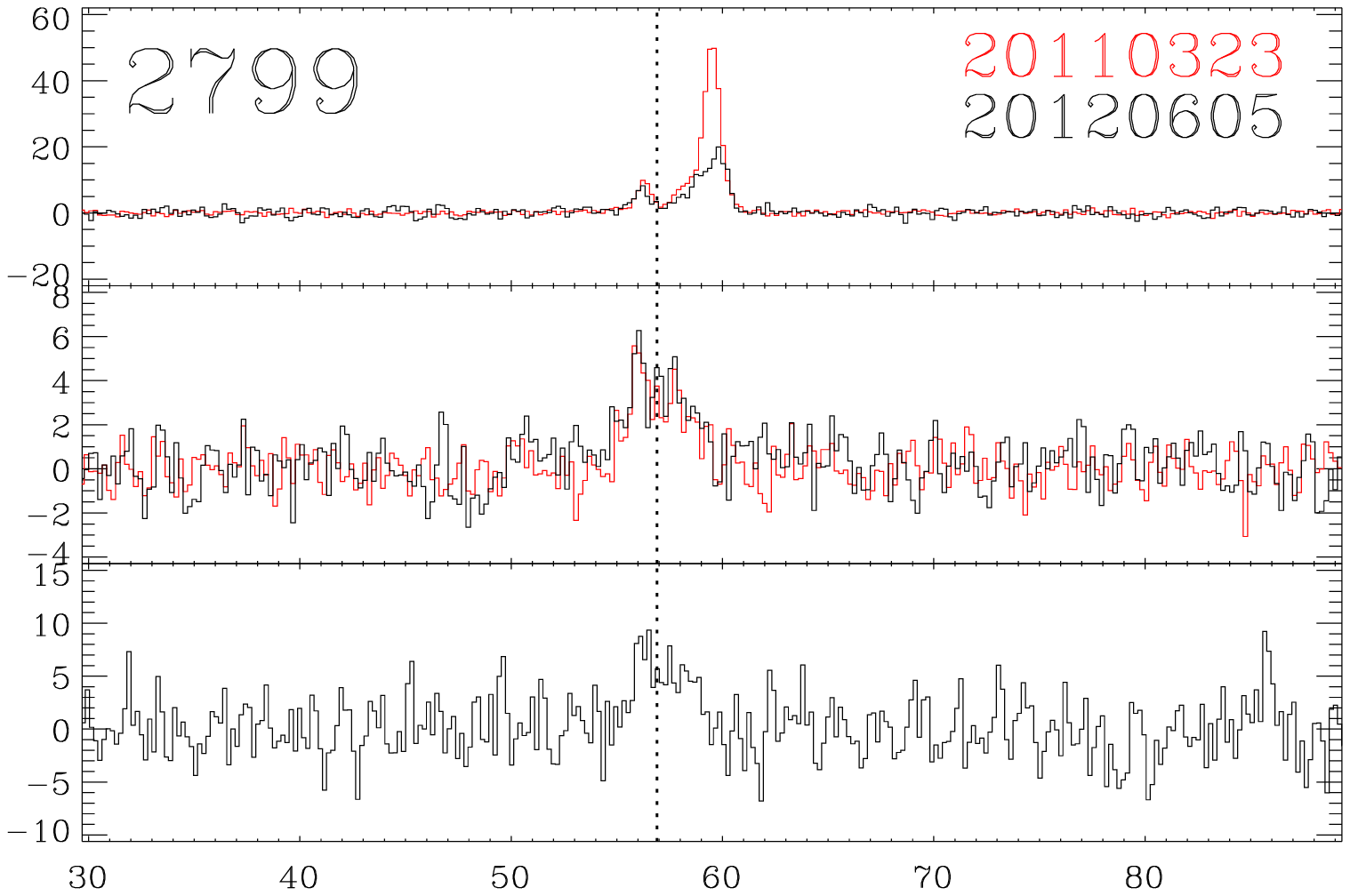} 
&
&
\includegraphics[width=50mm]{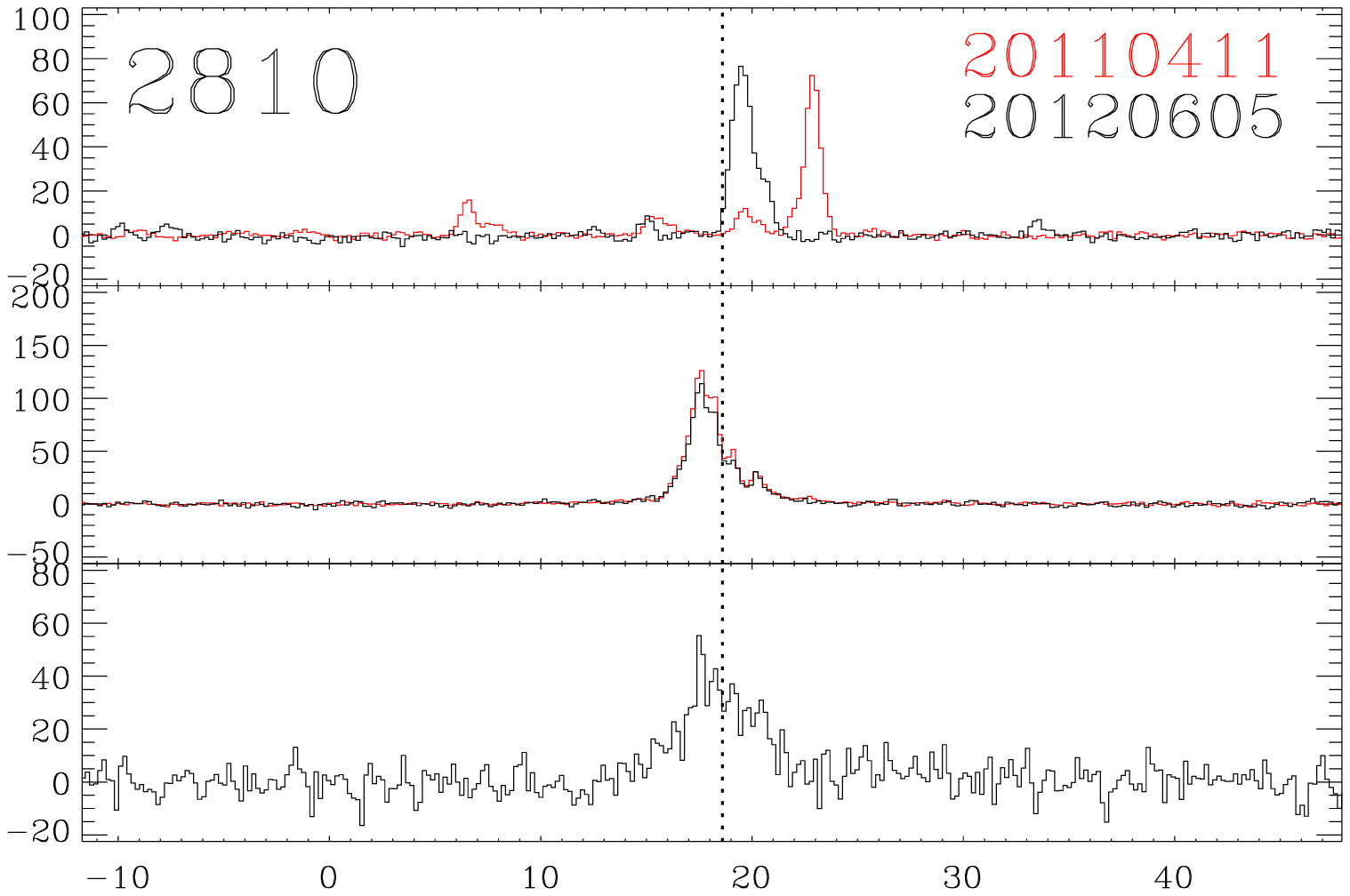} 
&
& 
\includegraphics[width=50mm]{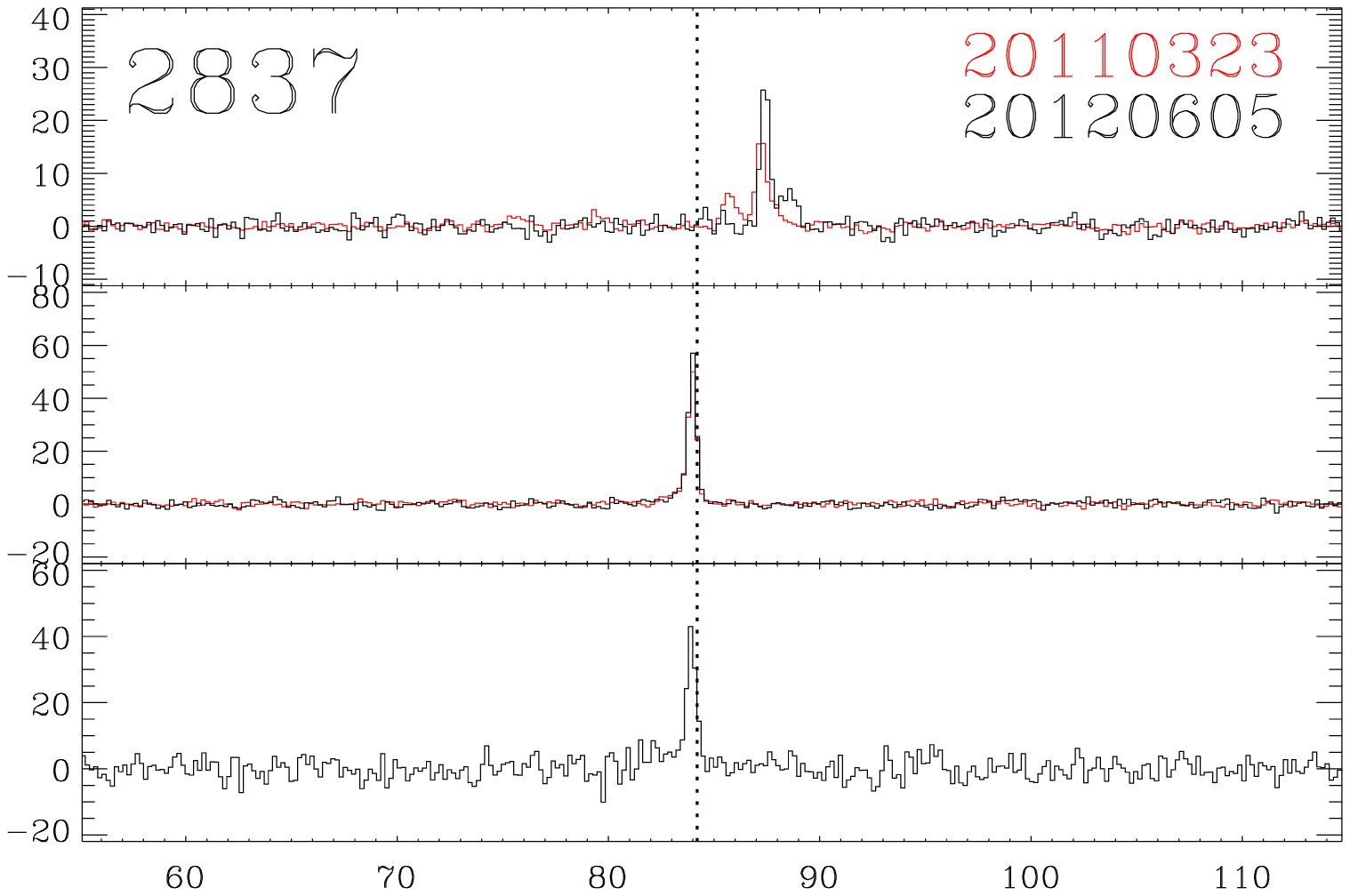} 
\\
&
\includegraphics[width=50mm]{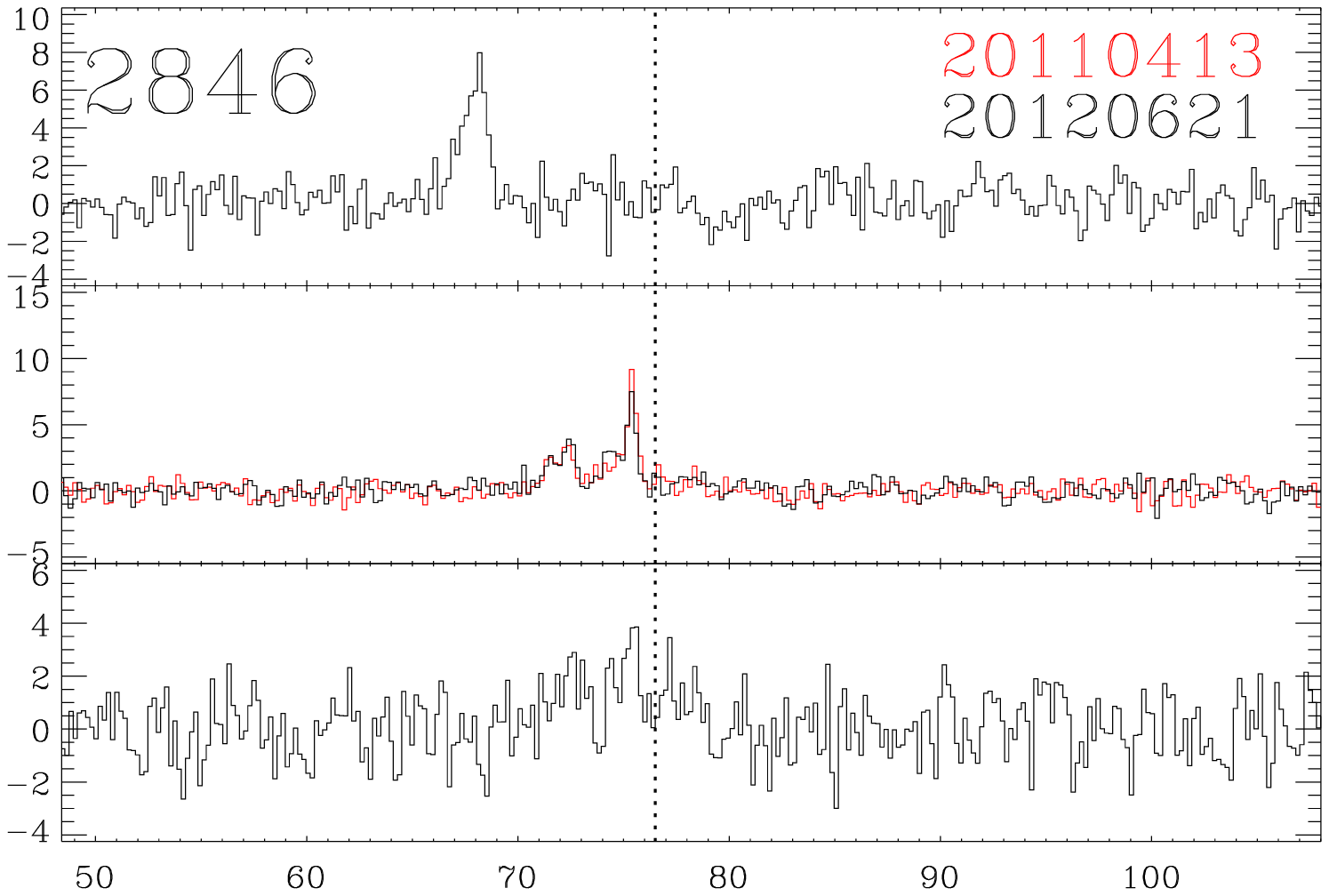} 
&
& 
\includegraphics[width=50mm]{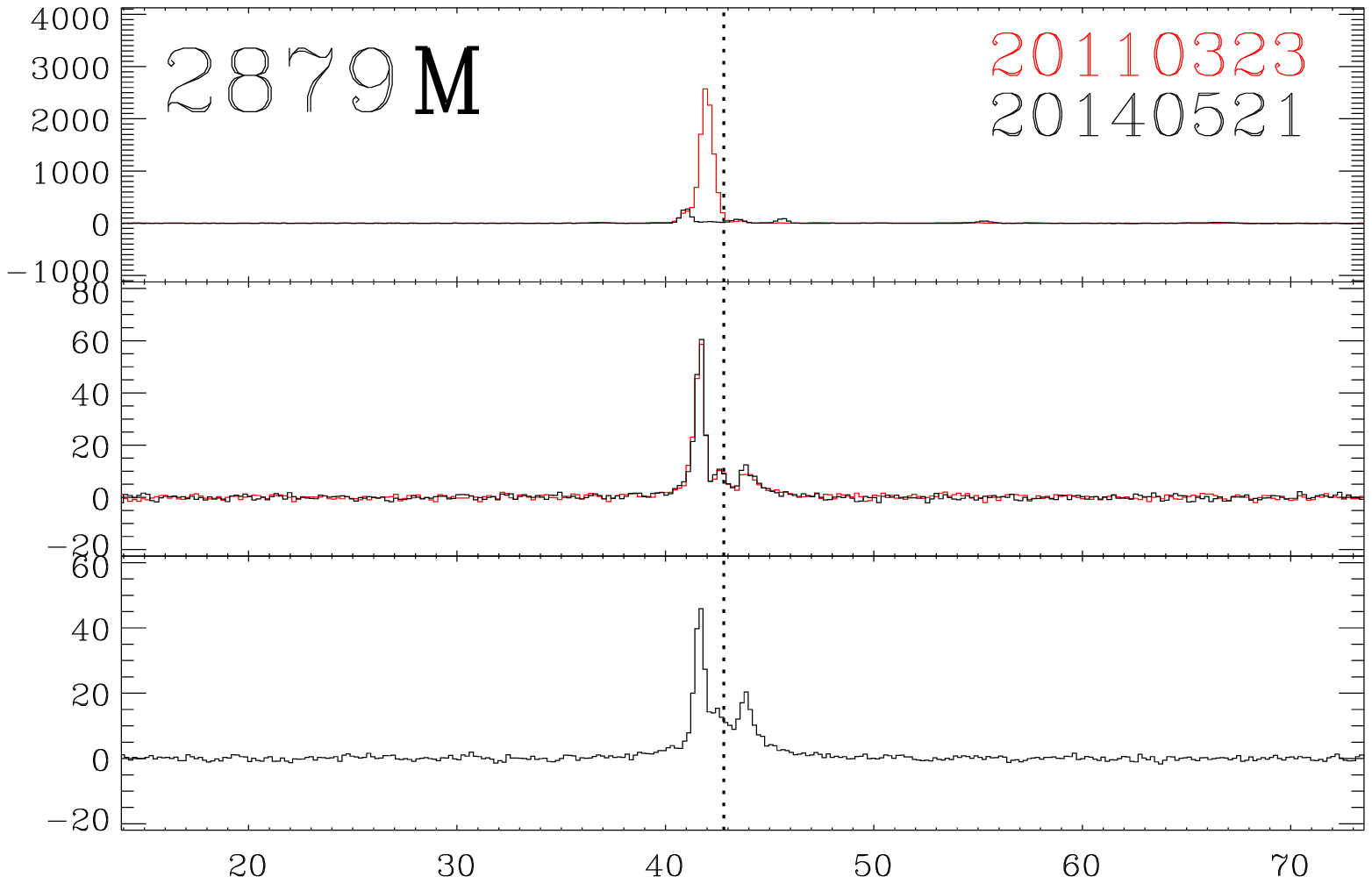} 
&
&
\includegraphics[width=50mm]{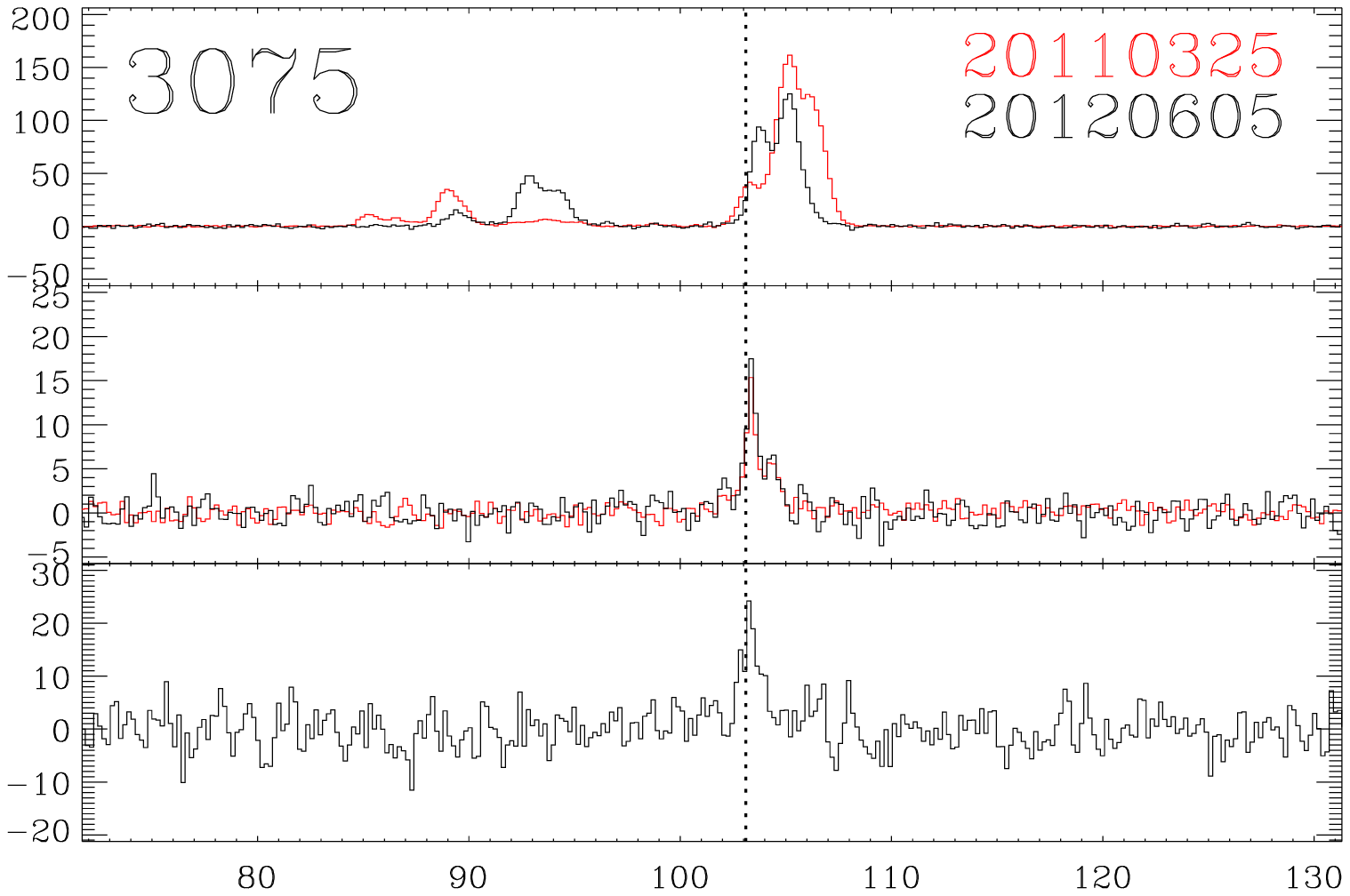} 
\\
&
\includegraphics[width=50mm]{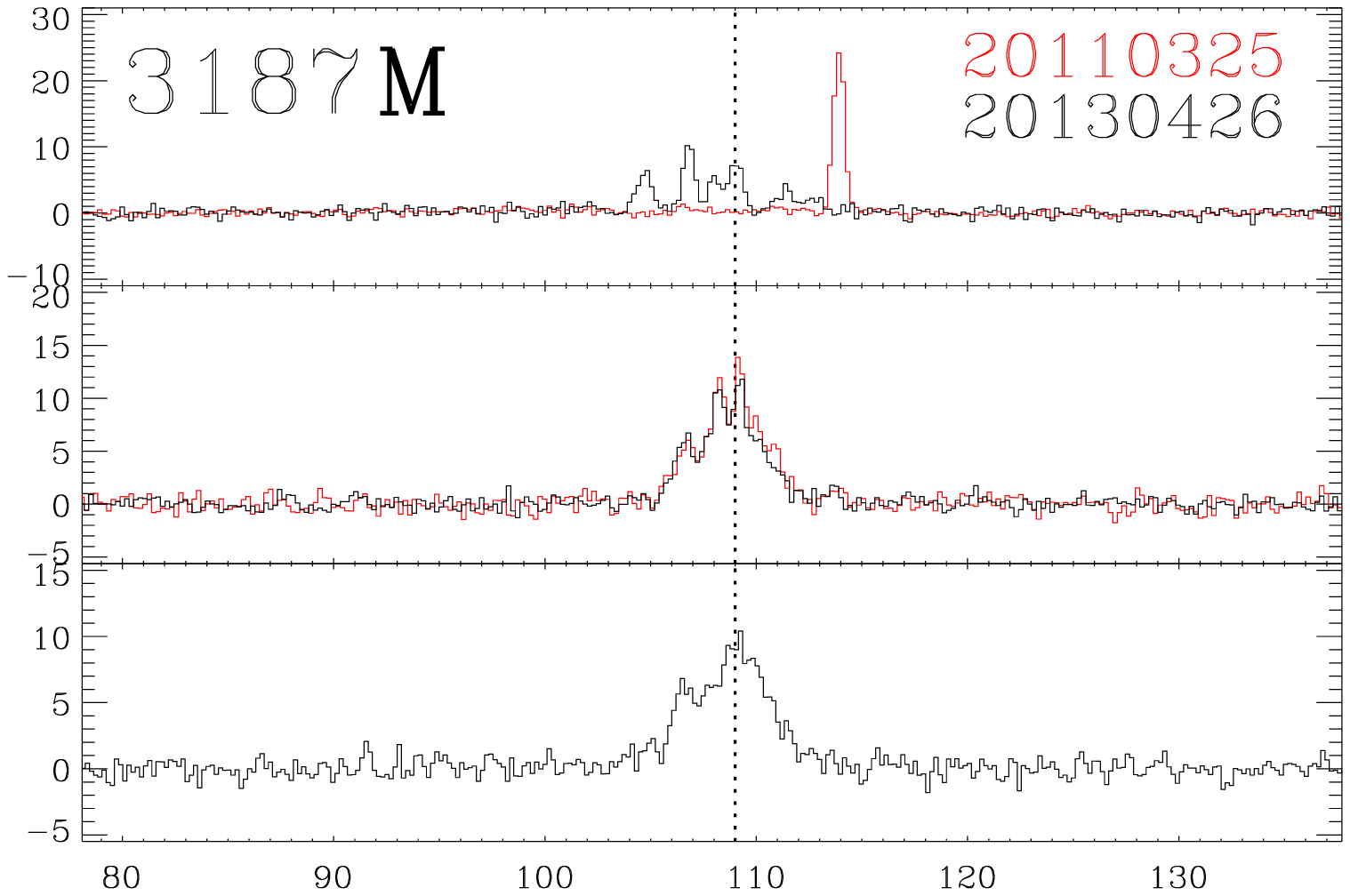} 
&
&  
\includegraphics[width=50mm]{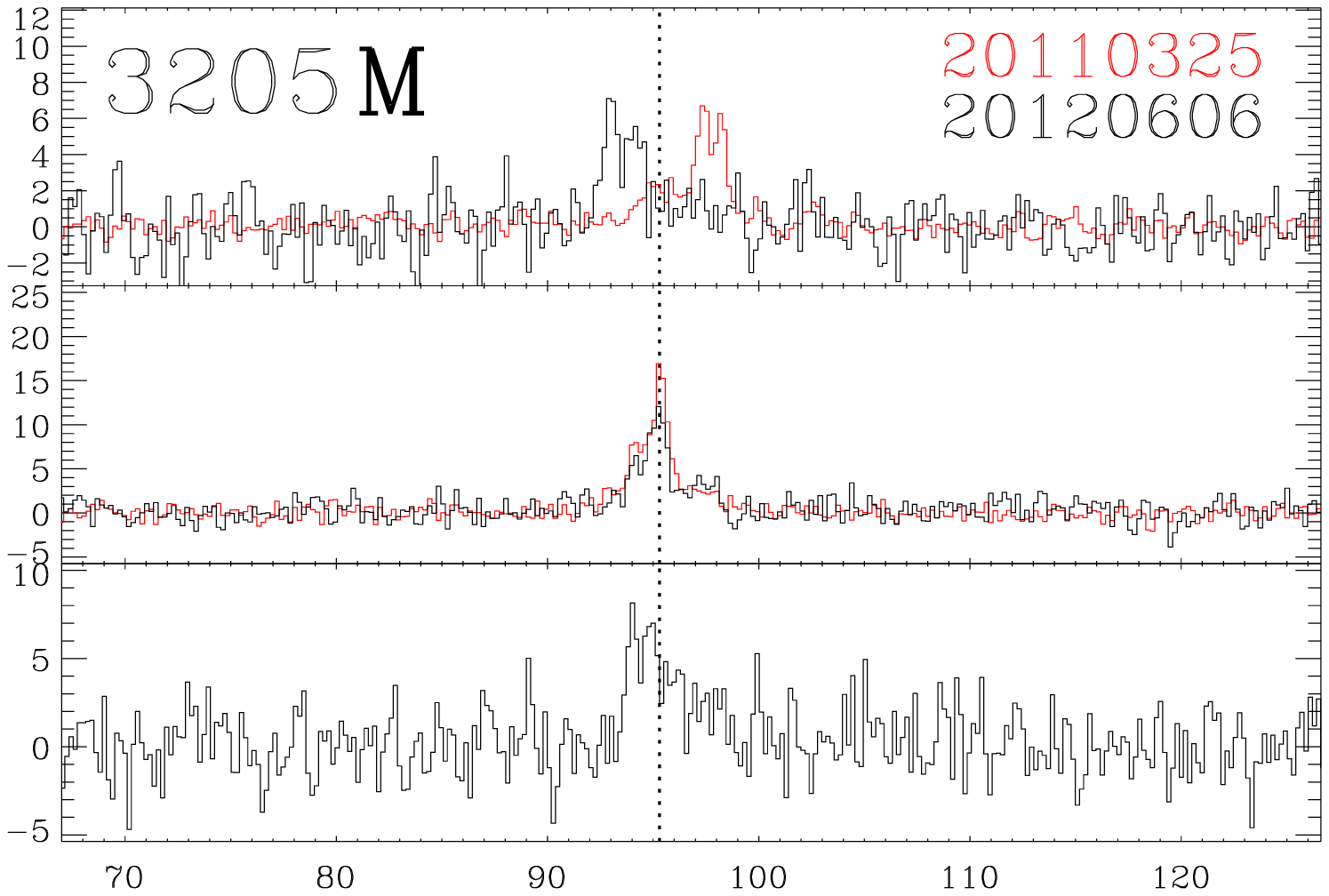} 
&
&
\includegraphics[width=50mm]{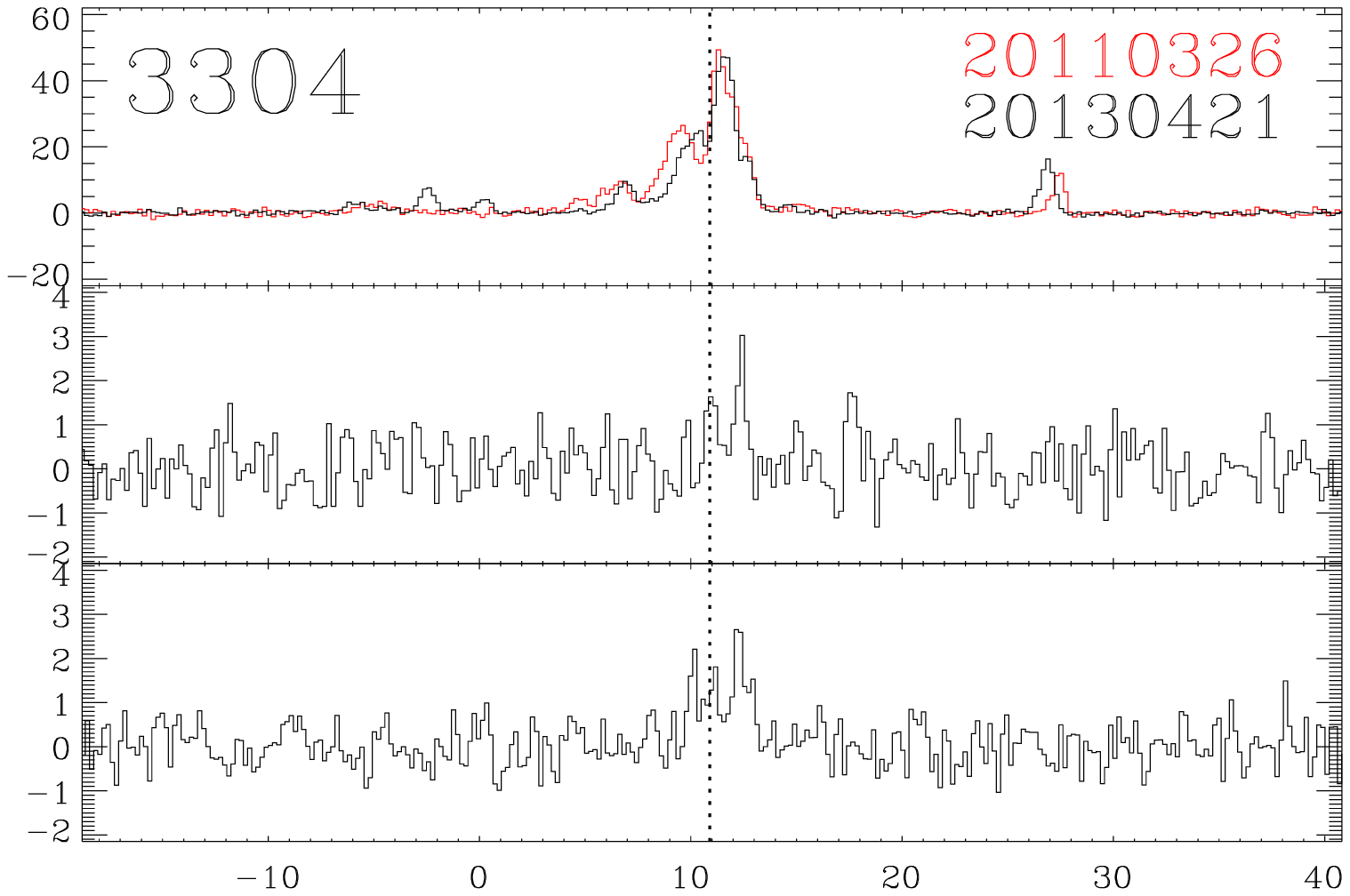} 
\\
&
\includegraphics[width=50mm]{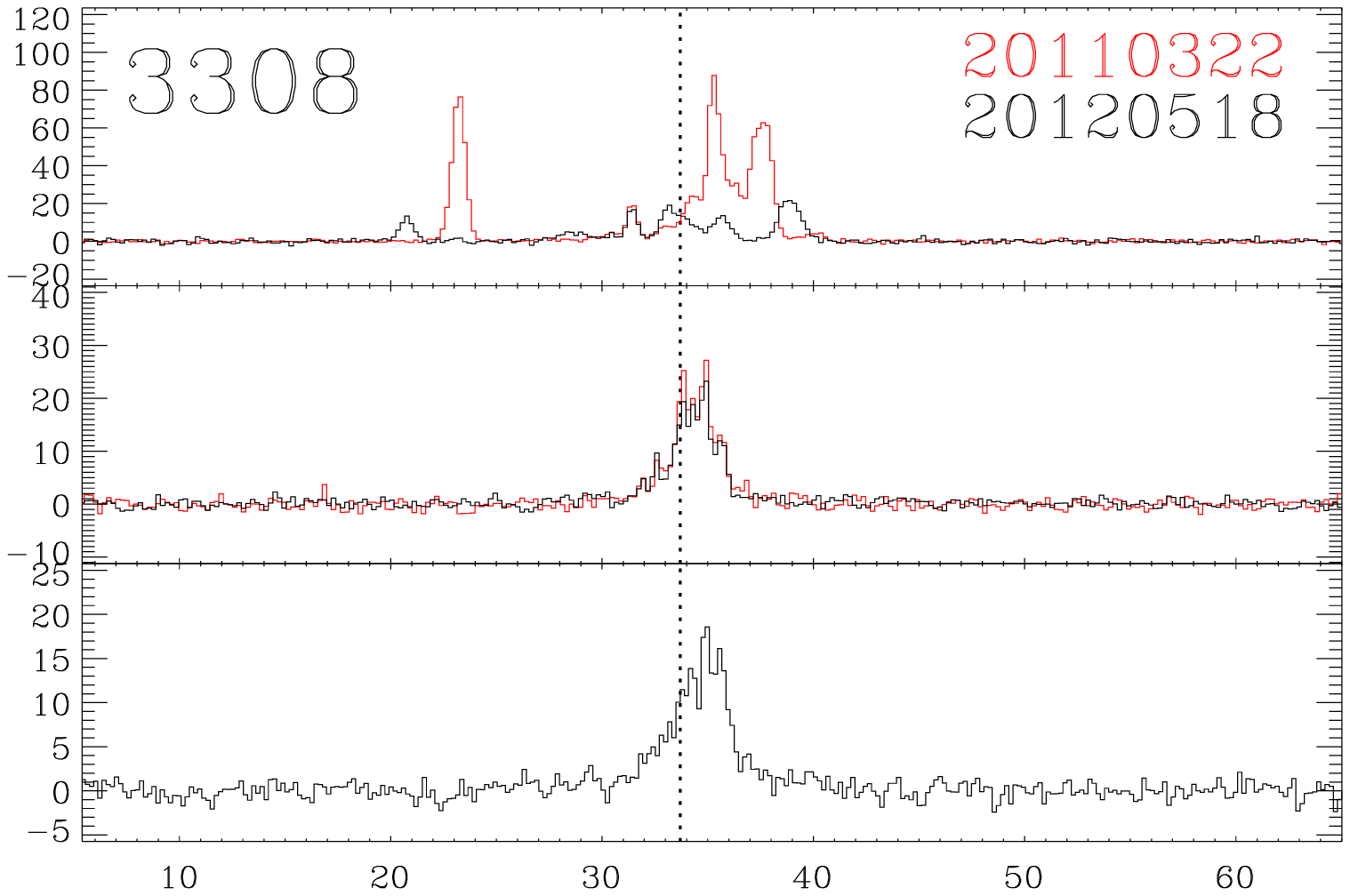} 
&
&  
\includegraphics[width=50mm]{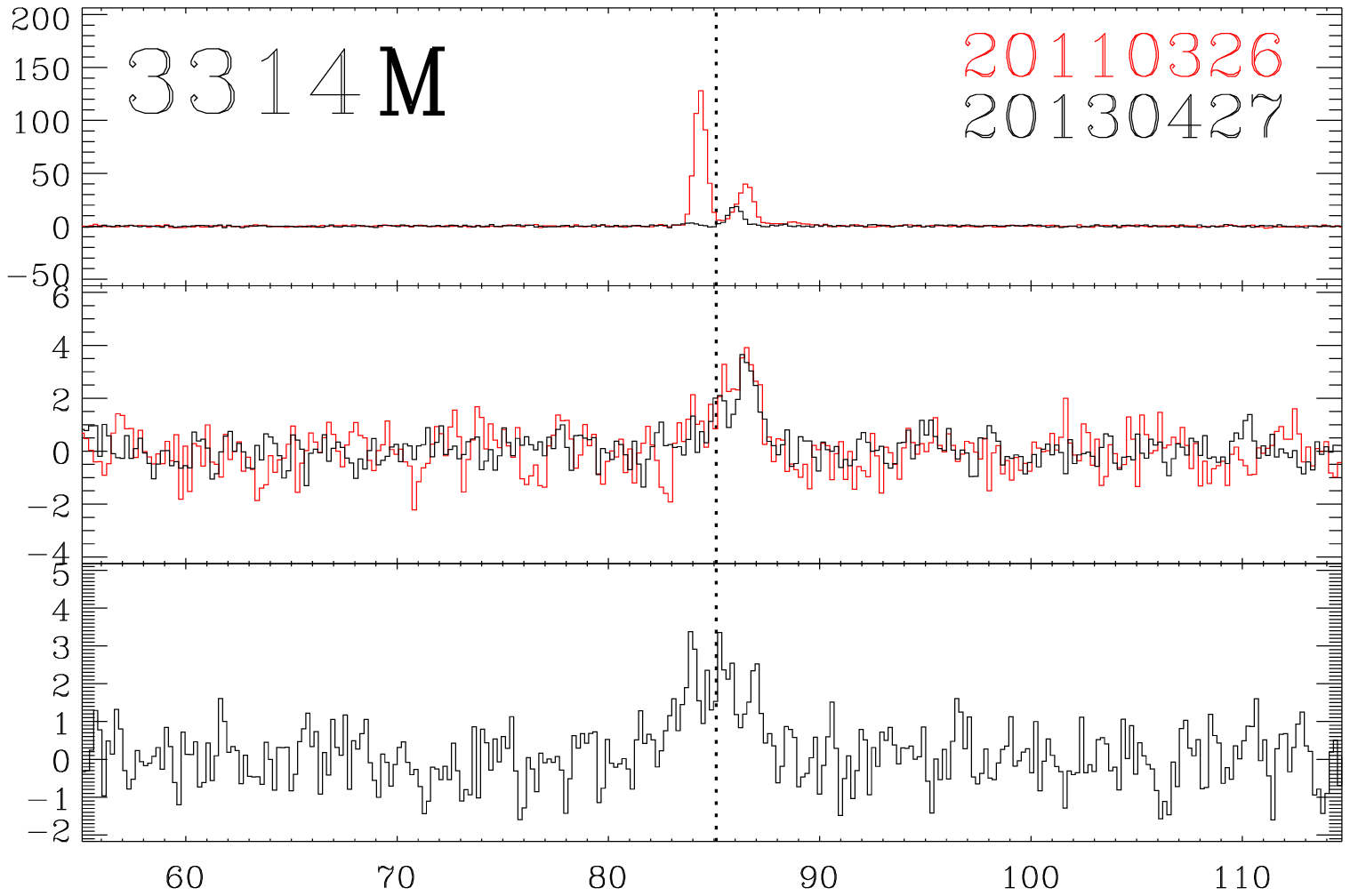} 
&
&
\includegraphics[width=50mm]{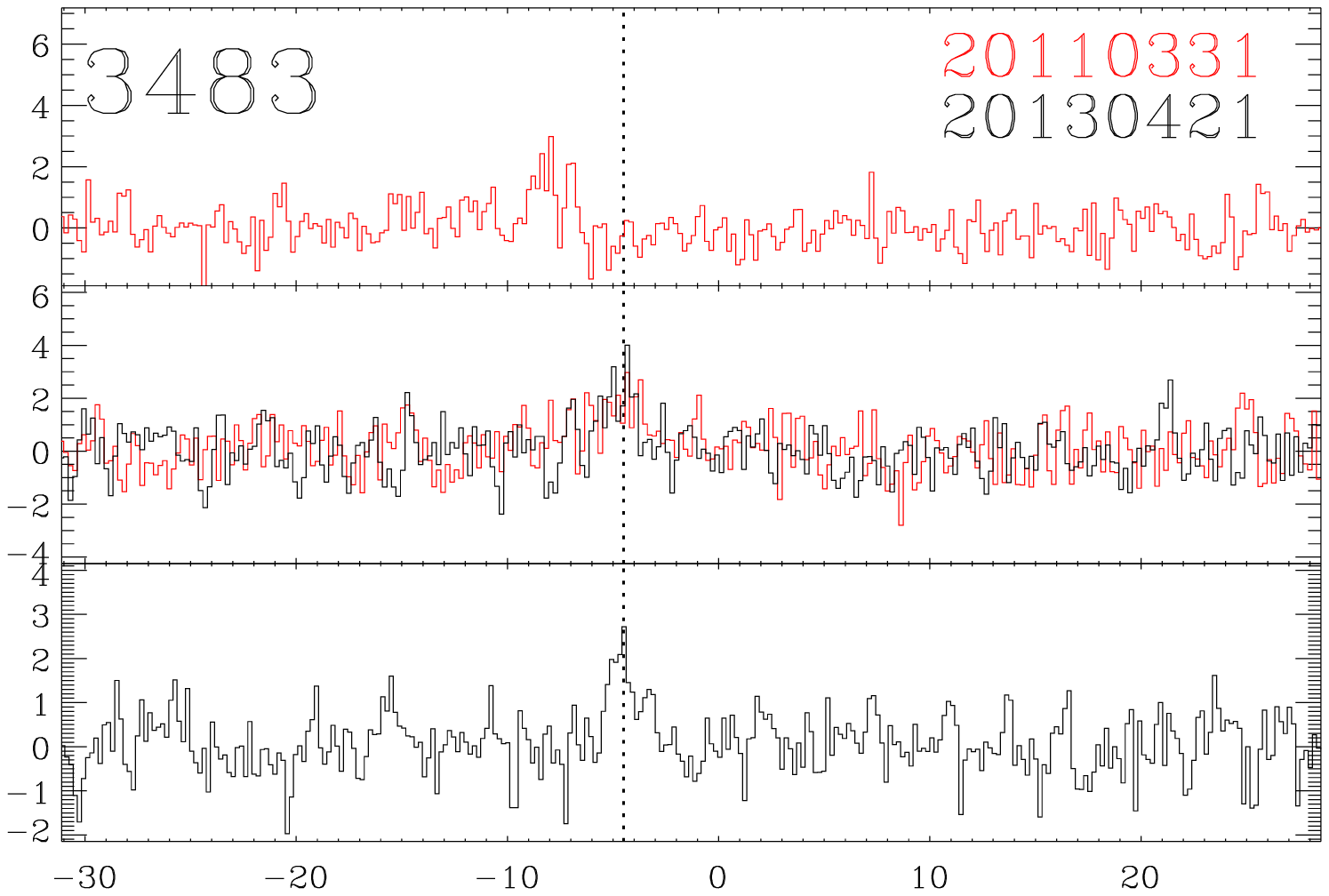} 
\\
&
\includegraphics[width=50mm]{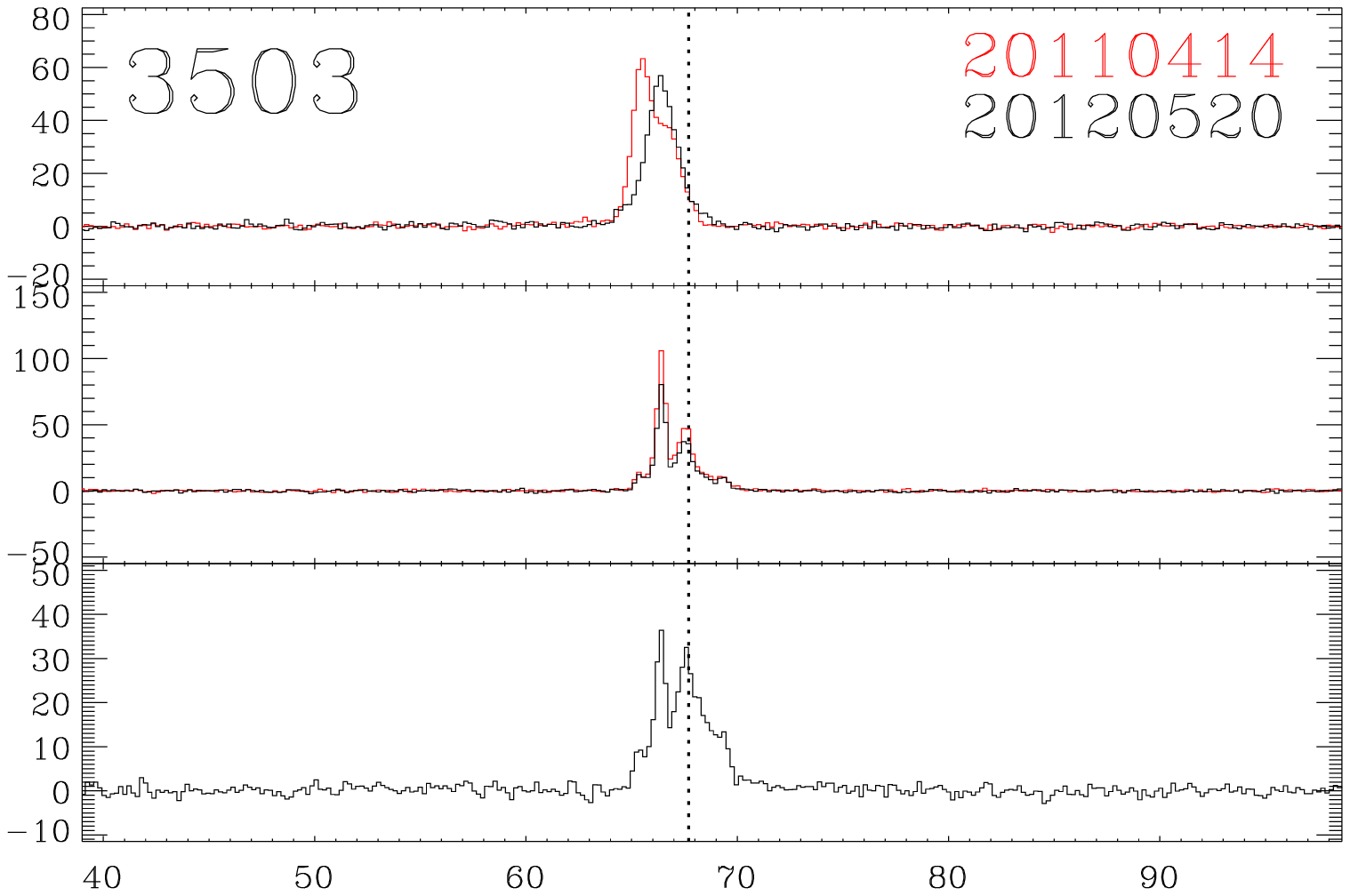} 
&
&  
\includegraphics[width=50mm]{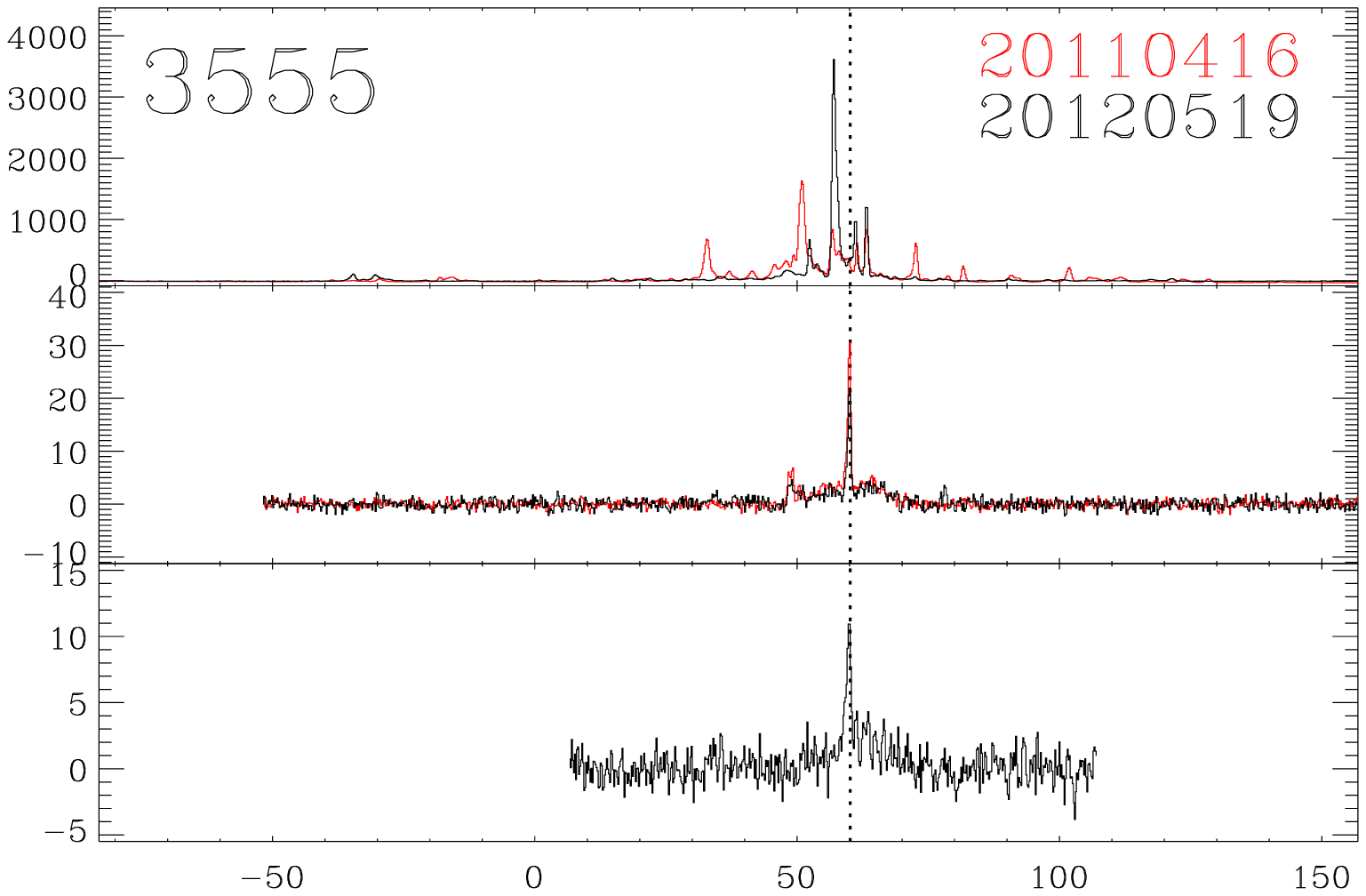} 
&
&
\includegraphics[width=50mm]{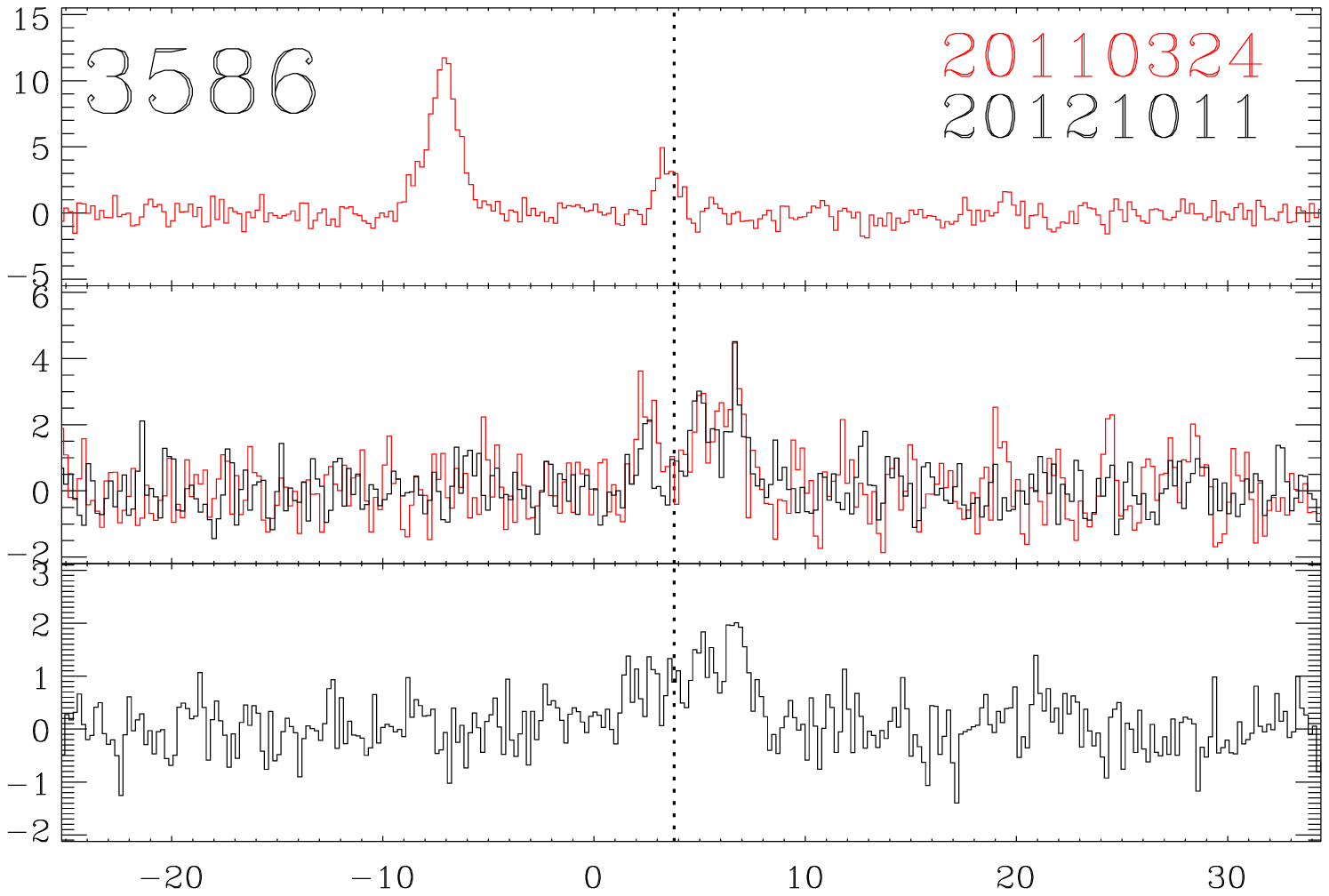} 
\\
&\multicolumn{5}{c}{LSR velocity (\kms)} \\
\end{tabular}
\caption{Continued}
\end{figure}
\clearpage

\begin{figure}
\figurenum{2}
\epsscale{1}
\begin{minipage}{0.1cm}
\rotatebox{90}{\textcolor{black}{Flux density (Jy)}}
\end{minipage}
\begin{tabular}{cccccc}
&
\includegraphics[width=50mm]{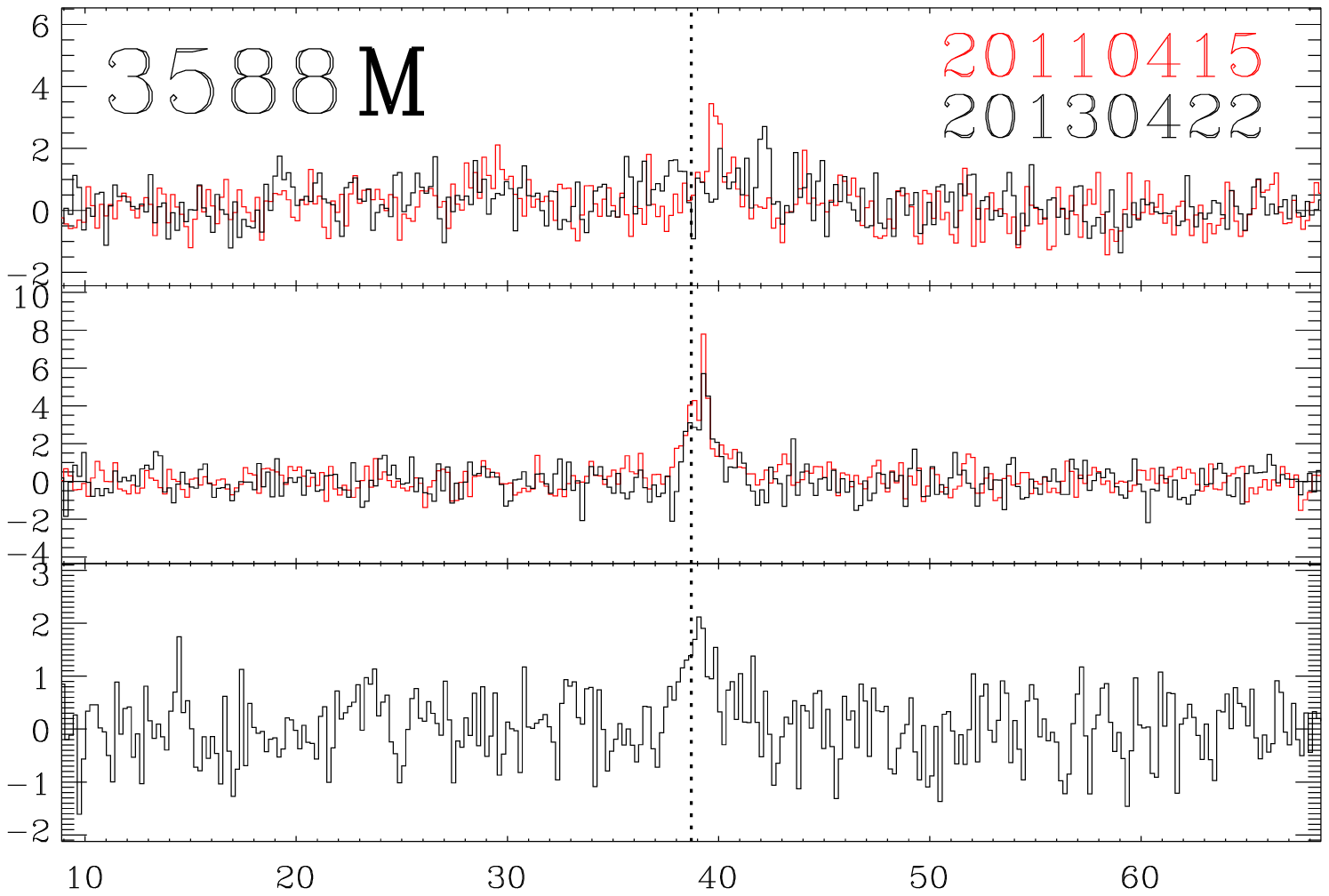} 
&
&
\includegraphics[width=50mm]{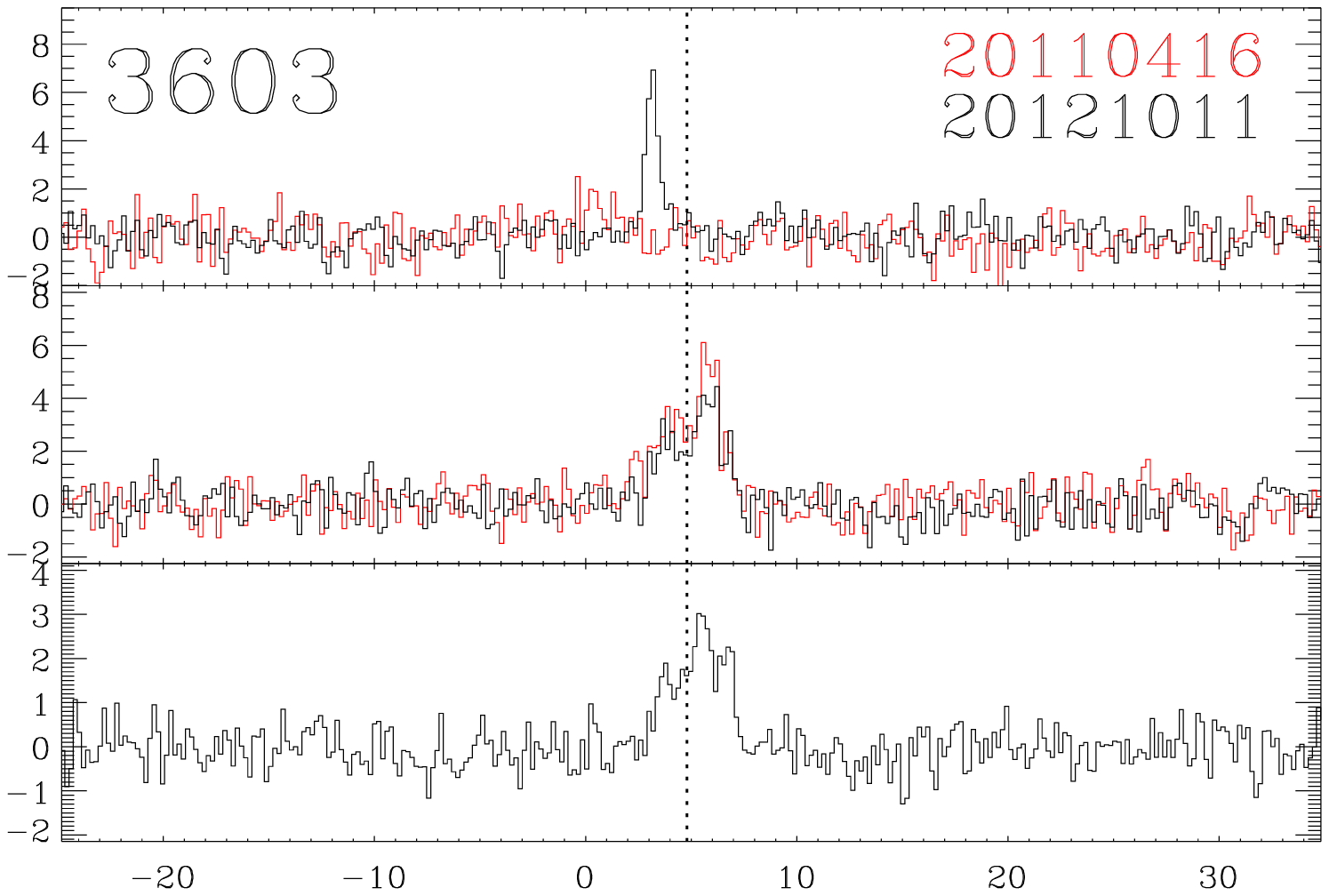} 
&
& 
\includegraphics[width=50mm]{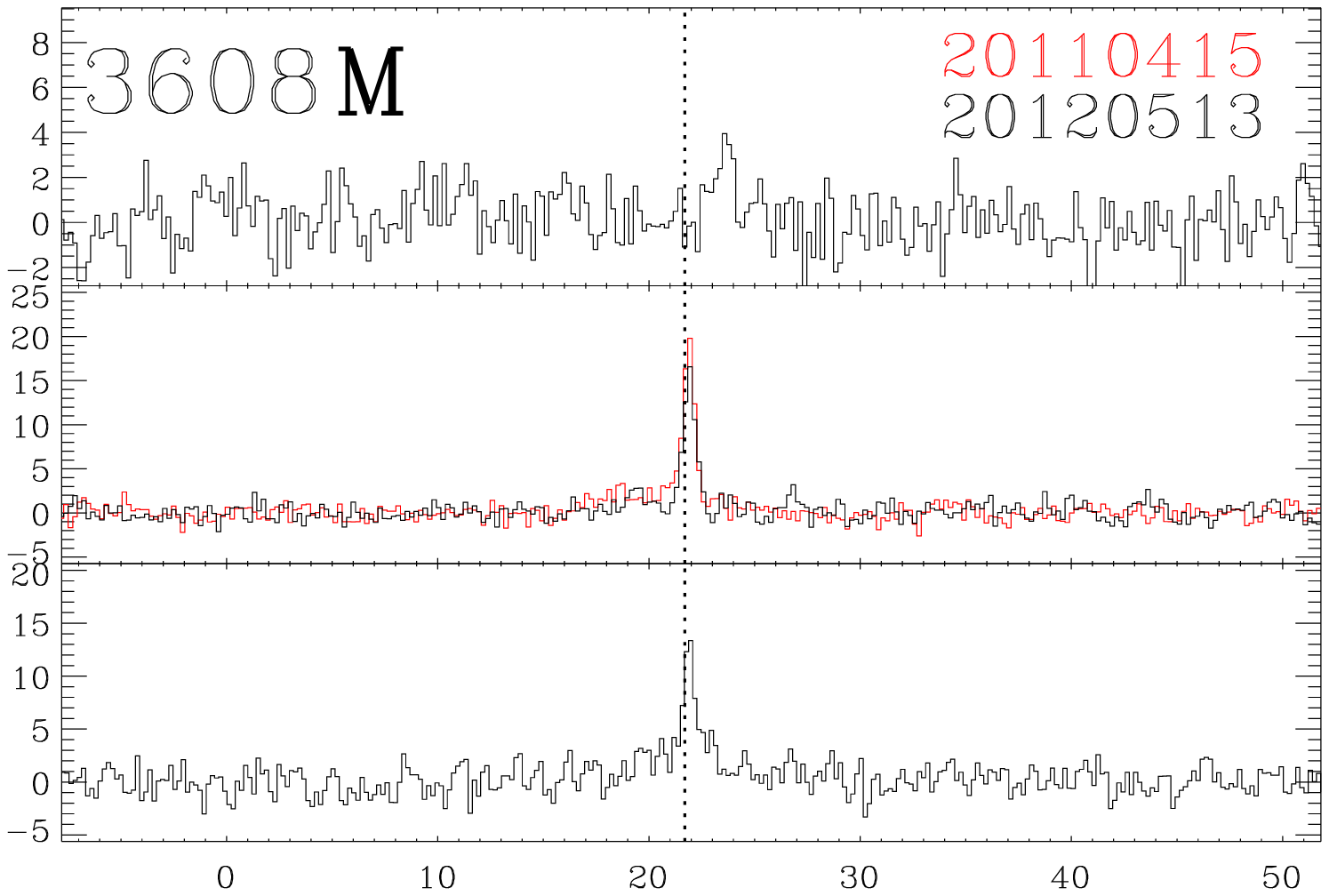} 
\\
&
\includegraphics[width=50mm]{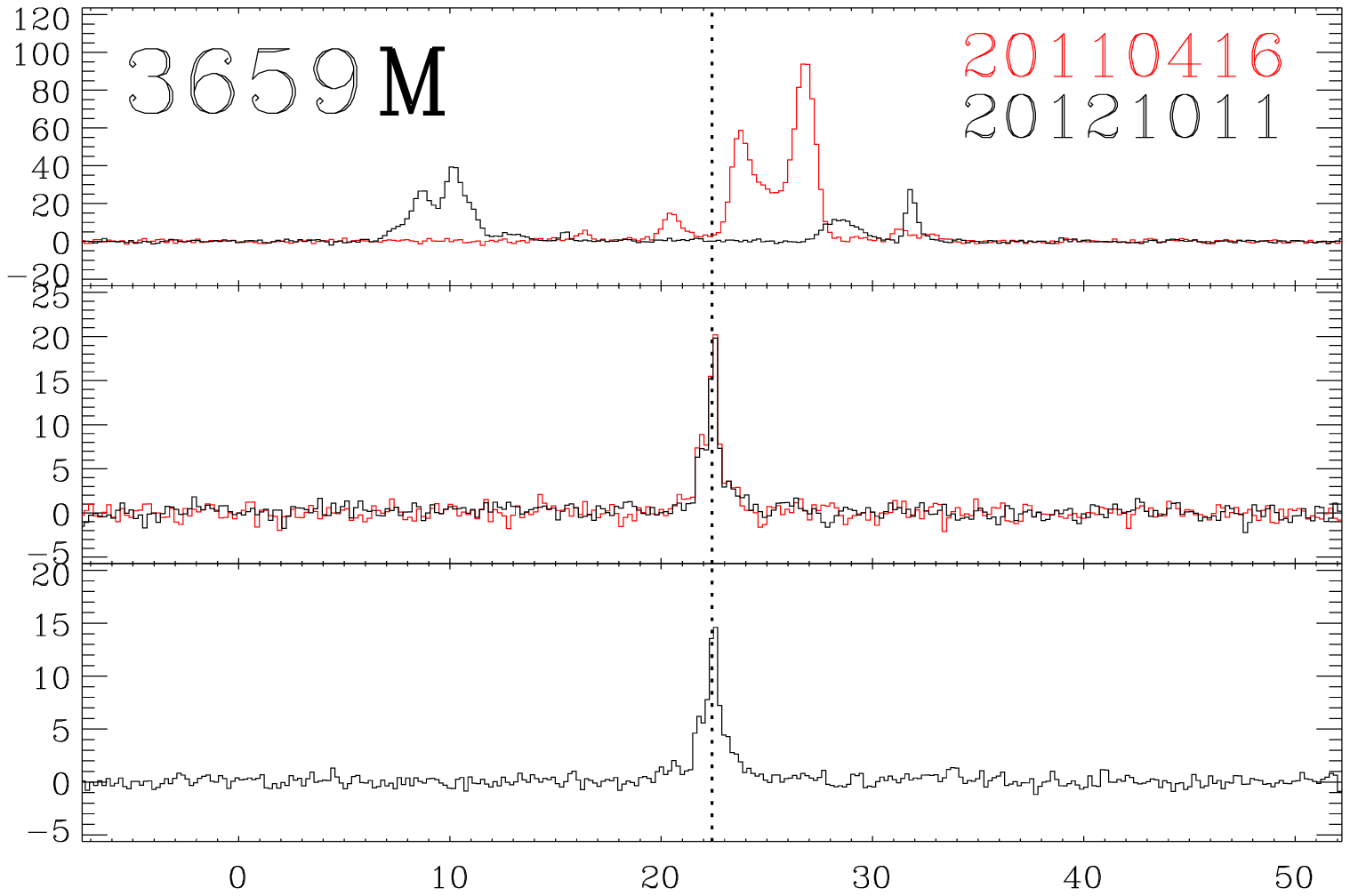} 
&
& 
\includegraphics[width=50mm]{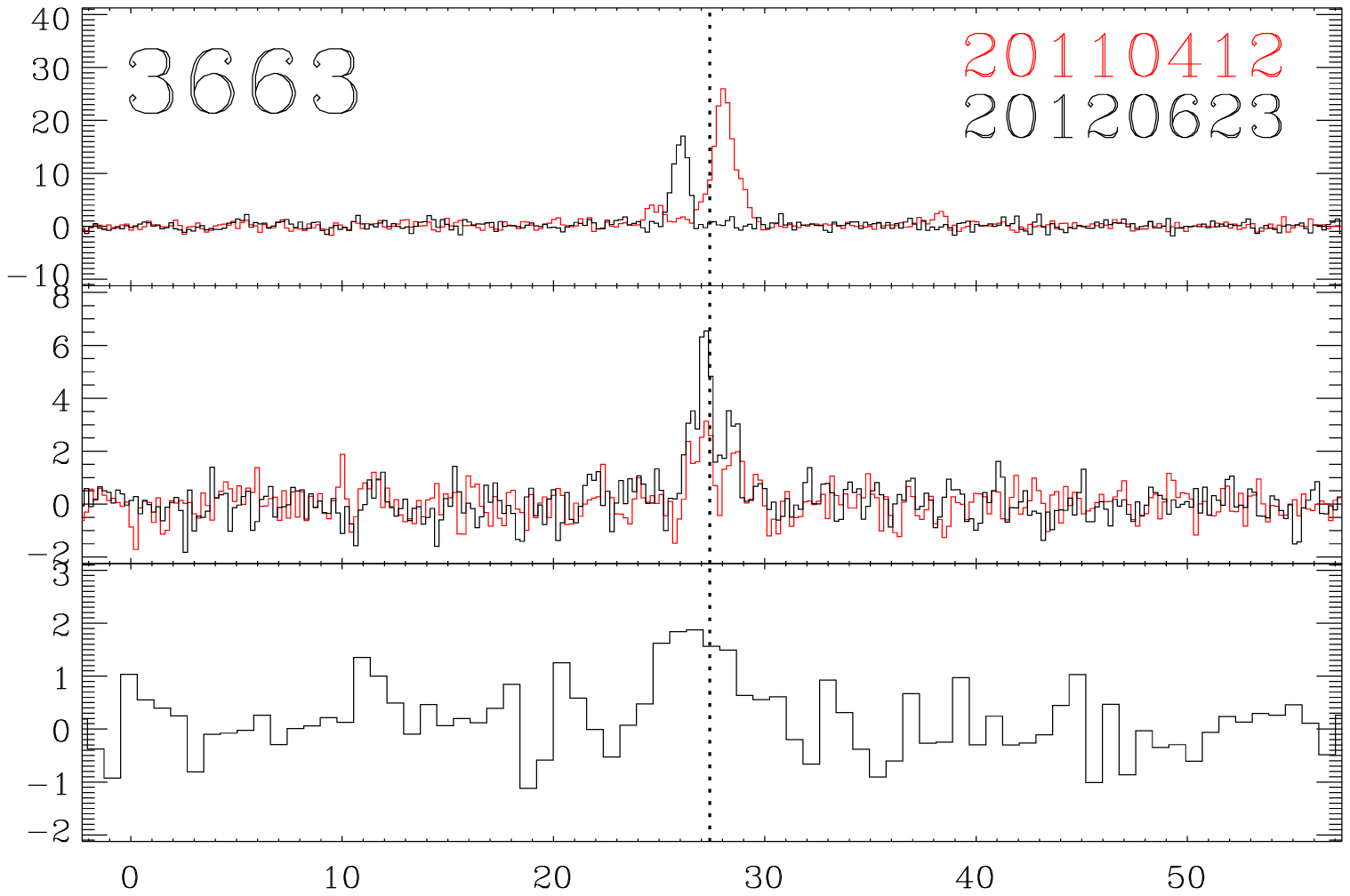} 
&
&
\includegraphics[width=50mm]{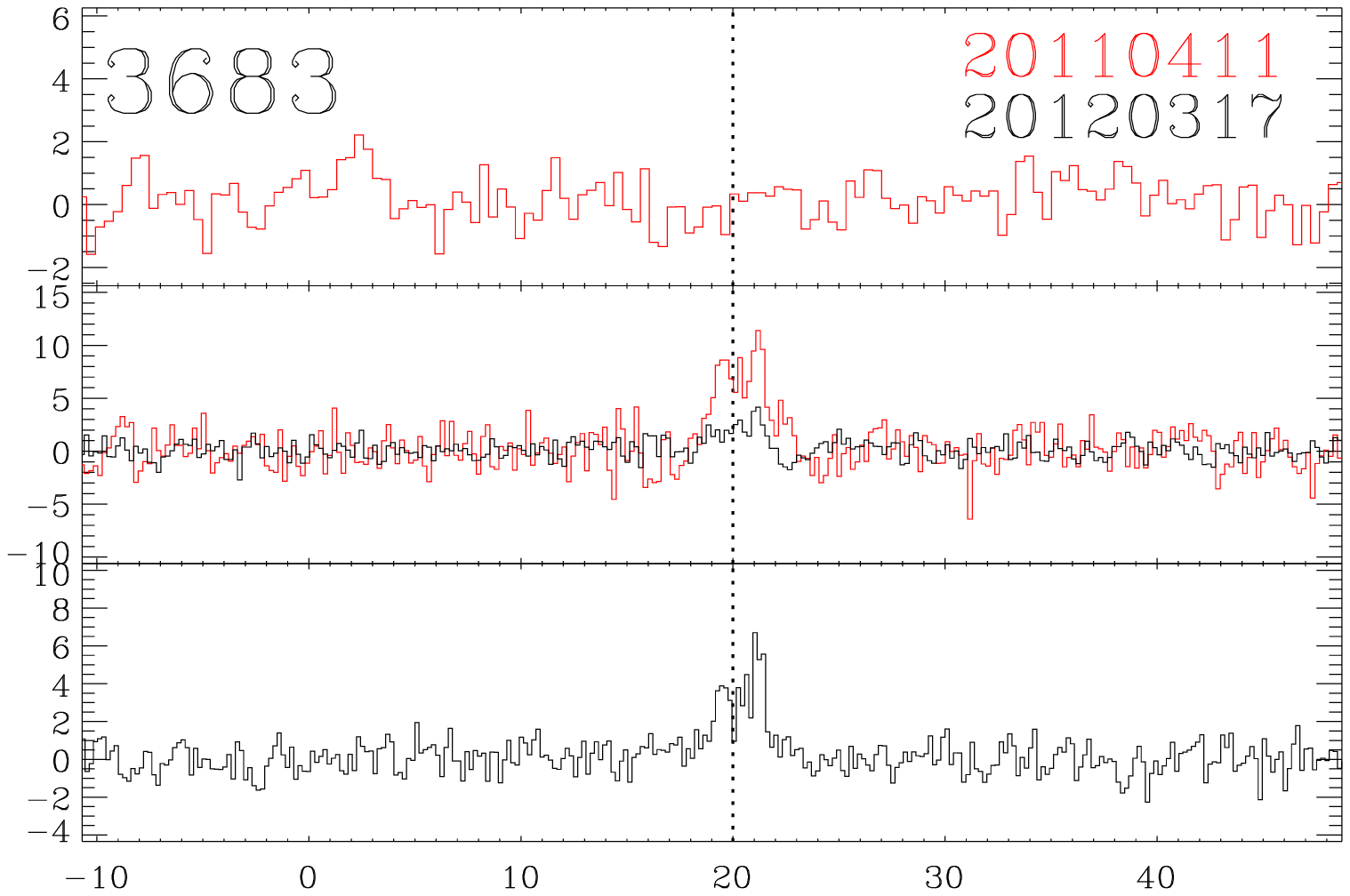} 
\\
&
\includegraphics[width=50mm]{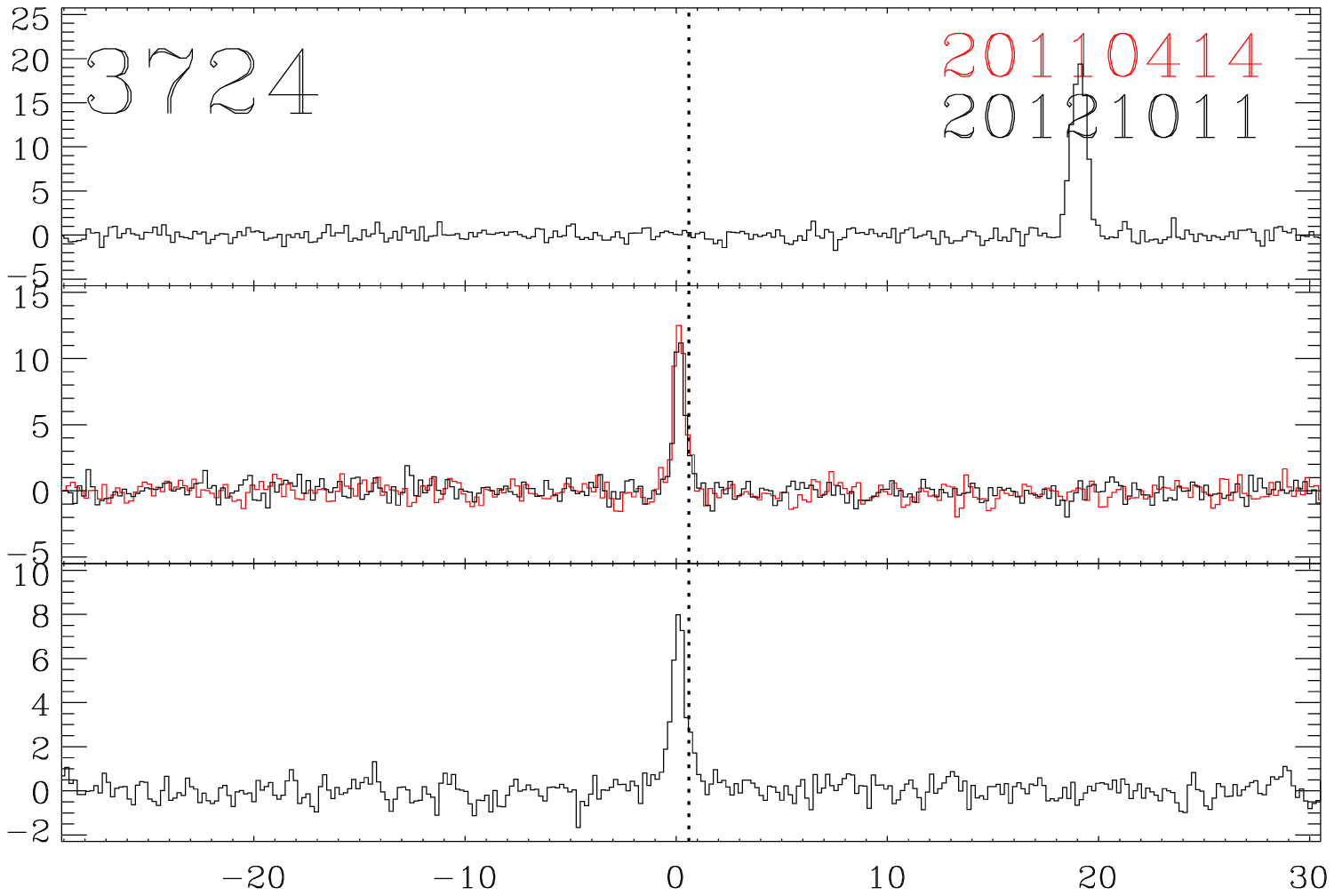} 
&
&  
\includegraphics[width=50mm]{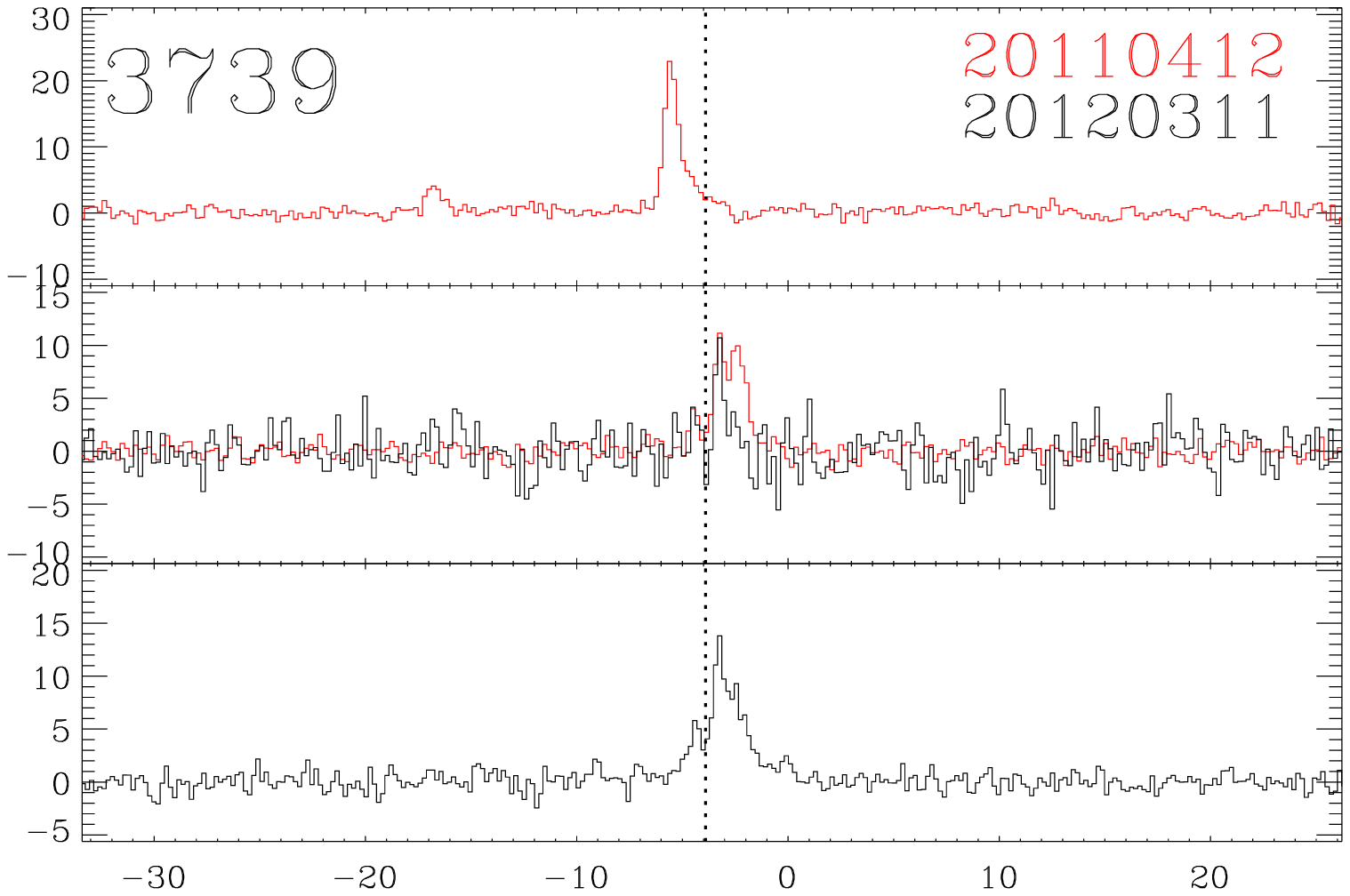} 
&
&
\includegraphics[width=50mm]{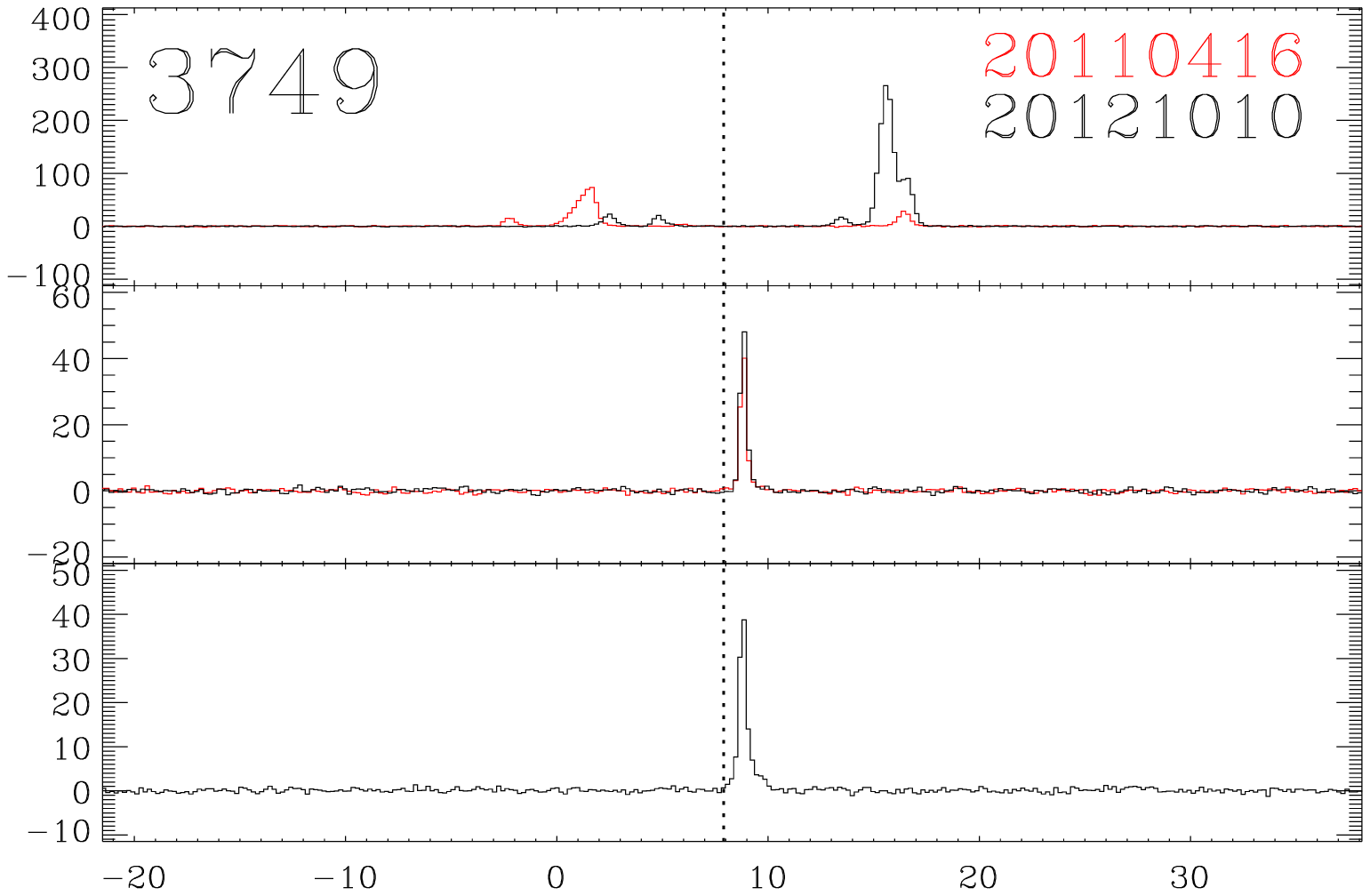} 
\\
&
\includegraphics[width=50mm]{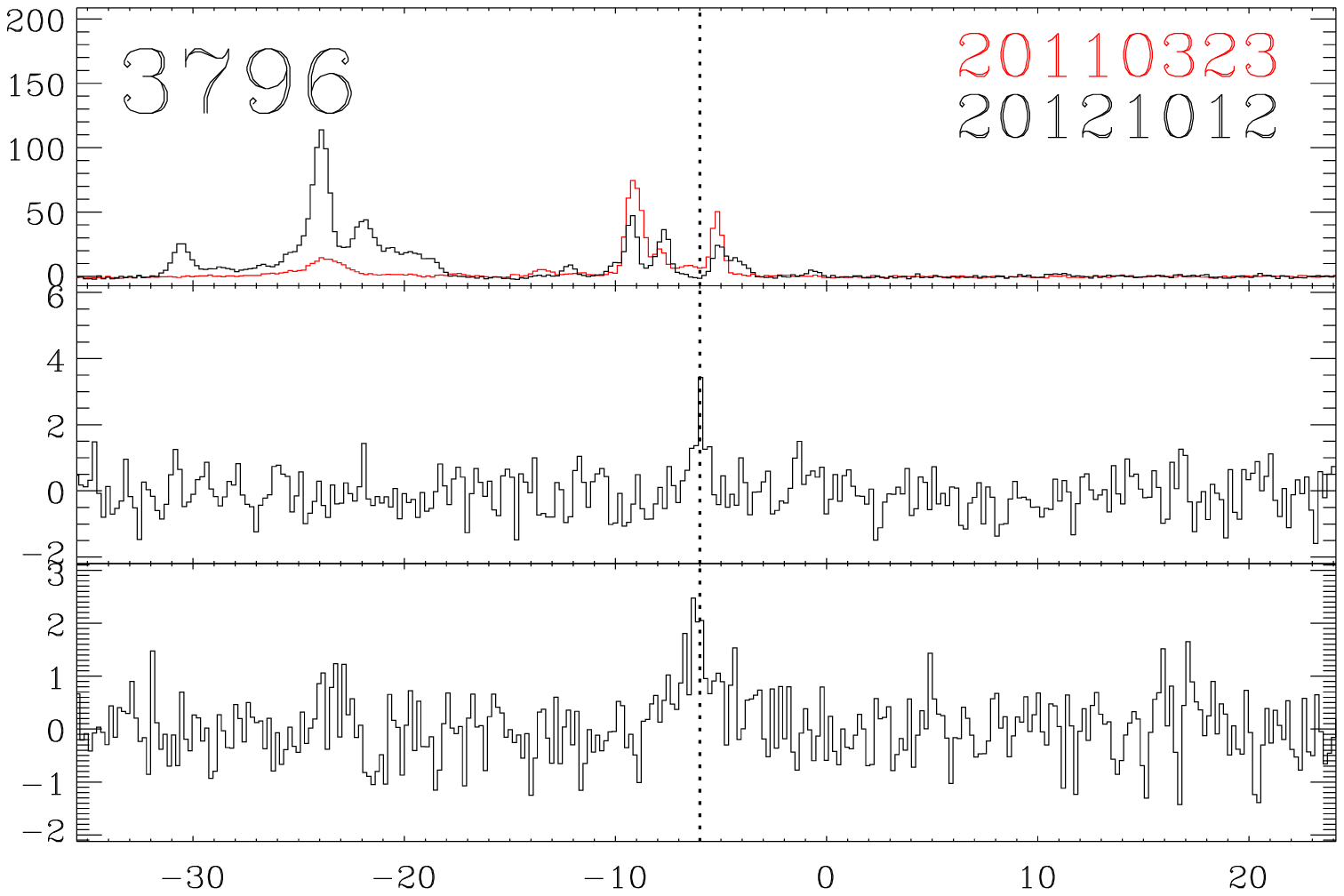} 
&
&  
\includegraphics[width=50mm]{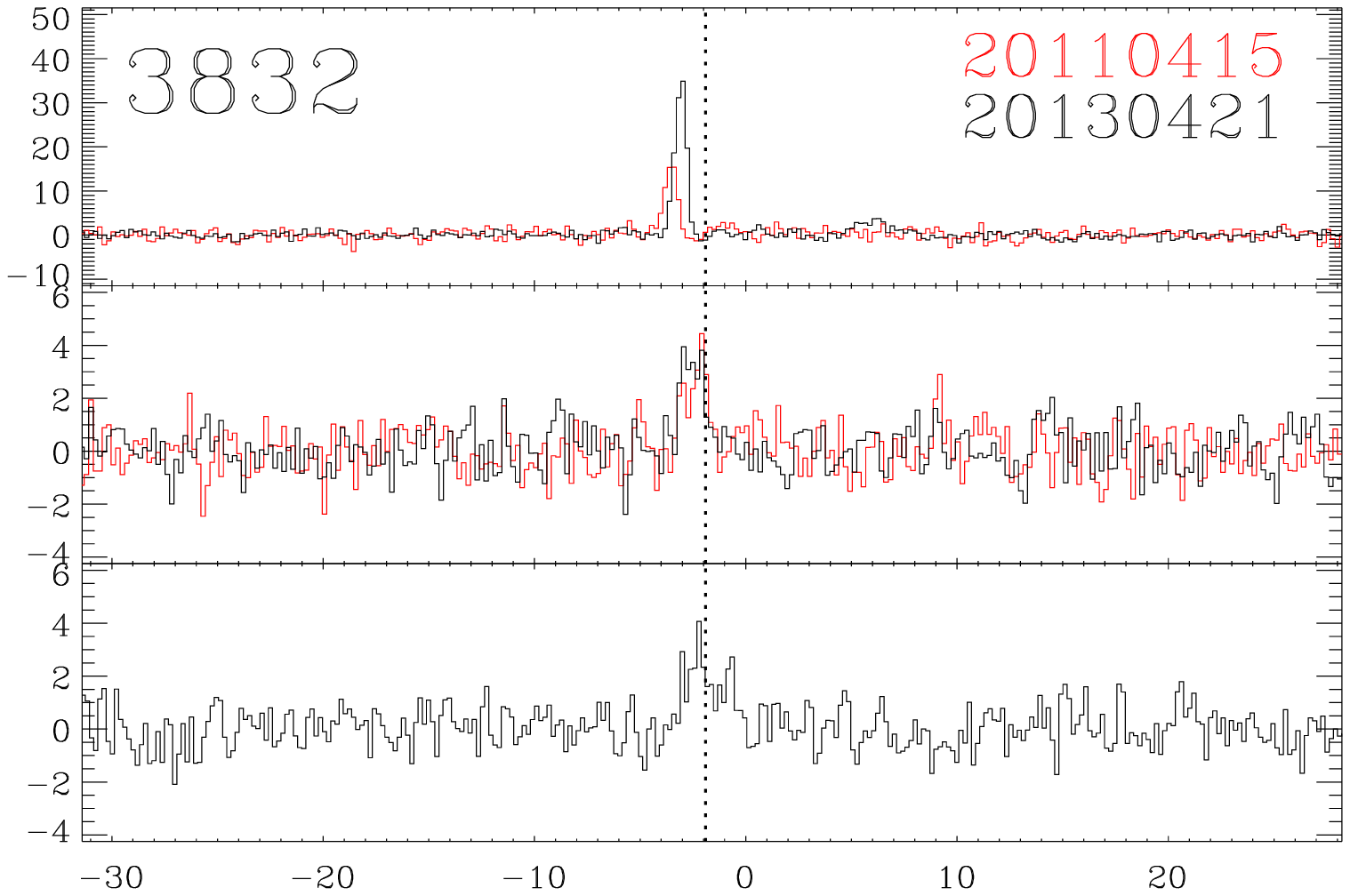} 
&
&
\includegraphics[width=50mm]{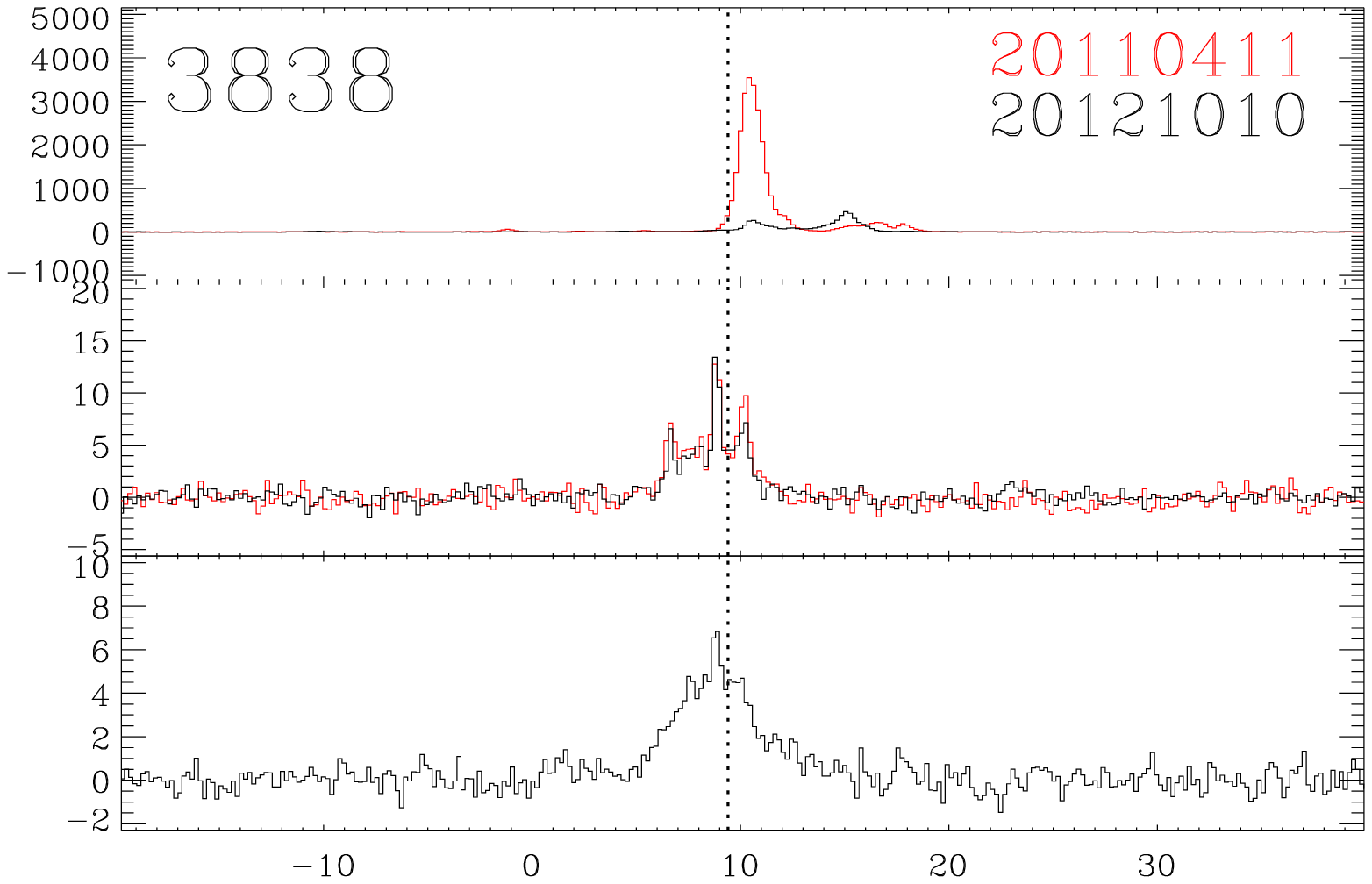} 
\\
&
\includegraphics[width=50mm]{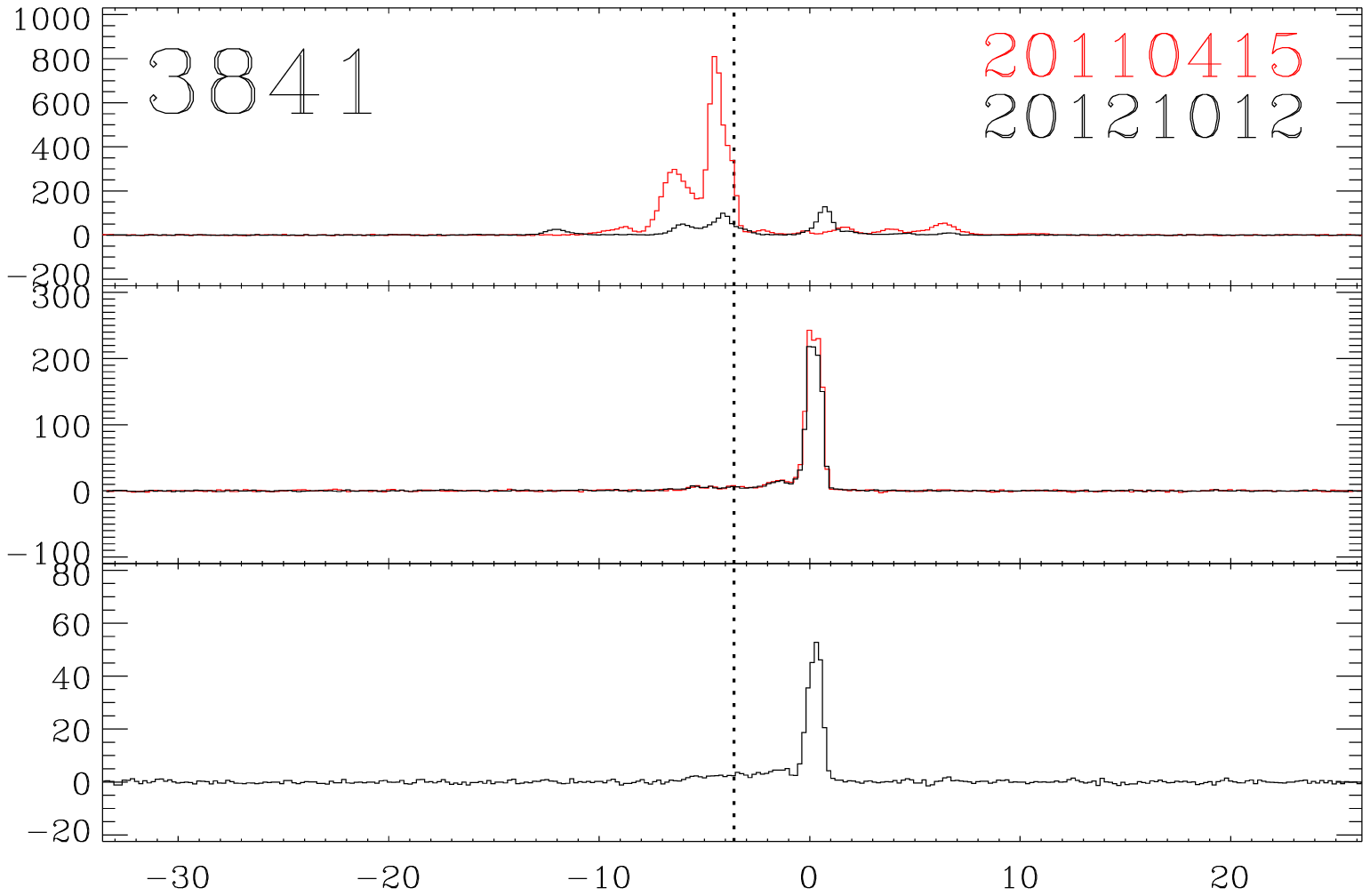} 
&
&  
\includegraphics[width=50mm]{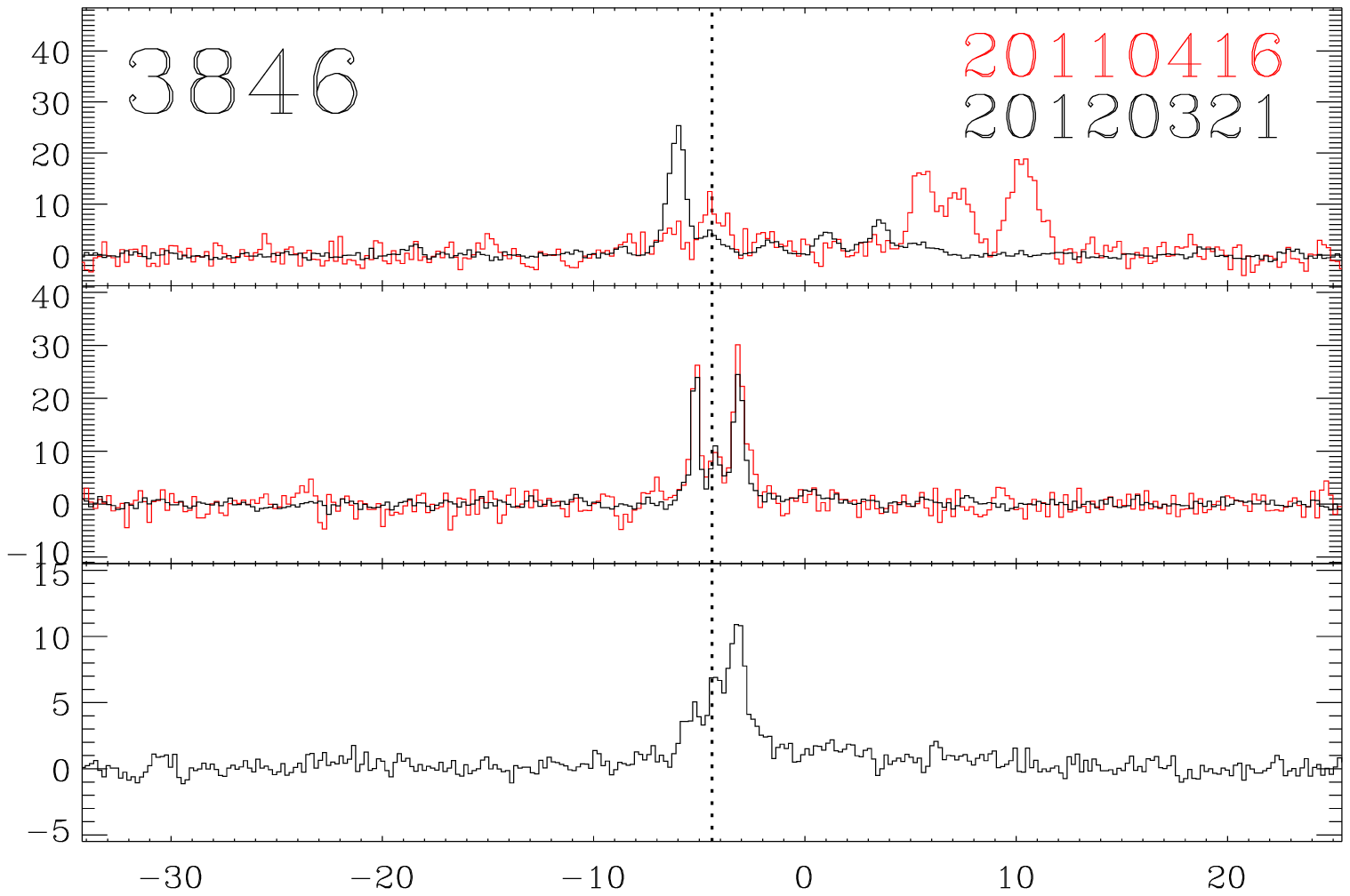} 
&
&
\includegraphics[width=50mm]{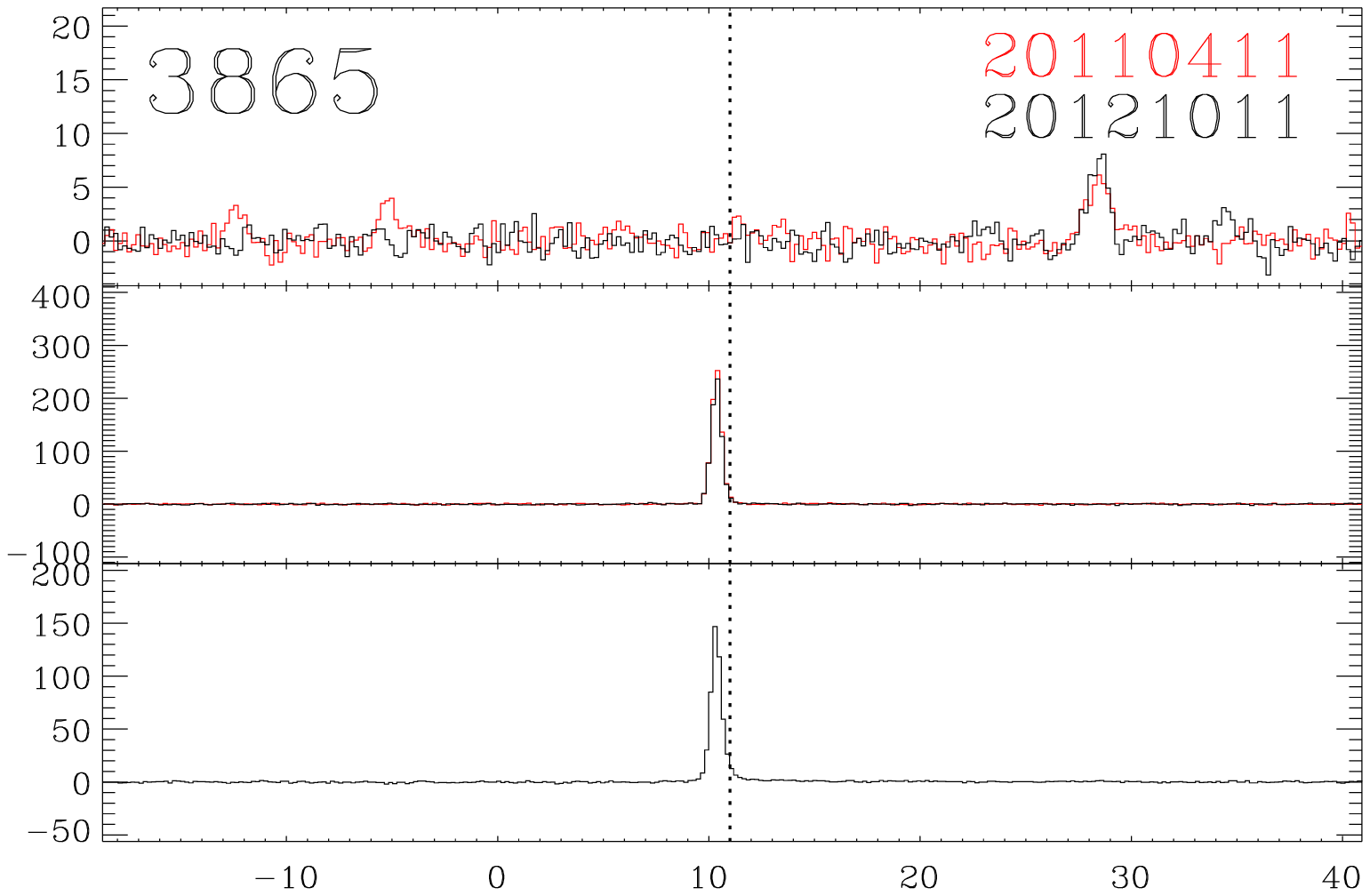} 
\\
&\multicolumn{5}{c}{LSR velocity (\kms)} \\
\end{tabular}
\caption{Continued}
\end{figure}
\clearpage

\begin{figure}
\figurenum{2}
\epsscale{1}
\begin{minipage}{0.1cm}
\rotatebox{90}{\textcolor{black}{Flux density (Jy)}}
\end{minipage}
\begin{tabular}{cccccc}
&
\includegraphics[width=50mm]{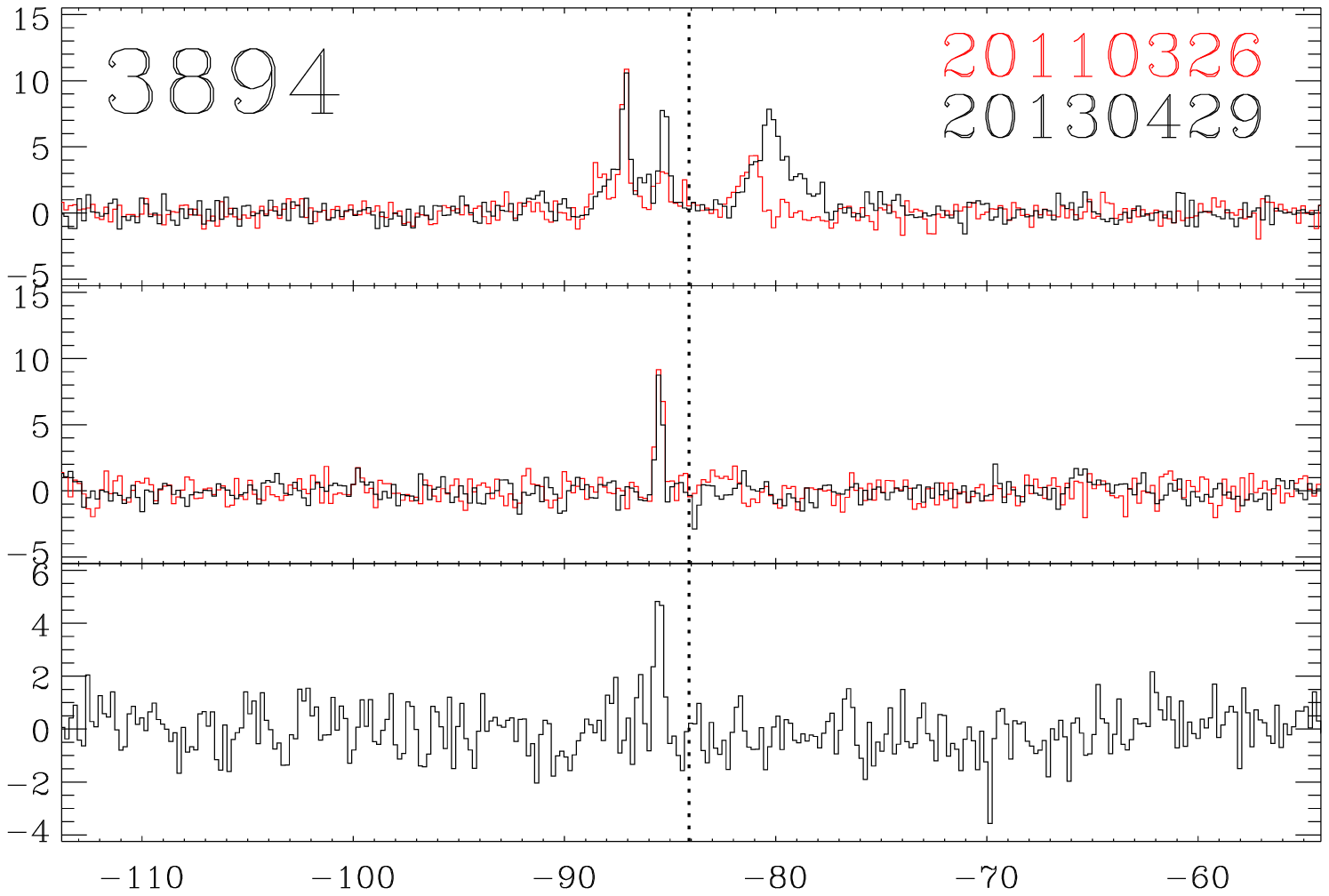} 
&
&
\includegraphics[width=50mm]{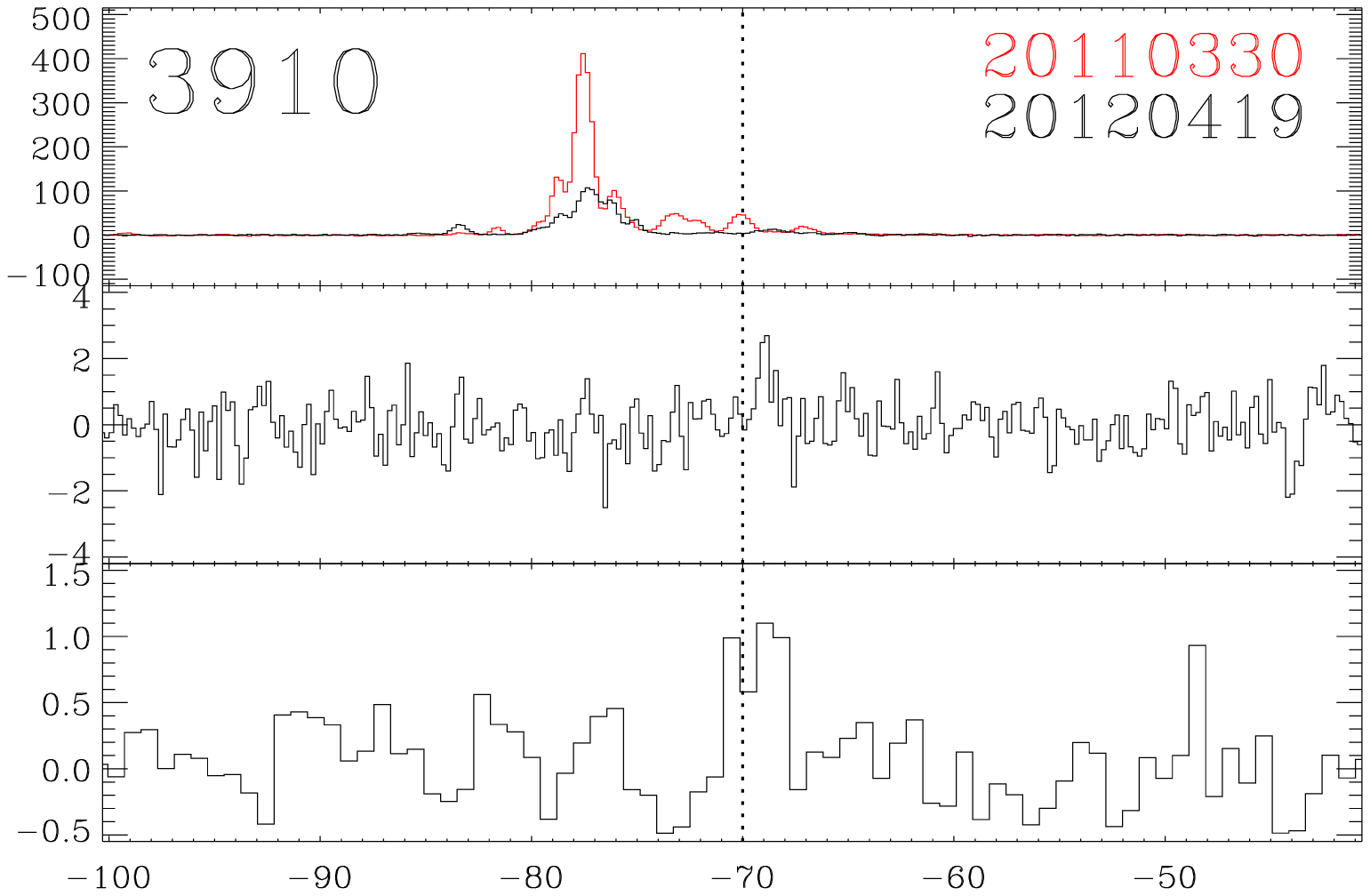} 
&
& 
\includegraphics[width=50mm]{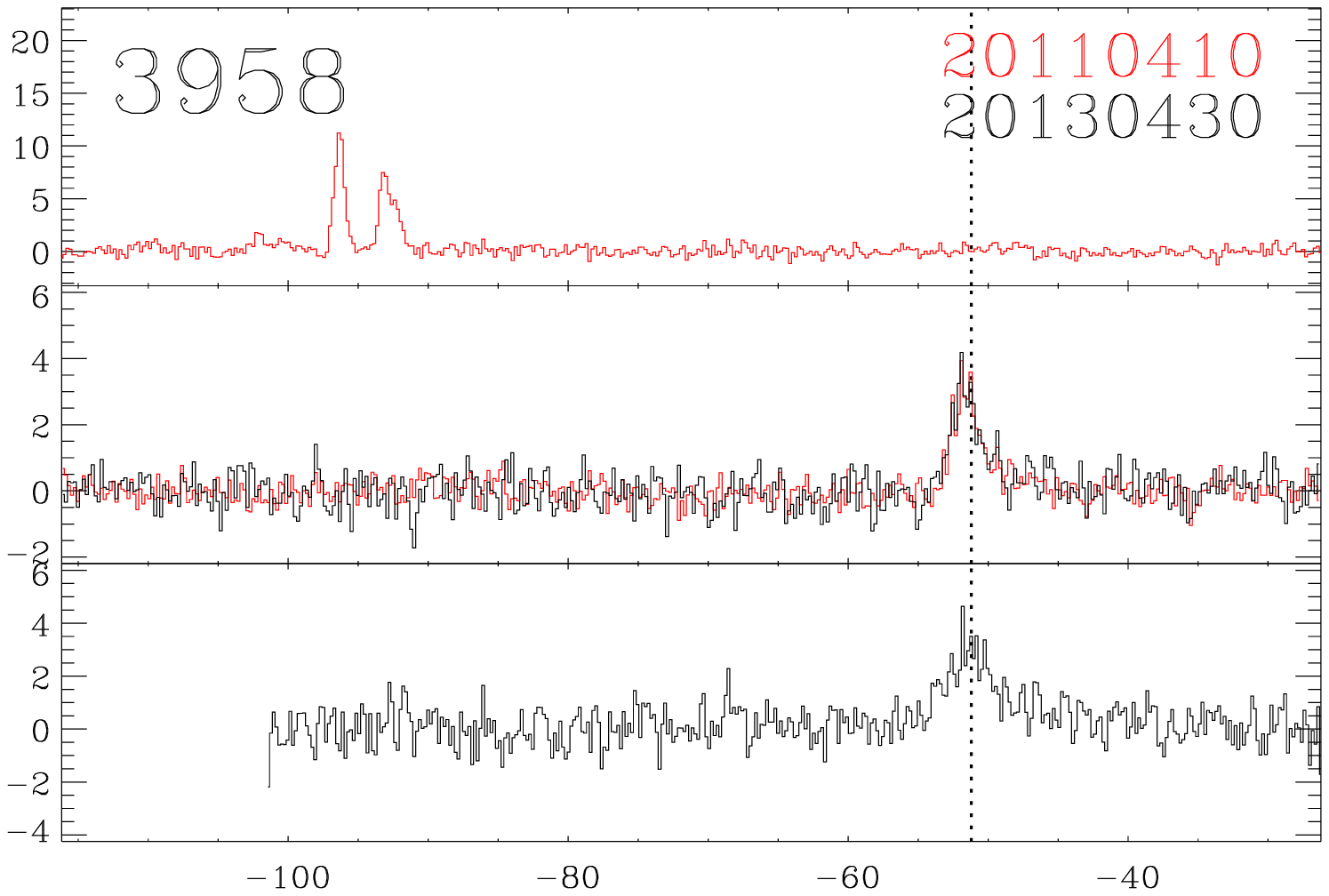} 
\\
&
\includegraphics[width=50mm]{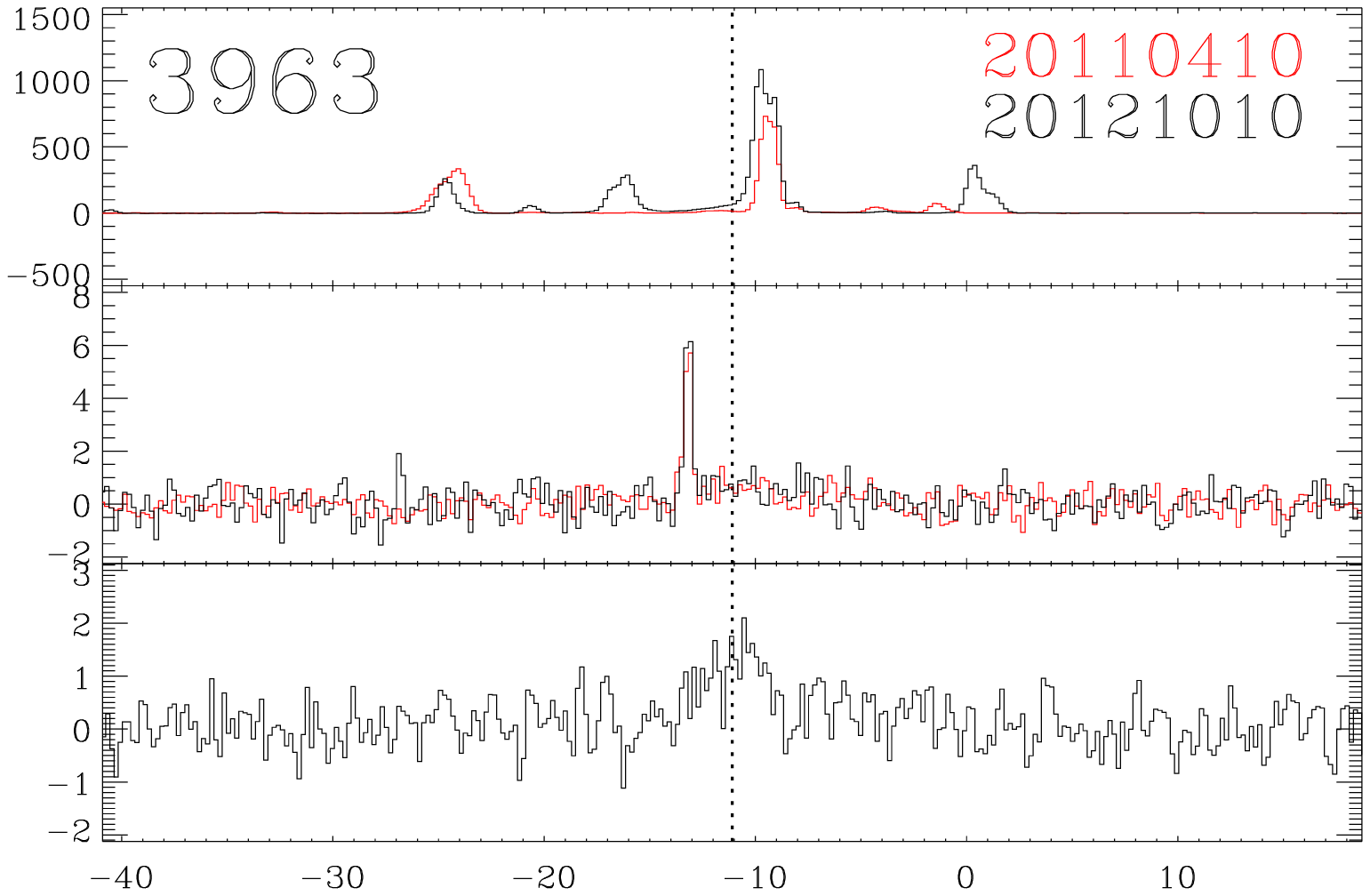} 
&
& 
\includegraphics[width=50mm]{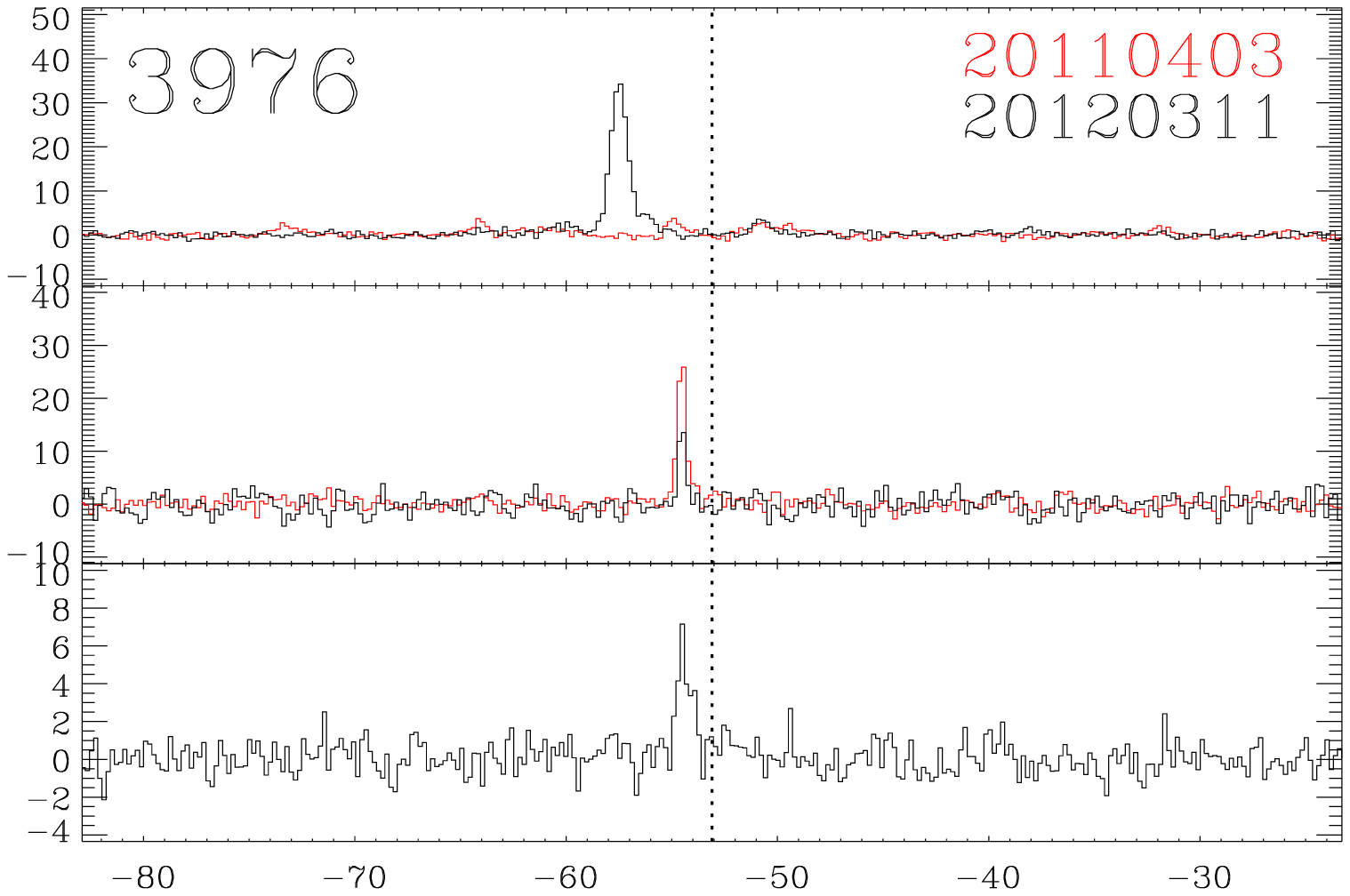} 
&
&
\includegraphics[width=50mm]{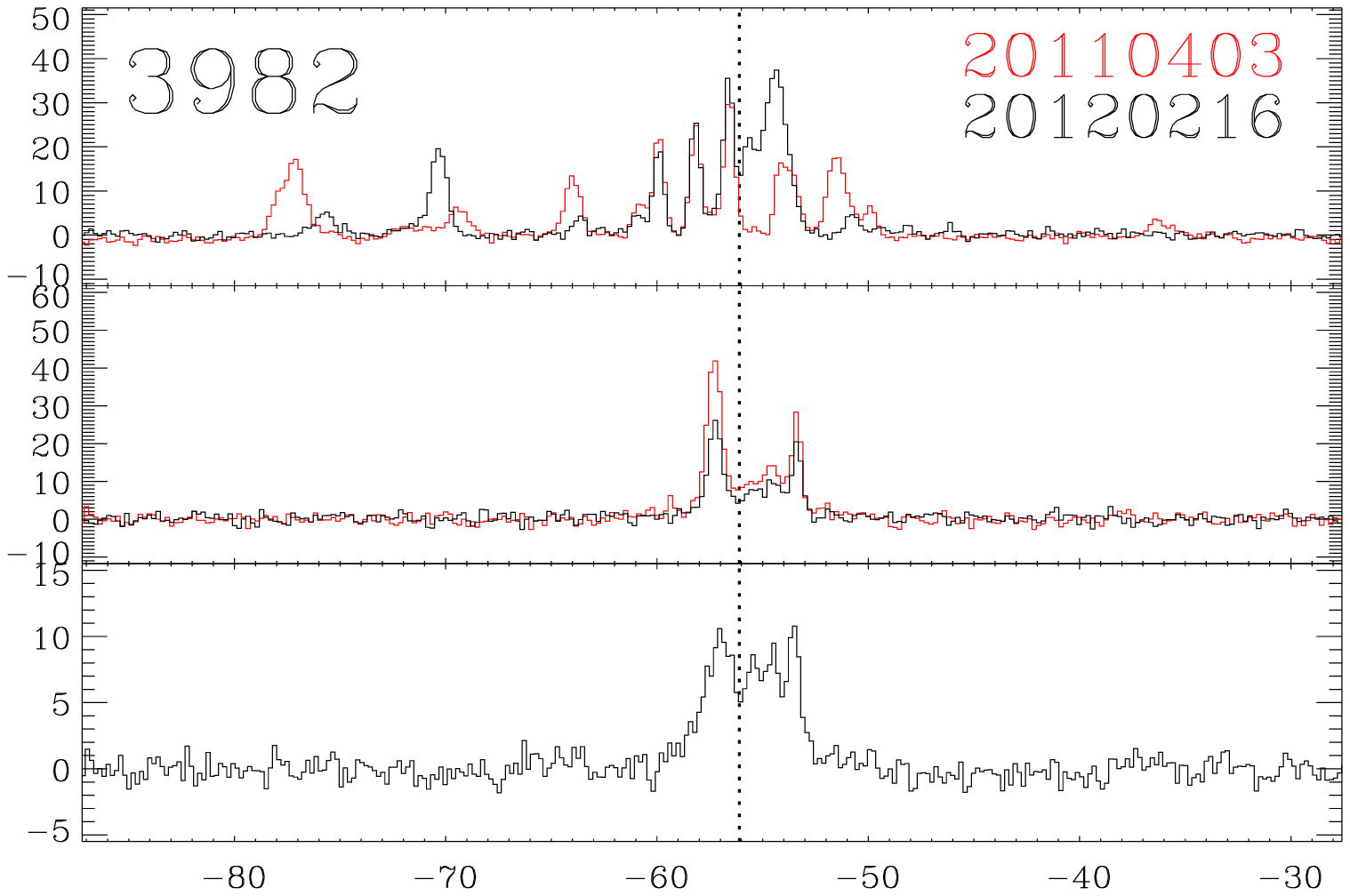} 
\\
&
\includegraphics[width=50mm]{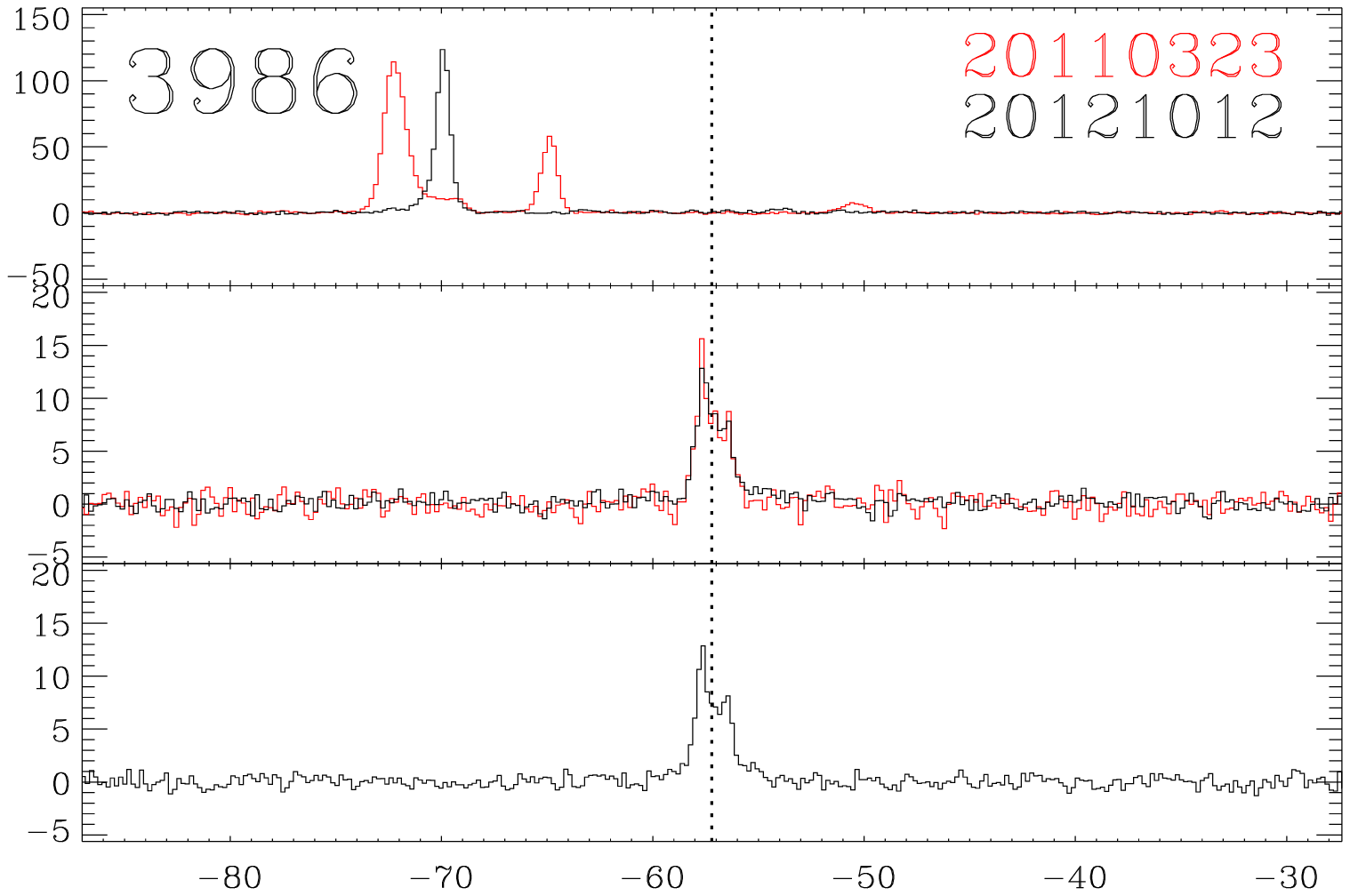} 
&
&  
\includegraphics[width=50mm]{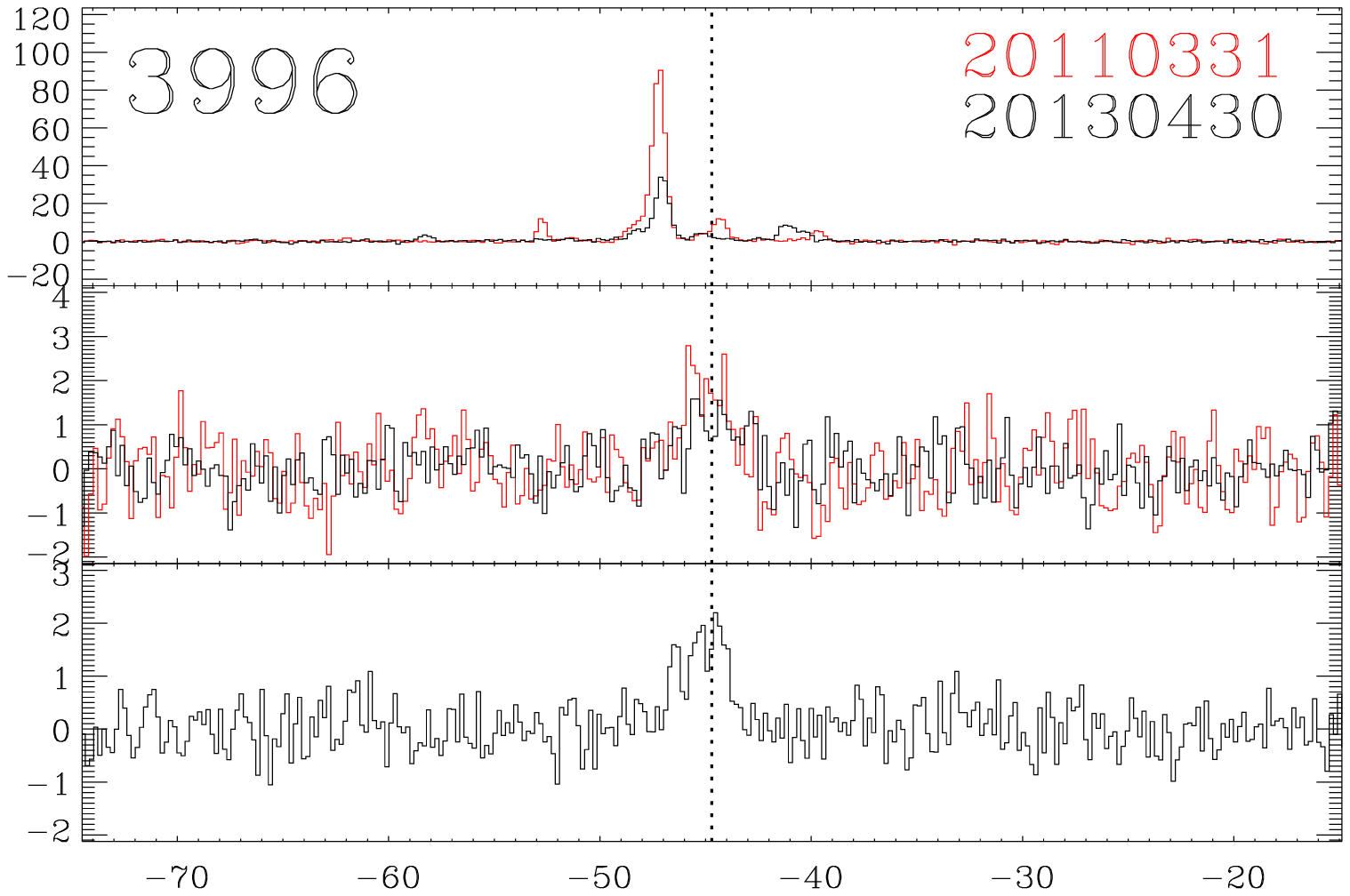} 
&
&
\includegraphics[width=50mm]{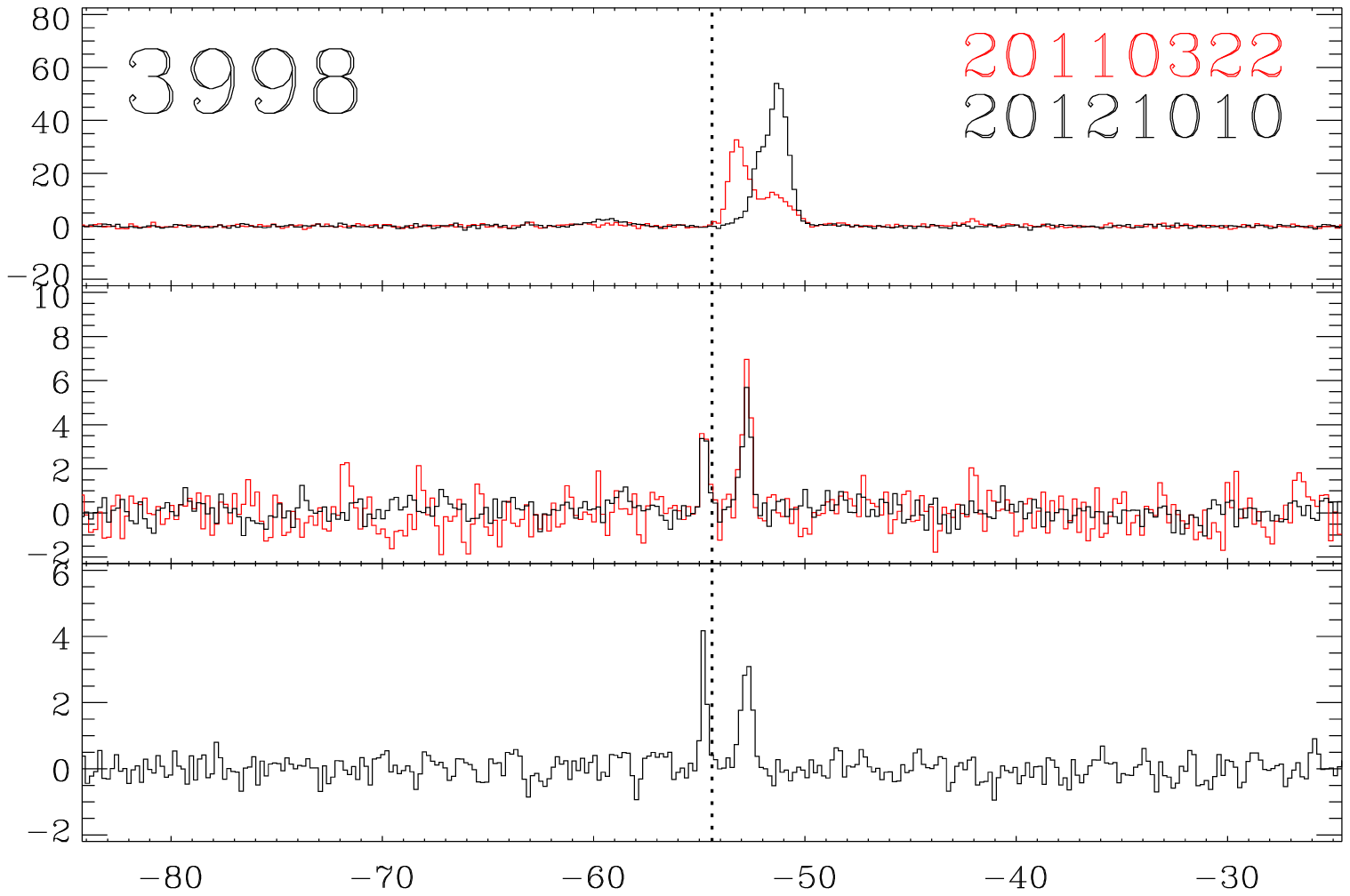} 
\\
&
\includegraphics[width=50mm]{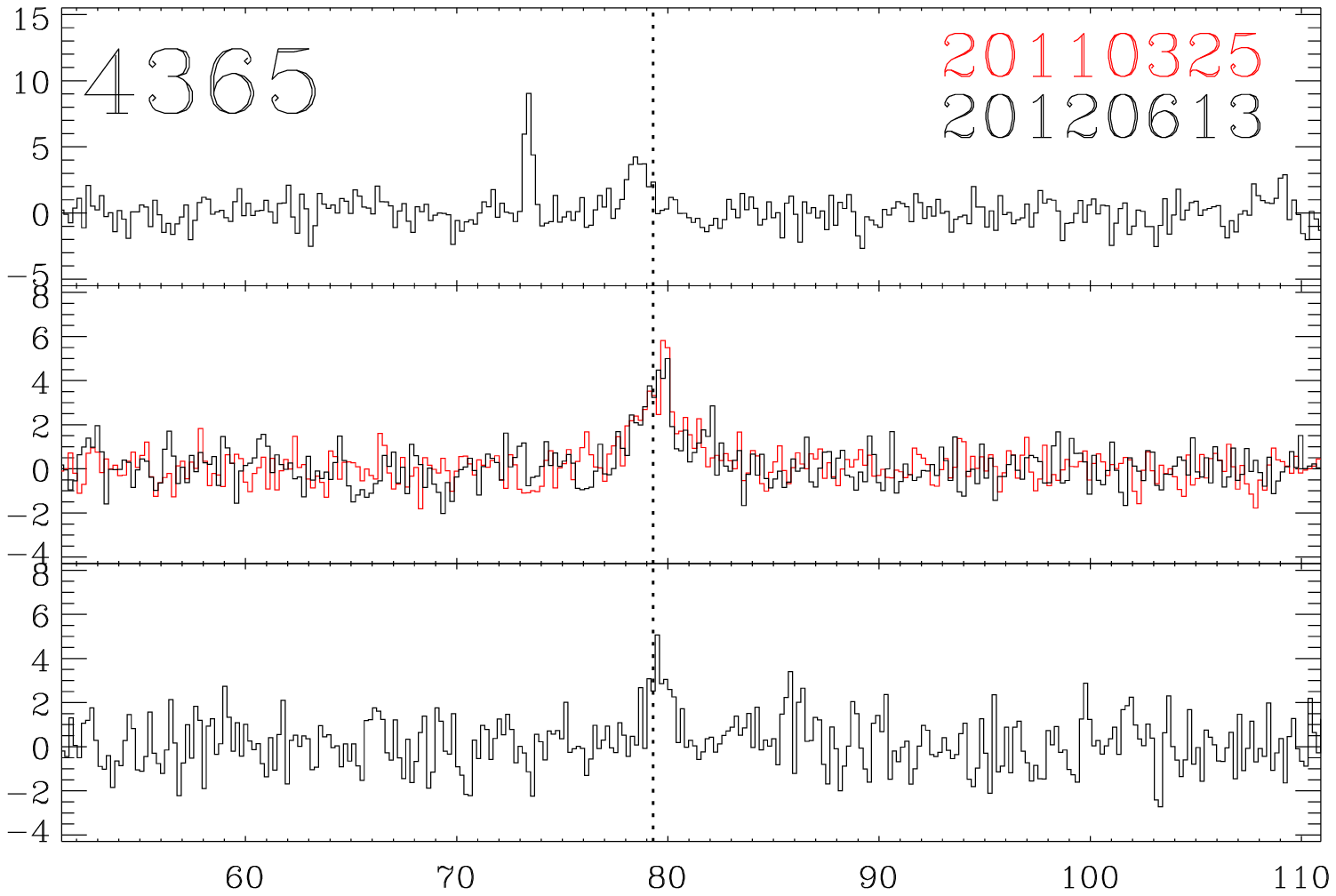} 
&
&  
\includegraphics[width=50mm]{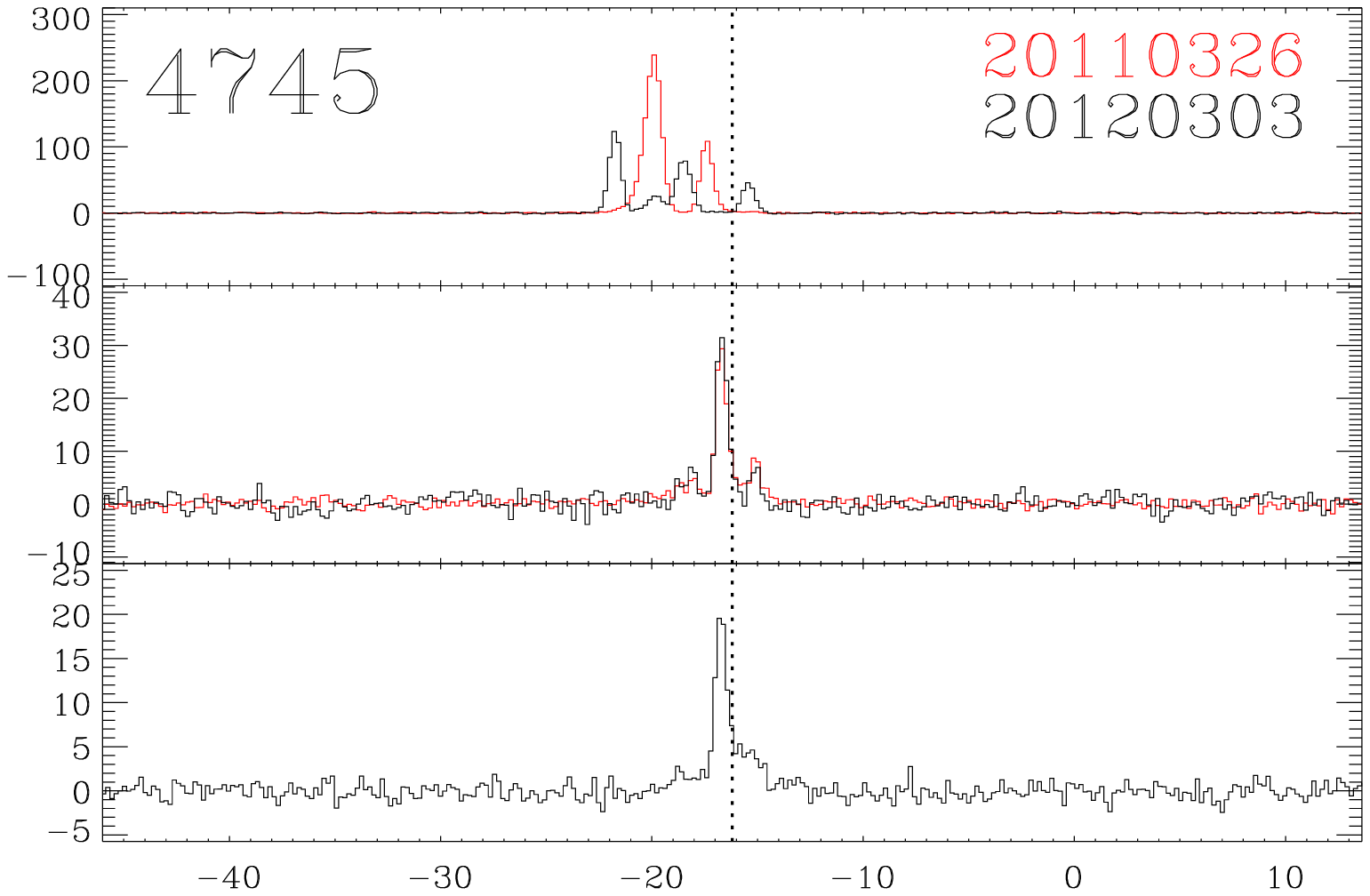} 
&
&
\includegraphics[width=50mm]{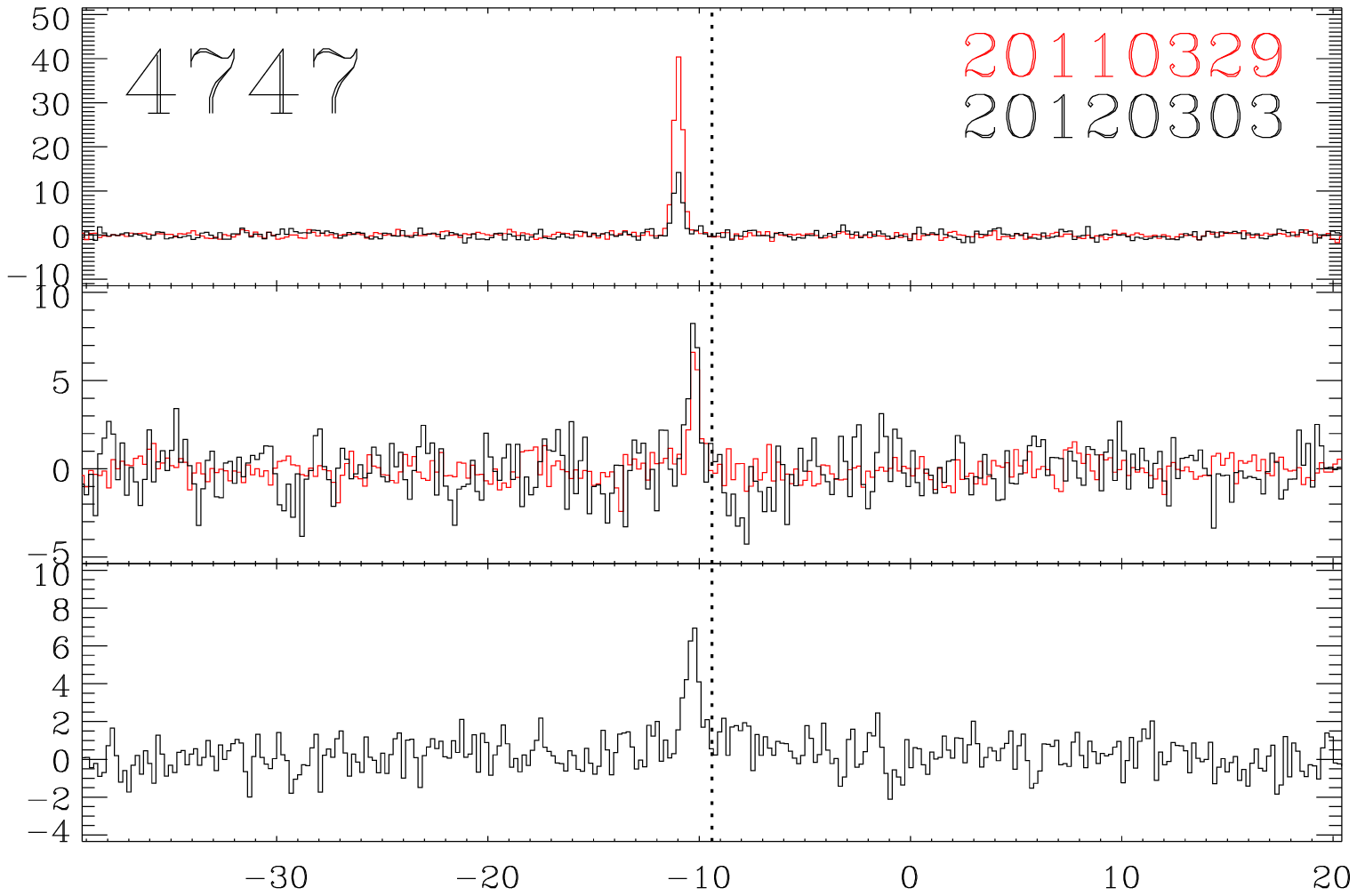} 
\\
&
\includegraphics[width=50mm]{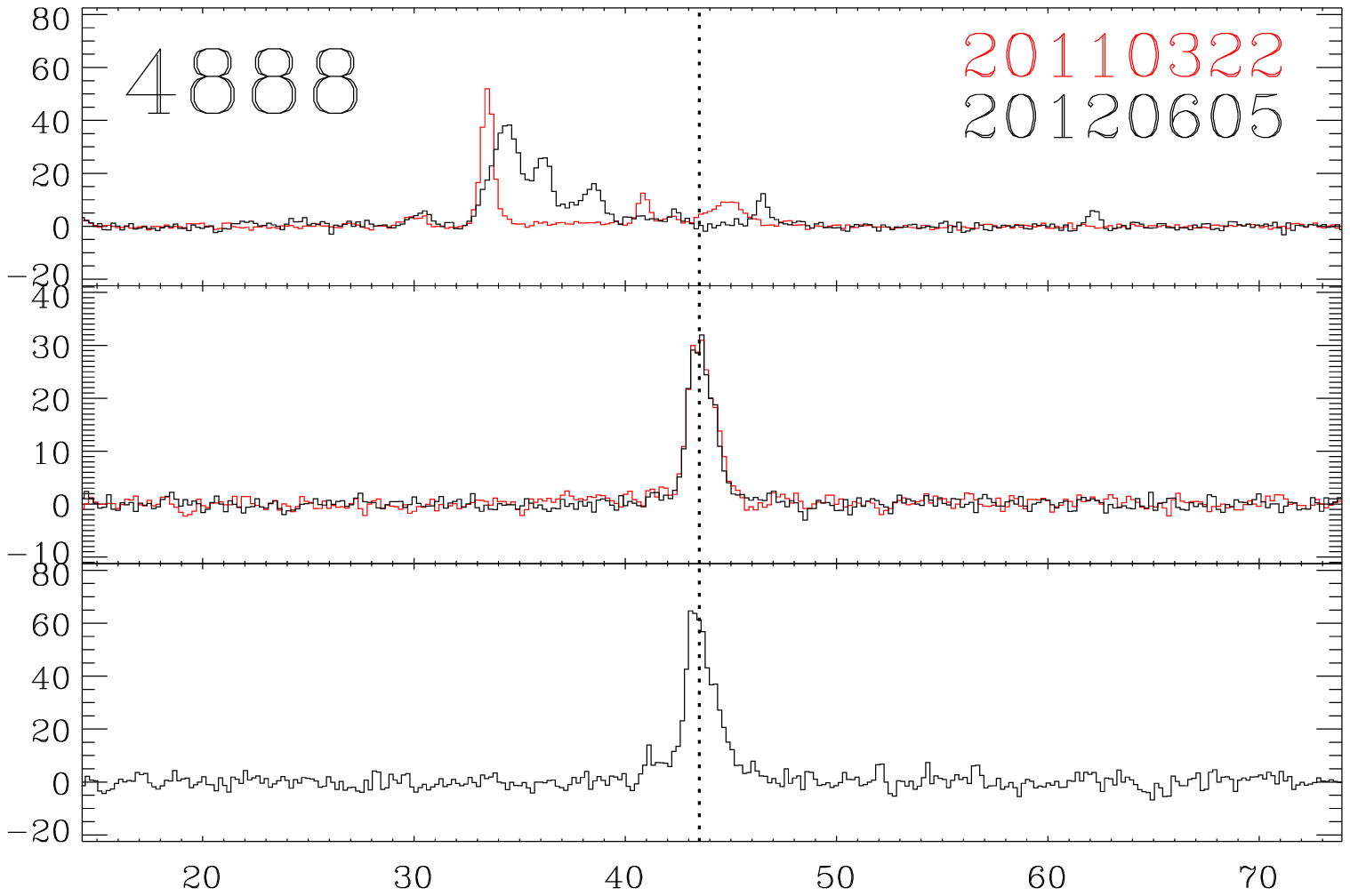} 
&
&  
\includegraphics[width=50mm]{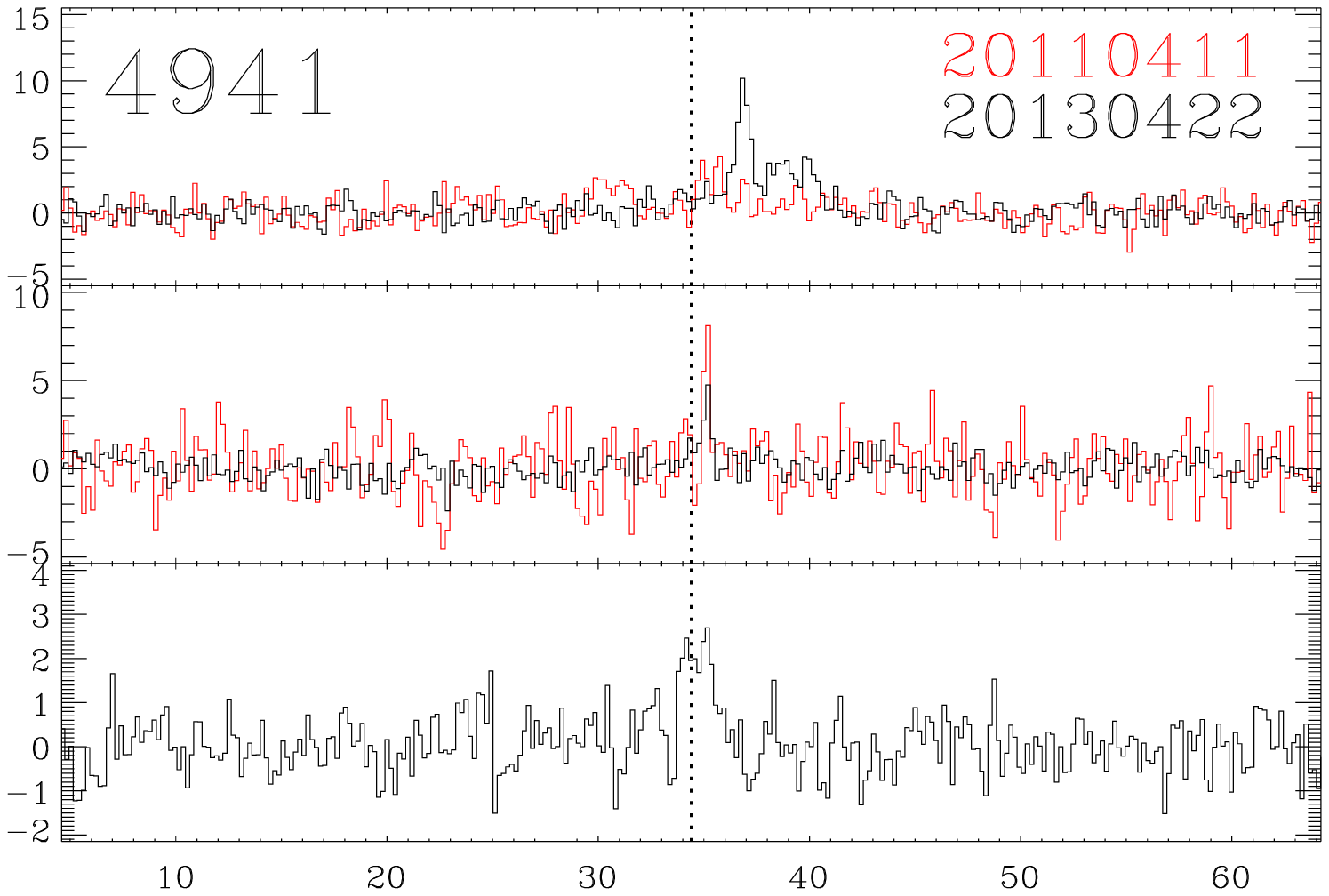} 
&
&
\\
&\multicolumn{5}{c}{LSR velocity (\kms)} \\
\end{tabular}
\caption{Continued}
\end{figure}
\clearpage

\begin{figure}
\epsscale{1}
\begin{minipage}{0.1cm}
\rotatebox{90}{\textcolor{black}{Flux density (Jy)}}
\end{minipage}
\begin{tabular}{cccccc}
&
\includegraphics[width=50mm]{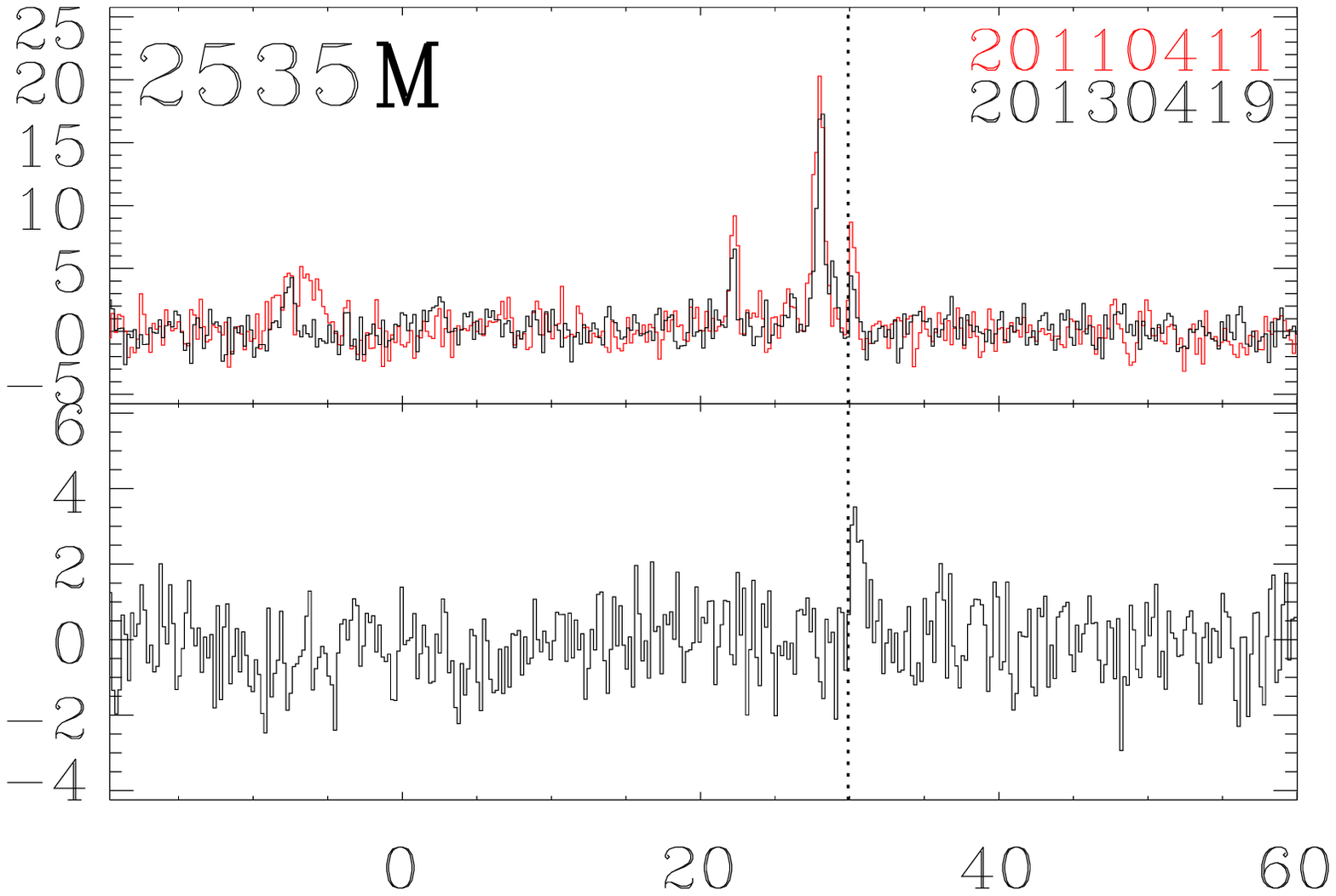} 
&
&
\includegraphics[width=50mm]{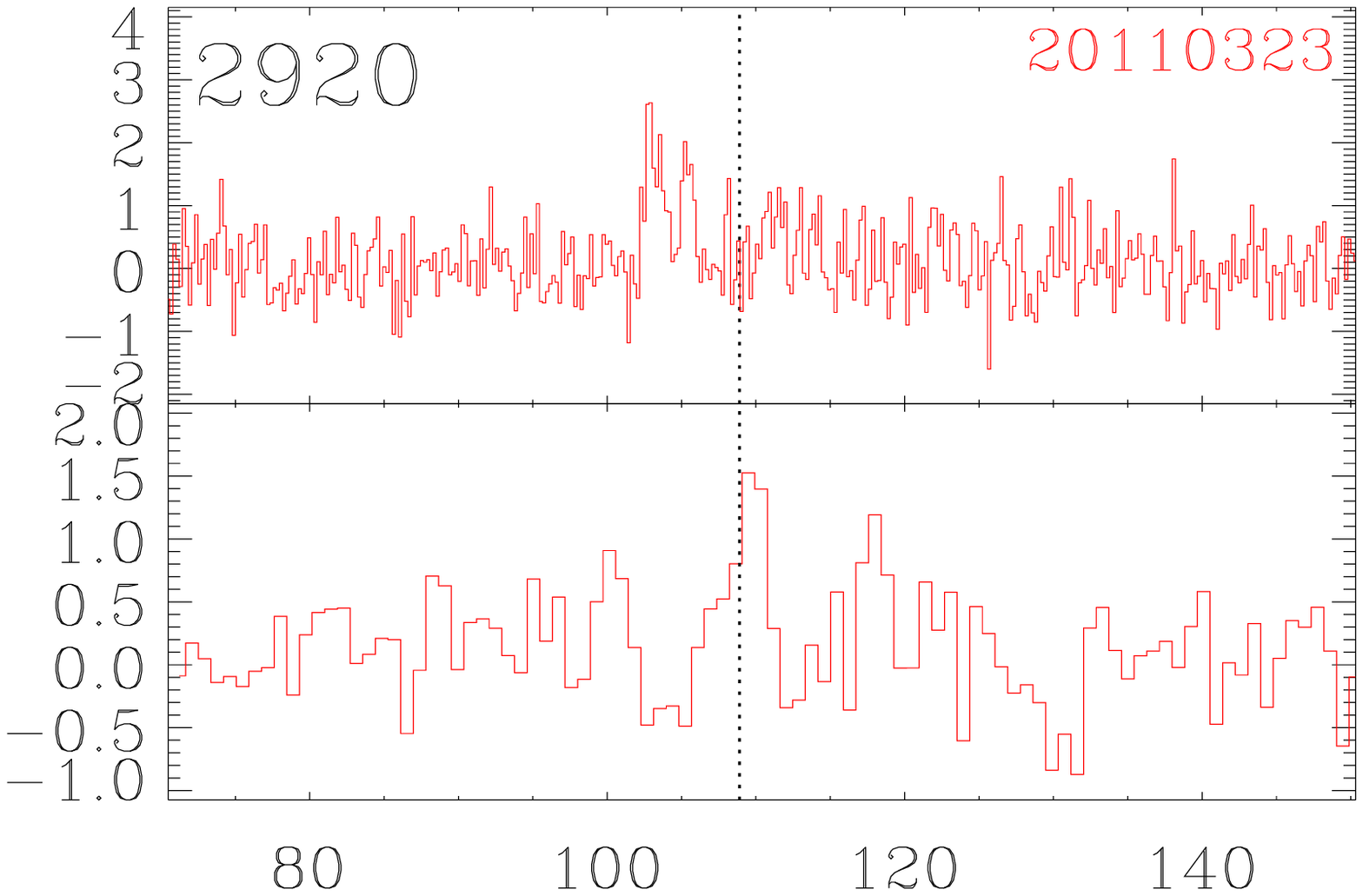} 
&
& 
\includegraphics[width=50mm]{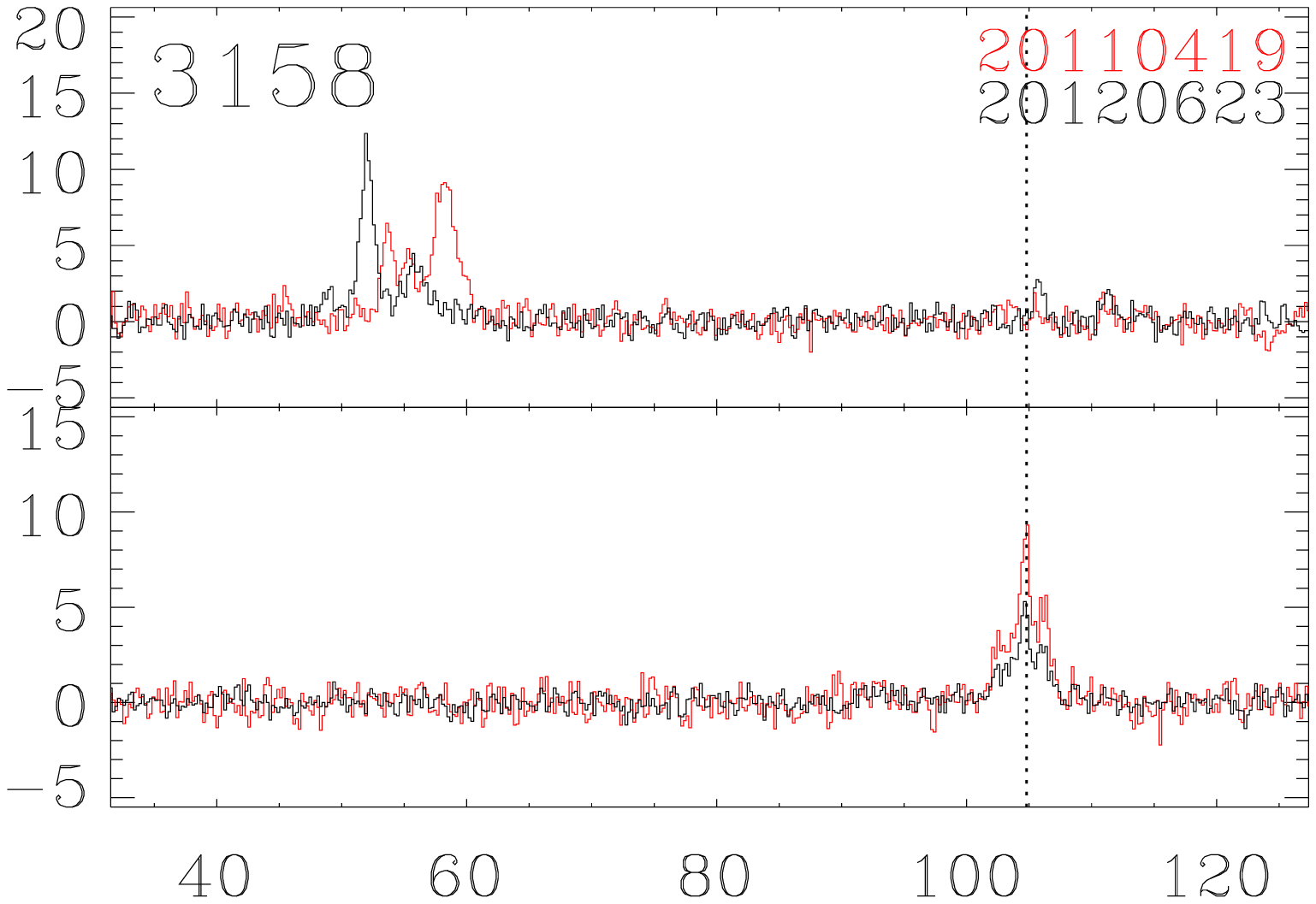} 
\\
&
\includegraphics[width=50mm]{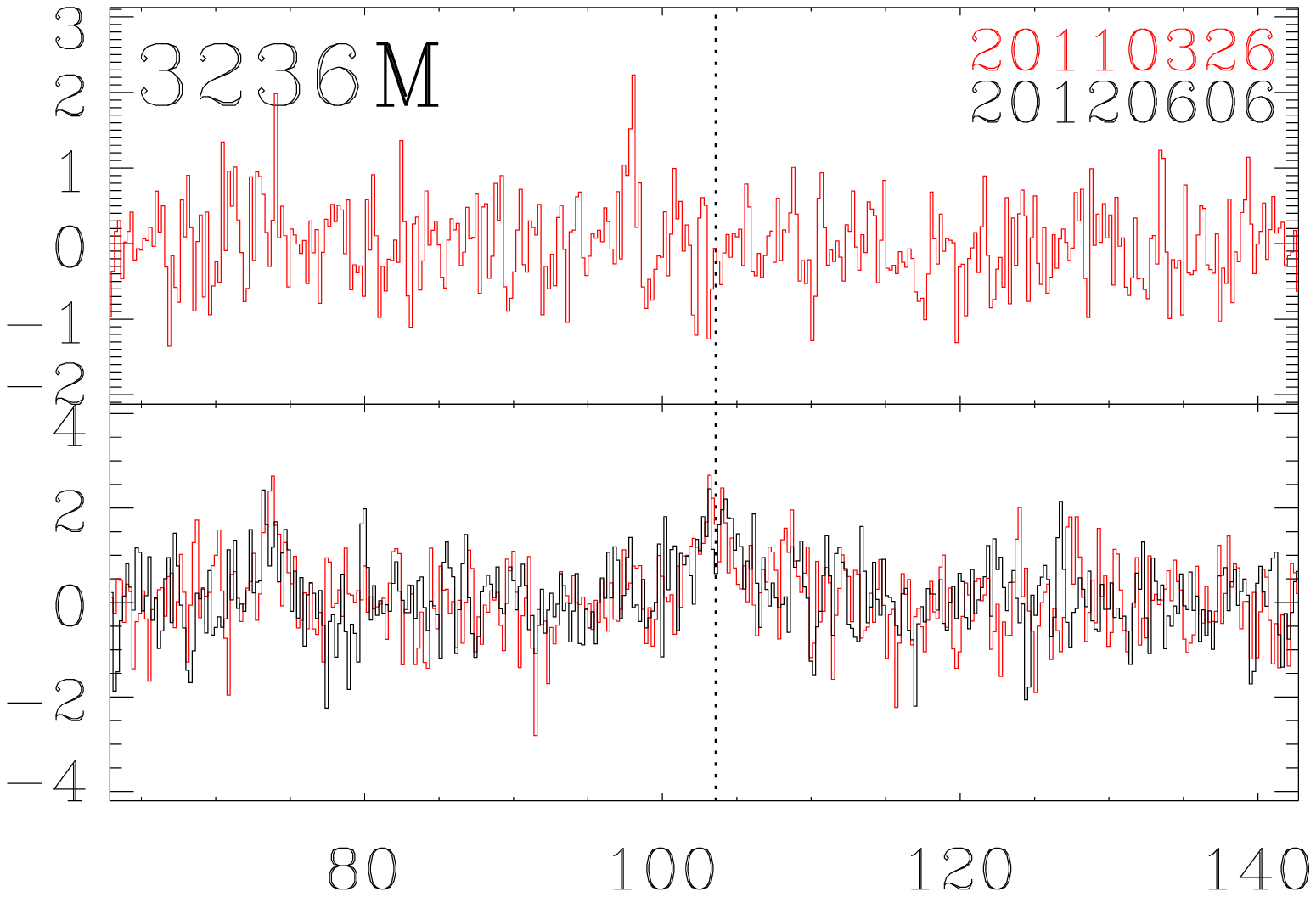} 
&
& 
\includegraphics[width=50mm]{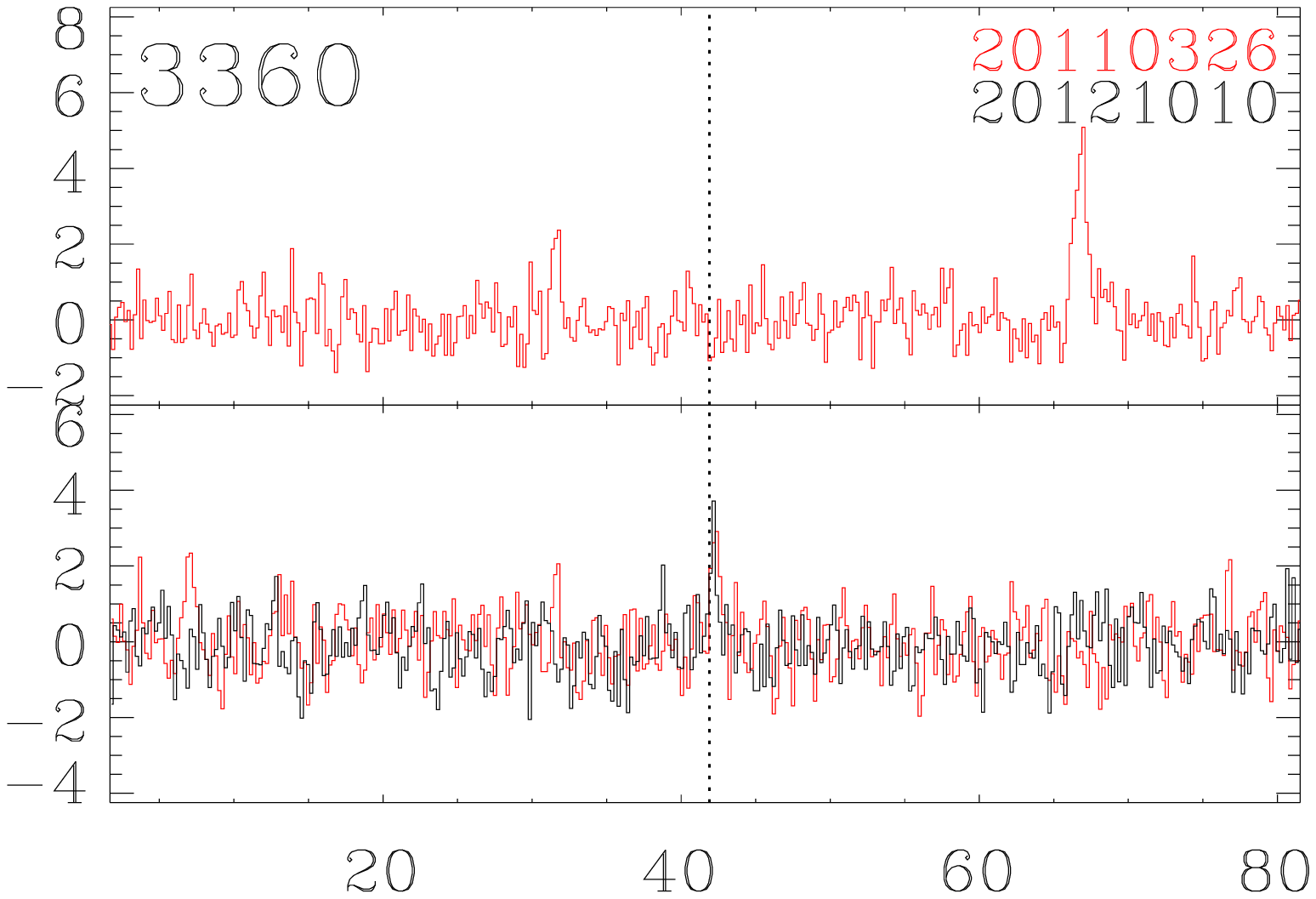} 
&
&
\includegraphics[width=50mm]{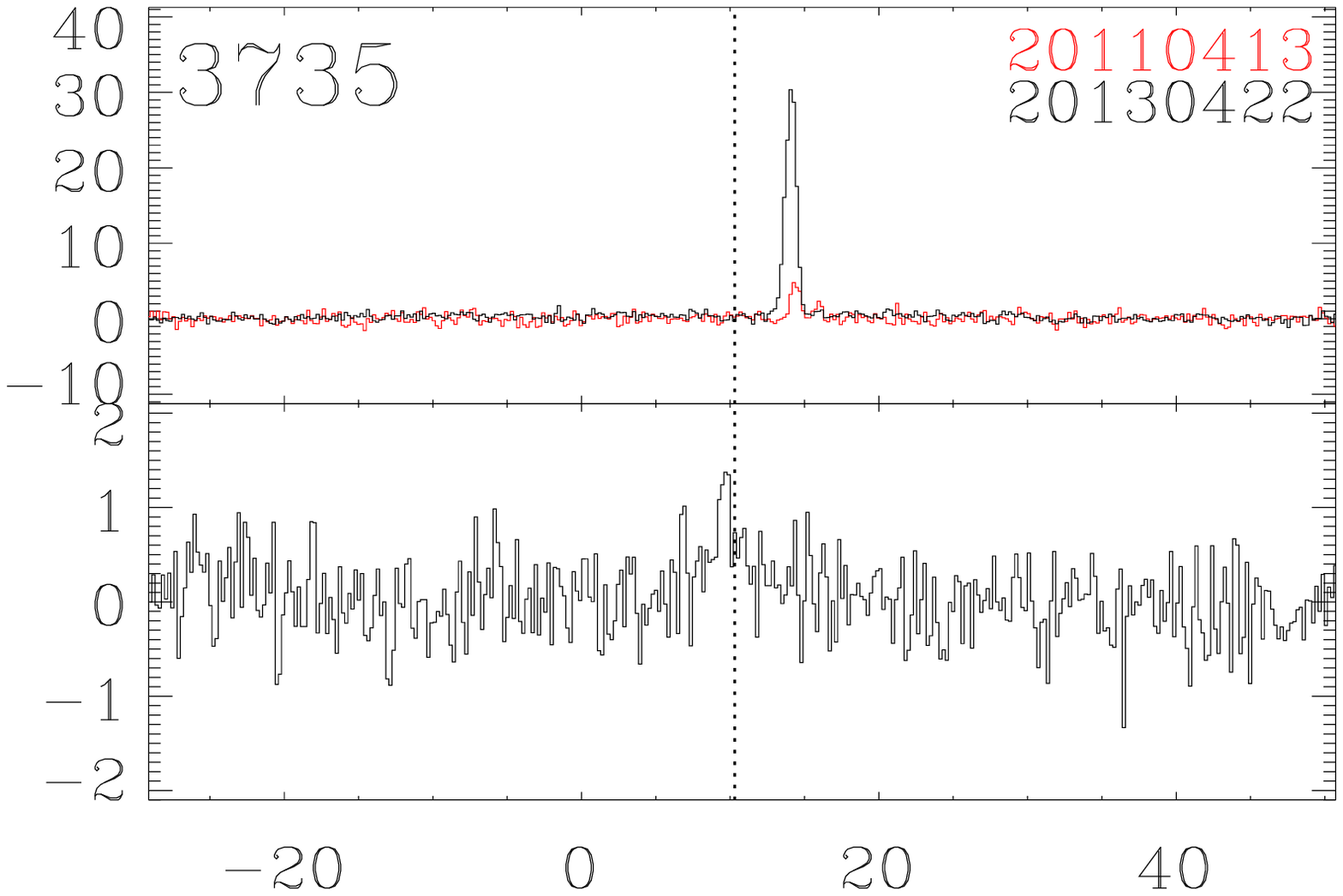} 
\\
&
\includegraphics[width=50mm]{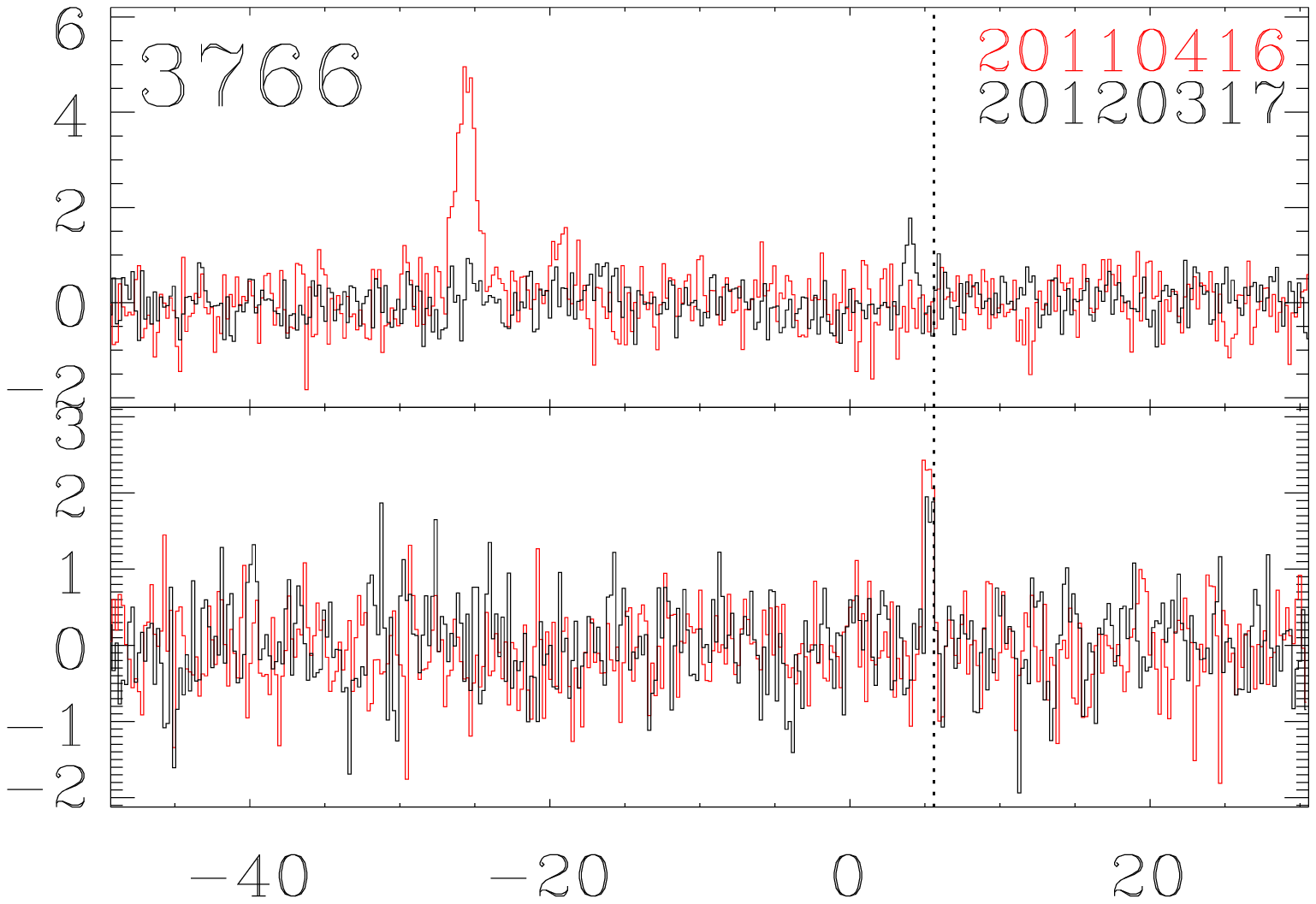} 
&
&
\includegraphics[width=50mm]{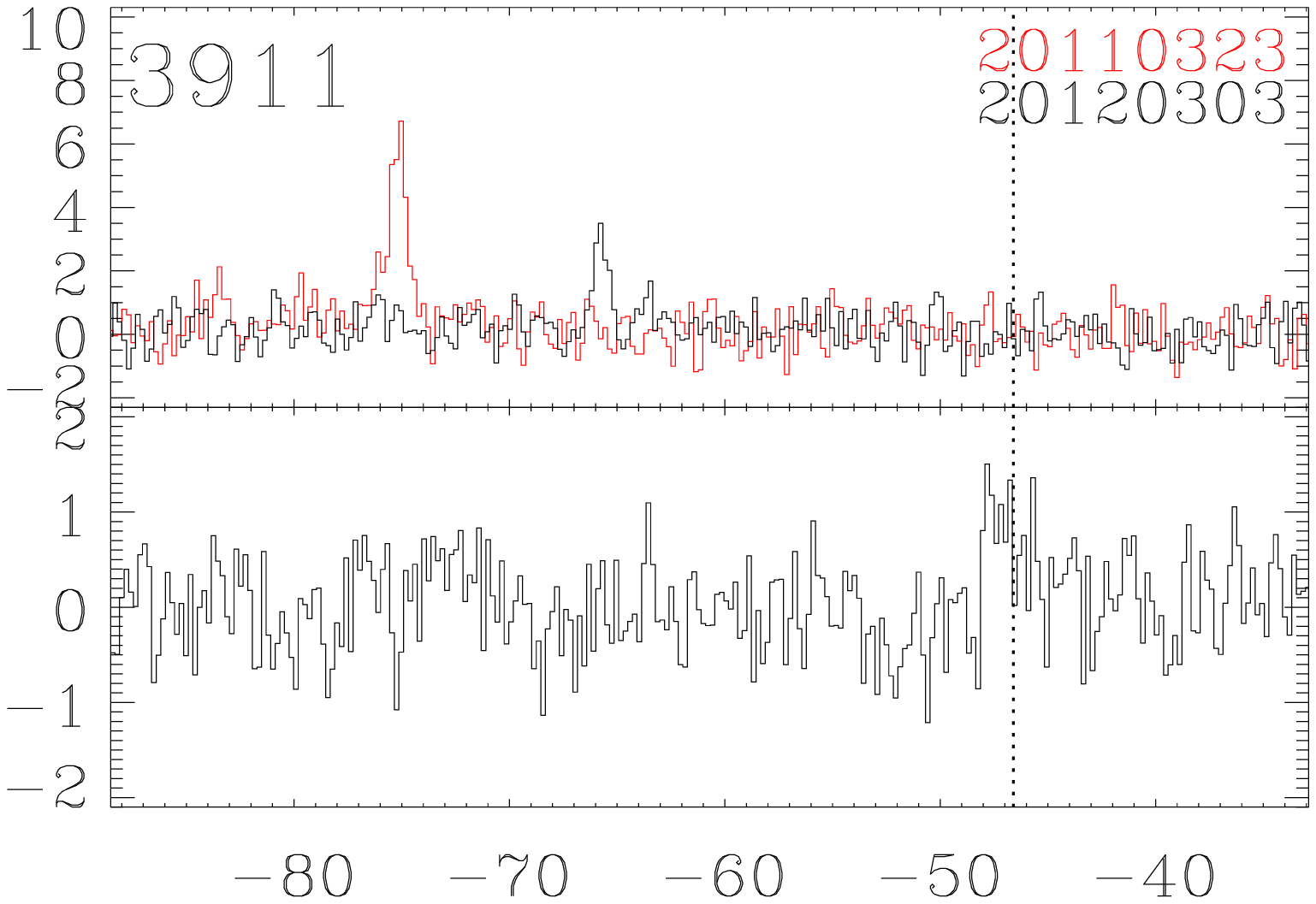}  
&
&
\includegraphics[width=50mm]{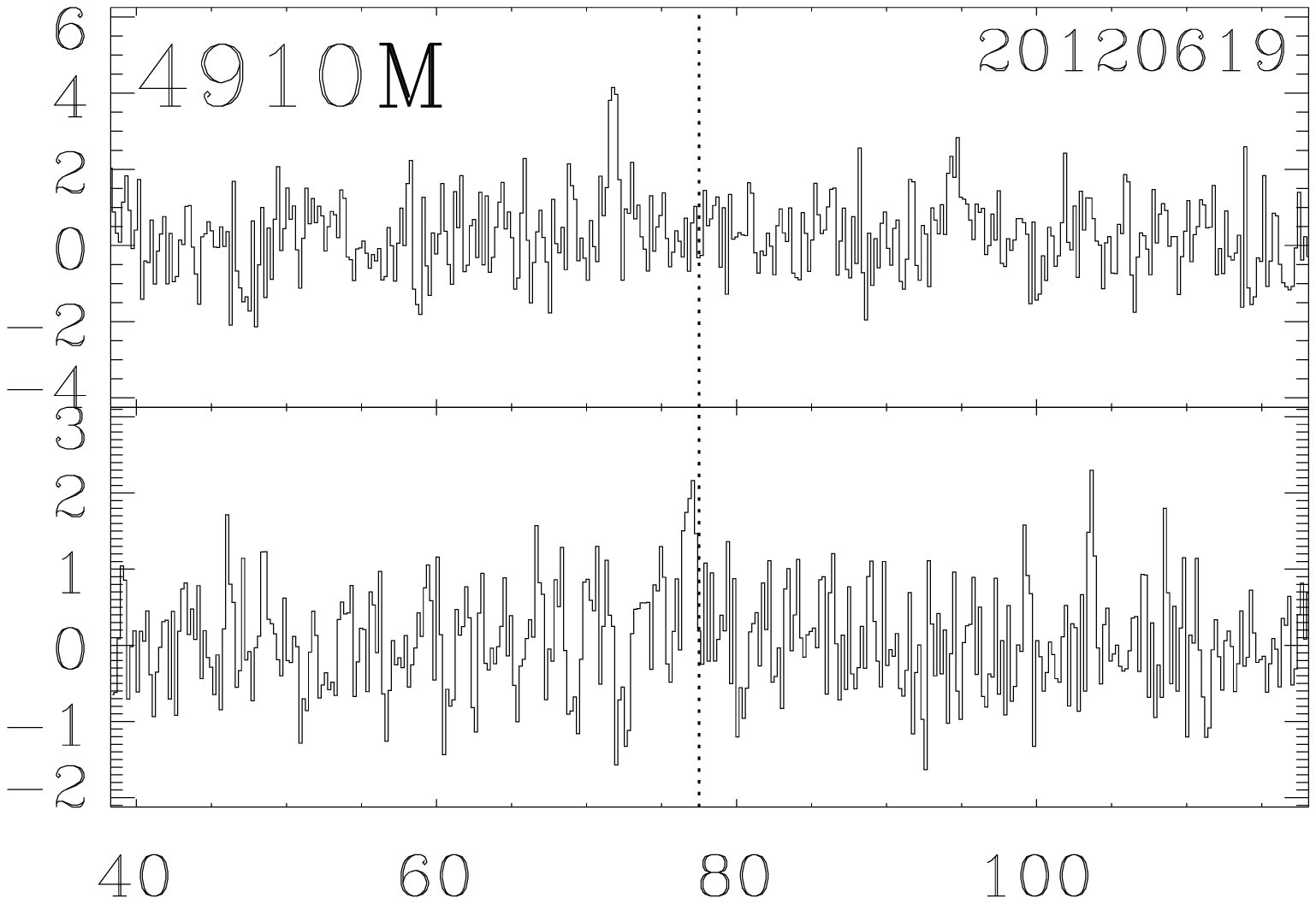} 
\\
\end{tabular}
\\
\begin{minipage}{\linewidth}
\center{LSR velocity (\kms)}
\end{minipage}
\caption{Same as in Figure~\ref{fig_224495} except for the source detected both at (upper panel) 22~GHz and (lower panel) 44~GHz.
\label{fig_2244}}
\end{figure}
\clearpage

\begin{figure}
\epsscale{1}
\begin{minipage}{0.1cm}
\rotatebox{90}{\textcolor{black}{Flux density (Jy)}}
\end{minipage}
\begin{tabular}{cccccc}
&
\includegraphics[width=50mm]{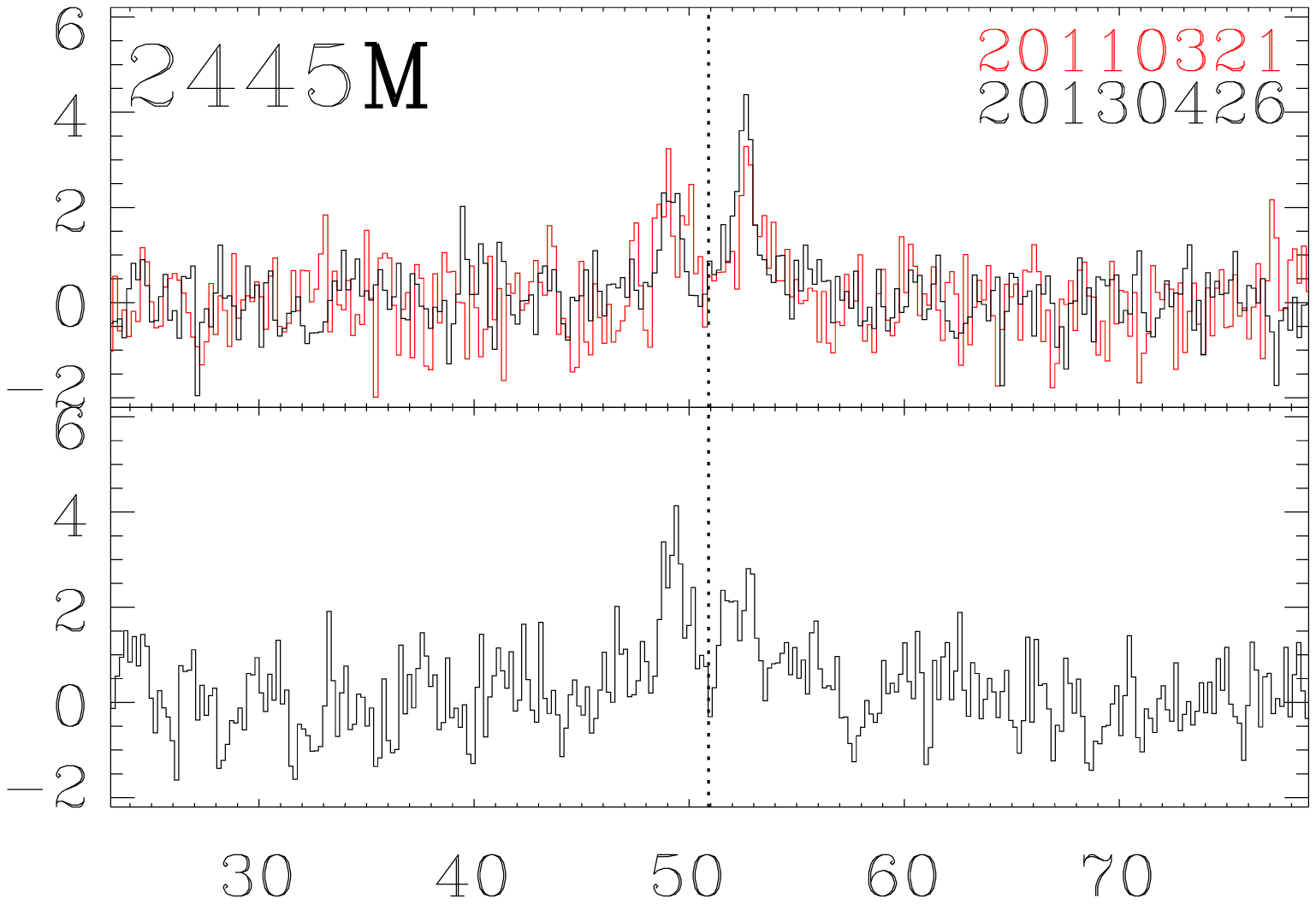} 
&
&
\includegraphics[width=50mm]{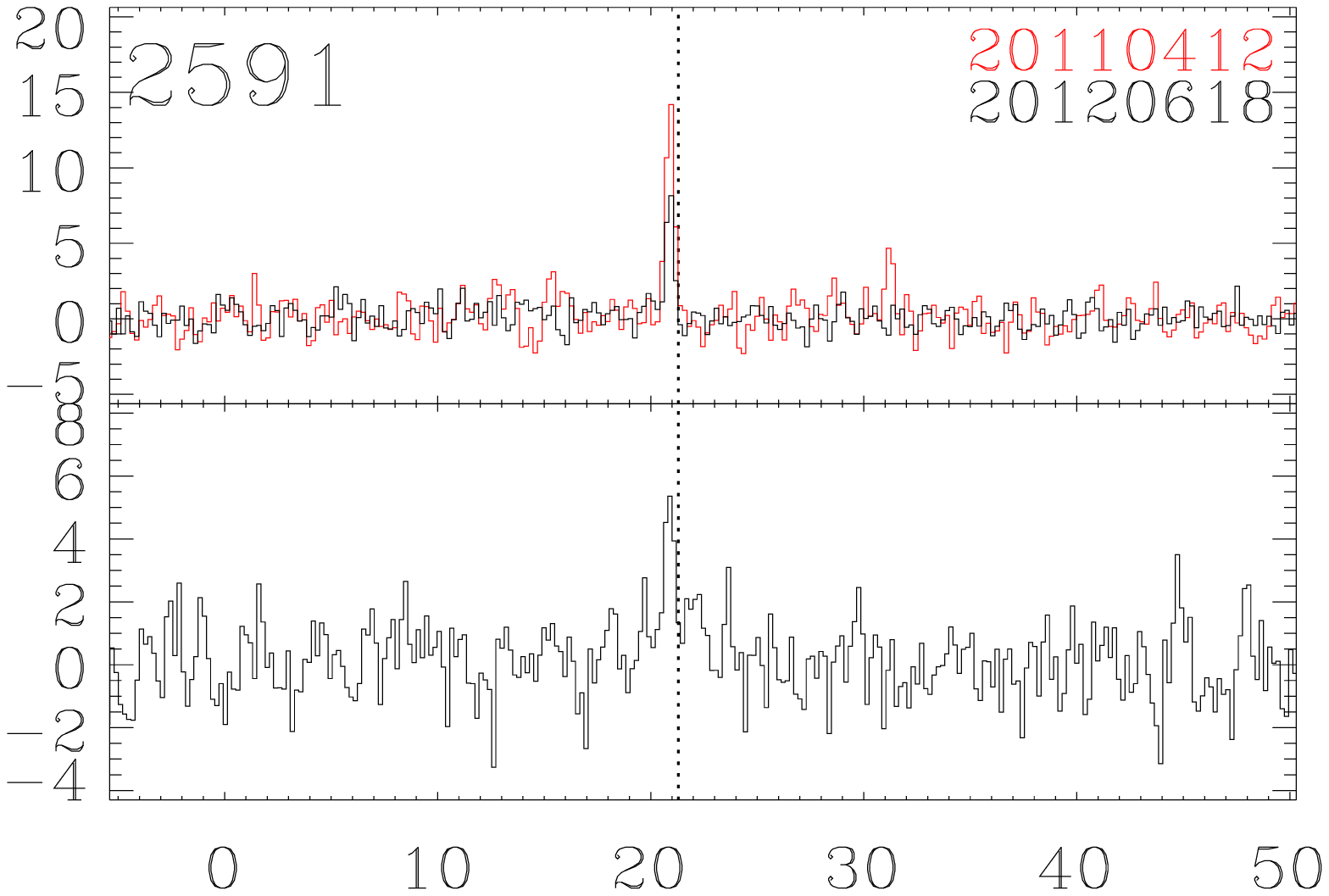} 
&
& 
\includegraphics[width=50mm]{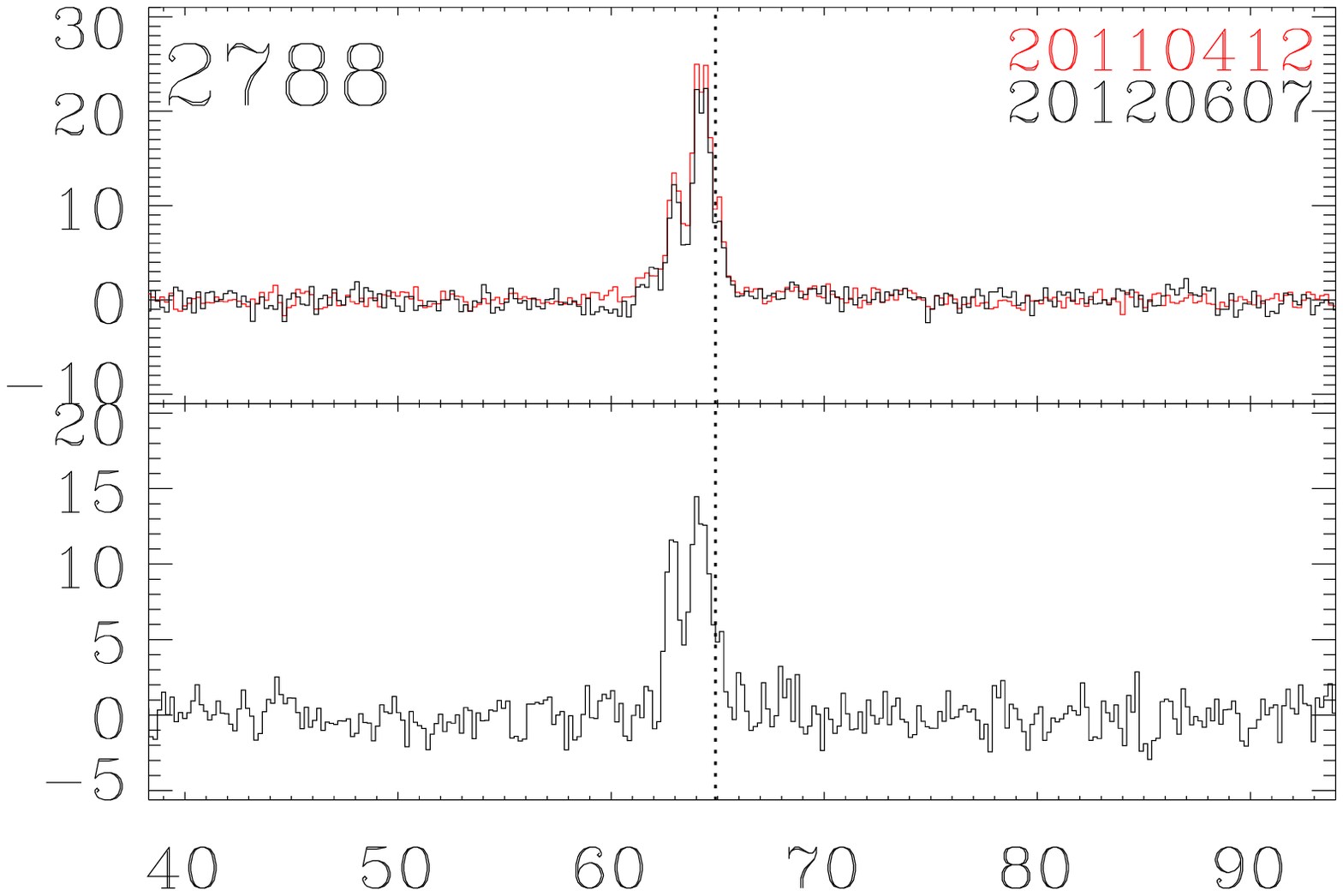} 
\\
&
\includegraphics[width=50mm]{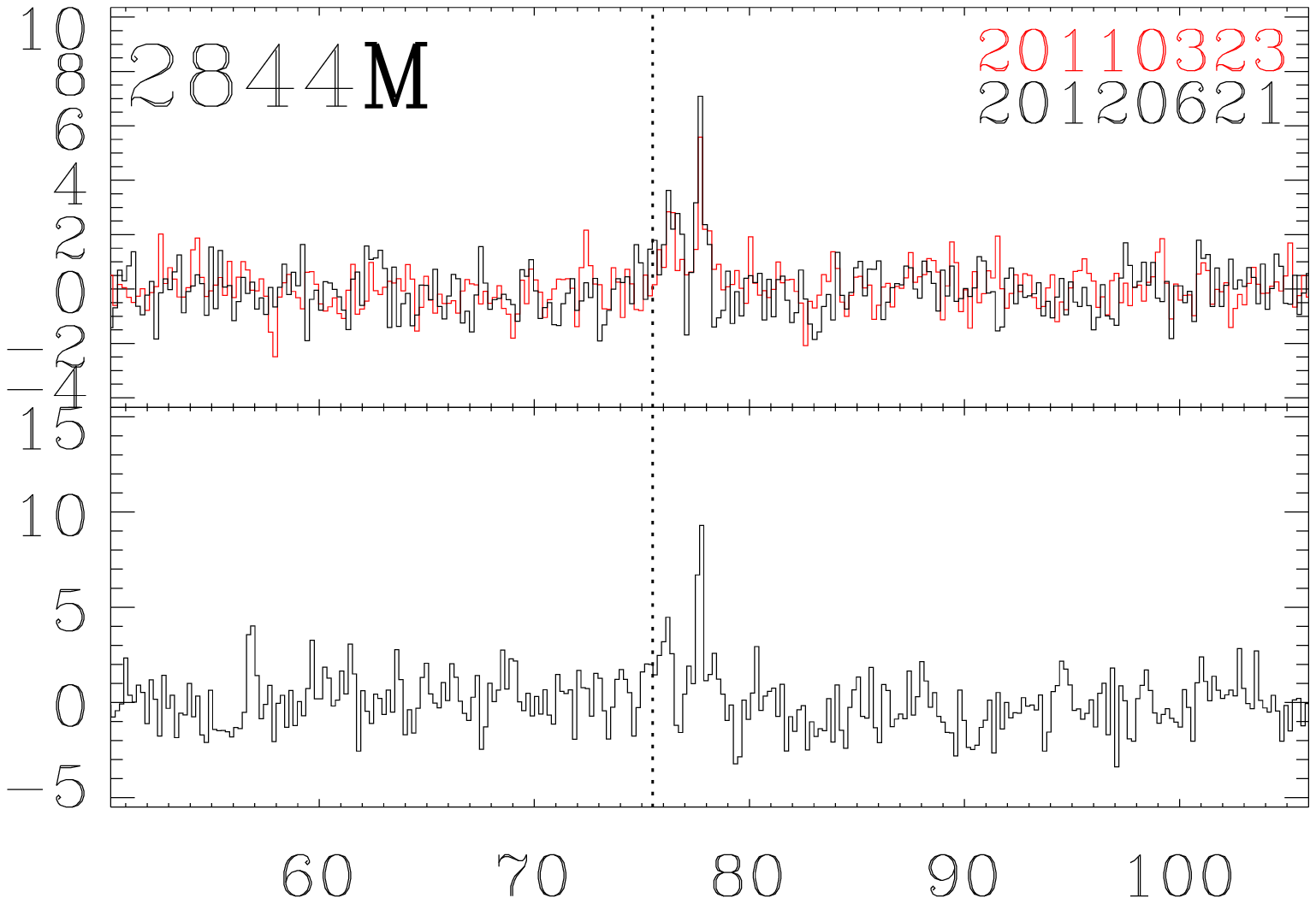} 
&
& 
\includegraphics[width=50mm]{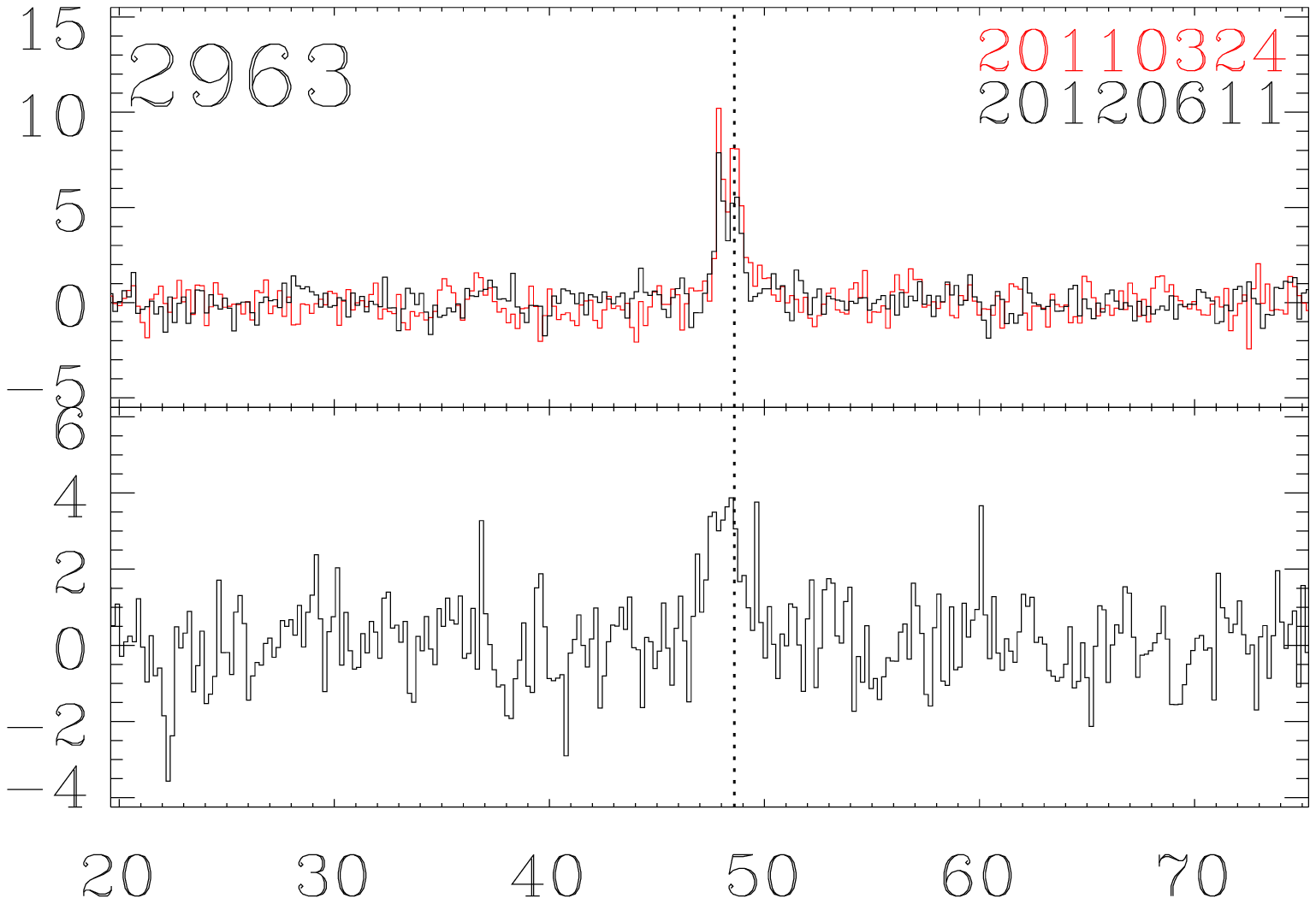} 
&
&
\includegraphics[width=50mm]{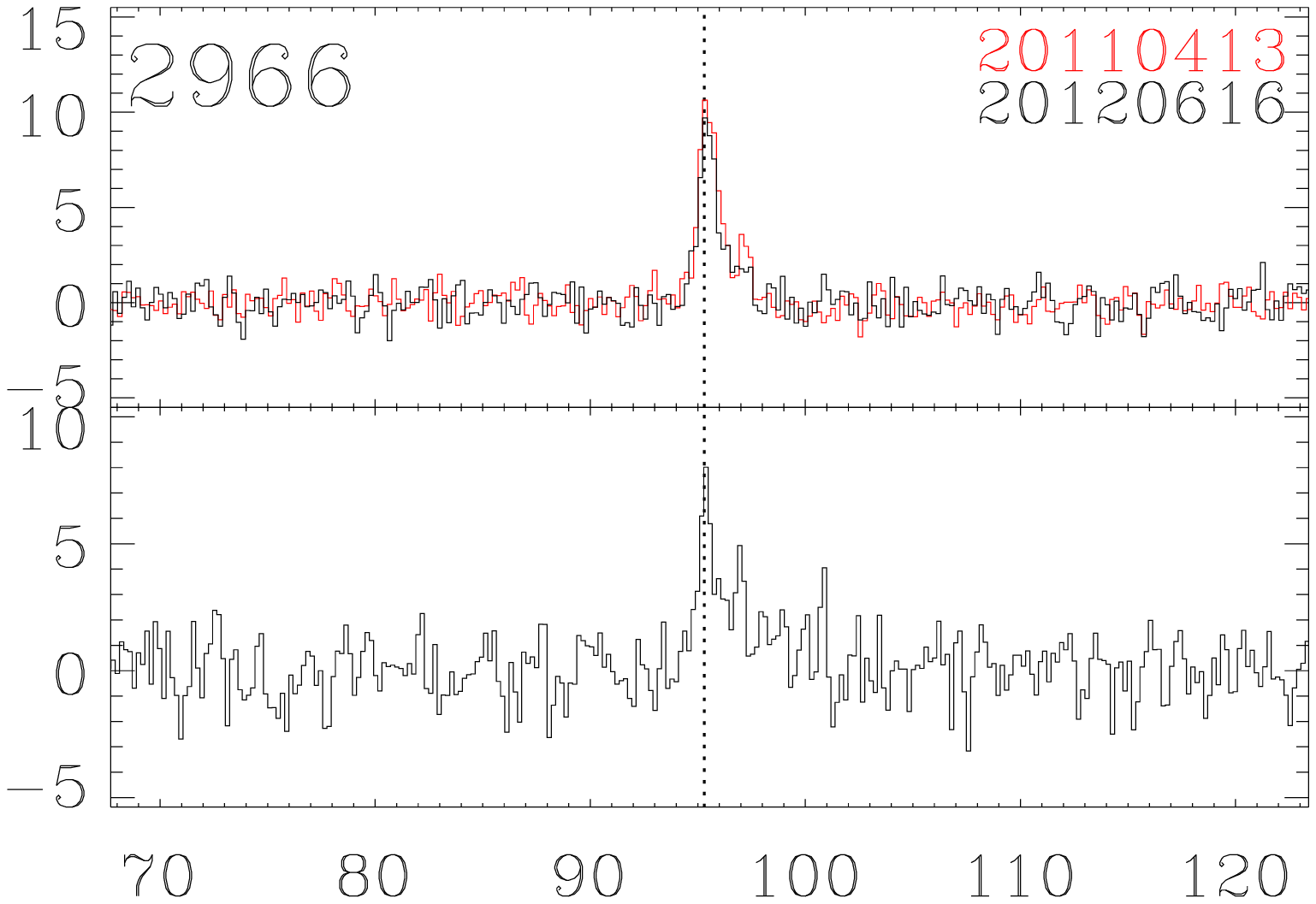} 
\\
&
\includegraphics[width=50mm]{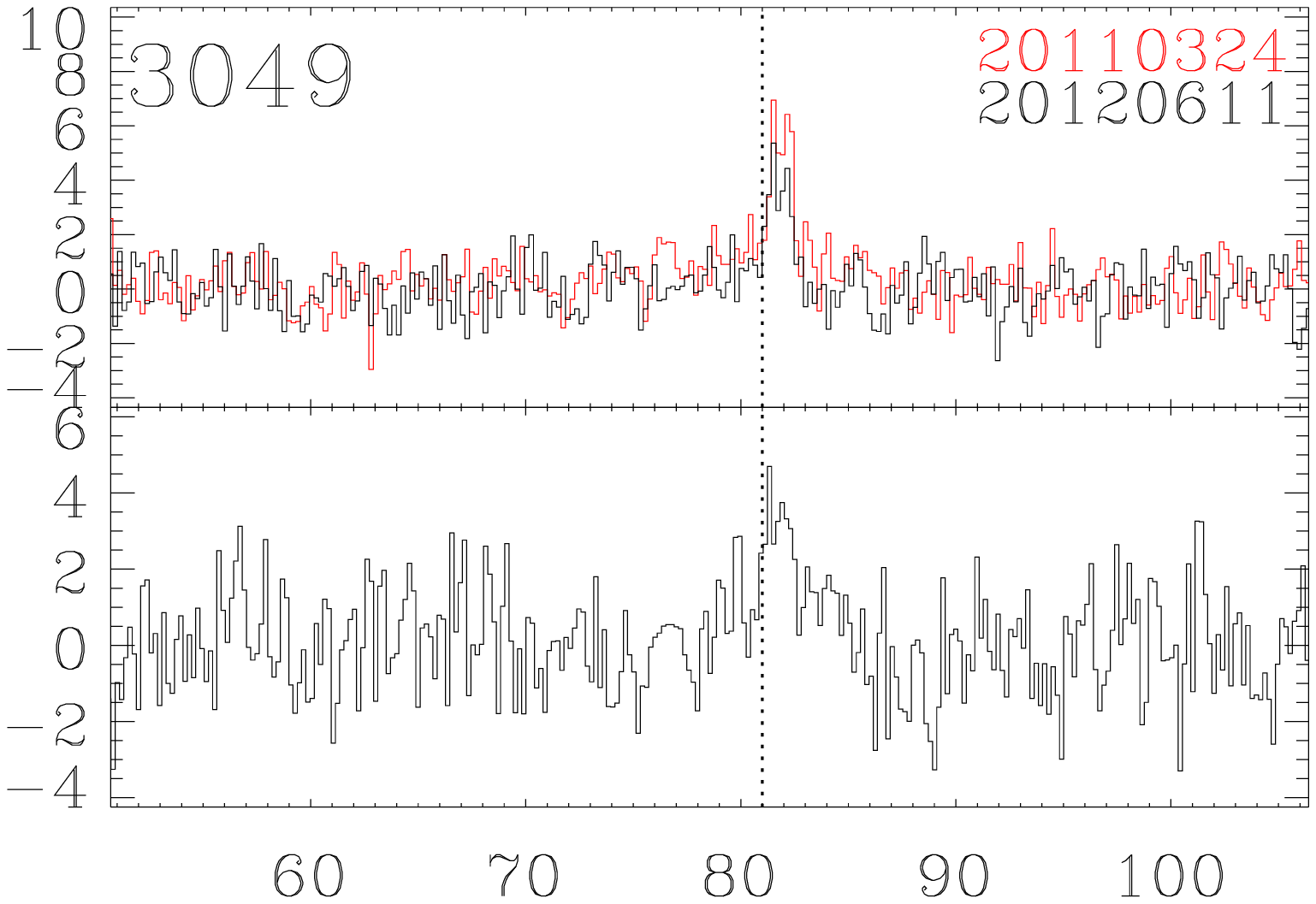} 
&
&  
\includegraphics[width=50mm]{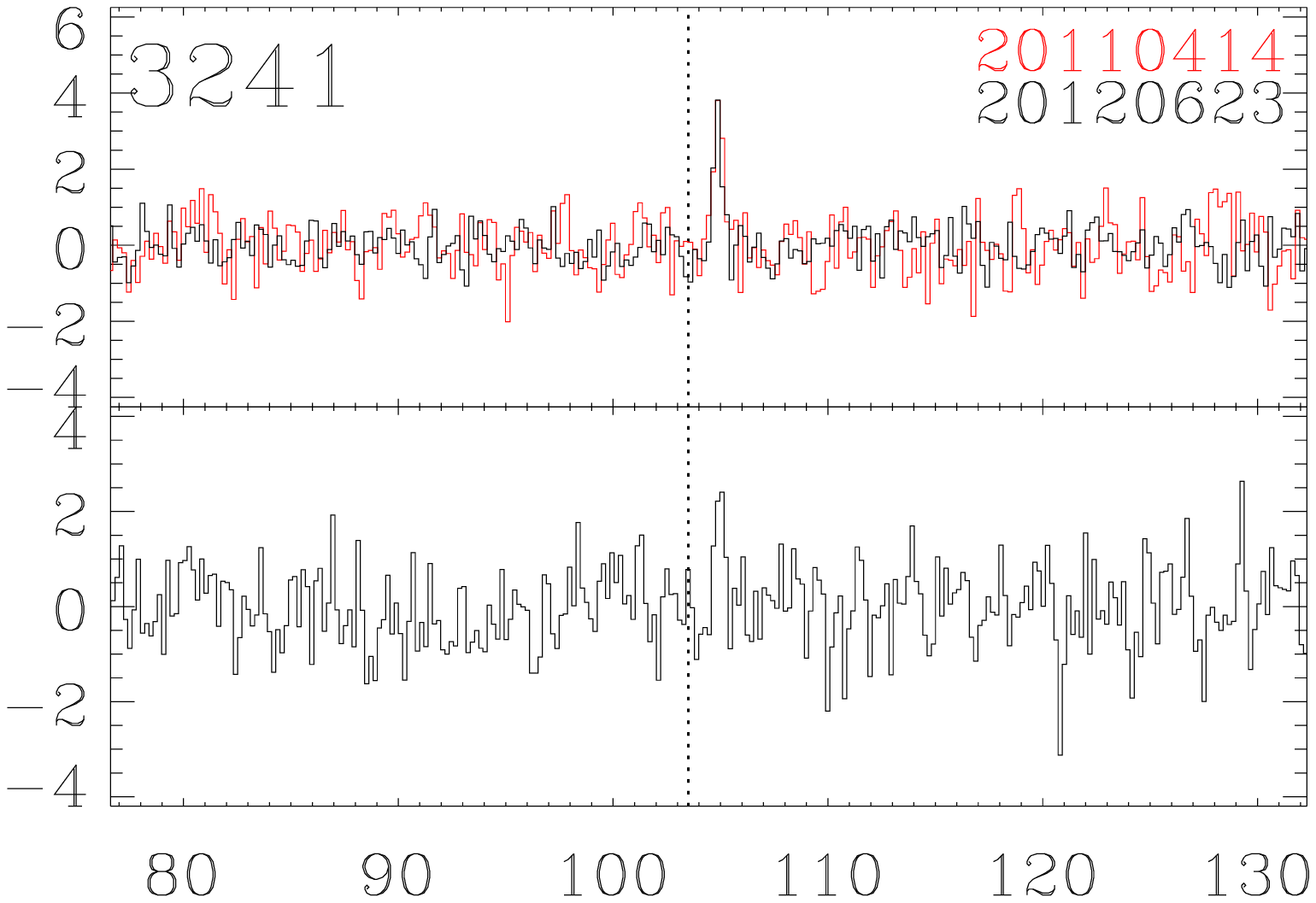} 
&
&
\includegraphics[width=50mm]{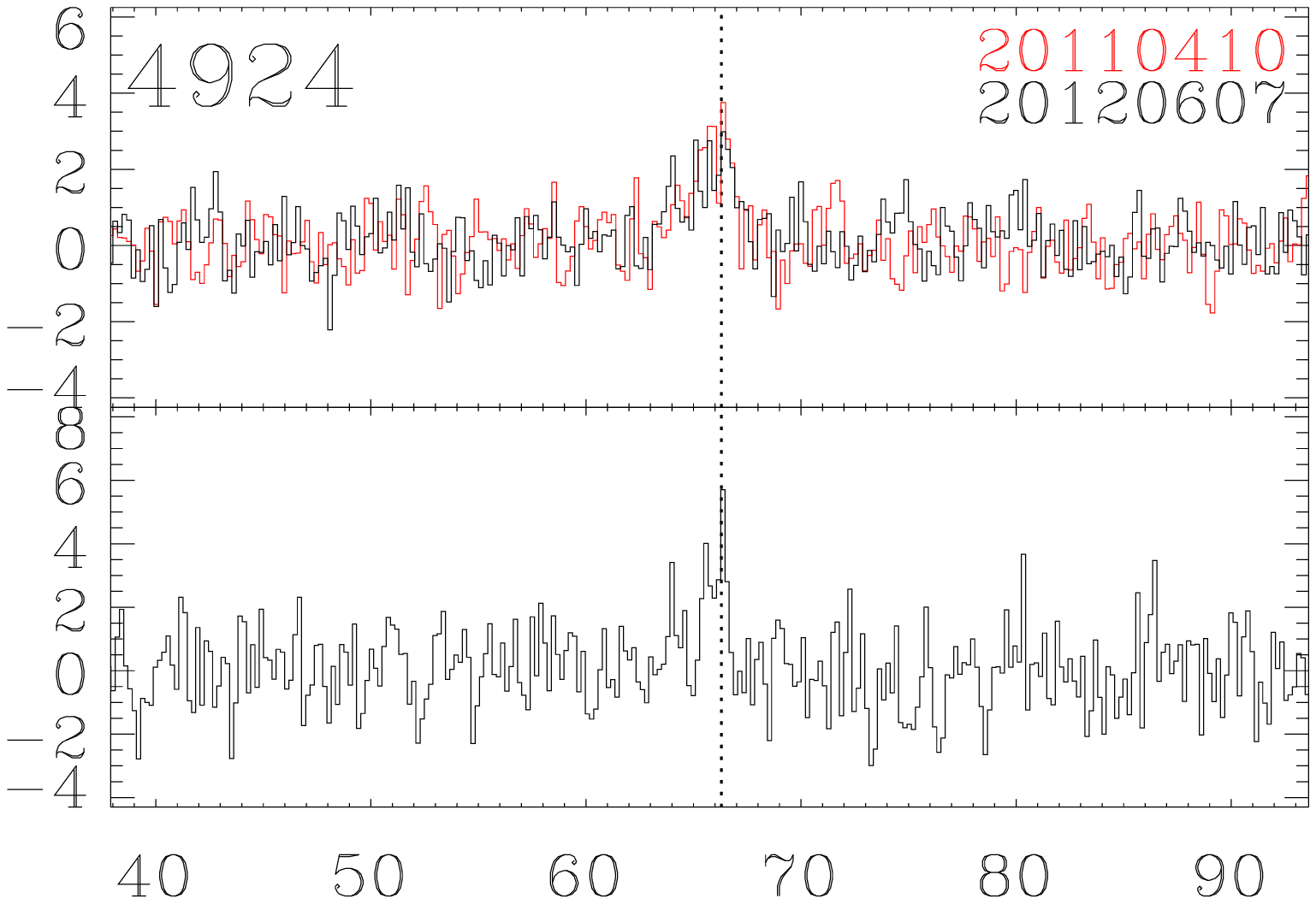} 
\\
\end{tabular}
\begin{minipage}{\linewidth}
\center{LSR velocity (\kms)}
\end{minipage}
\caption{Same as in Figure~\ref{fig_224495} except for the source detected both at (upper panel) 44~GHz and (lower panel) 95~GHz.
\label{fig_4495}}
\end{figure}
\clearpage

\begin{figure}
\epsscale{1}
\begin{minipage}{0.1cm}
\rotatebox{90}{\textcolor{black}{Flux density (Jy)}}
\end{minipage}
\begin{minipage}{\linewidth}
\plotone{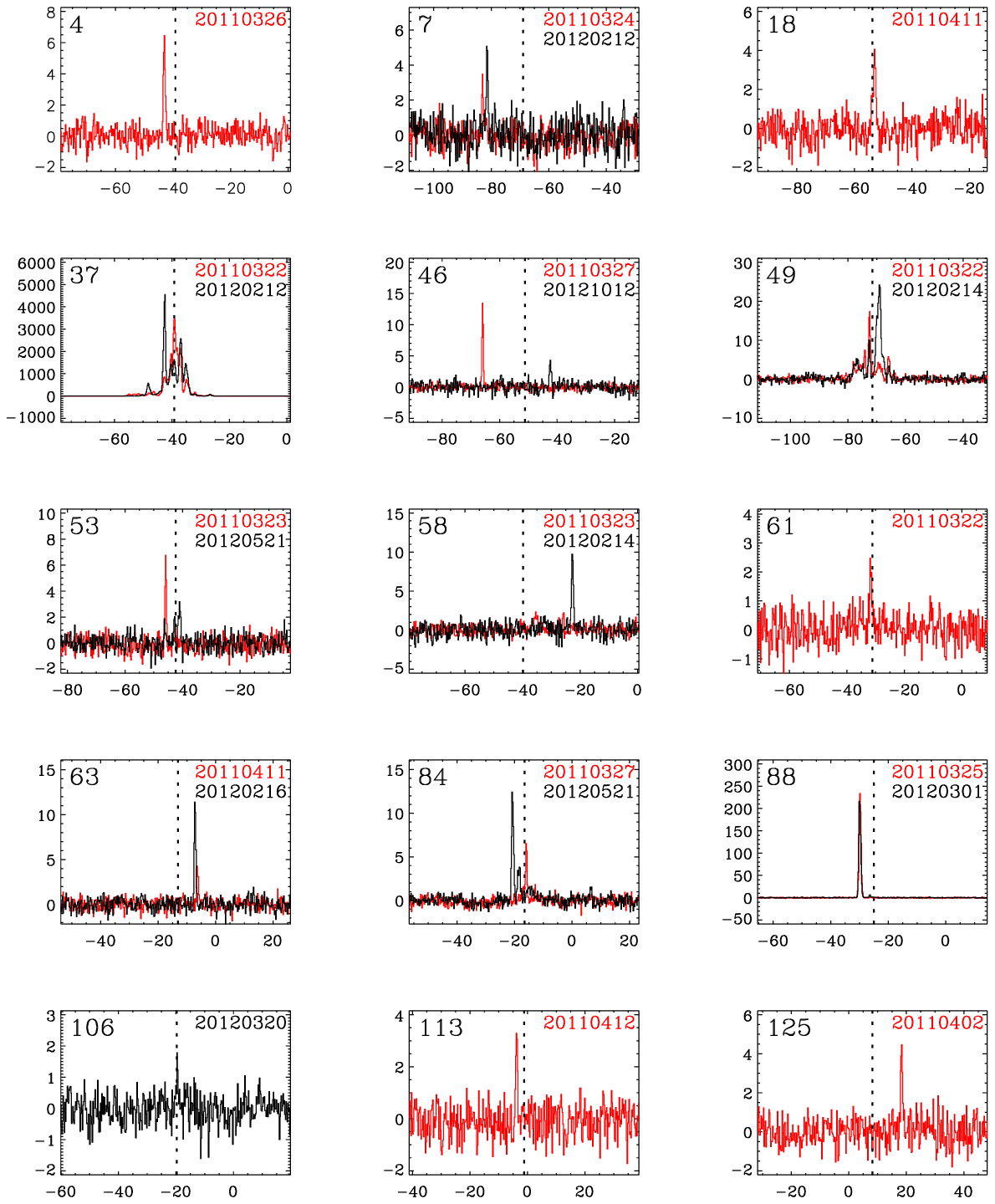}
{LSR velocity (\kms)}
\end{minipage}
\caption{Same as in Figure~\ref{fig_224495} except for the source detected only at 22~GHz.
\label{fig_22}}
\end{figure}
\clearpage

\begin{figure}
\figurenum{5}
\epsscale{1}
\begin{minipage}{0.1cm}
\rotatebox{90}{\textcolor{black}{Flux density (Jy)}}
\end{minipage}
\begin{minipage}{\linewidth}
\plotone{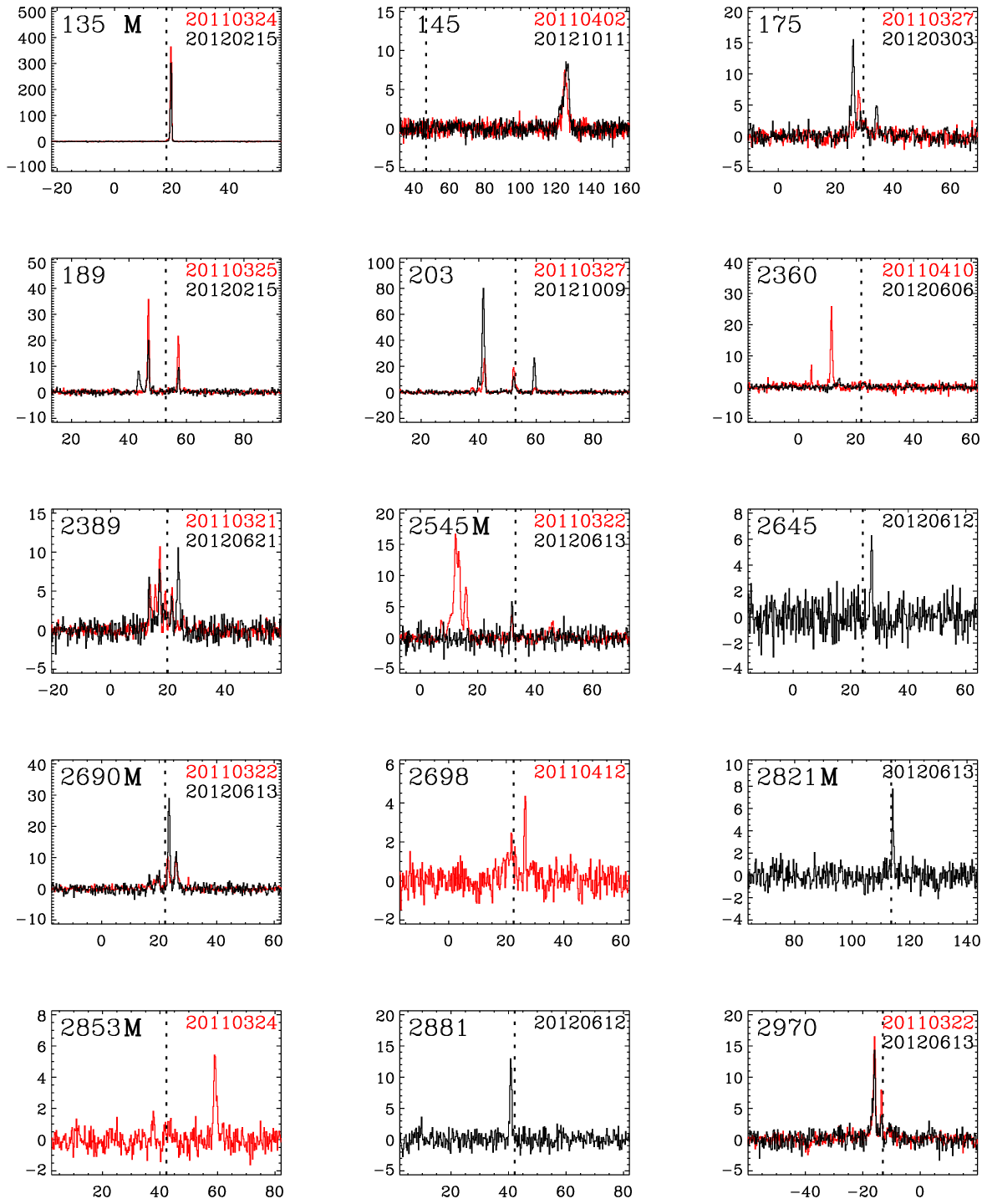}
{LSR velocity (\kms)}
\end{minipage}
\caption{Continued}
\end{figure}
\clearpage

\begin{figure}
\figurenum{5}
\epsscale{1}
\begin{minipage}{0.1cm}
\rotatebox{90}{\textcolor{black}{Flux density (Jy)}}
\end{minipage}
\begin{minipage}{\linewidth}
\plotone{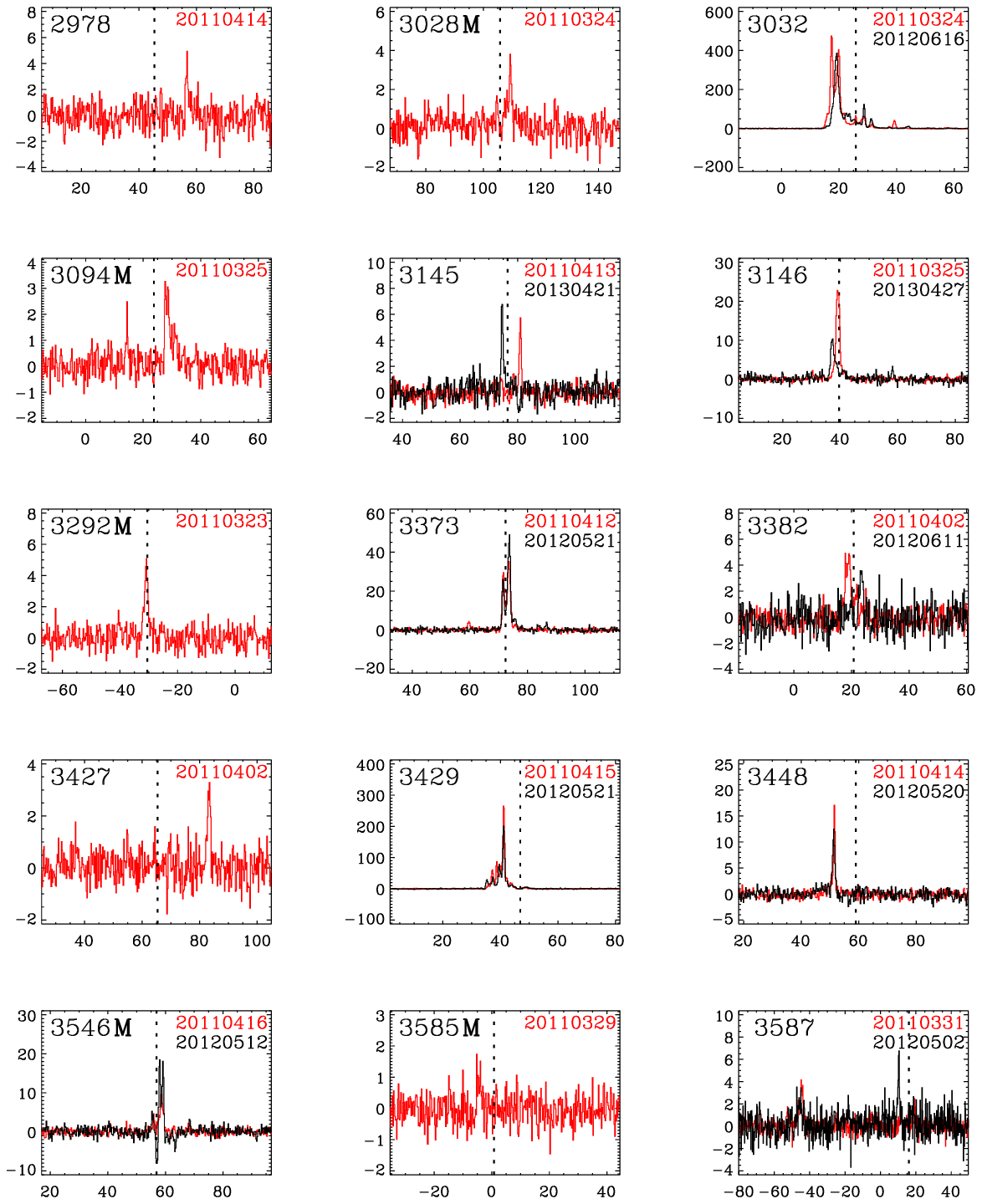}
{LSR velocity (\kms)}
\end{minipage}
\caption{Continued}
\end{figure}
\clearpage

\begin{figure}
\figurenum{5}
\epsscale{1}
\begin{minipage}{0.1cm}
\rotatebox{90}{\textcolor{black}{Flux density (Jy)}}
\end{minipage}
\begin{minipage}{\linewidth}
\plotone{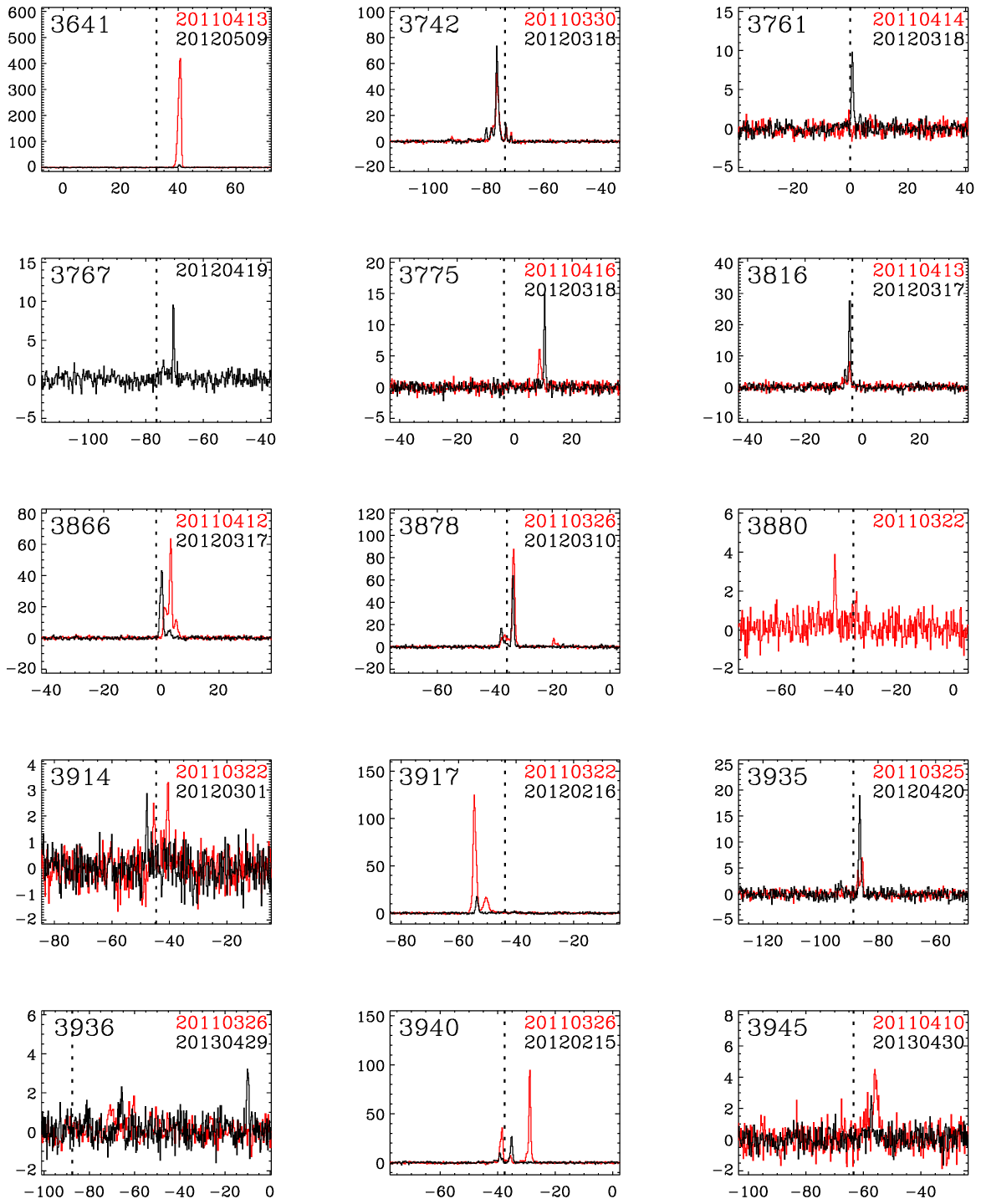}
{LSR velocity (\kms)}
\end{minipage}
\caption{Continued}
\end{figure}
\clearpage

\begin{figure}
\figurenum{5}
\epsscale{1}
\begin{minipage}{0.1cm}
\end{minipage}
\begin{minipage}{\linewidth}
\plotone{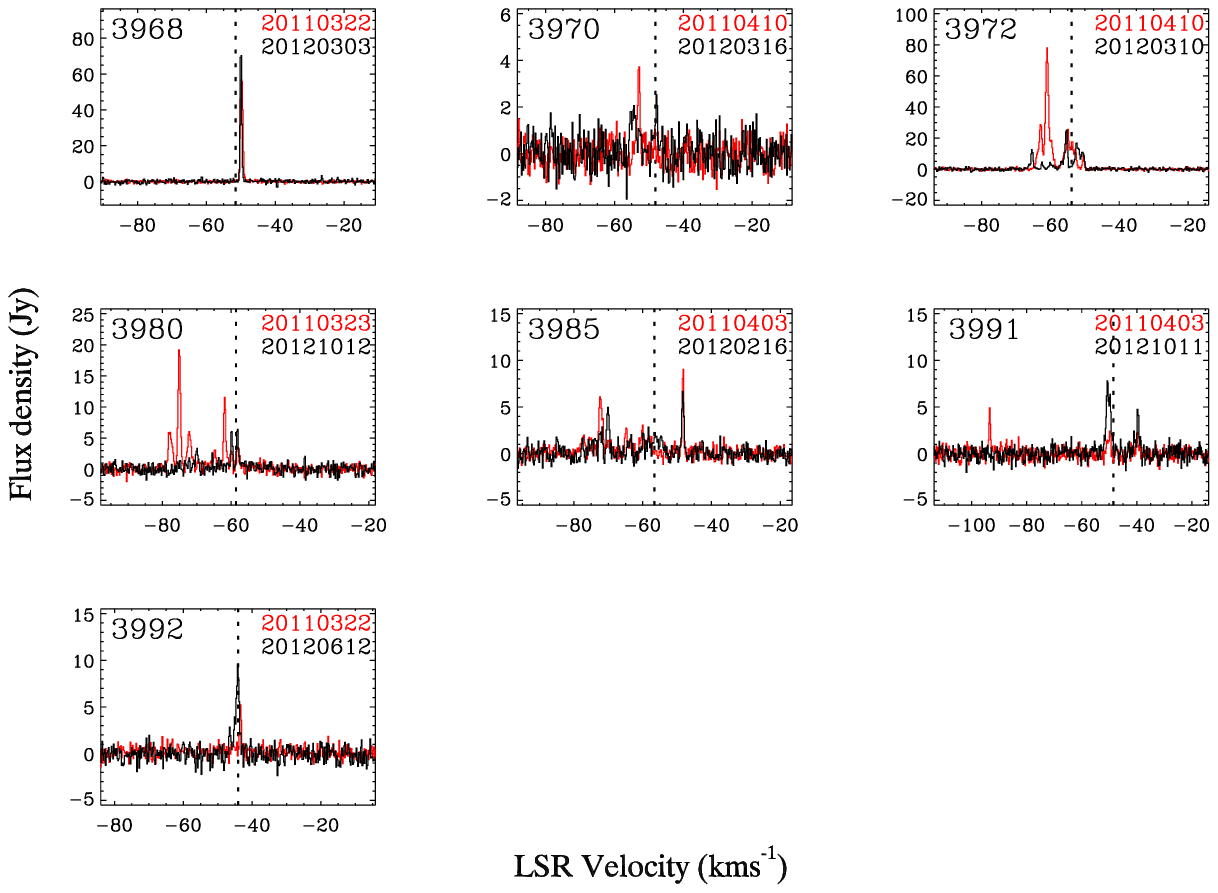}
\end{minipage}
\caption{Continued}
\end{figure}
\clearpage

\begin{figure}
\epsscale{1}
\plotone{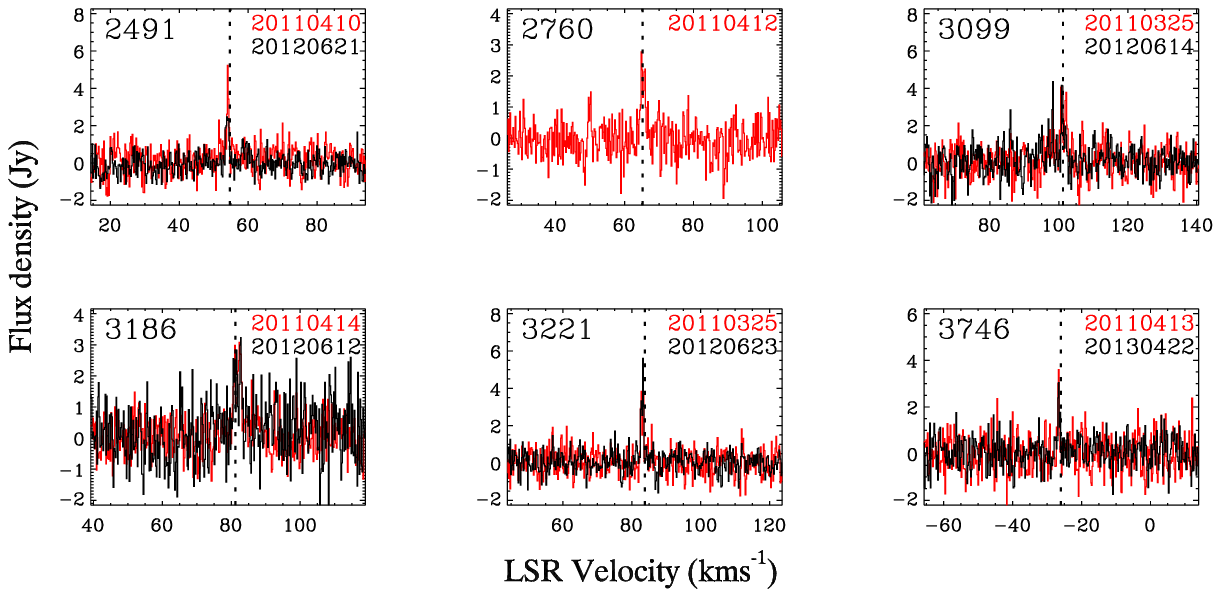}
\caption{Same as in Figure~\ref{fig_224495} except for the source detected only at 44~GHz.
\label{fig_44}}
\end{figure}
\clearpage

\begin{figure}
\epsscale{1}
\plotone{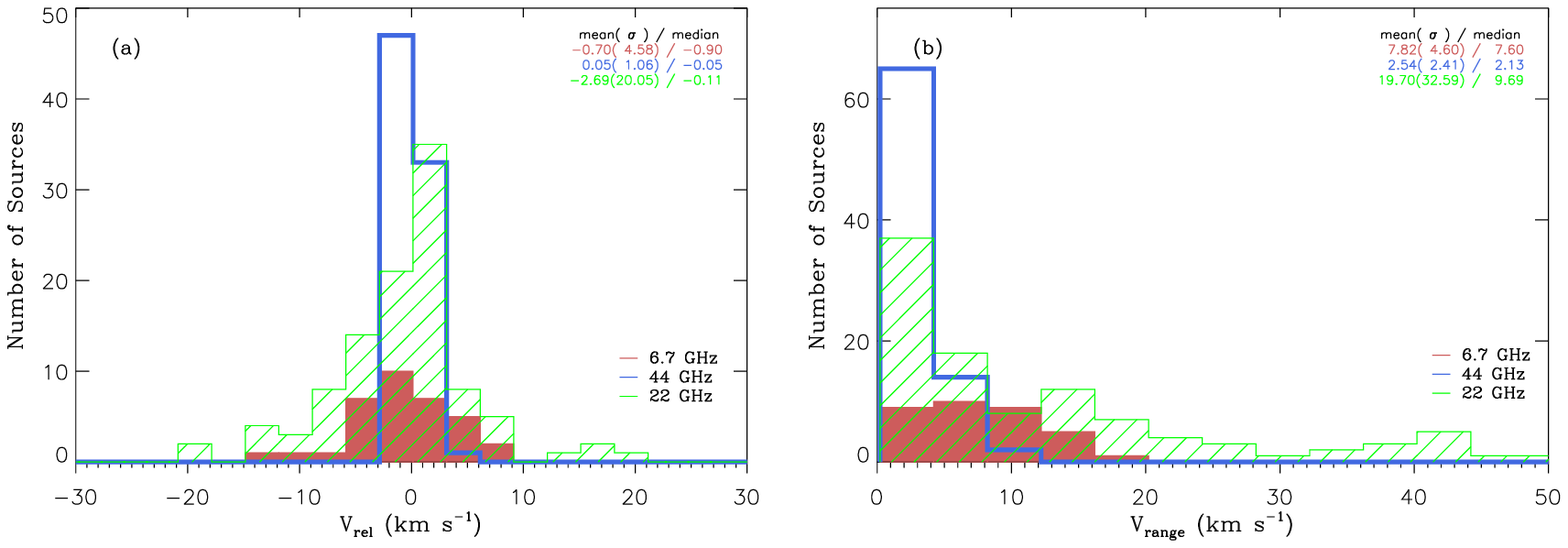}
\caption{
(a) Histogram of the relative peak velocities for the 22, 44, and 6.7~GHz maser sources in a bin size of 3~\kms. Green, blue, and brown colors represent 22, 44, and 6.7~GHz masers, respectively. The 22 and 44~GHz data are from the second-epoch survey of this study, while the 6.7~GHz data are from the MMB survey. The mean and median values of each distribution are listed in the top-right corner of each panel.
 (b) Same as in (a) except for the velocity ranges in 4 \kms\ bins. 
 \label{fig_Vrel_Vrange_hist}}
\end{figure}
\clearpage

\begin{figure}
\epsscale{1}
\begin{minipage}{0.1cm}
\rotatebox{90}{\textcolor{black}{Flux density (Jy) / T$_{\rm mb}$ (K)}}
\end{minipage}
\begin{tabular}{cccccc}
&
\includegraphics[width=50mm]{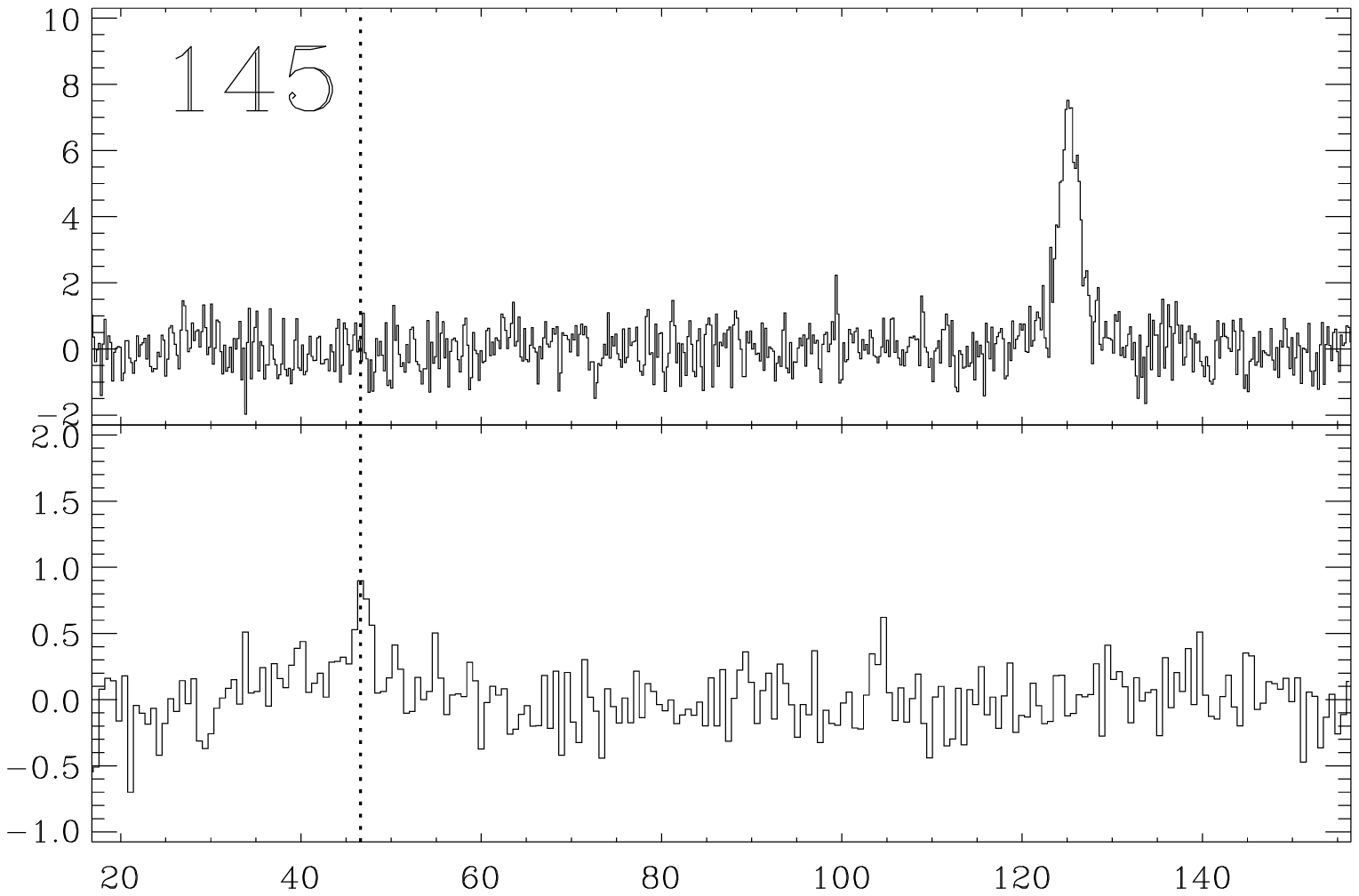} 
&
&
\includegraphics[width=50mm]{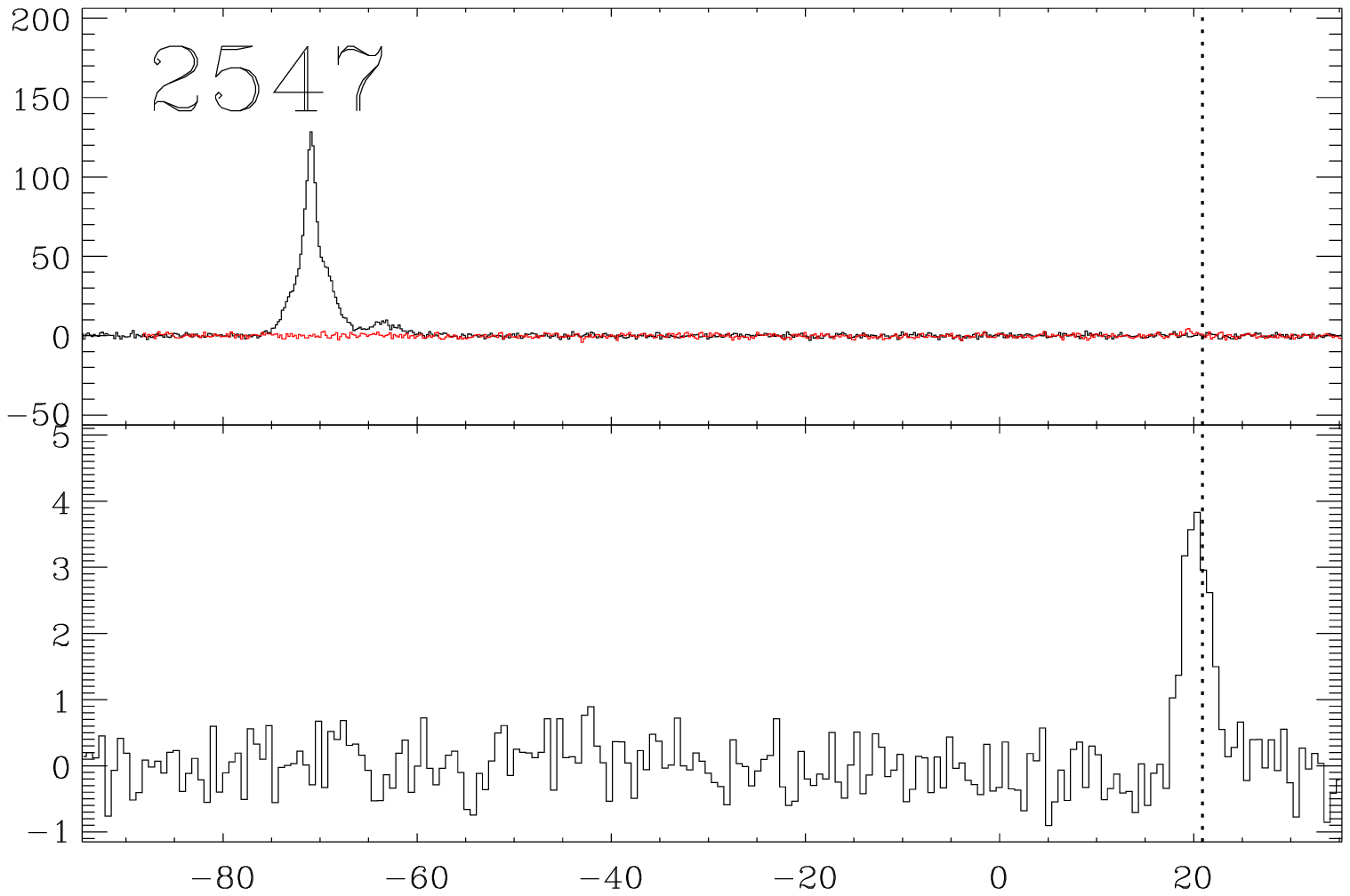} 
&
&
\includegraphics[width=50mm]{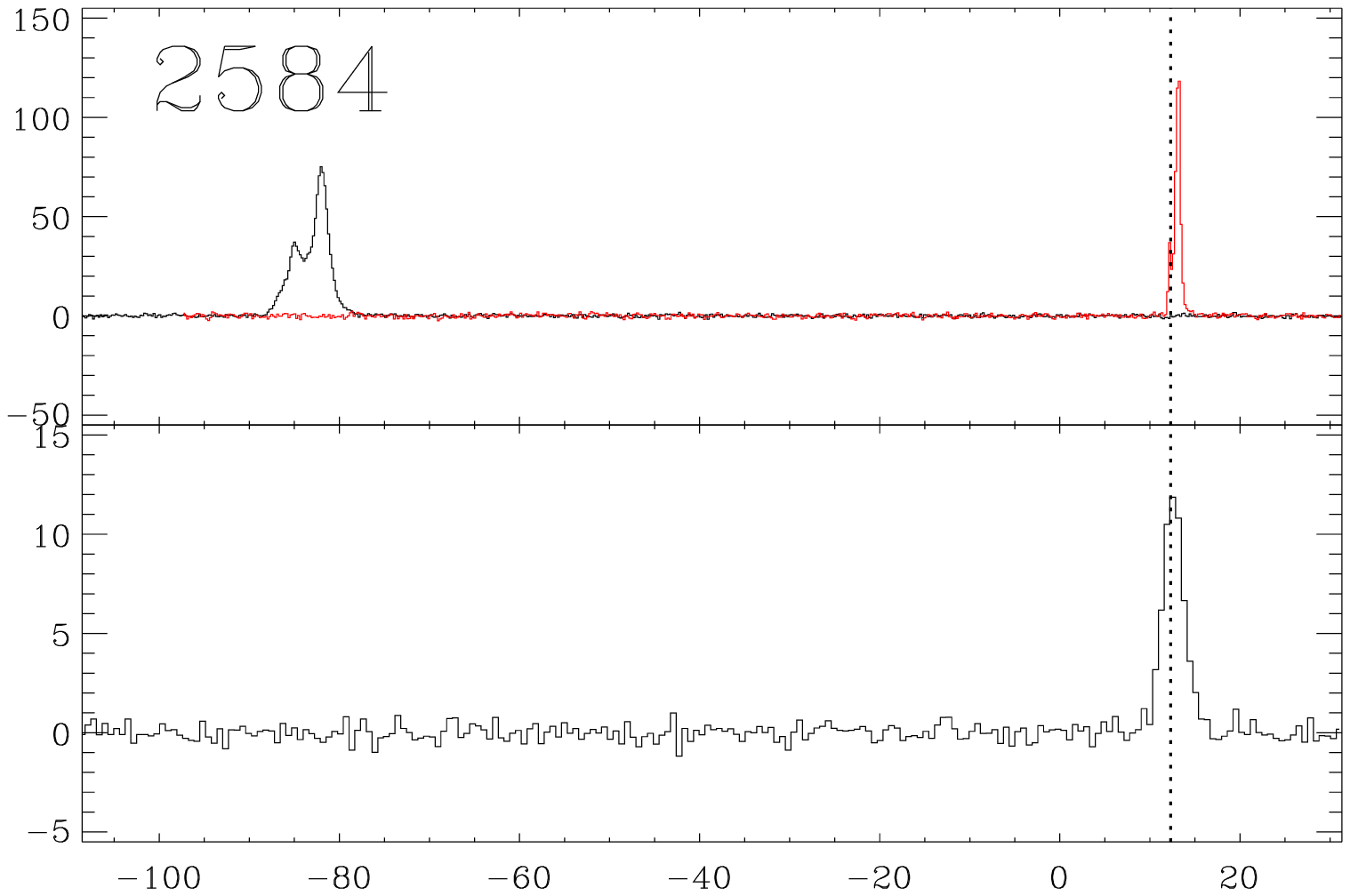}
\\
&
\includegraphics[width=50mm]{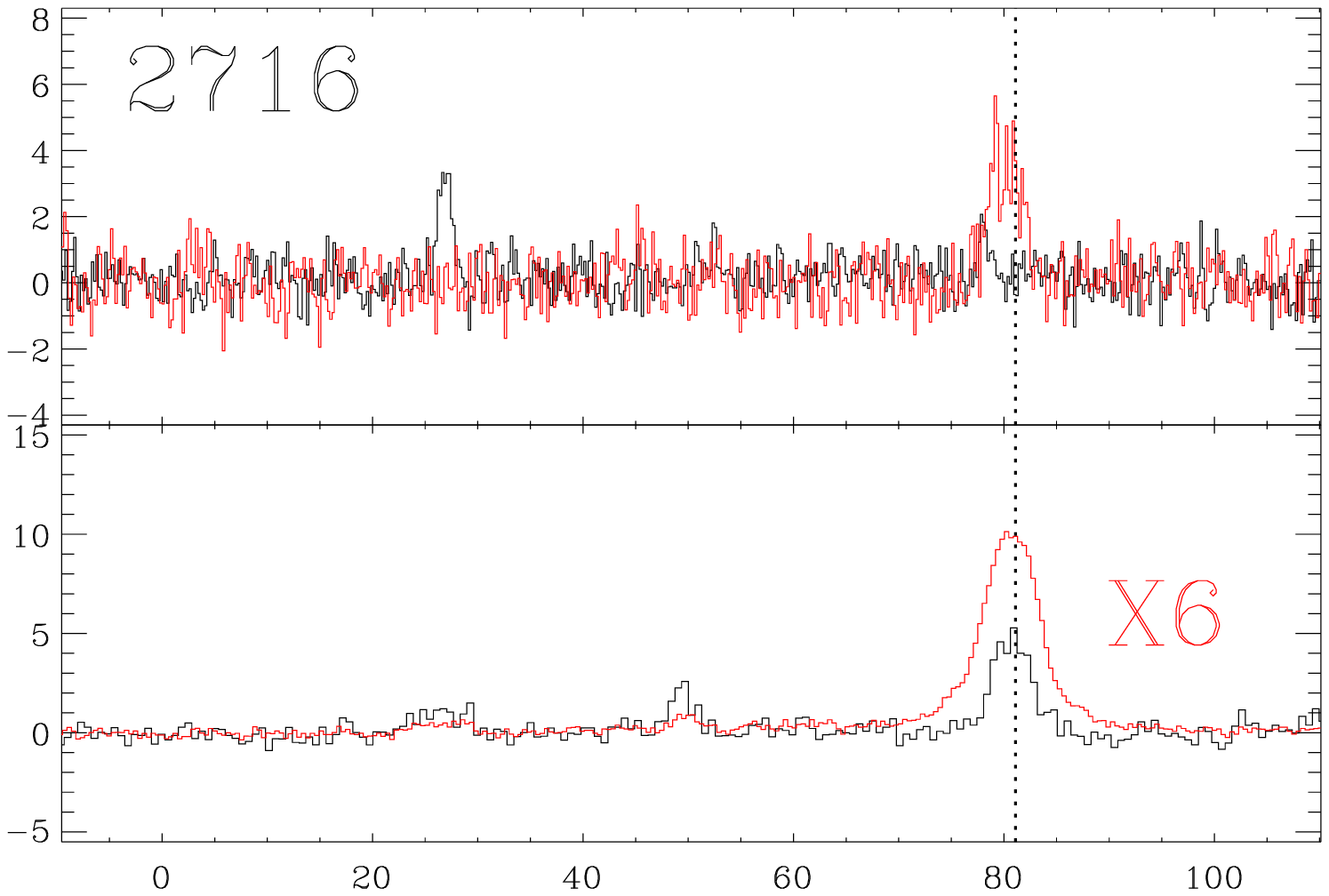} 
&
&
\includegraphics[width=50mm]{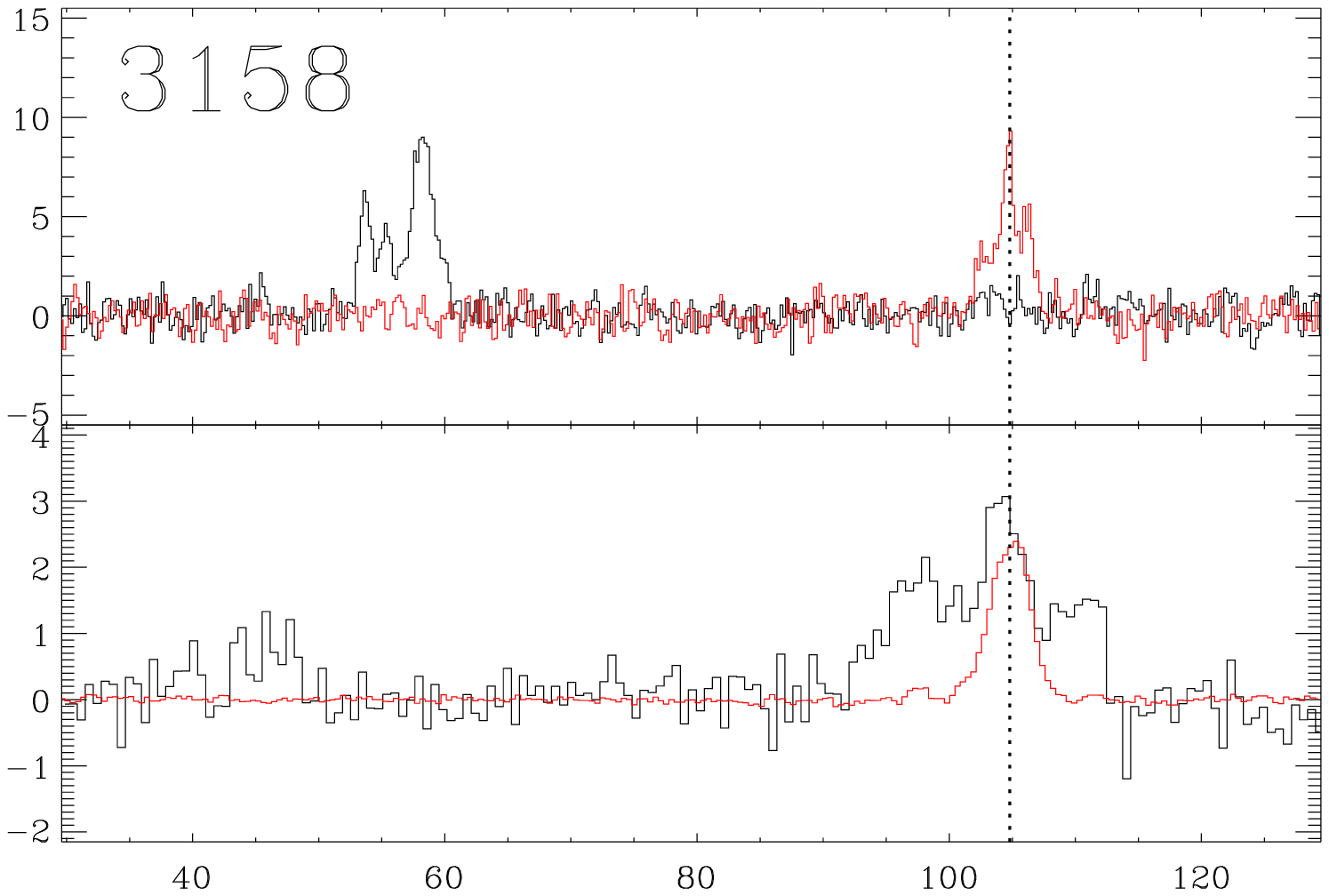}  
&
&
\includegraphics[width=50mm]{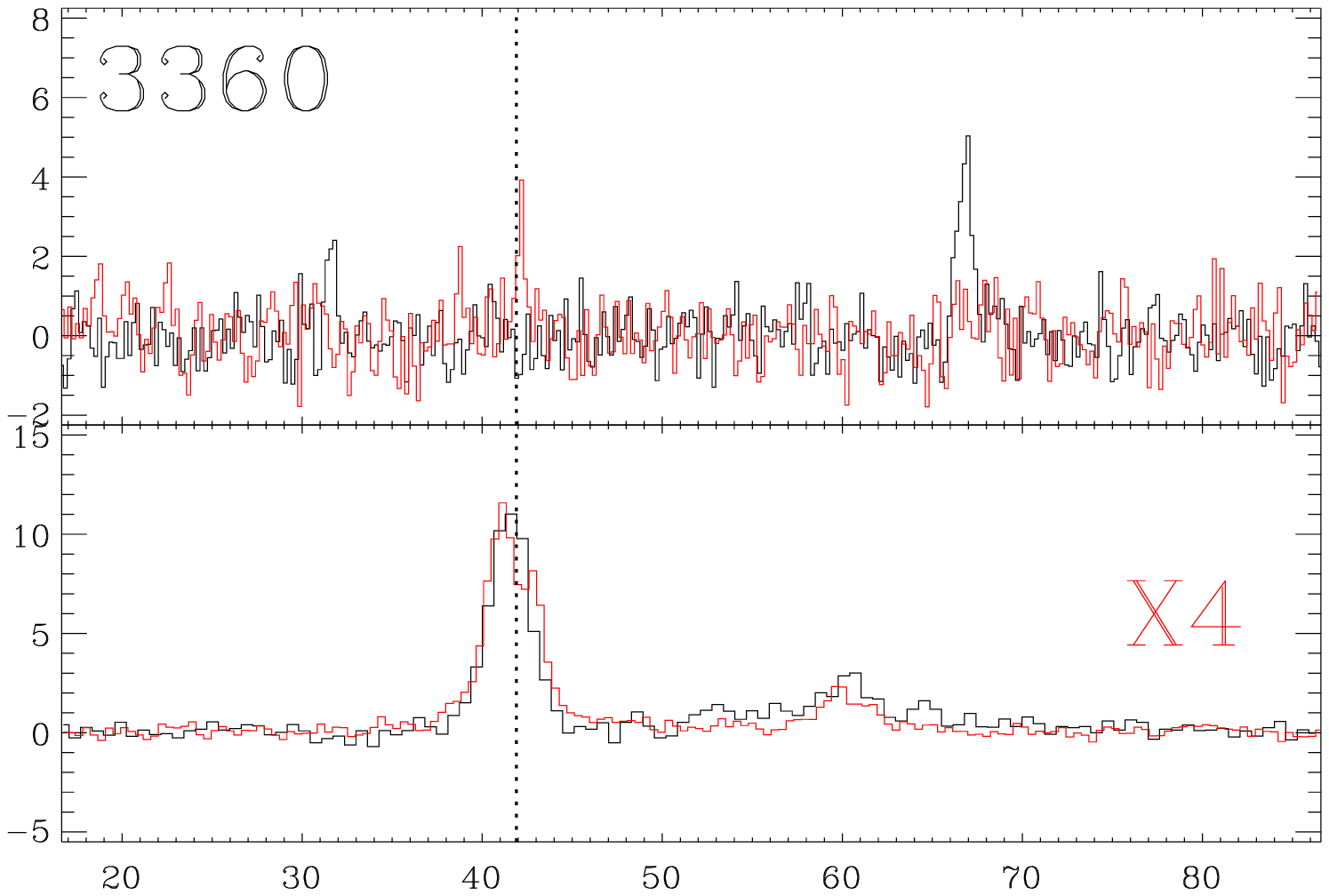} 
\\
&
\includegraphics[width=50mm]{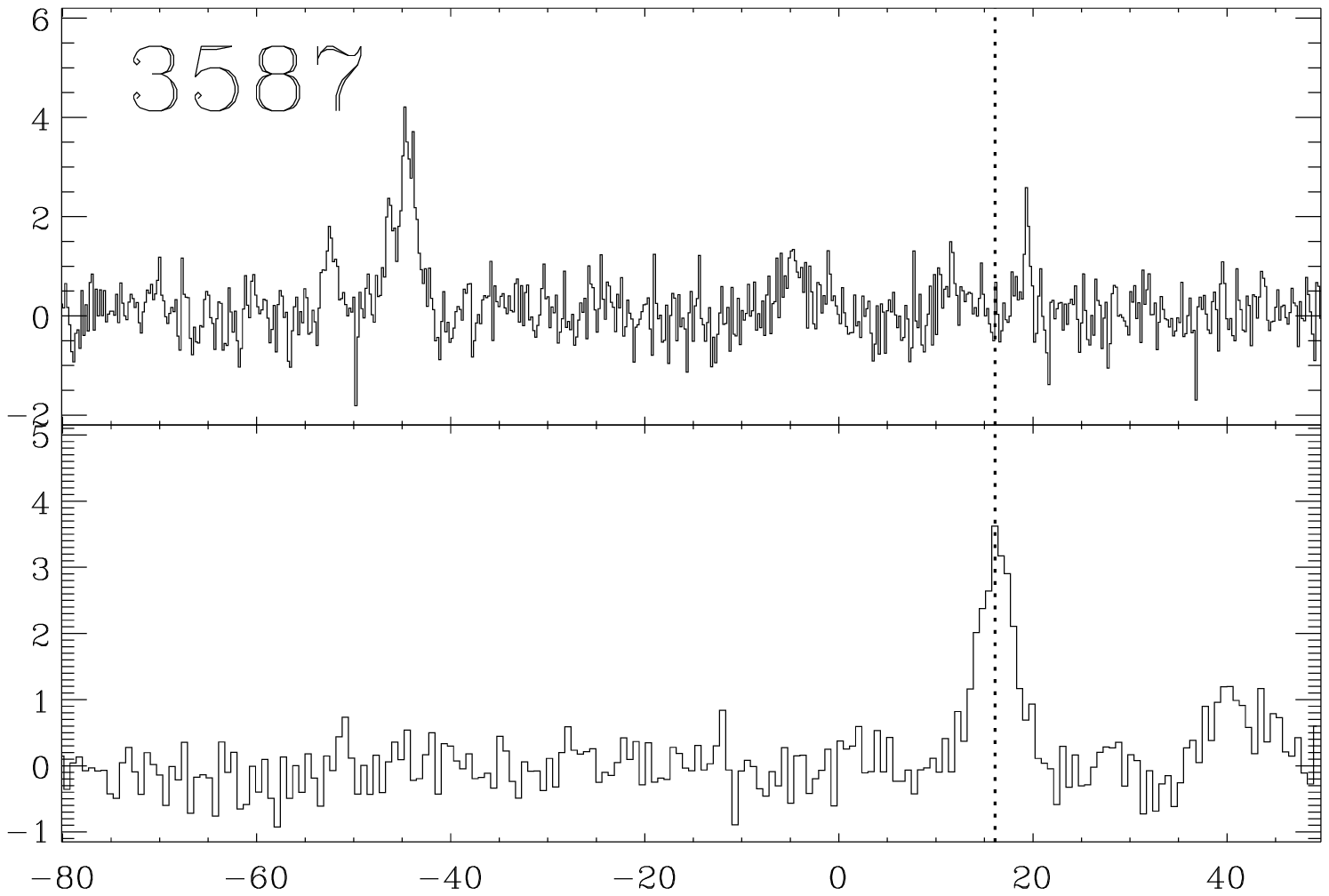}
&
&
\includegraphics[width=50mm]{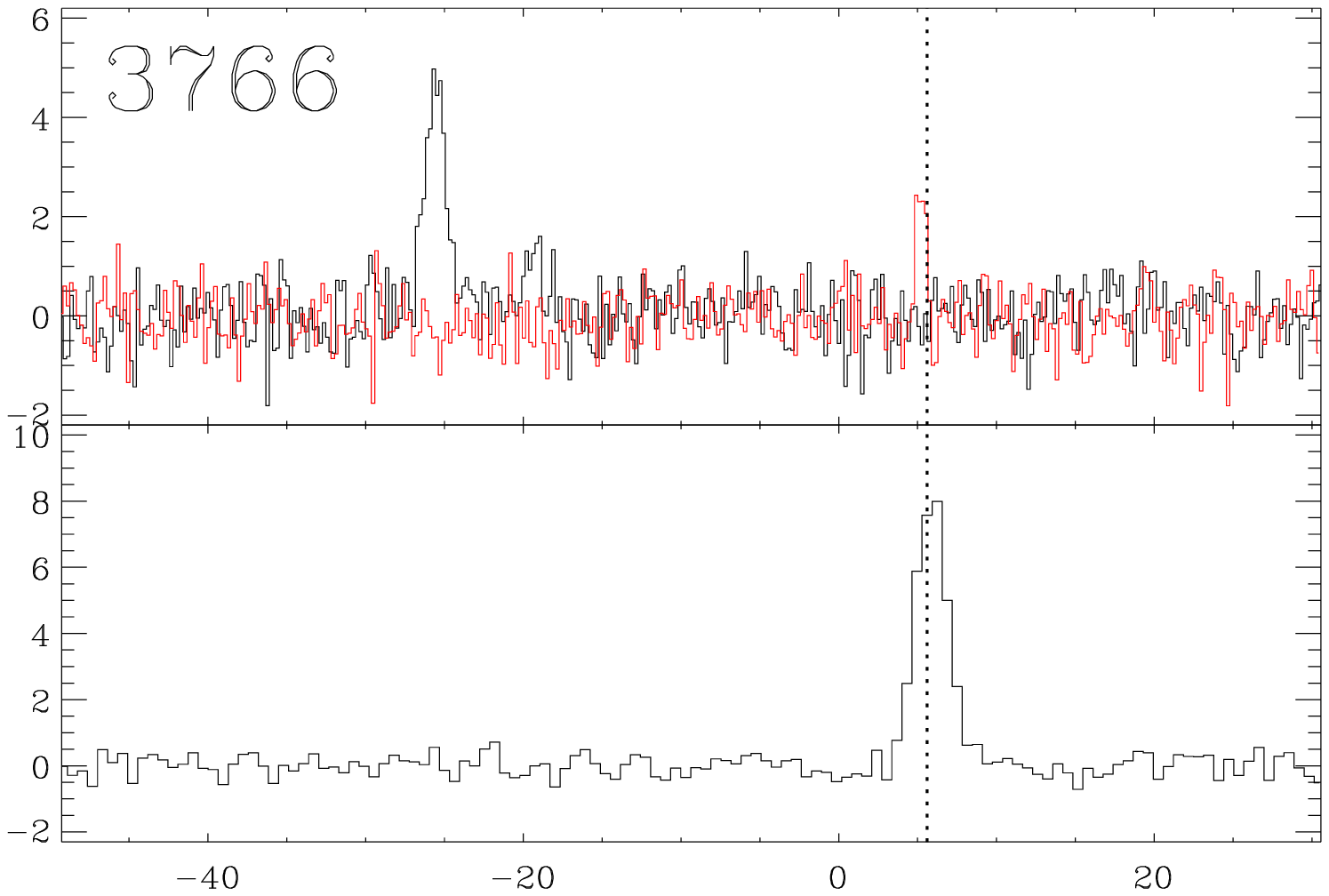} 
&
&
\includegraphics[width=50mm]{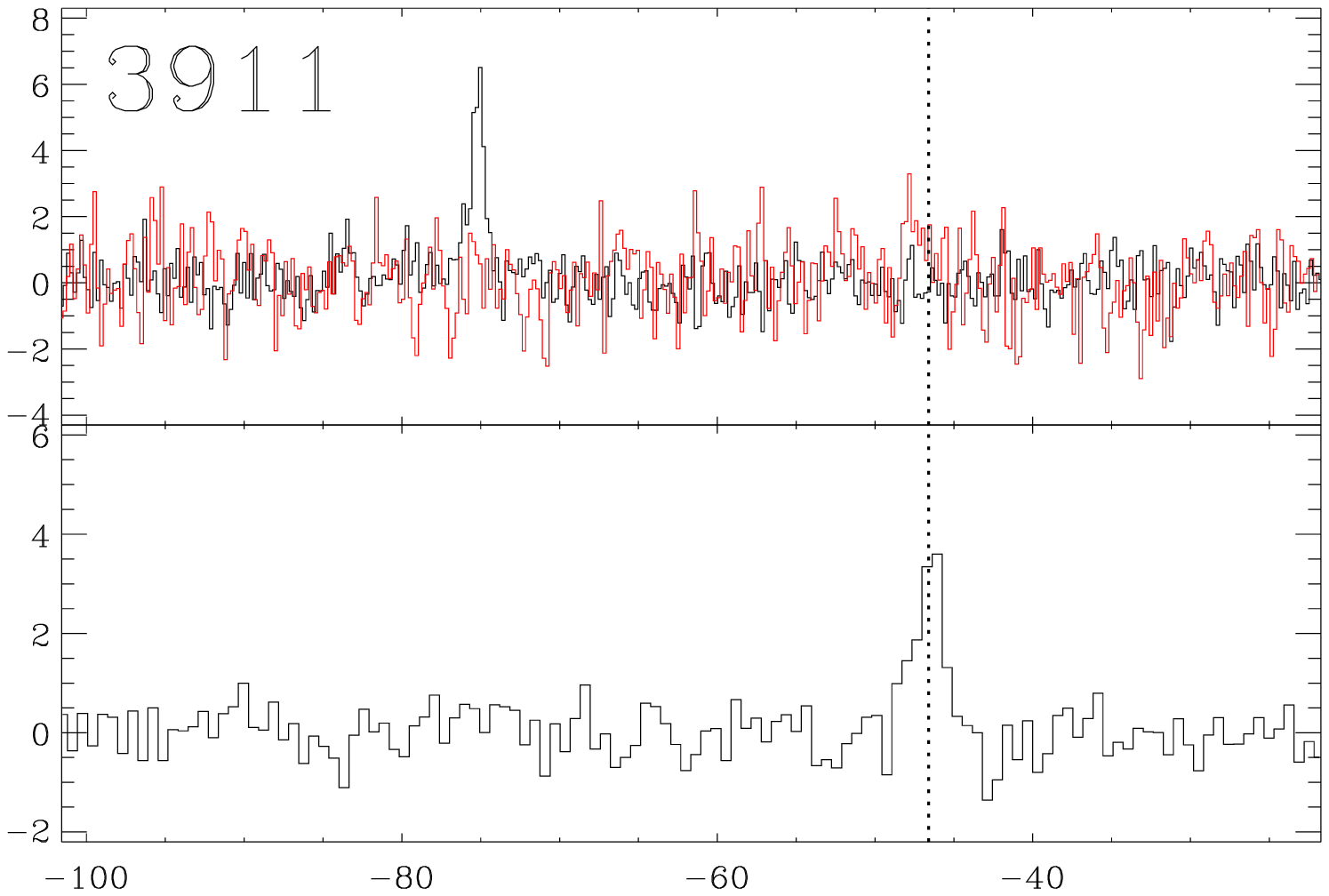}
\\
&
\includegraphics[width=50mm]{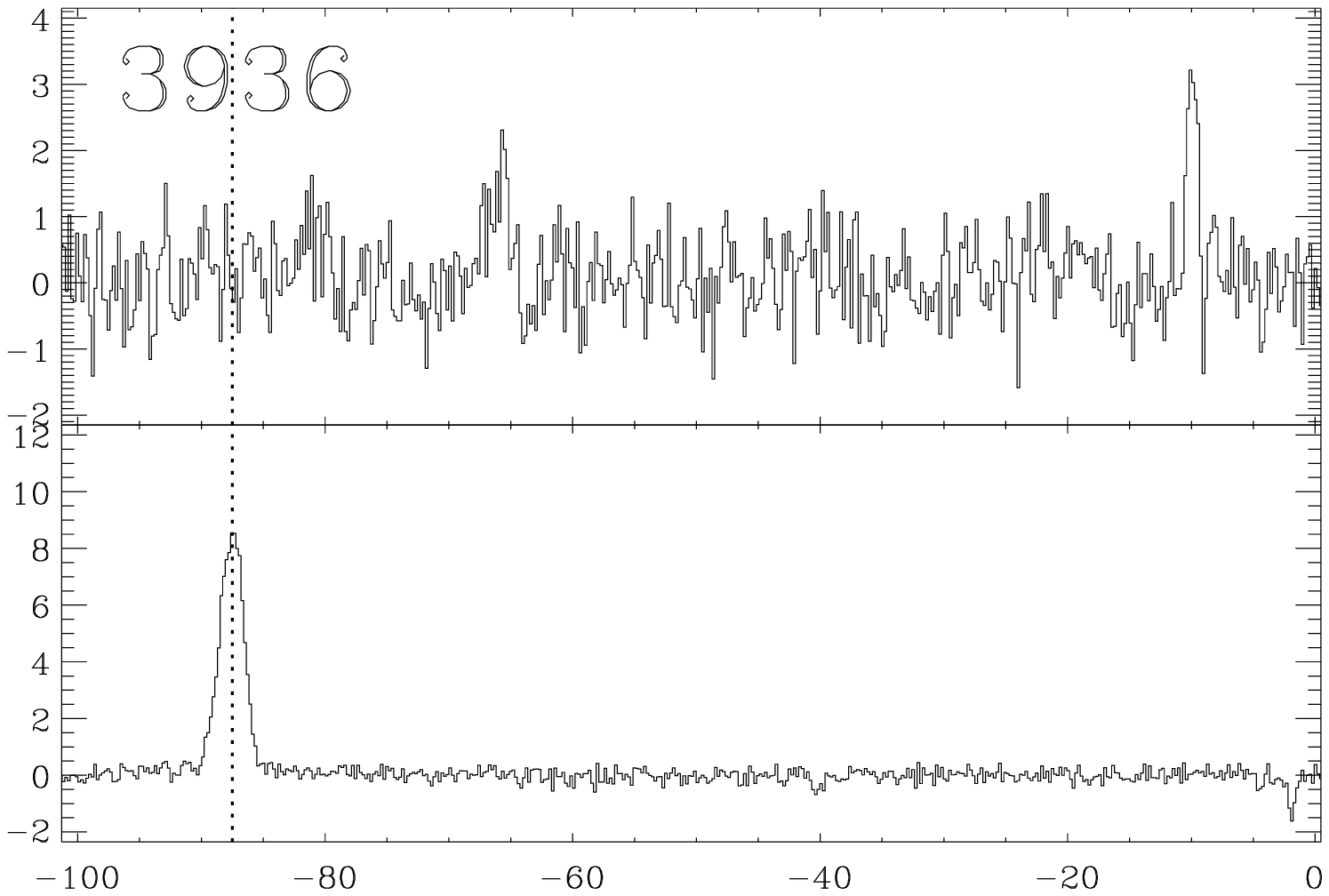}
&
&
\includegraphics[width=50mm]{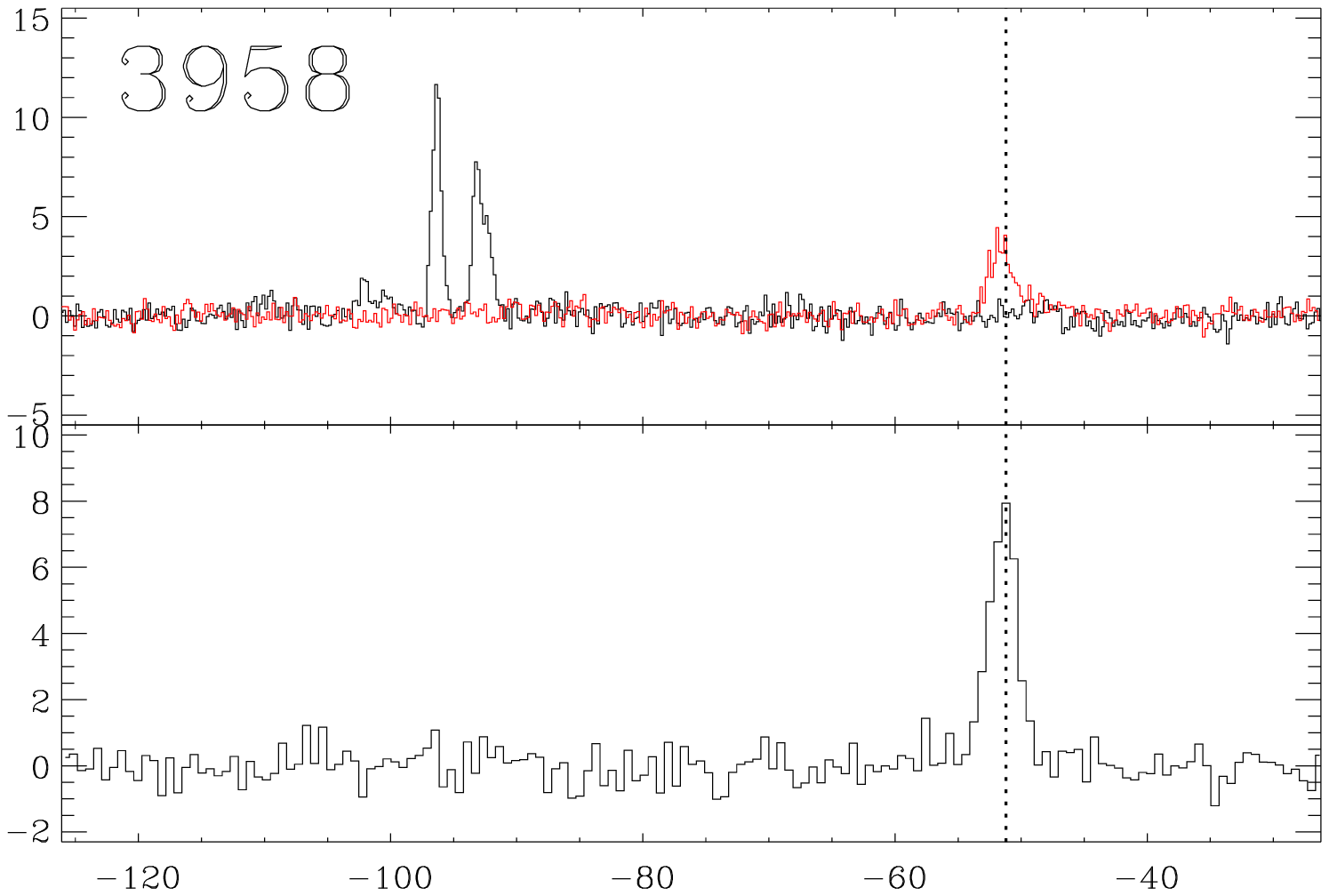}
\\
\end{tabular}
\begin{minipage}{\linewidth}
\center{LSR velocity (\kms)}
\end{minipage}
\caption{(Upper panel) The detected 22~GHz \water\ ($black$) and 44~GHz \methanol\ ($red$) maser spectra and 
(lower panel) $^{13}$CO~J=1$-$0 ($black$) and HCO$^+$~J=1$-$0 ($red$) line spectra of 
dominant shifted \water\ maser outflow source candidates ($^{13}$CO~J=2$-$1 line spectrum only for RMS~3936).
For each source, the source name is presented at the top-left corner of the upper panel and the systemic velocity is indicated by a vertical dotted line. 
\label{fig_HV}}
\end{figure}
\clearpage

\begin{figure}
\epsscale{1}
\plotone{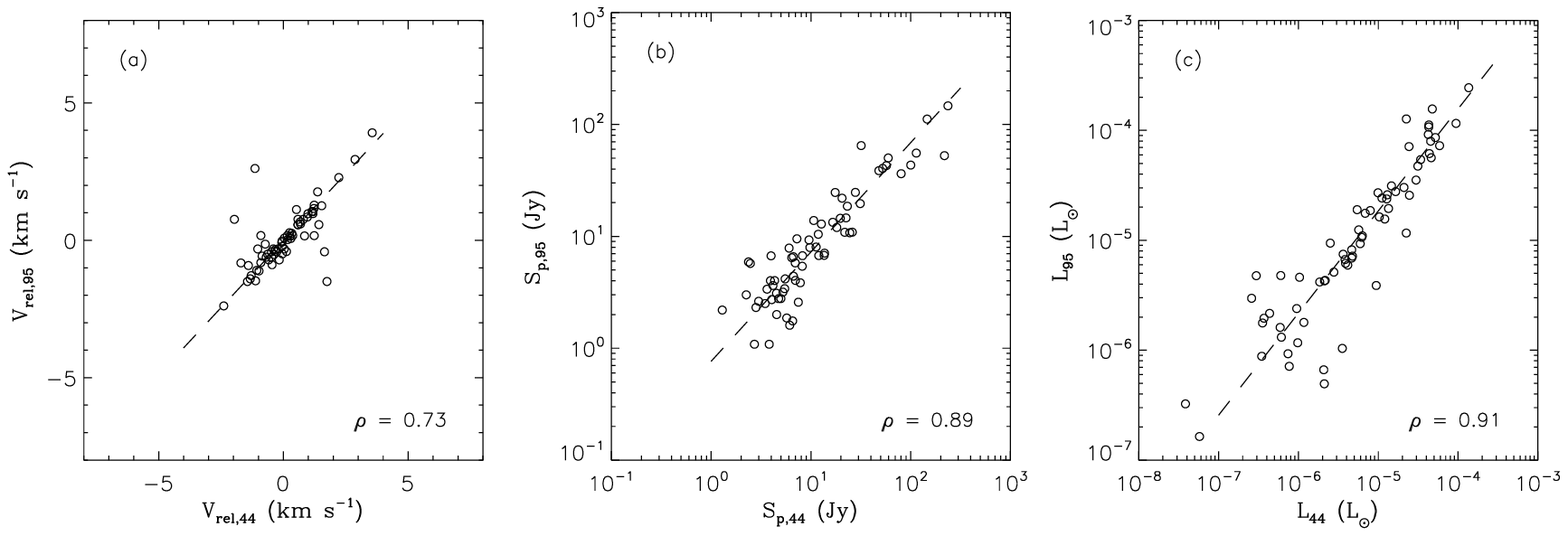}
\caption{
Comparison of (a) the relative peak velocities, (b) the peak flux densities, and (c) the isotropic luminosities of the 44 and 95~GHz class~I \methanol\ maser sources. 
In each panel the least-squares fitted relation is displayed by a dashed line with the correlation coefficient in the bottom-right corner.
\label{fig_Compare_4495}}
\end{figure}
\clearpage

\begin{figure}
\epsscale{0.6}
\plotone{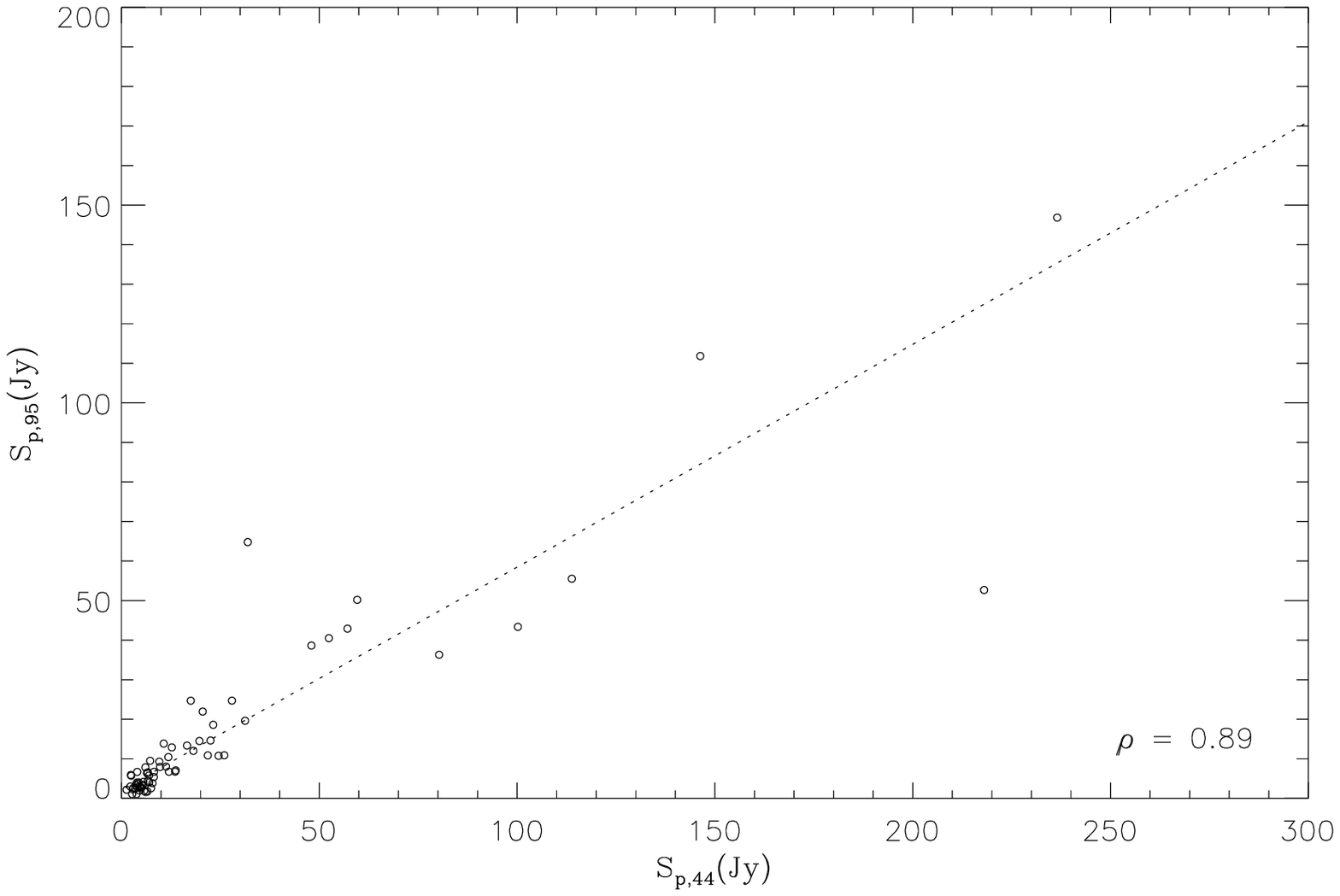}
\caption{
Same as in Figure~\ref{fig_Compare_4495} (b) but in linear scale.
\label{fig_Sp_4495_lin}}
\end{figure}
\clearpage

\begin{figure}
\epsscale{1}
\plotone{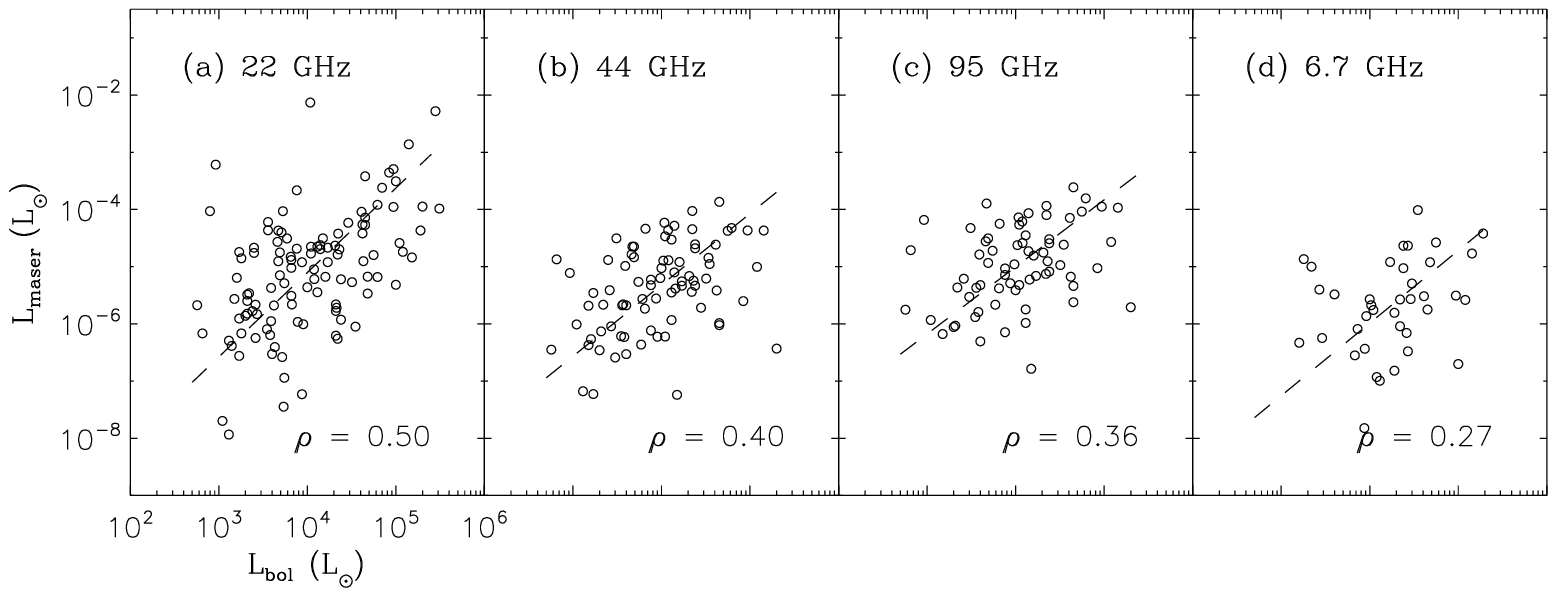}
\caption{Comparison of the isotropic maser luminosity with the bolometric luminosity for (a) the 22~GHz, (b) 44~GHz, (c) 95~GHz, and (d) 6.7~GHz maser sources. The 22, 44, and 95~GHz data are from the second-epoch survey of this study, while the 6.7~GHz data are from the MMB survey
\citep{gre10,gre12,bre15}.
The Pearson correlation coefficient is shown in the bottom-right corner of each panel.
\label{fig_lbol_lmaser}}
\end{figure}

\clearpage

\begin{figure}
\epsscale{1}
\plotone{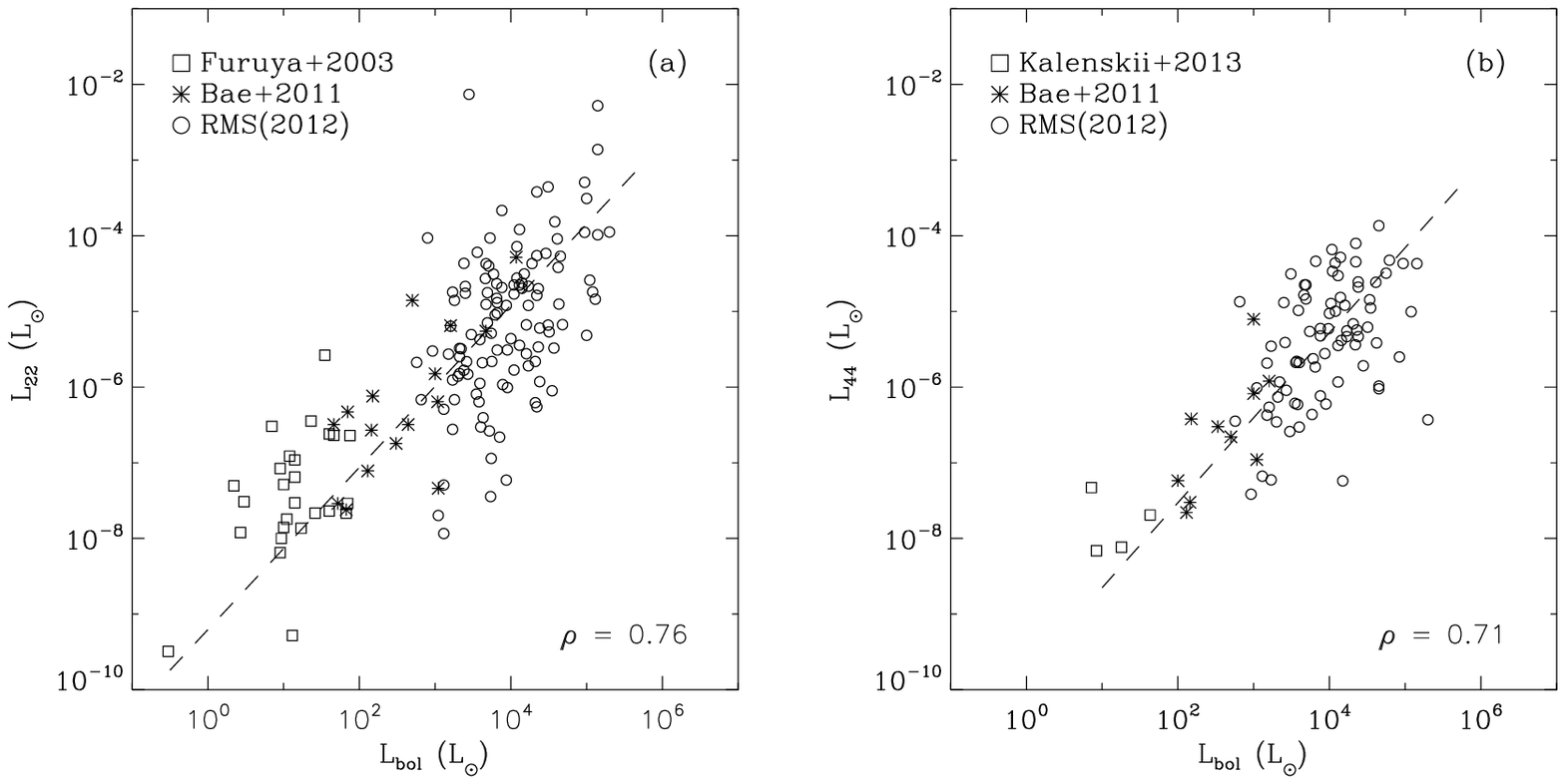}
\caption{Comparison of the isotropic maser luminosity with the bolometric luminosity for (a) the 22~GHz and (b) 44~GHz maser sources. The data points in low- and intermediate-mass regime are added from the literature. In both panels open circles and asterisks are data points of this study in the second epoch and \citet{bae11}, respectively. Open squares are data points from \citet{fur03} in (a) and \citet{kal13} in (b). The least-sqaures fitted relation is displayed by a dashed line with the correlation coefficient on the bottom-right corner in each panel.
\label{fig_lbol_lmaser_wide}}
\end{figure}

\clearpage

\begin{figure}
\epsscale{1}
\plotone{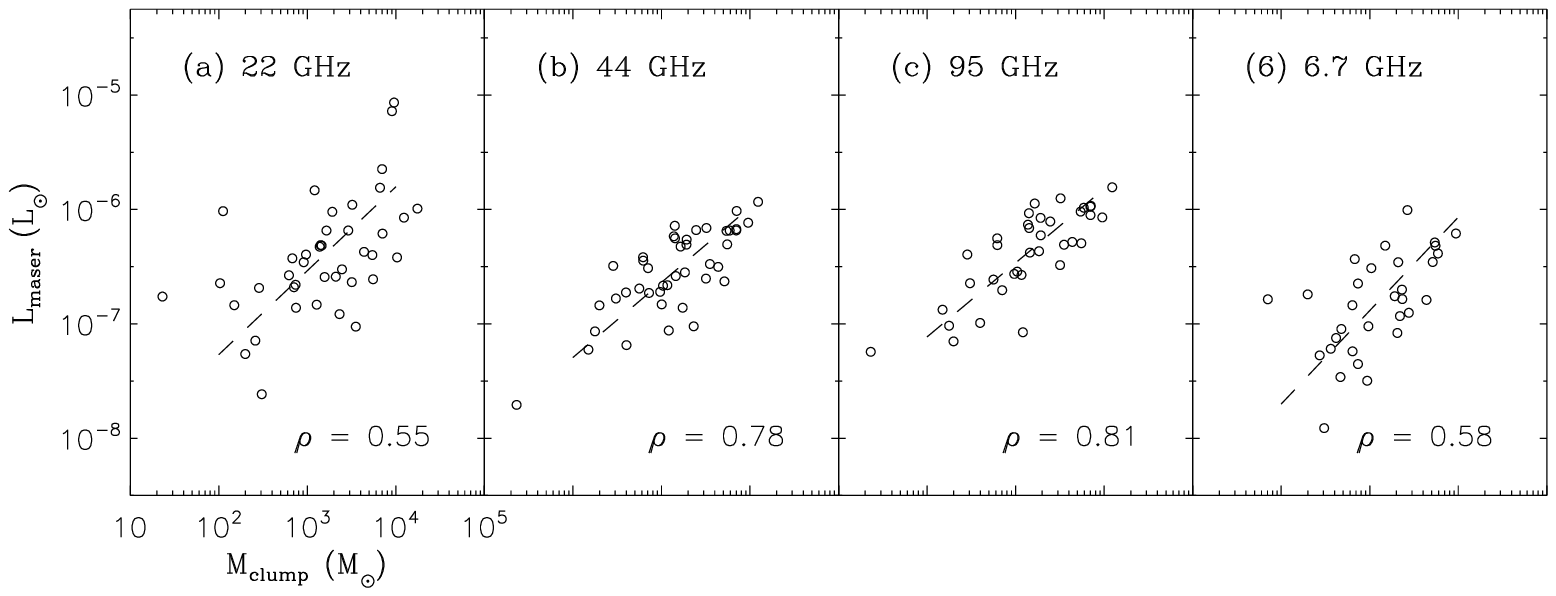}
\caption{Same as in Figure~\ref{fig_lbol_lmaser} except for the associated ATLASGAL clump mass.
In each panel the least-sqaures fitted relation is shown by a dashed line with the correlation coefficient on the bottom-right corner.
\label{fig_M_Lmaser}}
\end{figure}

\begin{figure}
\epsscale{1}
\plotone{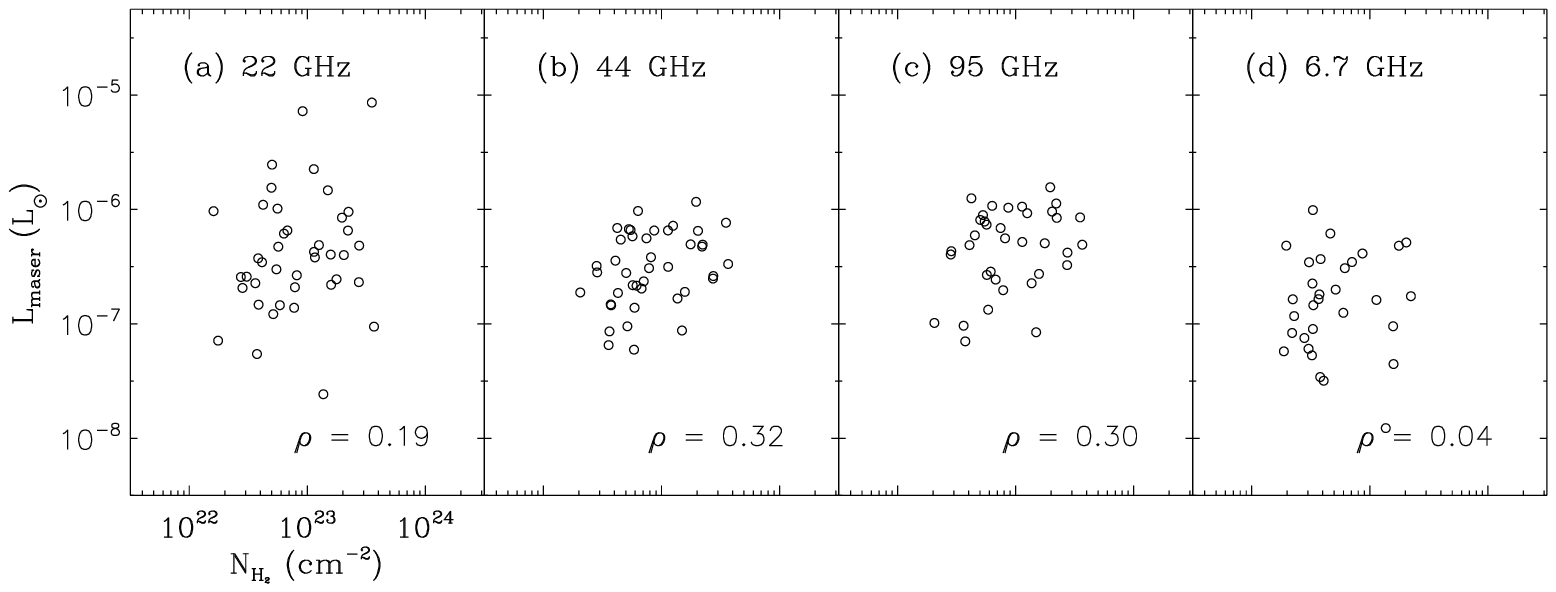}
\caption{Same as in Figure~\ref{fig_lbol_lmaser} except for the peak H$_2$ column density. 
\label{fig_NH2_Lmaser}}
\end{figure}
\clearpage

\begin{figure}
\epsscale{1}
\plotone{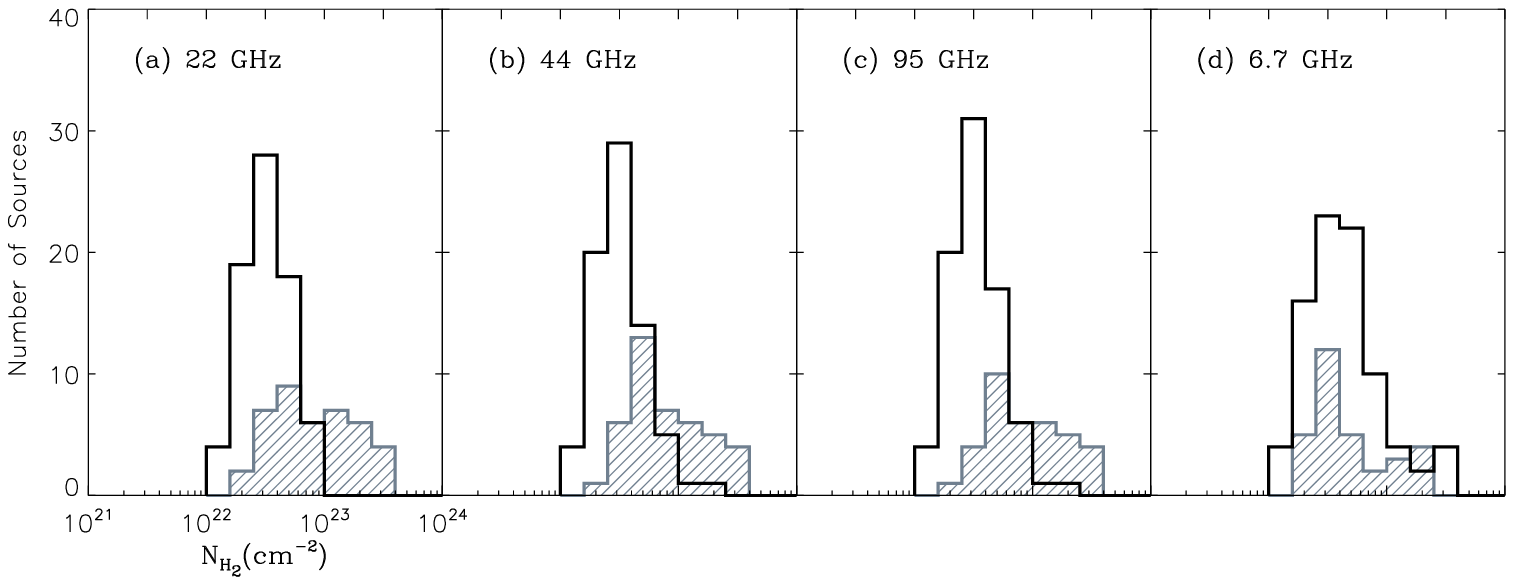}
\caption{The histograms of peak H$_2$ column density for detected and nondetected sources in each maser transition. The open black and hatched gray  lines represent the source with no emission and with emission for each of four masers, respectively. The size of bin is 0.2 dex.
\label{fig_NH2_hist}}
\end{figure}

\begin{figure}
\epsscale{1}
\plotone{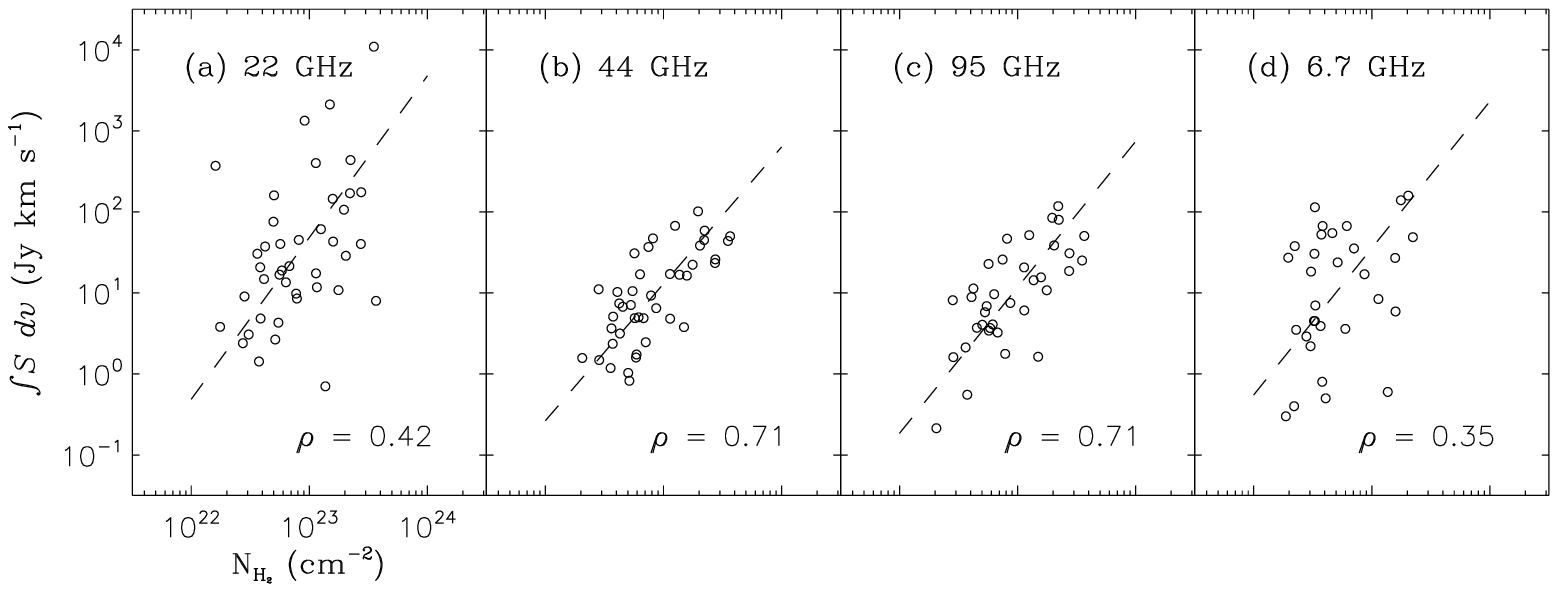}
\caption{The integrated maser flux density versus the peak H$_2$ column density. In each panel the least-sqaures fitted relation is shown by a dashed line with the correlation coefficient in the bottom-right corner.
\label{fig_NH2_Smint}}
\end{figure}

\begin{figure}
\epsscale{1}
\plotone{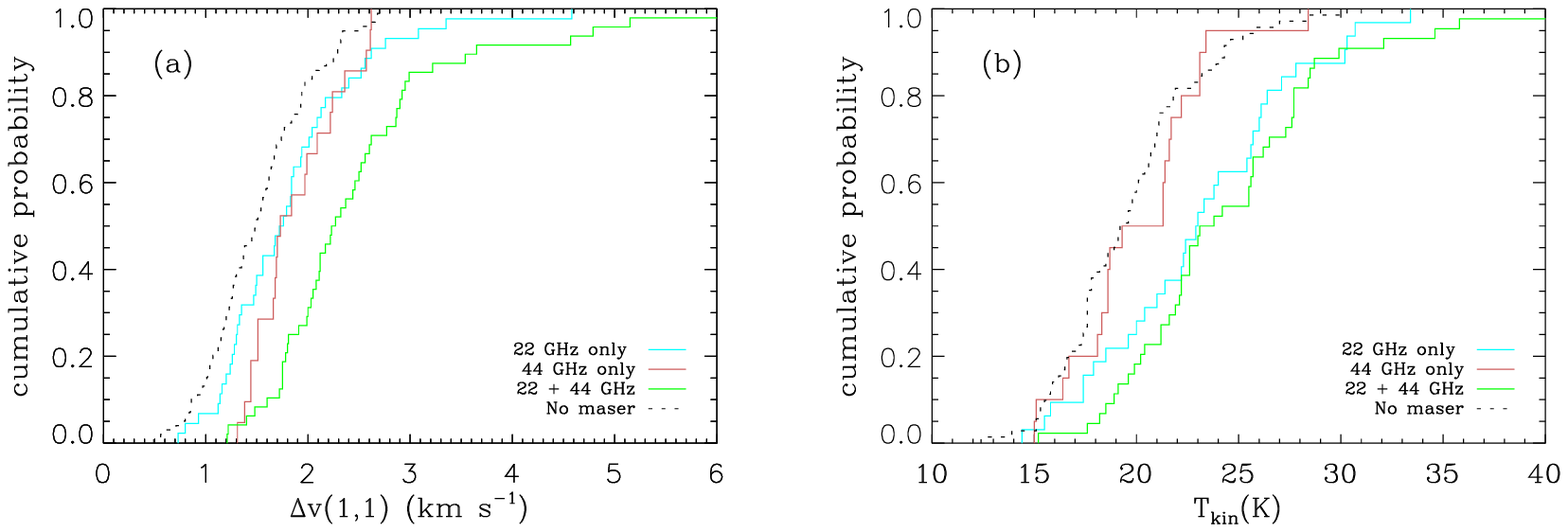}
\caption{Cumulative probability of (a) the NH$_3$ line width and (b) the kinetic temperature for 4 subsamples, which are divided on the bassis of the second-epoch survey results. The subsamples and corresponding colors are displayed on the bottom-right corner in each panel. 
\label{fig_NH3_cumul}}
\end{figure}

\clearpage

\begin{figure}
\epsscale{1}
\plotone{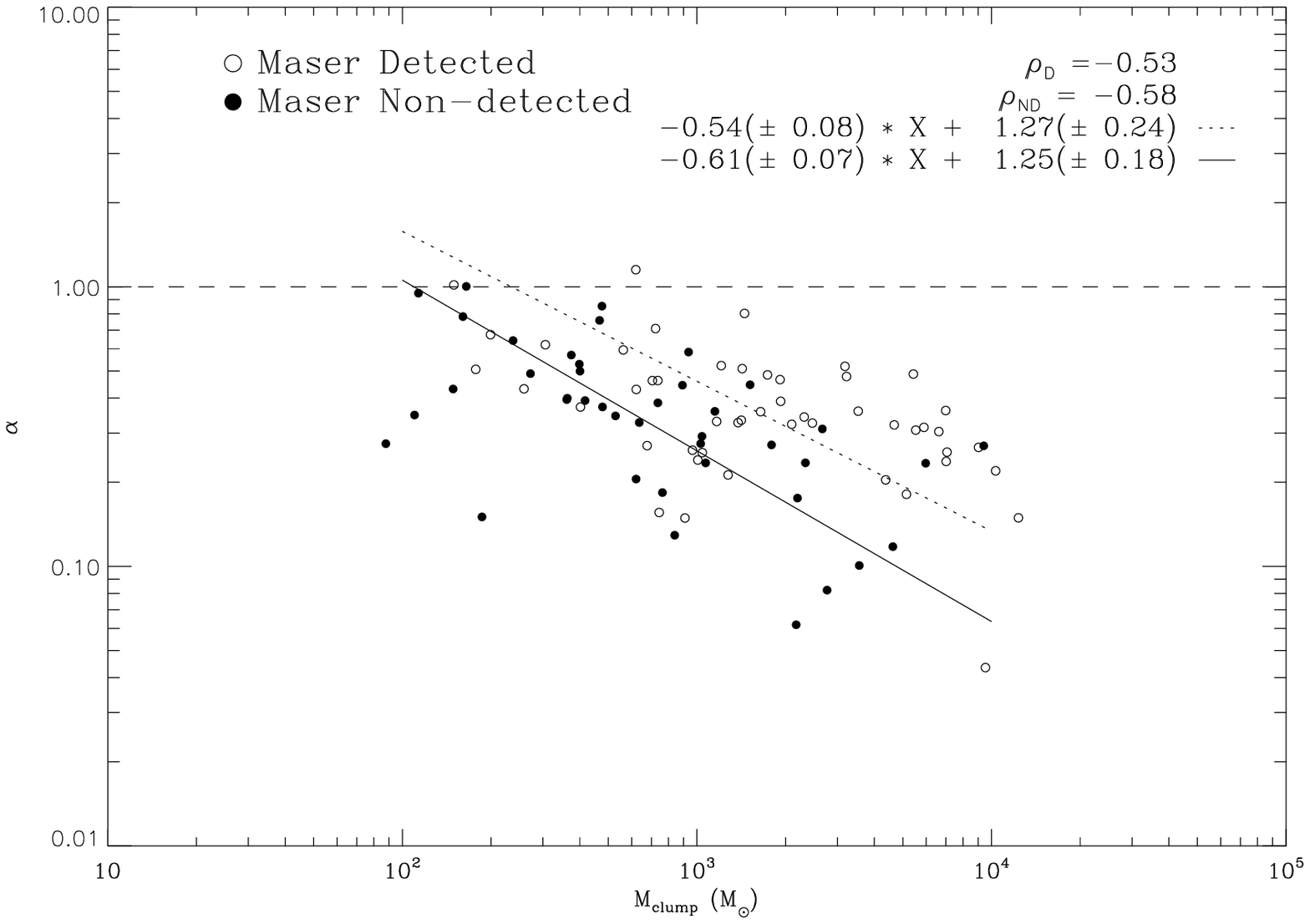}
\caption{The virial parameter ($\alpha$) versus the clump mass for the associated ATLASGAL clumps. Open and filled circles represent the sources with and without any maser emission, respectively.
The dashed line indicates the critical value ($\alpha$=1) for an isothermal sphere in hydrostatic equilibrium without magnetic support.
The area of $\alpha <$ 1 indicates the region where clumps are graviationally unstable and likely to be collapsing without additional support from strong magnetic field. 
The solid and dotted lines show the least-sqaures fitted relations for the filled and open circles, respectively.
The Pearson correlation coefficients are shown in the right upper corner. 
\label{fig_virial}}
\end{figure}

\clearpage

\begin{deluxetable}{ccccrrrcccc}
\rotate
\tabletypesize{\tiny}
\tablewidth{0pt}
\tablecaption{Summary of Target Sources \label{tbl_source}}
\tablehead{
\colhead{RMS\tablenotemark{a}} & \colhead{MSX} & \colhead{R.A.}   & \colhead{Decl.}  &
\colhead{Distance} & \colhead{$L_{\rm{bol}}$} & \colhead{$V_{\rm{sys}}$\tablenotemark{b}}& \multicolumn{3}{c}{Detection\tablenotemark{c}} & \colhead{MMB} \\
\cline{8-10}
\colhead{ID} & \colhead{name} & \colhead{(J2000.0)} & \colhead{(J2000.0)} & \colhead{(kpc)} &\colhead{($L_\odot$)} &\colhead{(\kms)} 
& \colhead{22 GHz} & \colhead{44 GHz} & \colhead{95 GHz} & \colhead{Coverage}}
\startdata
4  & G118.6172-01.3312  & 00:15:27.83  & +61:14:18.9  &   2.8 &    4900 & -39.20 & Y\tablenotemark{11,\rm{d}}  & n  & n  & ... \\
7  & G120.1483+03.3745  & 00:23:57.04  & +66:05:51.5  &   5.6 &   21000 & -68.90 & y\tablenotemark{4}  & n  & n  & ... \\
11  & G121.3479-03.3705  & 00:38:59.37  & +59:27:48.2  &   3.0 &    1100 & -41.90 & n  & n  & n  & ... \\
12  & G122.4459-00.7815  & 00:47:17.05  & +62:05:09.3  &   3.1 &    1600 & -42.80 & n\tablenotemark{e}  & n  & n  & ... \\
14  & G123.2836+03.0307  & 00:54:52.79  & +65:53:57.4  &   4.9 &    4300 & -62.10 & n  & n  & n  & ... \\
15  & G123.8059-01.7805  & 00:58:39.99  & +61:04:43.0  &   2.1 &    2400 & -31.90 & n  & n  & n  & ... \\
16  & G124.0144-00.0267  & 01:00:55.41  & +62:49:30.4  &   3.1 &    1800 & -43.40 & n  & n  & n  & ... \\
18\tablenotemark{H}  & G125.6045+02.1038B  & 01:16:36.21  & +64:50:38.4  &   4.1 &    8400 & -53.70 & y\tablenotemark{4,\rm{d}}  & n  & n  & ... \\
\enddata
\tablenotetext{a}{For multiple RMS sources within a 20$\arcsec$ radius, the superscript Y, H, D, and E indicate the type of neaby sources as YSO, HII region, Diffuse HII region and Evolved stars, respectively.
 Dagger symbol represents the single source finally classified with HII region, of which it's type was YSO at the observation time.}
\tablenotetext{b}{The systemic velocity from the RMS archives.}
\tablenotetext{c}{Y: New detection, y: detection, n: non-detection}
\tablenotetext{d}{Detected in this survey but not detected in \citet{urq11}.}
\tablenotetext{e}{Detected in \citet{urq11} but not detected in this survey.}
\tablenotetext{f}{Maser emission was detected by the first sidelobe.}
\tablenotetext{g}{Adopted distance from nearby known YSO, VV Mon\citep{tes98}.}
\tablenotetext{h}{Maser emission was detected not only by the main beam but also by the first sidelobe at different velocities.}
\tablerefs{1. \citet{han95}, 2. \citet{valt95}, 3. \citet{han98}, 4. \citet{vald01}, 5. \citet{kur04}, 
 6. \citet{sun07}, 7. \citet{valt07}, 8. \citet{urq09}, 9. \citet{bre11}, 10. \citet{chen11}, 11. \citet{urq11}, 12. \citet{bay12}, 13. \citet{chen12}, 14 \citet{tit14}}
\end{deluxetable}

\begin{deluxetable}{llrcccccc}
\tabletypesize{\tiny}
\tablecaption{Summary of Observations}
\tablewidth{0pt}
\tablehead{
\colhead{} & \colhead{Maser} & \colhead{Frequency} & \colhead{FWHM} & \colhead{} & \colhead{Velocity Resolution} & \colhead{Velocity Coverage} & \colhead{T$_{\textnormal{sys}}$} &\colhead{$f$\tablenotemark{a}} \\
\colhead{Telescope} & \colhead{Transition} & \colhead{(GHz)} &\colhead{($''$)} & \colhead{$\eta_A$} & \colhead{(\kms)} & \colhead{(\kms)} & \colhead{(K)} & \colhead{(Jy K$^{-1}$)}
}
\startdata
 KVN~21m & H$_2$O 6$_{16}$--5$_{23}$ & 22.235080 & 120 & 0.6 & 0.105 & 431 & 60--200 & 13.3 \\
 & CH$_3$OH 7$_0$--6$_1 A^+$ & 44.069430 & 62 & 0.6 & 0.053 & 218 & 120--350 & 13.3 \\
 & CH$_3$OH 8$_0$--7$_1 A^+$ & 95.169463 & 32 & 0.5 & 0.025 & 101 & 160--470 & 16.0 \\
 & HCO$^{+}$~J=1--0 & 89.188523 & 32 & 0.4\tablenotemark{b} & 0.026 & 108 & 170--200 &  \\
 TRAO~14m & $^{13}$CO~J=1--0 & 110.201353 & 48 & 0.5\tablenotemark{b} & 0.703 & 300 & 600--700 & \\
\enddata
\tablenotetext{a}{The conversion factor from T$_A^*$ to flux density.}
\tablenotetext{b}{The main-beam efficiency.}
\label{OBS_summary}
\end{deluxetable}

\begin{deluxetable}{lccc}
\tabletypesize{\tiny}
\tablecaption{Summary of Detections \label{tbl_DR} }
\tablewidth{0pt}
\tablehead{
\colhead{Maser} & \colhead{Epoch} & \colhead{Number} & \colhead{Detection Rate (\%)}
}
\startdata
                       22 GHz  &      2011  &   126 &    42 \\ 
                               &      2012  &   112 &    37 \\ 
                               &  Subtotal  &   135 &    45 \\ 
\hline
                       44 GHz  &      2011  &    76 &    25 \\ 
                               &      2012  &    81 &    27 \\ 
                               &  Subtotal  &    83 &    28 \\ 
\hline 
                       95 GHz  &      2012  &    68 &    23 \\ 
\hline 
                      Any    & Both         &   151 &    51 \\ 
\enddata
\end{deluxetable}

\begin{deluxetable}{ccccrrrrrrrc}
\tabletypesize{\tiny}
\tablewidth{0pt}
\tablecaption{Detected Maser Line Parameters\label{tbl_linepara}}
\tablehead{
\colhead{RMS} &  & \colhead{Epoch} &  & \colhead{$V_{\rm p}$} & \colhead{$S_{\rm p}$} &
\colhead{$\int S_{\rm \nu}dv$} & \colhead{$V_{\rm min}$} & \colhead{$V_{\rm max}$} & \colhead{$V_{\rm range}$} & \colhead{$L_{\rm maser}$} &  \\
 \colhead{ID} & \colhead{Maser}  &\colhead{(yyyymmdd)} & \colhead{Station\tablenotemark{a}} & \colhead{(\kms)} & 
 \colhead{(Jy)} & \colhead{(Jy \kms)} & \colhead{(\kms)} &\colhead{(\kms)} &\colhead{(\kms)} & \colhead{($L_{\odot}$)} & Notes\tablenotemark{b}}
\startdata
4 & 22 & 20110326 & YS &  -43.00 &      6.4 &      4.7 &  -43.42 &  -42.58 &    0.84 &   8.55E-07 &          \\
7 & 22 & 20110324 & YS &  -83.03 &      3.6 &      1.7 &  -83.24 &  -82.82 &    0.42 &   1.26E-06 &          \\
 &  & 20120212 & YS &  -81.56 &      5.1 &      3.0 &  -81.56 &  -81.13 &    0.42 &   2.18E-06 &          \\
18 & 22 & 20110411 & US &  -52.96 &      4.1 &      2.9 &  -53.17 &  -52.54 &    0.63 &   1.12E-06 &          \\
35 & 22 & 20110322 & YS &  -39.04 &   1221.6 &   4519.8 &  -56.10 &  -24.71 &   31.39 &   4.19E-04 &          \\
 &  & 20120521 & YS &  -42.62 &    841.7 &   4096.1 &  -62.21 &  -25.56 &   36.66 &   3.80E-04 &          \\
 & 44 & 20110322 & YS &  -42.55 &      1.2 &      1.0 &  -42.55 &  -42.55 &    0.21 &   1.92E-07 &          \\
 &  & 20120521 & YS &  -42.55 &      2.2 &      5.2 &  -43.62 &  -40.64 &    2.98 &   9.50E-07 &          \\
\enddata
\tablenotetext{a}{YS, US, and TN are the KVN Yonsei, Ulsan, and Tamna stations, respectively. MMB is the MMB survey of 6.7~GHz class~II \methanol\ masers.}
\tablenotetext{b}{D: \water ~maser with a dominant shifted feature at that epoch, H: presence of \water ~maser emission with high-velocity($>$30 \kms) component}
\end{deluxetable}

\begin{deluxetable}{cccrrrccrrrccrrr}
\rotate
\tabletypesize{\tiny}
\tablecaption{Statistics of Maser Line Parameters\label{tbl_line_summary}}
\tablewidth{0pt}
\tablehead{
& & \multicolumn{4}{c}{$S_{\rm{p}}$ (Jy)} & & \multicolumn{4}{c}{$V_{\rm rel}$ (\kms)} & & \multicolumn{4}{c}{$V_{\rm{range}}$ (\kms)} \\
\cline{3-6} \cline{8-11} \cline{13-16}
\colhead{Maser} & \colhead{Number} & \colhead{mean} & \colhead{median} & \colhead{min} & \colhead{max}
& & \colhead{mean}  & \colhead{median} & \colhead{min} & \colhead{max}
& & \colhead{mean}  & \colhead{median} & \colhead{min} & \colhead{max}}
\startdata
1st Epoch \\ 
\cline{1-1} 
       22  &   126 &    156.6 $\pm$    525.5 &     13.5 &      0.8 &   3540.6 &  &    -3.73 $\pm$    18.63 &    -0.93 &   -94.33 &    78.47 &  &    17.37 $\pm$    24.21 &    12.01 &     0.21 &   188.13 \\
       44  &    76 &     24.9 $\pm$     46.0 &      9.1 &      1.2 &    252.5 &  &     0.02 $\pm$     1.05 &    -0.05 &    -2.39 &     3.59 &  &     2.92 $\pm$     2.60 &     2.34 &     0.21 &    15.52 \\
2nd Epoch \\
\cline{1-1} 
       22  &   112 &    144.4 $\pm$    563.6 &     17.2 &      1.2 &   4552.1 &  &    -2.69 $\pm$    20.05 &    -0.11 &   -92.86 &    79.11 &  &    19.70 $\pm$    32.59 &     9.69 &     0.21 &   230.68 \\
       44  &    81 &     21.5 $\pm$     41.6 &      6.7 &      1.1 &    236.5 &  &     0.05 $\pm$     1.06 &    -0.05 &    -2.39 &     3.56 &  &     2.54 $\pm$     2.41 &     2.13 &     0.21 &    11.27 \\
       95  &    68 &     16.2 $\pm$     25.0 &      6.7 &      1.1 &    146.9 &  &     0.09 $\pm$     1.08 &    -0.02 &    -2.39 &     3.91 &  &     2.96 $\pm$     2.31 &     2.36 &     0.20 &    13.39 \\
Either Epoch \\
\enddata
\tablecomments{Upper : Maser sources for KVN observations \\ 
     Lower : Maser sources combining the current work with 6.7 GHz in an region covered with MMB survey} 
\end{deluxetable}

\begin{deluxetable}{ccccccccc}
\tabletypesize{\tiny}
\tablecaption{Dominant Shifted \water ~Maser Source Candidates}
\tablewidth{0pt}
\tablehead{
\colhead{RMS} & \colhead{$^{13}$CO\tablenotemark{a}} & \colhead{HCO$^{+}$\tablenotemark{a}} & \colhead{} \\
\colhead{ID} & \colhead{J=1--0} & \colhead{J=1-0} & \colhead{Note\tablenotemark{b}} 
}
\startdata
 145 & n & -- & R \\
 2547 & n & -- & B \\
 2584 & n & -- & B \\
 2716 & y & y & N \\
 3158 & y & n & B \\
 3360 & y & y & N \\
 3587 & n & -- & B \\
 3766 & n & -- & B \\
 3911 & n & -- & B \\
 3936 & n\tablenotemark{c} & -- & R \\
 3958 & n & -- & B \\
\enddata
\tablenotetext{a}{The presence of molecular line emission associated with the dominant blue- or redshifted \water\ maser feature candidates.}
\tablenotetext{b}{B: dominant blueshifted, R: dominant redshifted, N: no dominant maser emission}
\tablenotetext{c}{The $^{13}$CO~J=2--1 line data from the JCMT archives.}
\label{tab_HV}
\end{deluxetable}

\end{document}